\documentclass[preprint]{aa}
\pdfoutput=1

\usepackage{booktabs}
\usepackage{csvsimple}
\usepackage{pgfplotstable}
\usepackage{array}
\usepackage{amsmath}
\usepackage{amsfonts}
\usepackage{dsfont}
\usepackage{amsxtra}
\usepackage{amssymb}
\usepackage{upgreek}
\usepackage{comment}

\usepackage{multirow}
\usepackage{url}

\usepackage{graphicx,epsfig}

\usepackage{xspace} 

\usepackage{color}
\usepackage{xcolor}

\usepackage{lineno}
\usepackage[export]{adjustbox}

\newcommand{\be}{\begin{equation}}
\newcommand{\ee}{\end{equation}}
\newcommand{\bea}{\begin{eqnarray}}
\newcommand{\eea}{\end{eqnarray}}
\newcommand{\beaa}{\begin{eqnarray*}}
\newcommand{\eeaa}{\end{eqnarray*}}
\newcommand{\ba}{\begin{array}} 
\newcommand{\ea}{\end{array}}
\newcommand{\bi}{\begin{itemize}}
\newcommand{\ei}{\end{itemize}}
\newcommand{\ben}{\begin{enumerate}}
\newcommand{\een}{\end{enumerate}}

\newcommand{\lb}{\label}

\newcommand{\al}{\alpha}
\newcommand{\bt}{\beta}

\newcommand{\Fermi}{\textit{Fermi}\xspace}

\newcommand{\twopicsp}{0.45}

\definecolor{darkgreen}{rgb}{0.0, 0.7, 0.0}

\begin{document}

   \title{Machine learning methods for constructing \\ probabilistic \Fermi-LAT catalogs%
   \thanks{The data in tables \ref{tab:prob_cat},
\ref{tab:parkes},
\ref{tab:psr_candidates_3class}, and
\ref{tab:other_candidates_3class} and probabilistic catalogs are available in electronic form
at the CDS via anonymous ftp to cdsarc.u-strasbg.fr (130.79.128.5)
or via \url{http://cdsarc.u-strasbg.fr/viz-bin/cat/J/A+A/660/A87}}
}

   \author{A. Bhat \thanks{\email{aakash.bhat@fau.de}}
          \inst{1}
          \and
          D. Malyshev \thanks{on leave of absence from NRC ``Kurchatov Institute'' - ITEP, B. Cheremushkinskaya st. 25, Moscow, Russia 117218, 
          \email{dmitry.malyshev@fau.de}}
          \inst{1}
          }

   \institute{
             Erlangen Centre for Astroparticle Physics, Erwin-Rommel-Str. 1, Erlangen, Germany
             }

\titlerunning{ML methods for probabilistic \Fermi-LAT catalogs}
\authorrunning{A. Bhat and D. Malyshev}
 
\abstract
{
Classification of sources is one of the most important tasks in astronomy.
Sources detected in one wavelength band, for example using gamma rays, may have several possible associations in other wavebands, or
there may be no plausible association candidates.
}
{
In this work we aim to determine the probabilistic classification of unassociated sources in the third \Fermi Large Area Telescope (LAT) point source catalog (3FGL) and the fourth \Fermi LAT data release 2 point source catalog (4FGL-DR2) using two classes  -- pulsars and active galactic nuclei (AGNs) -- or three classes -- pulsars, AGNs, and ``OTHER'' sources.
}
{
We use several machine learning (ML) methods to determine a probabilistic classification of \Fermi-LAT sources.
We evaluate the dependence of results on the meta-parameters of the ML methods, such as the maximal depth of the trees in tree-based classification methods and the number of neurons in neural networks.
}
{
We determine a probabilistic classification of both associated and unassociated sources in the 3FGL and 4FGL-DR2 catalogs.
We cross-check the accuracy by comparing the predicted classes of unassociated sources in 3FGL with their associations in 4FGL-DR2 for cases where such associations exist.
We find that in the two-class 
case it is important to correct for the presence of OTHER sources among the unassociated ones in order to realistically estimate the number of pulsars and AGNs.
We find that the three-class classification, despite different types of sources in the OTHER class, has a similar performance as the two-class classification in terms of reliability diagrams and, at the same time, it does not require adjustment due to presence of the OTHER sources among the unassociated sources.
We show an example of the use of the probabilistic catalogs for population studies, which include associated and 
unassociated sources.
}
{}


\maketitle


\section{Introduction}

The multiwavelength association of astronomical sources is important for understanding their nature.
Unfortunately, in many cases a firm association of sources at different wavelengths is not possible.
For example, about one-third of the gamma-ray sources in \Fermi Large Area Telescope (LAT) catalogs are unassociated
\citep{2010ApJS..188..405A, 2012ApJS..199...31N, 2015ApJS..218...23A, 2020ApJS..247...33A}.
It is at least useful to know the classes to which the unassociated sources belong or, as is more typical,
the probabilities of the sources belonging to various classes.
In this paper we used several machine learning (ML) algorithms to find a probabilistic classification of
 sources in the third \Fermi-LAT catalog \citep[3FGL;][]{2015ApJS..218...23A} and the fourth data release two catalog
\citep[4FGL-DR2;][]{2020ApJS..247...33A, 2020arXiv200511208B}.%
\footnote{The 4FGL-DR2 catalog \citep{2020arXiv200511208B} is an update of the 4FGL catalog \citep{2020ApJS..247...33A}
based on 10 years of the \Fermi-LAT data compared to 8 years of the \Fermi-LAT data in the 4FGL catalog.
We use the 4FGL notation when we talk about general features of the 4FGL catalog, which are also present in the 4FGL-DR2.
The 4FGL-DR2 notation is used for specific calculations in this paper.}
We used the versions gll\_psc\_v16.fit for 3FGL and gll\_psc\_v27.fit for 4FGL-DR2.

We refer to the catalogs where the classification of the sources is given in terms of probabilities as ``probabilistic catalogs.''
In general, the classes may include the possibility that a source is not a real source but a fluctuation of the background 
\citep{2021A&A...656A..62P}
or that a source is an overlay of two sources.
Probabilistic catalogs
have previously been introduced for optical sources 
\citep[e.g.,][]{2010EAS....45..351H, 2013AJ....146....7B}
and for gamma-ray sources \citep{2017ApJ...839....4D}.
Bayesian association probabilities were also included in the 4FGL catalog \citep{2020ApJS..247...33A} for faint sources.
Probabilistic classification of unassociated \Fermi-LAT sources was performed by, for example,
\cite{2012ApJ...753...83A}, \cite{2016ApJ...820....8S}, \cite{2016ApJ...825...69M}, \cite{2017A&A...602A..86L}, \cite{2020MNRAS.492.5377L}, 
\cite{2021MNRAS.507.4061F}, and \cite{2021RAA....21...15Z},
or in the application for the subclassification of blazars by
\cite{2013MNRAS.428..220H}, \cite{2014ApJ...782...41D},
\cite{2016MNRAS.462.3180C}, \cite{2017MNRAS.470.1291S}, and \cite{2019MNRAS.490.4770K, 2020MNRAS.493.1926K},
and in subclassification of pulsars 
by \cite{2012MNRAS.424.2832L} and \cite{2016ApJ...820....8S}.
In this work we considered the classification of gamma-ray sources into two classes -- active galactic nuclei (AGNs) and pulsars -- as well as into three classes 
-- AGNs, pulsars, and other associated sources (``OTHER'').
We revisited the probabilistic classification of 3FGL sources and compared the results of the classification of unassociated sources
with their respective associations in 4FGL-DR2.
We also determined a probabilistic classification of the 4FGL-DR2 sources.

Catalogs of gamma-ray point sources are typically designed to have low false detection rates. 
Nevertheless, 469 sources out of the 3033 in the 3FGL catalog \citep{2015ApJS..218...23A} have no counterparts 
in the 4FGL catalog \citep{2020ApJS..247...33A}.
This is much larger than the expected false detection rate in 3FGL arising from statistical fluctuations.
For the majority of sources in the 3FGL catalog without counterparts in the 4FGL catalog, the problem is not the false detection 
but rather the association.
For example, some sources can be detected due to deficiencies in the Galactic diffuse emission model.
In this case, the statistical significance of the detection is high, but the association is wrong: the sources should be classified as
a part of the Galactic diffuse emission rather than point-like sources.
Another reason could be that two (or more) point-like sources in 3FGL are associated with a single extended source in 4FGL,
or a single source is resolved into two sources.
Again, this is a problem of classification (or association) rather than false detection.

The absence of a previously detected source in a new catalog may also be due to variability.
In particular, flat spectrum radio quasars (FSRQs) are highly variable AGNs.
If a source was active during the observation time of 3FGL but inactive afterward, 
then its significance in the 4FGL can be below the detection threshold.
This problem is connected to the selection of a hard detection threshold of $TS = 25$ for the 3FGL and 4FGL 
catalogs. Selection of a lower detection threshold could help to keep the variable sources inside the catalog, 
but it will not solve the problem since the variable sources near the lower threshold can also disappear in the new catalog.
Moreover, a lower threshold would lead to more false detections due to fluctuations of the background.
Thus, on the one hand, a lower threshold can be useful in studies where a more complete list of sources is desirable and a higher false detection rate is admissible.
On the other hand, a lower threshold can be problematic for studies where 
a clean sample is necessary. 
This problem of detection threshold selection can be ameliorated with the development of a probabilistic catalog.
In such a catalog, each point-like object detected above a certain relatively low confidence level
is probabilistically classified into classes, which include the statistical fluctuation class.
At low confidence, the probability of a source coming from a background fluctuation is high.
This probability decreases as the significance of sources increases.
Apart from the statistical fluctuation class, classes can include various types of Galactic and extra-galactic sources, diffuse emission deficiencies, extended sources, and so on.
Any user of such a catalog has the freedom to choose the probability threshold for the class that he or she is interested in.
In this paper we make a first step in this direction by providing a probabilistic classification of \Fermi-LAT sources into two or three classes.
We also show how the probabilistic catalogs can be used for population studies of sources, for example,
as a function of their flux or position on the sky, where one includes not only associated sources but also unassociated ones according to their class probabilities.

The paper is organized as follows.
In Sect. \ref{sec:methods} we discuss general questions about the construction of the probabilistic catalogs and the choices of the ML methods.
In Sect. \ref{sec:training} we construct the classification algorithms using the associated sources in the 3FGL catalog for training. We consider several aspects: (1) feature selection, (2) training of the algorithms and selection of meta-parameters,
and (3) oversampling of the data sets in order to have equal numbers of pulsars and AGNs in training (there are many more AGNs observed than pulsars).

In Sect. \ref{sec:prob_cats} we apply the classification algorithms determined in Sect. \ref{sec:training} to the classification of 3FGL and 4FGL-DR2 sources.
We compare our predictions for the unassociated sources in 3FGL with the respective associations in 4FGL-DR2.
In Sect. \ref{sec:3class} we classify sources in the 3FGL and 4FGL-DR2 catalogs into three classes (AGNs, pulsars, and OTHER).
In Sect. \ref{sec:pop_studies} we show applications of the probabilistic catalogs for predicting the number of AGNs, pulsars, and OTHER sources among the unassociated sources and in the construction of the source counts as a function of their flux, $N(S)$, and as a function of 
Galactic latitude and longitude, $N(b)$ and $N(\ell)$.
We compare the $N(S)$, $N(b)$, and $N(\ell)$ distributions for associated and unassociated sources in the 3FGL and 4FGL-DR2 catalogs.
In Sect. \ref{sec:conclusions} we present our conclusions.

In Appendix \ref{sec:app} we perform further studies of the meta-parameters of some of the ML algorithms, 
in Appendix \ref{sec:app_O_vs_S} we compare the oversampling method used in the paper with Synthetic Minority Over-sampling Technique (SMOTE),
in Appendix \ref{sec:thres} we study the effect of the choice of the probability threshold on the precision and recall of classification,
and in Appendix \ref{sec:reliability} we discuss the reliability diagrams.

\section{Choice of methods}
\lb{sec:methods}

\subsection{General methodology}

The first choices that must be made to construct probabilistic catalogs are the choices of the input data and the ML methods to be used.
For the input data we took associated point sources in the 3FGL or 4FGL-DR2 catalogs, which we then split into training and testing subsets.
We considered four ML algorithms: random forests \citep[RF;][]{709601, Breiman:2001hzm}, 
boosted decision trees \citep[BDT;][]{friedman2001},  
logistic regression \citep[LR;][]{cox1958}, 
and neural networks \citep[NN;][]{Hopfield:1982pe}.
Although the performance of algorithms on testing data is slightly different, 
we report the classification probabilities for all four algorithms.
The difference among the predictions serves as a measure of modeling uncertainty related 
to the choice of the classification algorithm.

\subsection{Discussion of the classification algorithms}
\lb{sec:class_alg}

One of the simplest and most transparent algorithms for classification is decision trees.
In this algorithm, at each step the sample is split into two subsets using one of the input features.
The choice of the feature and the separating value are determined by minimizing an objective function, such as misclassification
error, Gini index, or cross-entropy.
This method is very intuitive, since at each step the results can be described in words. 
For example, at the first step, the sources can be split into mostly Galactic and extragalactic sources by a cut on the Galactic latitude.
At the next step, the high latitude sources can be further sub-split into millisecond pulsars and OTHER sources via a cut on the spectral index around 1 GeV (pulsars have a hard spectrum below a few GeV) and so on.
One of the main problems with decision trees is either overfitting or bias: if a tree is too deep, then it will pick up particular cases of the training sample resulting in overfitting, while if the trees are too shallow they will not be able to describe the data well, thereby leading to a bias. 
As a result, one needs to be very careful when selecting the depth of the tree.
This problem can be avoided if a random subset of features is used to find a division at each node. This is the basis of the RF algorithm,
where the final classification is given by an average of several trees with random subsets of features used at each node.
Another problem with the simple trees algorithm is that it can miss the classification of some subsets of data. This is rectified in the BDT algorithm, where the final classification is given by a collection of trees, where each new tree is created by increasing the weights of misclassified samples of the previous step. 
Finally, simple trees predict classes for the data samples, while we would like to have probabilities for these classes (also known as soft classification).
RF and BDT algorithms, by virtue of averaging, provide probabilities. As a result, we use RF and BDT algorithms rather than simple decision trees in this paper.

Tree-based algorithms, even after averaging in RF and BDT methods, have sharp edges among domains with different probabilities.
In LR algorithm, the probabilities of classes are by construction smooth functions of input features.
In particular, for two-class classification the probability of class 1, given the set of features $x$, is modeled by the sigmoid (logit) function
\bea
\lb{eq:logit}
p_1(x) = \frac{e^{m(x)}}{1 + e^{m(x)}}.
\eea
The probability of class 0 is then modeled as $p_0(x) = 1 - p_1(x)$.
Therefore, if $m(x)$ is a linear function of features, then the boundary between the domains, defined, for example, as $p_1(x) = 0.5$, will also be linear
at $m(x) = 0$.
More complicated boundaries can be modeled by taking nonlinear functions $m(x)$.
Unknown parameters of the function $m(x)$ are determined by maximizing the log likelihood of the model given the known classes of the data in the training sample.
A useful feature of the LR method is that it, by construction, provides probabilities of classes with smooth transitions among domains of different classes.
A limitation is that the form of the probability function is fixed to the sigmoid function in Eq. (\ref{eq:logit}).

We notice that if $m(x)$ is a linear function of features $x$, then the LR model is obtained by an application of sigmoid function to a linear combination of input features.
This is in fact a single layer perceptron, or a NN, with several input nodes (each node corresponding to a feature) and one output node, which corresponds to $p_0(x)$, but without any hidden layers.
The output value is obtained by a nonlinear transformation (sigmoid) of a linear combination of features.
A neural network with several hidden layers is obtained by a sequence of nonlinear transformations of linear combinations of features.
In particular, the values in the first hidden layer are obtained by a nonlinear transformation of linear combinations of input features.
Then the values in the second hidden layer are obtained by a nonlinear transformation of linear combinations of values in the first hidden layer and so on till the required number of hidden layers is reached.
In the context of NN, the nonlinear transformations are also called activation functions.
If the activation function for the output layer is sigmoid, then the output values can be interpreted as probabilities.

\section{Construction of probabilistic catalogs}
\lb{sec:training}

One of the first problems with the 3FGL and 4FGL-DR2 catalogs is that
some of the sources in the catalogs have missing or unphysical values (e.g., infinity).
In order to avoid a bias in predictions, we include sources with missing or unphysical values only in testing or in predictions (for unassociated sources), but not in training.
If the value is infinity, then we formally substitute it by the largest value found in the sample multiplied by 10.
An unphysical zero (e.g., in significance) is substituted by the smallest value in the sample divided by 10,
while a missing value is substituted by the average of the sample.
There can be other ways to replace the missing or unphysical values, for example, by using k nearest neighbors regression, 
but since the number of such sources is relatively small (13 for 3FGL and 14 for 4FGL-DR2), 
the choice of the method to replace the missing values does not significantly affect the results.
In the final probabilistic catalogs, we use a column ``Missing\_Values\_Flag'' to mark 
the sources with missing or unphysical values.

As an example of the construction of a probabilistic catalog, we use the 3FGL catalog.
In this section we perform a two-class classification to separate point sources into pulsars and AGNs.
Thus for training and testing, we sub-selected the sources, which are associated with pulsars and AGNs.
The three-class classification into pulsars, AGNs, and OTHER sources is discussed in Sect. \ref{sec:3class}.
After training the algorithms, we tested the performance with the test sources and predicted the classes of the unassociated sources.
The general workflow had the following three steps: (1) Select data for training and testing. (2) Optimize algorithms using training data sets.
We selected meta-parameters of the algorithms by optimizing the accuracy of the classification and tested for overfitting using the test data sets.
In order to get stable results, we repeated the separation of the data into training and testing samples 100 times and 
averaged the accuracy. (3)
Make predictions for unassociated point sources in the 3FGL catalog.
We also applied the classification to associated sources, which we use for consistency checks.

\
As a result of the analysis in this section, we selected meta-parameters for the four ML algorithms,
which we then use in the following section to construct probabilistic catalogs
based on the \Fermi-LAT 3FGL and 4FGL-DR2 catalogs.

\subsection{Data and feature selection}

For training of the algorithms we used the associated sources without missing or unphysical values, 
which were classified as either AGNs (classification labels in the 3FGL catalog: agn, FSRQ, fsrq, BLL, bll, BCU, bcu, RDG, rdg, NLSY1, nlsy1, ssrq, and sey) or pulsars (classification labels in 3FGL: PSR, psr).%
\footnote{In the following we use PSR as a shorthand notation for pulsar in tables, figures, and in the labels of probabilistic catalogs.}
There are 1905 such sources in the 3FGL catalog. 

There are several dozen features of point sources quoted in the catalog, such as the position, photon, and energy fluxes integrated in different energy bands, spectral parameters, and variability index, as well as the corresponding uncertainties. 
We took some of the main features and also added four hardness ratios defined as 
\bea
\lb{eq:hr}
HR_{ij} = \frac{EF_j - EF_i}{EF_j + EF_i},
\eea
where $EF_i$ is the energy flux in bin $i$ and $j = i + 1$ (i.e., the bins are consecutive).

Spectral index is one of the most important characteristic of sources. 
Unfortunately in the 3FGL catalog, the definition of the spectral index is different for associated and unassociated sources.
In particular, the gamma-ray flux of pulsars is described by a power-law with a (super)exponential cut-off $\propto E^{-\Gamma} e^{-(E / E_c)^b}$, where the ``Spectral\_Index'' feature in the catalog is the parameter $\Gamma$.
On the other hand, gamma-ray flux of unassociated sources with significant curvature is represented by the log-parabola function $\propto (E/E_0)^{-\al - \bt \ln (E/E_0)}$,
where the Spectral\_Index 
feature is the parameter $\al$, that is, the tilt in the spectrum at the pivot energy $E_0$ (which also varies for different sources).
Since the Spectral\_Index feature has different definitions for associated pulsars and for possible pulsars among unassociated sources,
its use for training the algorithms to separate pulsars from AGNs is problematic.
If one fits all spectra of sources in the catalog by a power-law function, then the corresponding indices of the power laws are represented by the
``PowerLaw\_Index'' feature in the catalog.
This feature is defined uniformly for all associated and unassociated sources (i.e., it is safe to use for training).
Unfortunately, the power-law function is not a good description of the gamma-ray flux from pulsars.
Consequently, in the classification of the 3FGL sources we constructed a new feature: the index at 500 MeV (denoted in the following as ``500MeV\_Index''), defined as minus the derivative of the log flux:
\bea
\lb{eq:n500_def}
n({\rm 500\,MeV}) = - \left. \frac{d \ln F}{d \ln E} \right|_{E = \rm 500\,MeV}
.\eea
For log-parabola and for a power law with a (super)exponential cutoff, it is respectively
\bea
n(\rm 500\,MeV) &=& \al + 2 \bt \ln(\rm 500\,MeV / E_0);    \\
n({\rm 500\,MeV}) &=& \Gamma + b\,({\rm 500\,MeV} / E_c)^b
.\eea
This feature has a more uniform definition for all sources in the 3FGL catalog than the Spectral\_Index. It also has a better separating power 
than PowerLaw\_Index, provided that pulsars have typically harder spectra at energies below 1 GeV than AGNs.

\begin{figure*}[h]
\centering
\includegraphics[width=0.75\textwidth]{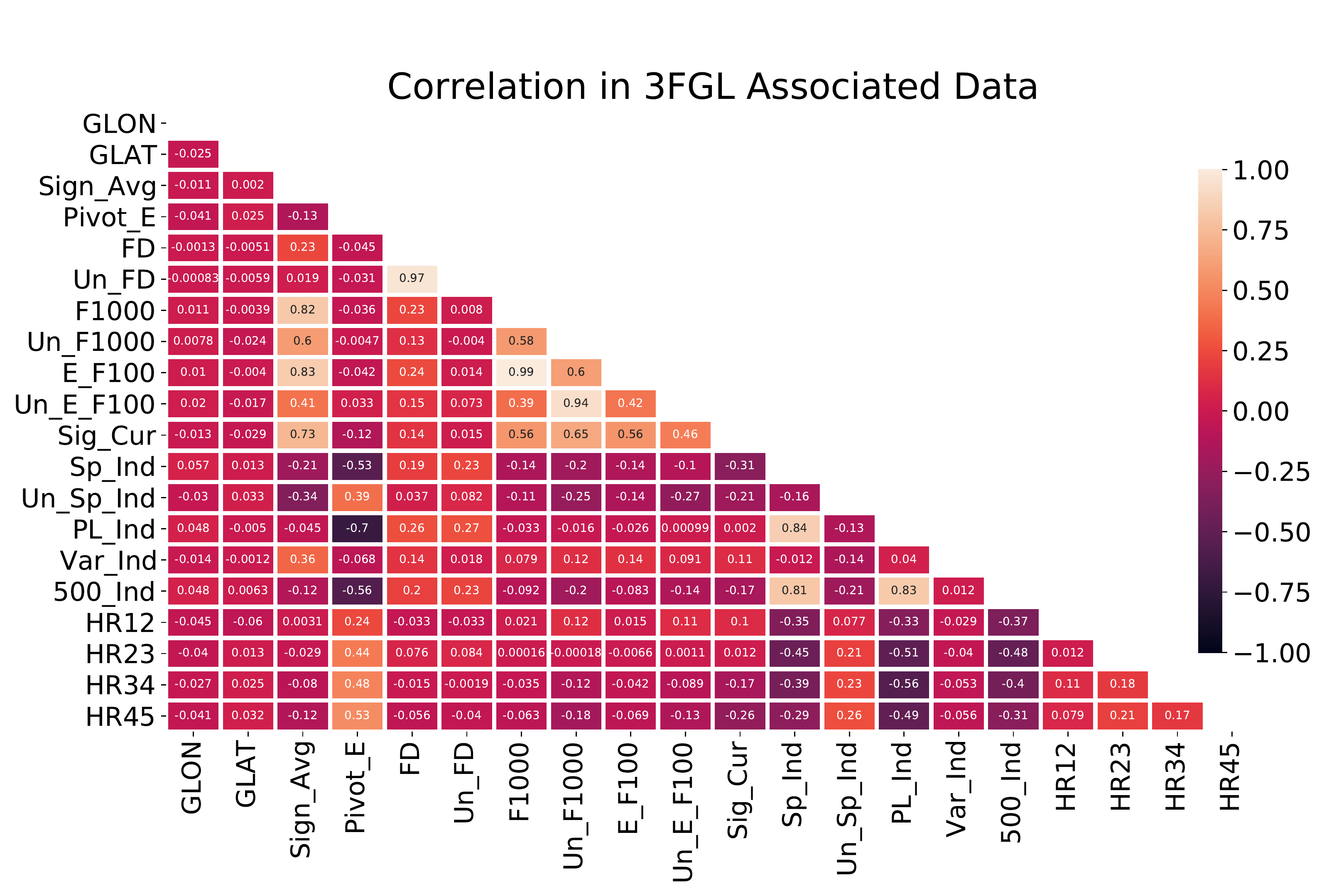}
\caption{Correlation matrix of features for the associated sources in the 3FGL catalog.
The 500MeV\_Index is defined in Eq. (\ref{eq:n500_def}).
The hardness ratios (HR12, HR23, etc.) are defined in Eq. (\ref{eq:hr}).
All other features are taken directly from the 3FGL catalog (see text for the description of labels).
}
\label{fig:assoc_corr_3fgli}
\end{figure*}

In order to select independent features, we calculated the Pearson correlation coefficients for the features in the 3FGL catalog.
We shortened the notation of some features for brevity; the full 3FGL names of the features (in case of shortened notations) and some explanations are given in parentheses
(for the full explanation we refer to the 3FGL catalog \citep{2015ApJS..218...23A}):
GLON (Galactic longitude), 
GLAT (Galactic latitude),
Sign\_Avg (Signif\_Avg -- average significance), 
Pivot\_E (Pivot\_Energy), 
FD (Flux\_Density -- differential flux at Pivot\_Energy), 
Un\_FD (Unc\_Flux\_Density -- uncertainty of Flux\_Density), 
F1000 (Flux\_1000 -- flux above 1000 MeV), 
Un\_F1000 (Unc\_Flux\_1000 -- uncertainty of Flux\_1000),
E\_F100 (Energy\_Flux100 -- energy flux above 100 MeV),  
Un\_E\_F100 (Unc\_Energy\_Flux100 -- uncertainty of Energy\_Flux100), 
Sig\_Cur (Signif\_Curve -- significance of spectral curvature), 
Sp\_Ind (Spectral\_Index), 
Un\_Sp\_Ind (Unc\_Spectral\_Index -- uncertainty of Spectral\_Index),
PL\_Ind (PowerLaw\_Index), 
Var\_Ind (Variability\_Index), 
500\_Ind (500MeV\_Index -- spectral index at 500 MeV defined in Eq. (\ref{eq:n500_def})), 
and HR$_{ij}$ for the hardness ratios defined above.
A graphical representation of the correlations is shown in Fig.~\ref{fig:assoc_corr_3fgli}. 
In the following, if two features have (anti)correlation $\gtrsim 0.75$ 
($\lesssim -0.75$), then we keep only one of the features for classification.
Taking into account the correlation among the features and the above discussion of the spectral index definition,
we selected the following 11 features for the classification of the 3FGL sources:
GLAT, GLON, ln(Energy\_Flux\_100), $\ln$(Unc\_Energy\_Flux100), 500MeV\_Index, $\ln$(Signif\_Curve), 
$\ln$(Variability\_Index), and the four hardness ratios ${\rm HR}_{ij}$.  
The table of features and their statistics can be found in Appendix \ref{sec:app}.

\subsection{Construction of classification algorithms}

The number of tunable parameters in the classification algorithms is not fixed a priori. 
Moreover, there is a certain freedom in the choice of the architecture of the algorithms, such as
the number of hidden layers and the number of neurons in NN.
In general, one starts with a simple model and increases the complexity (the number of tunable parameters)
until the model can describe the data well without overfitting it.
The overfitting is tested by splitting the input data into training and testing samples.
The training sample is used for optimizing the parameters,
while the test sample is used to check that the model is not overtrained (for overtrained models the accuracy on the test
sample is significantly worse than the performance on the training sample).
For our catalogs we split the data randomly into 70\% training and 30\% testing samples.

In this paper we determined the probabilistic classification of a source with an algorithm by the class with the maximal probability (as estimated by this algorithm). In the case of two classes, this is the class with probability larger than 0.5. In the case of three classes, the largest probability can be smaller than 0.5 but always larger than 1/3. Although the classification probabilities for some sources are not very large; for example, a significant fraction of sources classified as pulsars may turn out to be AGNs or OTHER sources, the main goal of our analysis is not to determine a list of sources, which are classified as pulsars or AGNs with high probabilities, but to determine the probabilities themselves and to estimate the uncertainties on the probabilities. In other words, our main goal is the construction of the probabilistic catalogs, which we make available online \citep{SOM_material}. A user of these probabilistic catalogs can choose a smaller or a larger probability threshold for a particular class depending on the purpose of their analysis.

\subsubsection{Random forests}
\lb{sec:rf}

The two main parameters characterizing the RF algorithm are the number of trees and the maximum depth allowed in the trees. 
We used the Gini index as the objective function for the optimization of parameters (split values of features in the nodes).

\begin{figure}[h]
\centering
\includegraphics[width=0.5\textwidth]{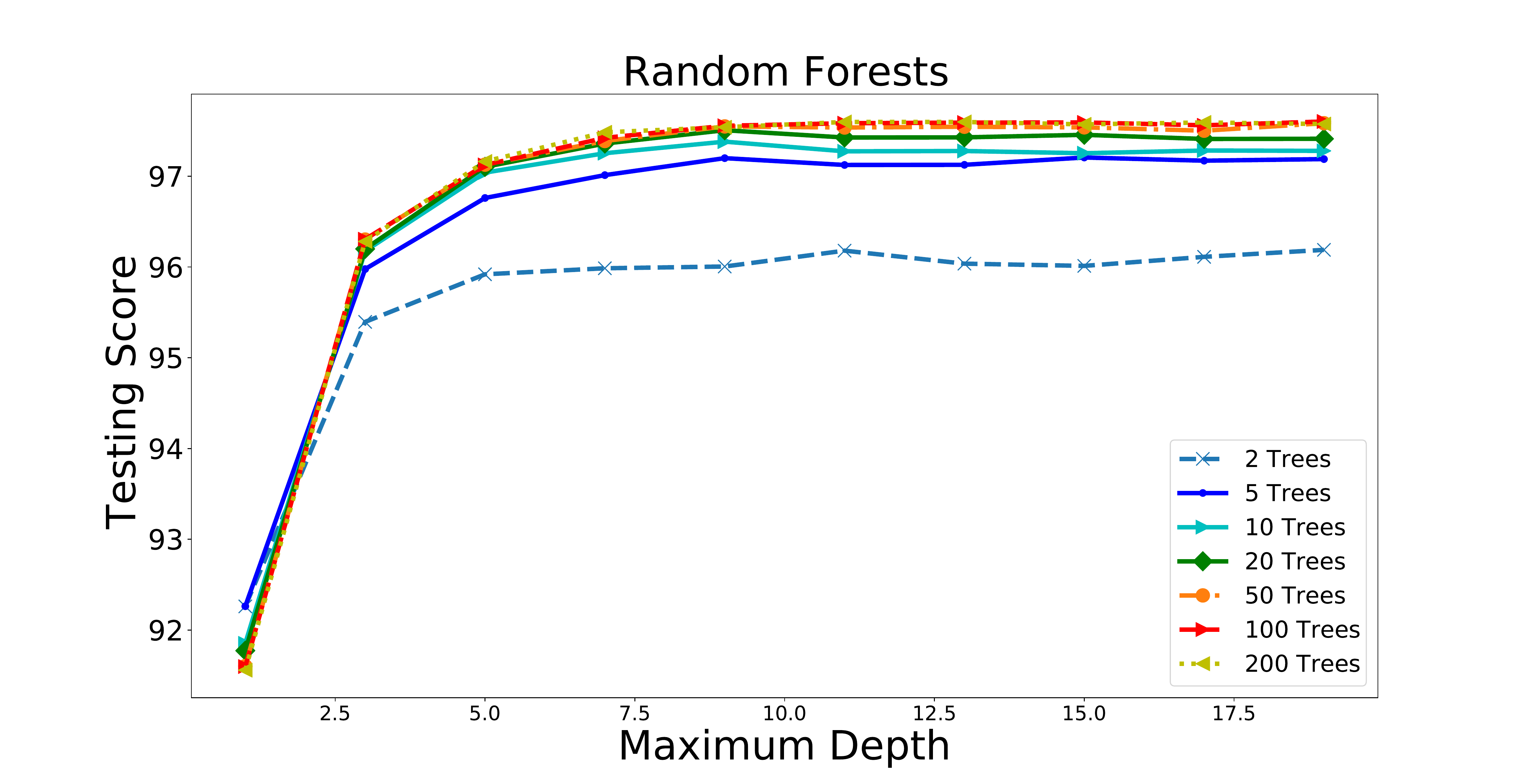}
\caption{
Test score (accuracy) of RF classification as a function of the number of trees and 
the maximal depth of trees.
}
\label{fig:RF_complexity}
\end{figure}

Figure \ref{fig:RF_complexity} shows the dependence of the accuracy of the test sample as a function of maximum depth and the number of trees. 
The results for each point are averaged over 100 realizations of the split into training and testing samples.
We notice that the accuracy does not decrease as the maximal depth of the trees increases (i.e., there is no overfitting as the complexity of the model increases with increased maximum depth).
This is due to the random choice of a subset of features at each node (maximal number of allowed features is $\sqrt{\text{\# features}}$).
It is also insensitive to the number of trees above approximately 20 trees.
For classification we used 50 trees with a maximum depth of 6.

\begin{figure}[h]
\centering
\includegraphics[width=0.5\textwidth]{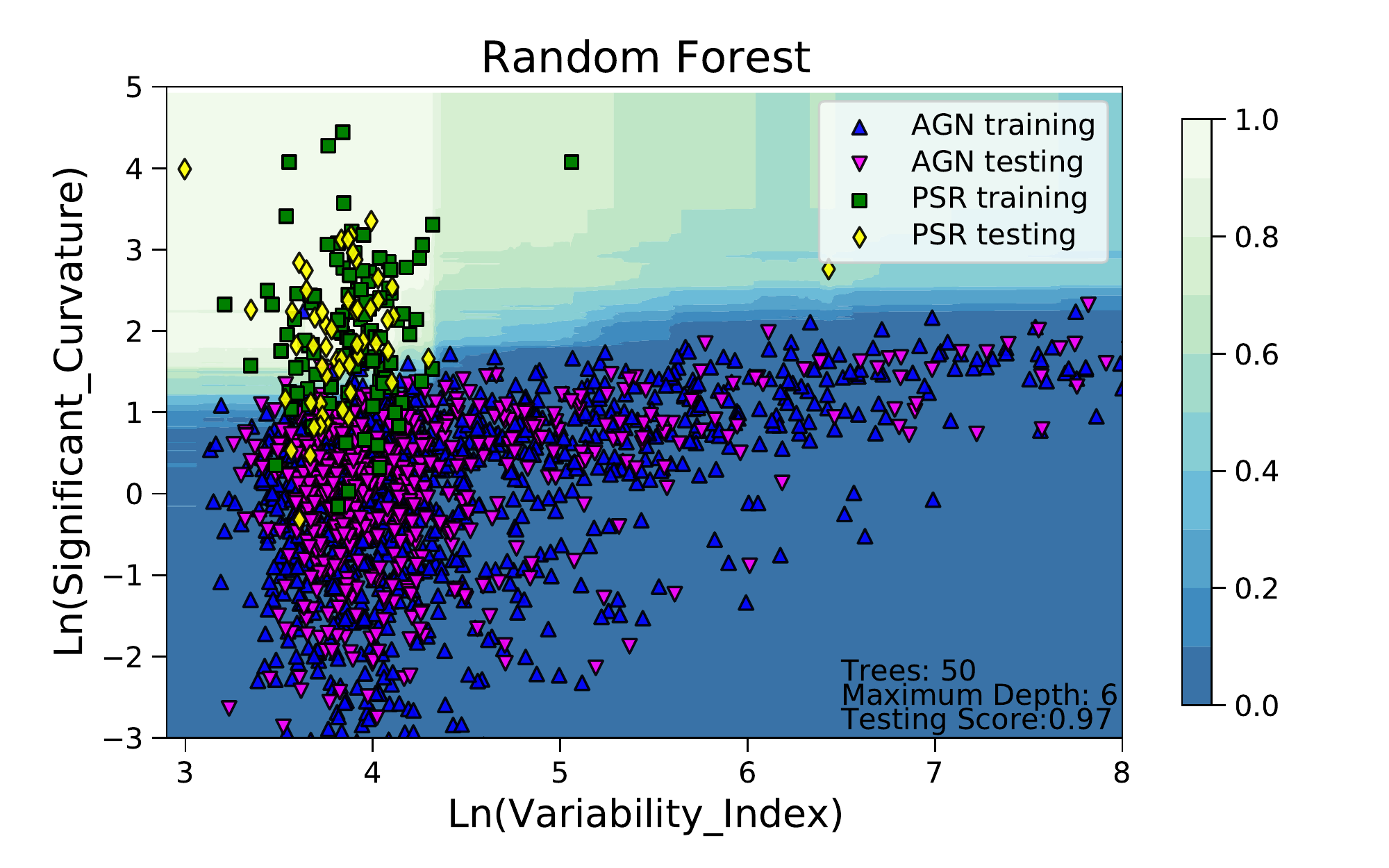}
\caption{RF classification domains showing class probabilities for training with two features
averaged over 100 random splits into training and testing samples.
One of these splits is shown for illustration.
The color scale describes the probability of a source being a pulsar.
}
\label{fig:RF_domains}
\end{figure}

In order to illustrate the separation of point sources into AGNs and pulsars, we retrain the RF algorithm using only two features: log of curvature significance and log of the variability index, and plot the resulting probabilities of classes in Fig. \ref{fig:RF_domains}
for the model with 50 trees with a maximum depth of 6.
The probabilities are averaged over 100 splits into training and testing samples.
It is important to note that in this plot the model is trained on only two features. Nevertheless a good testing accuracy of 97\% is reached, 
which is similar to the accuracy of the RF classification with all 11 features.
For the final classification with RF, we used 11 features and averaged over 1000 splits into training and testing samples.

\subsubsection{Boosted decision trees}

\begin{figure}[h]
\centering
\includegraphics[width=\twopicsp\textwidth]{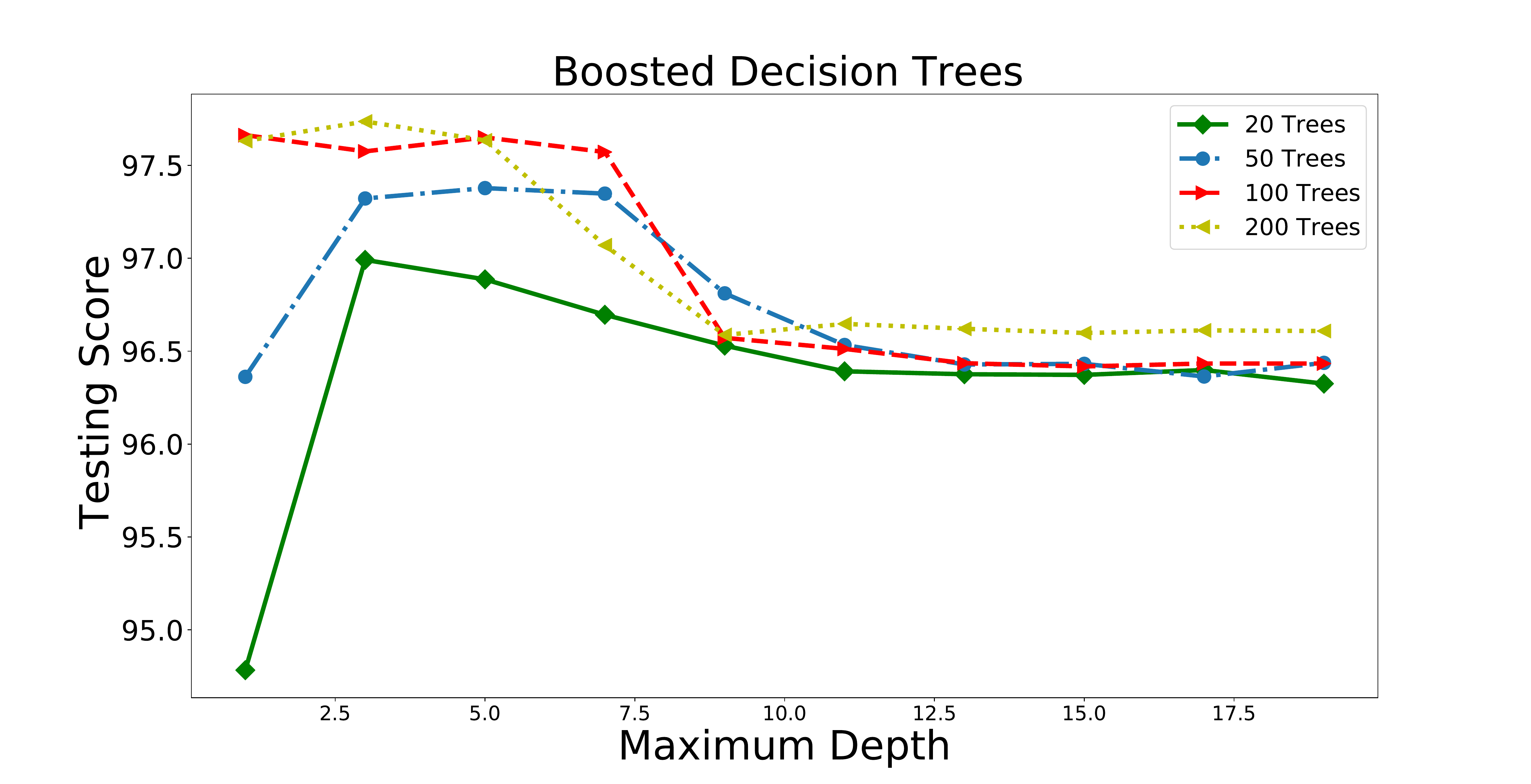}
\caption{Dependence of BDT accuracy on maximum depth and the numbers of trees.}
\label{fig:BDT_depth}
\end{figure}

The meta-parameters for BDT algorithms are similar to RF algorithms: the number of trees and the maximal depth.
We used the gradient boosting algorithm for the construction of BDT \citep{gb}.
The classification is performed by a weighted average of trees, where the trees are constructed recursively in order to better address 
misclassifications from the previous step. 
Dependence of the accuracy on tree depth is shown in Fig.~\ref{fig:BDT_depth}. 
Unlike the RF, which is also an ensemble-based method, the testing accuracy drops for the maximal depths larger than 7.

\begin{figure}[h]
\centering
\includegraphics[width=0.5\textwidth]{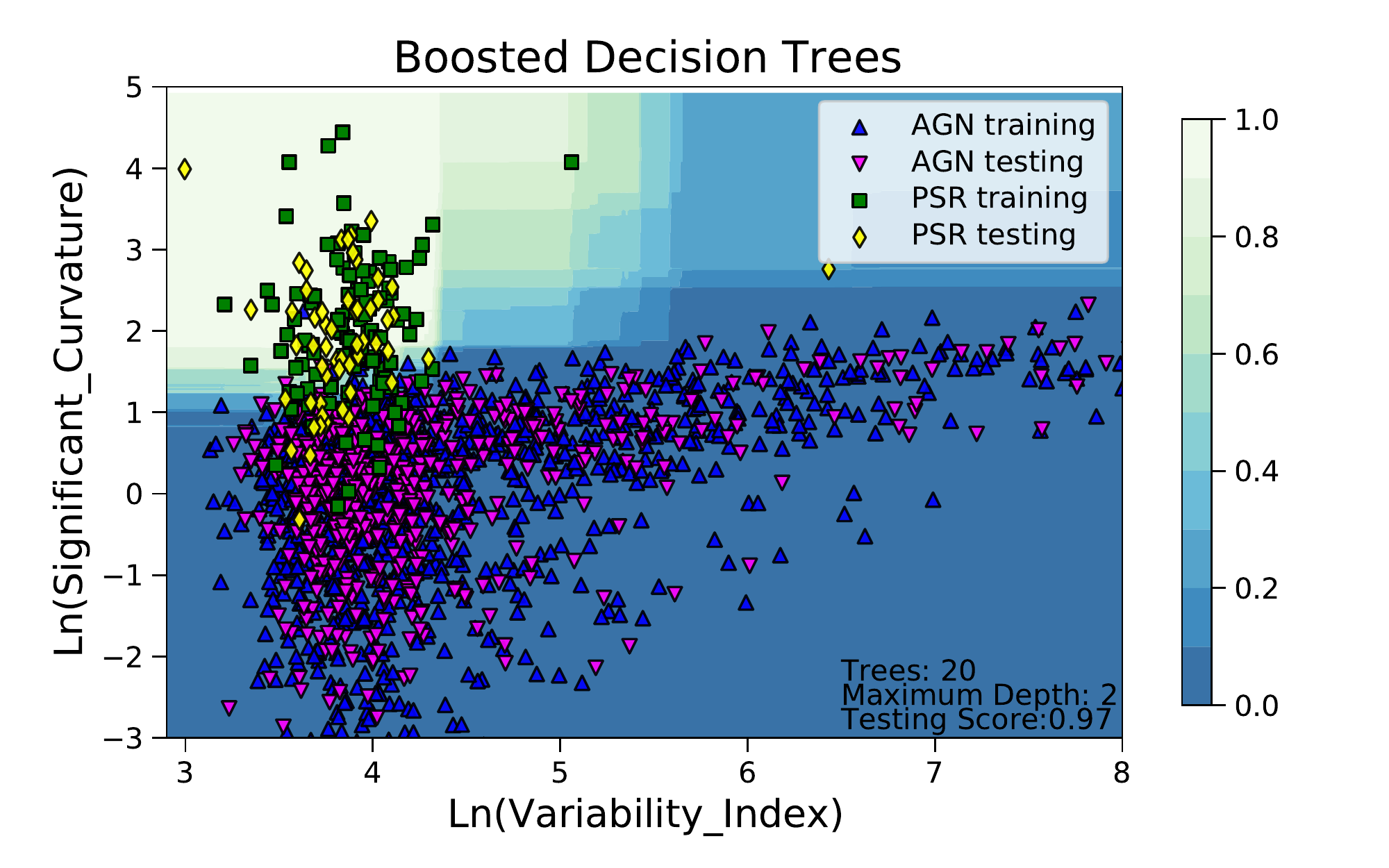}
\caption{Classification domains for BDT for training with two features 
averaged over 100 splits into training and testing samples.
}
\label{fig:BDT_domains}
\end{figure}

The classification domains in case of two features for 20 trees and the maximum depth of 2 are presented in Fig.~\ref{fig:BDT_domains}. 
For the classification we used BDT with 100 trees and the maximum depth of 2.

In tree-based algorithms, one can calculate the feature importance by using the averaged reduction in impurity for nodes (the Gini index in our case) involving the different features. 
The importance of features for the case of two different algorithms: RF with 50 trees and maximum depth of 6, and BDT with 100 trees with maximum depth of 2,  are shown in Table \ref{tab:feat_imp}.
We find that the most important feature for both cases is the significance of curvature.
Other significant features are the hardness ratio of the last two energy bins, uncertainty of the energy flux at 100 MeV, and the variability index.

\begin{table}[!h]
\caption{Feature importances for RF and BDT algorithms.
}
\label{tab:feat_imp}
\tiny
\centering
\renewcommand{\tabcolsep}{1mm}
\renewcommand{\arraystretch}{1}

\begin{tabular}{c c c}
\hline
\hline
Feature & RF: 50, 6 & BDT: 100, 2\\
\hline
{ $\ln$(Signif\_Curve)}&  0.331  & 0.518   \\
{ HR45}&0.137&0.071\\
{ $\ln$(Unc\_Energy\_Flux100)} &0.122& 0.050   \\
$\ln$(Variability\_Index)& 0.098&0.225  \\
$\ln$(Energy\_Flux100) & 0.071&0.019   \\
500MeV\_Index&0.065& 0.028  \\
HR23 & 0.062&0.052  \\
HR12& 0.052&0.012  \\
HR34&0.025&0.005\\
GLAT &0.017& 0.002     \\
GLON & 0.014&0.011  \\
\hline
\end{tabular}
\tablefoot{RF algorithm: 50 trees with maximal depth 6; BDT algorithm: 100 trees with maximal depth 2.
The features are ordered by decreasing importance in the case of the RF algorithm.
}
\end{table}

It is interesting to note that Galactic latitude is among the least significant features.
We also used sin(GLAT) to check that this is not due to scaling (i.e., the large range of values of GLAT),
but the significance is similar to the GLAT itself.
We further discuss the dependence on GLAT in Sect. \ref{sec:lat-lon-profiles}, 
where we compare the latitude and longitude profiles of the associated and unassociated source counts.%
\footnote{Feature importances for the classification of 4FGL-DR2 sources with RF and BDT algorithms are reported in Appendix \ref{sec:app}.}

\subsubsection{Neural networks}

In the case of NN, the number of free parameters depends on the number of hidden layers and on the number of neurons in the hidden layers. The final model accuracy also depends on the number of epochs that the network is allowed to be trained for and on the optimization algorithm. 

\begin{figure}[h]
\centering
\includegraphics[width=0.45\textwidth]{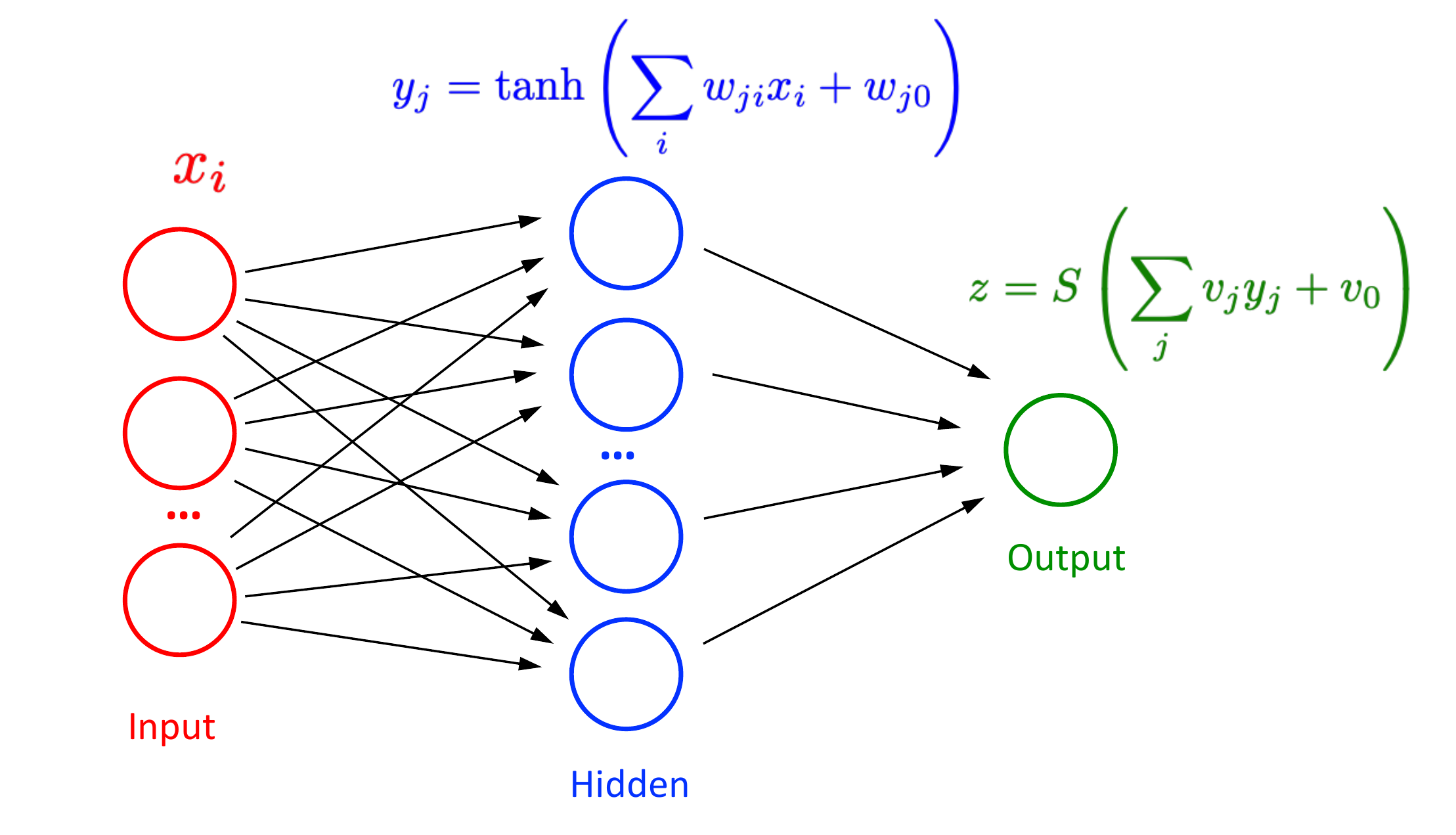}
\caption{
NN architecture that we used in the construction of the probabilistic catalogs.
The activation function in the output layer is sigmoid $S(x) = {e^{x}}/{(1 + e^{x})}$.
}
\label{fig:NN_structure}
\end{figure}

The general architecture of the NN that we used in this paper is shown in Fig. \ref{fig:NN_structure}.
It is a fully connected NN with 11 input nodes (shown by red circles with input features $x_i$), one hidden layer (shown by blue circles),
and an output layer (shown by the green circle).
The hidden layer consists of several nodes with values $y_j$. 
For the activation function at the hidden layer we used either hyperbolic tangent (tanh - shown on the plot) or rectified linear unit (relu).
The activation function for the output layer is sigmoid, which we used to make sure that the output value can be interpreted as a class probability.
The unknown parameters are weights of features in the hidden layer $w_{ji}$ and in the output layer $v_j$ including
offsets $w_{j0}$ and $v_0$.
The unknown parameters were optimized by minimizing a loss function, which we chose to be
the cross entropy
$-\text{log}L = - \sum_i (y_i\text{log}(p_i)+(1-y_i)\text{log}(1 - p_i))$, 
where $y_i = 0,\,1$ are the true labels of the sources and $p_i$ are the predicted class probabilities.
We also used NN with two hidden layers, but the accuracy was similar to the networks with one hidden layer (Appendix \ref{sec:app}). For the final classification model, we chose to use one hidden layer.

\begin{figure}[h]
\centering
\includegraphics[width=0.5\textwidth]{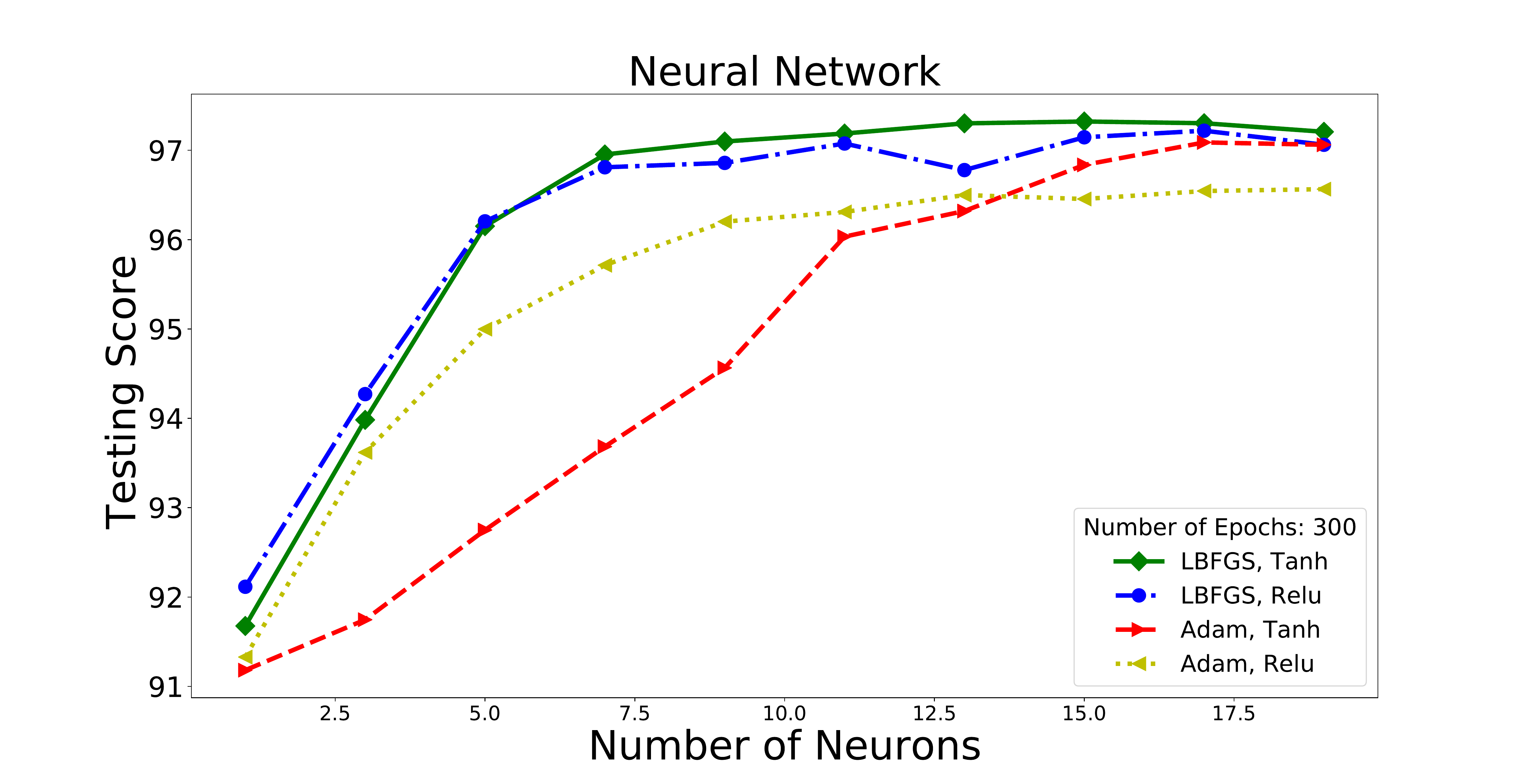}
\caption{Dependence of accuracy on the number of neurons for different NN models.}
\label{fig:NN_neurons}
\end{figure}

Dependence of the testing accuracy on the number of neurons in the hidden layer, on the activation function, 
and on the optimization algorithm is shown in Fig. \ref{fig:NN_neurons}. 
We compared two activation functions at the hidden layer (tanh and relu) and two optimization algorithms: 
limited memory Broyden-Fletcher-Goldfarb-Shanno \citep[LBFGS;][]{lbfgs} 
and the stochastic gradient descent algorithm Adam \citep{2014arXiv1412.6980K}.
We used 300 epochs for training.
Around 11 neurons in the hidden layer appears to be an optimal choice, since increasing the number of neurons led to no significant increase in accuracy for all models. 

\begin{figure}[h]
\centering
\includegraphics[width=0.5\textwidth]{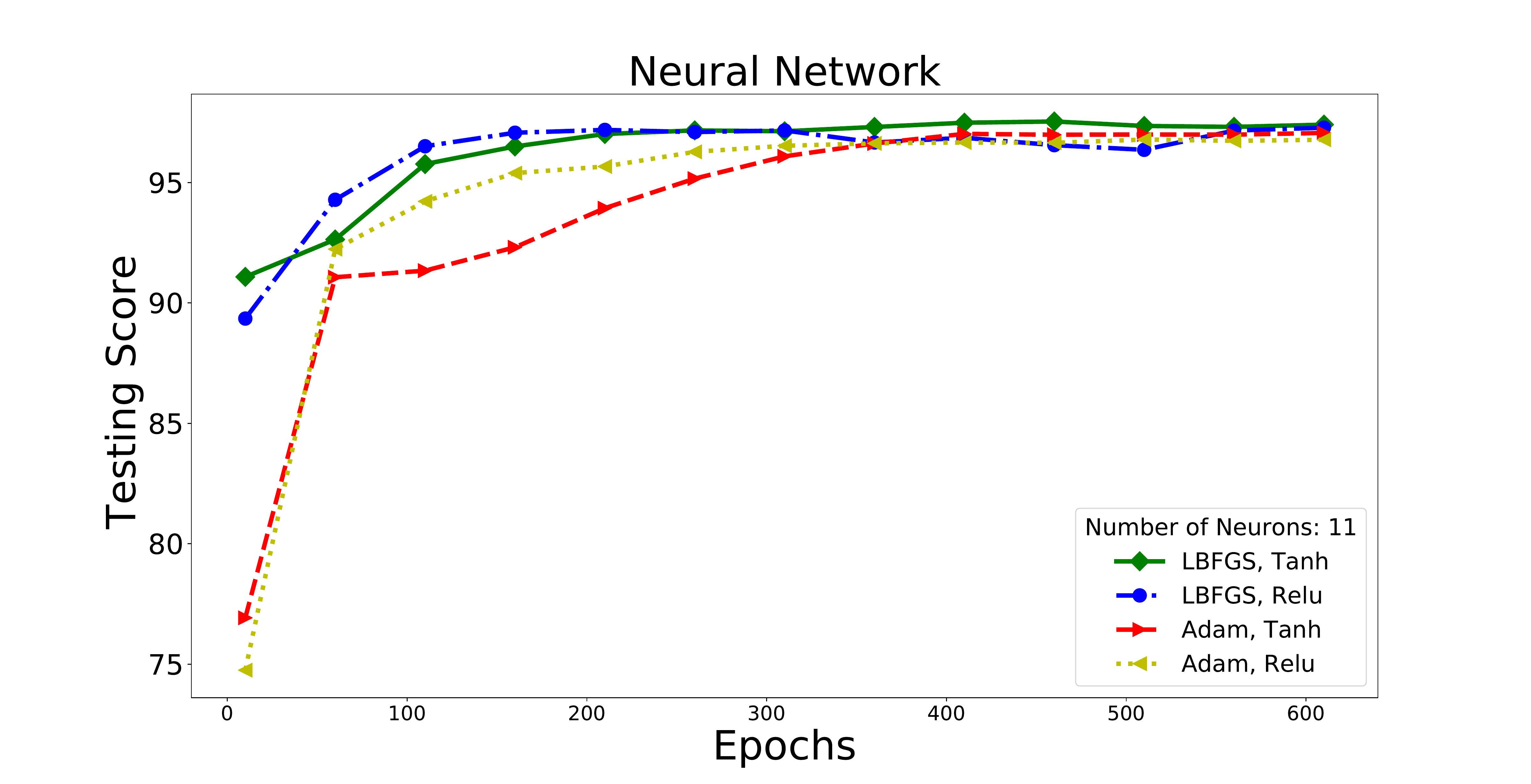}
\caption{
Dependence of testing accuracy on the number of epochs in training for different solvers and activation functions.
}
\label{fig:NN_epochs}
\end{figure}

Dependence on the number of epochs (number of iterations in fitting) is presented in Fig. \ref{fig:NN_epochs}. 
The accuracy increases with higher number of epochs and saturates at around 200 for LBFGS and 300 for Adam.

\begin{figure}[h]
\centering
\includegraphics[width=0.5\textwidth]{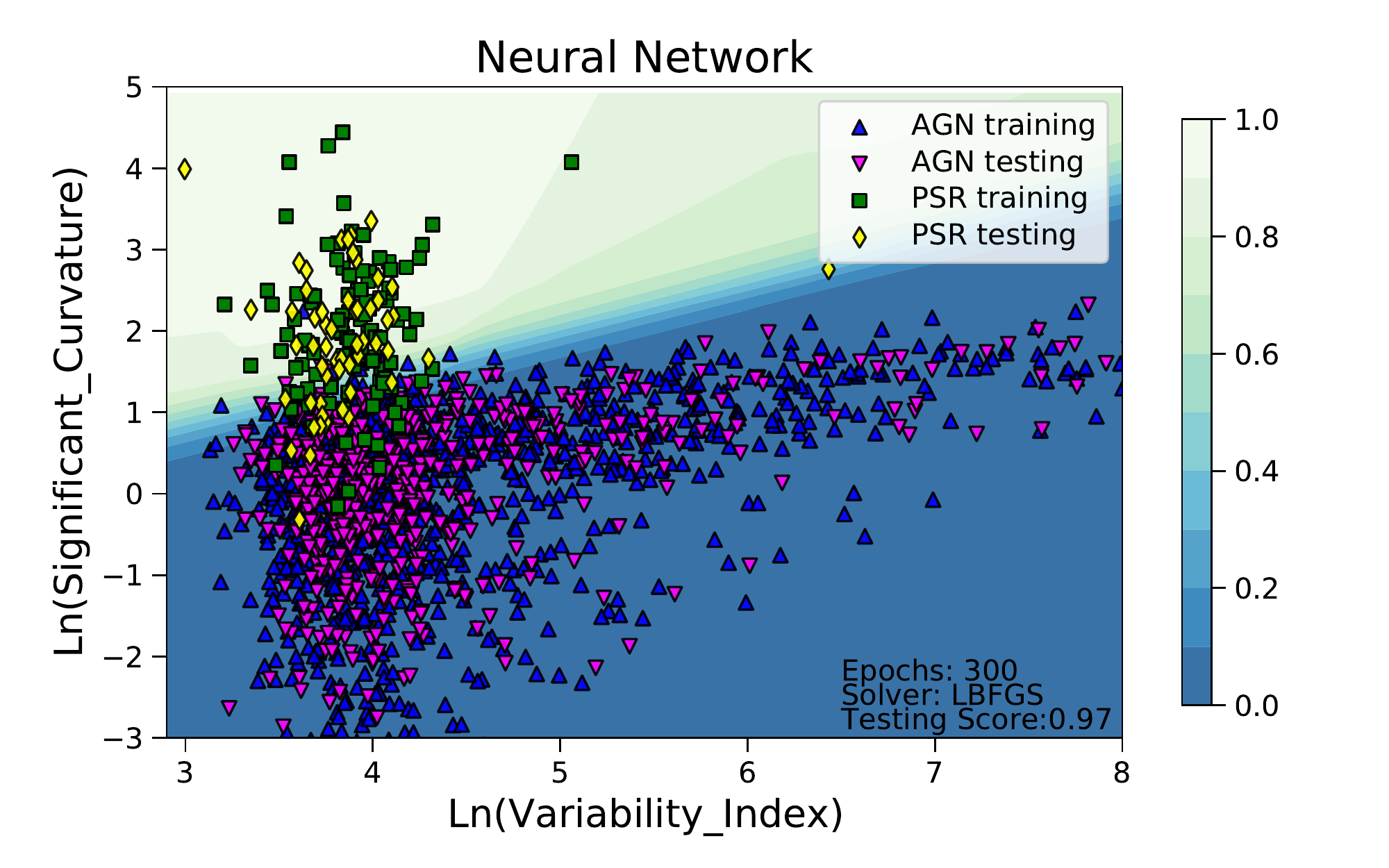}
\caption{NN classification domains for two input features
averaged over 100 random splits into training and testing samples.
We used two neurons in the hidden layer, the tanh activation function, and the LBFGS solver. 
}
\label{fig:NN_domains}
\end{figure}

We illustrate the classification domains for NN with two input features in Fig. \ref{fig:NN_domains}. 
In this case we also used only two neurons in the hidden layer.
One can see that the separation boundary is smoother compared to the RF domains in Fig. \ref{fig:RF_domains} or BDT domains in Fig. \ref{fig:BDT_domains}.
For our final model we chose one hidden layer with eleven neurons, 300 training epochs, LBFGS solver, and tanh activation function at the hidden layer.

\subsubsection{Logistic regression}

\begin{figure}[h]
\centering
\includegraphics[width=\twopicsp\textwidth]{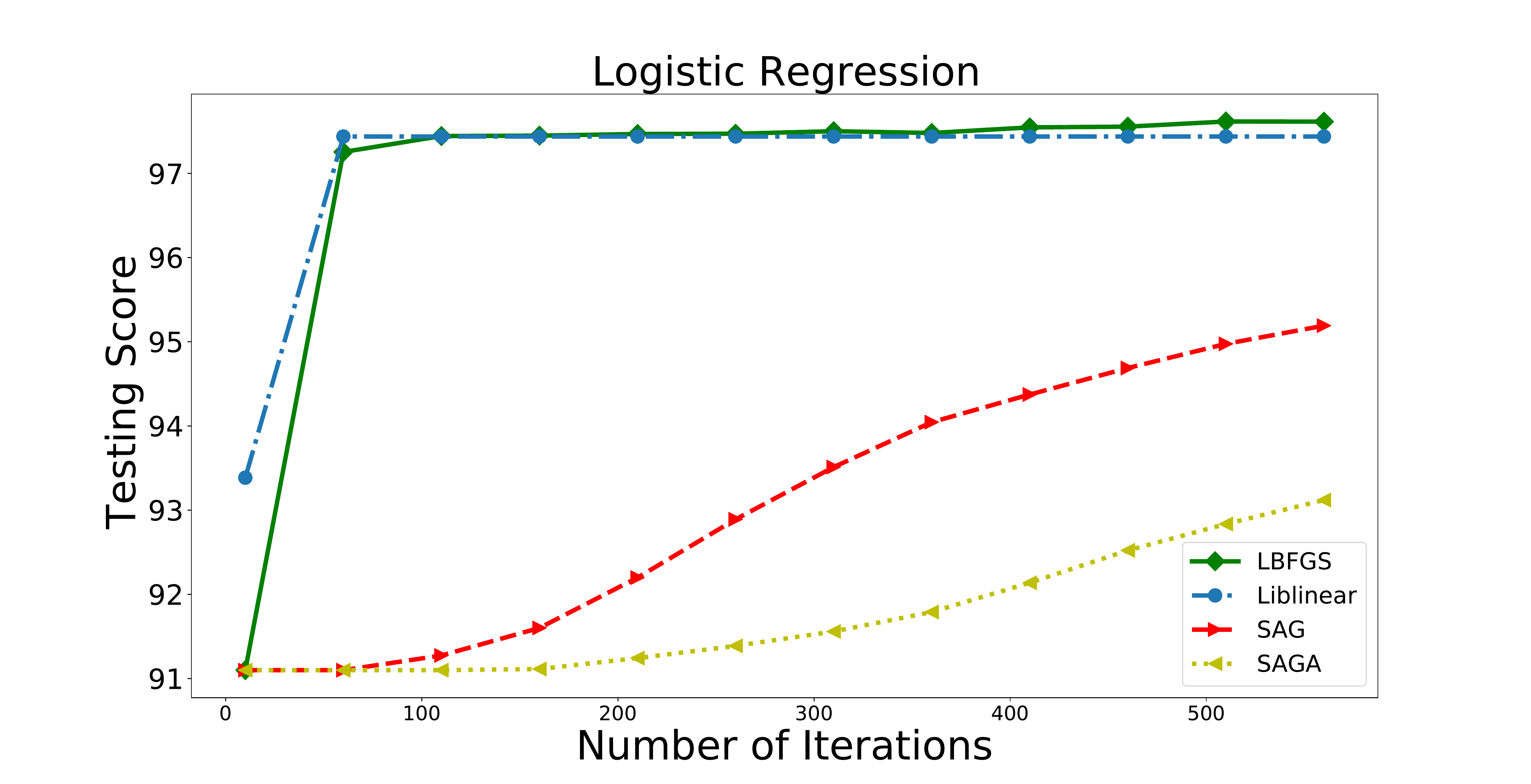}
\caption{Dependence of LR testing accuracy on the number of iterations for different solvers.}
\label{fig:LR_accuracy}
\end{figure}

\begin{figure}[h]
\centering
\includegraphics[width=0.5\textwidth]{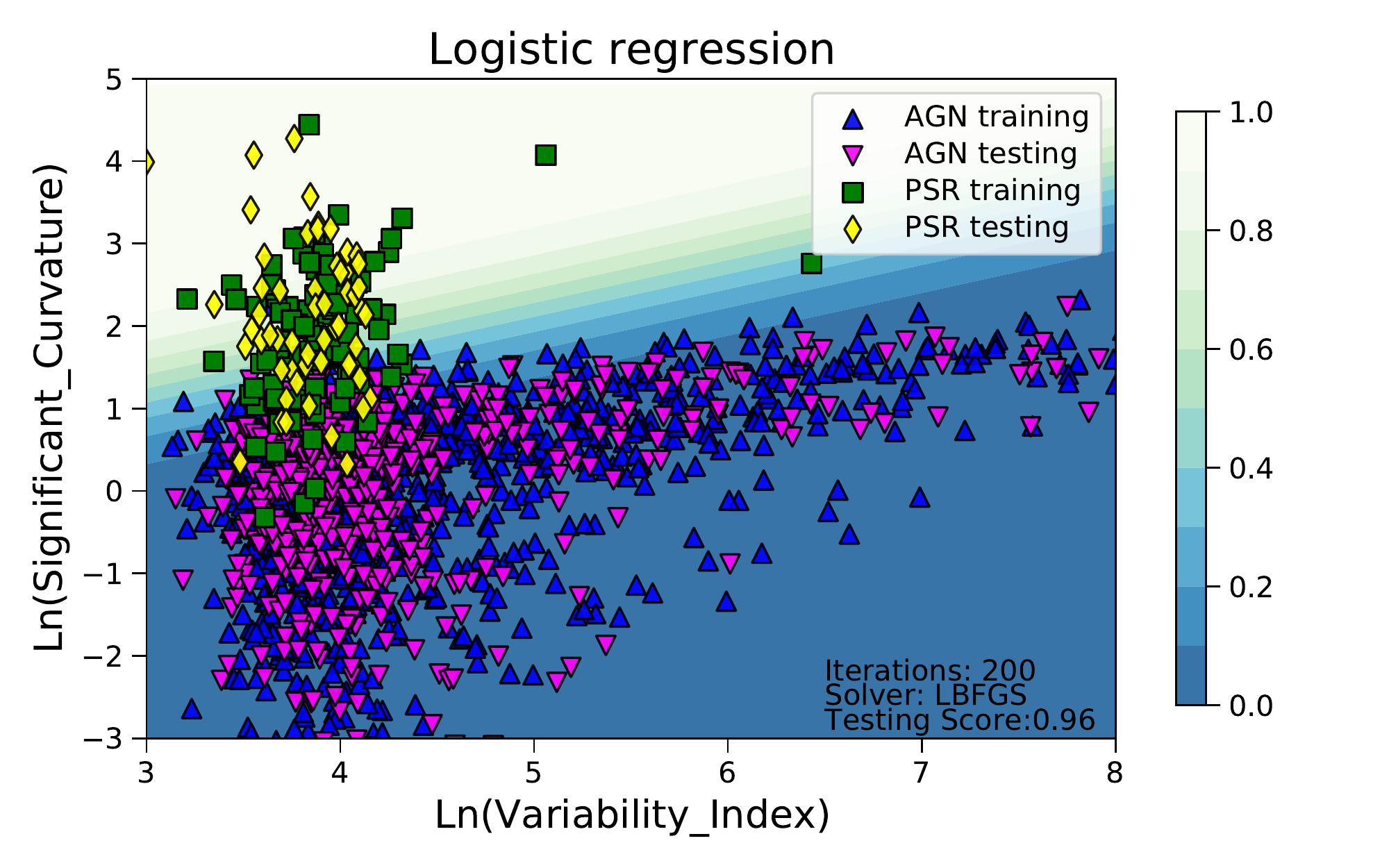}
\caption{Classification domains for LR with two features 
averaged over 100 random splits into training and testing samples.}
\label{fig:LR_domains}
\end{figure}

As we have discussed in Sect. \ref{sec:class_alg}, 
the probability of belonging to class 1 or 0 in LR is represented by the sigmoid function
$p_1(x) = 1 - p_0(x) = \frac{e^{m(x)}}{1 + e^{m(x)}}$ (see Eq. (\ref{eq:logit})),
where $m(x)$ is a function of input features $x$.
The complexity of the model is given by the number of parameters in $m(x)$.
We considered two cases for $m(x)$: linear and quadratic function of the input features $x$.
Quadratic $m(x)$ resulted in a similar accuracy as linear $m(x)$.
Consequently, we restricted our attention to linear functions $m(x) = f_0 + \sum_{k = 1}^{11} f_k x_k$.
In Fig. \ref{fig:LR_accuracy} we show the accuracy of the LR method as a function of the number of iterations
for different solvers, for example, LBFGS \citep{lbfgs}, the stochastic average gradient \citep[SAG;][]{sag}, SAGA \citep[a variant of SAG,][]{saga},
and LIBLINEAR \citep[a special solver for LR and support vector machine classifications,][]{ll}.
As one can see from Fig. \ref{fig:LR_accuracy}, LBFGS and LIBLINEAR outperform the other two solvers and converge much faster.
In order to illustrate the probability domains in LR, we show the classification with two features (LBFGs, 200 iterations)
in Fig. \ref{fig:LR_domains}. The domains look similar to the domains in the NN case (Fig. \ref{fig:NN_domains}).
For the final classification we used LBFGs solver with 200 iterations.

\subsection{Oversampling}
\lb{sec:oversampling}

\Fermi-LAT catalogs have many more AGNs than pulsars, (i.e., the data sets are imbalanced).
For example, the 3FGL catalog has 1744 associated AGNs (1739 without missing or unphysical values)
and 167 associated pulsars (166 without missing or unphysical values).
In the previous subsections we have optimized the overall accuracy. In this case, the algorithms try to identify AGNs rather than pulsars,
since it gives better accuracy. As a result, in the region of parameter space, where both pulsars and AGNs are present, the algorithms
will give a higher probability of a source being an AGN.

The problem of classification of imbalanced data sets can be quantitatively described in terms of precision and recall.
If we denote by ``\# true'' the number of pulsars in the data set, by ``\# positive'' -- the number of sources predicted to be pulsars, and by 
``\# true positive'' -- the number of pulsars predicted to be pulsars, then  $precision \rm = \frac{\#\ true\ positive}{\#\ positive}$ is a measure of how clean the prediction is, while $recall \rm = \frac{\#\ true\ positive}{\#\ true}$ is a measure of how well the algorithm can detect pulsars (i.e., how complete the list of predicted pulsars is).
If we reduce the pulsar domain by attributing uncertain sources predominantly to AGNs, then for pulsars the precision will increase, but the recall will decrease.

\begin{figure}[h]
\centering
\includegraphics[width=\twopicsp\textwidth]{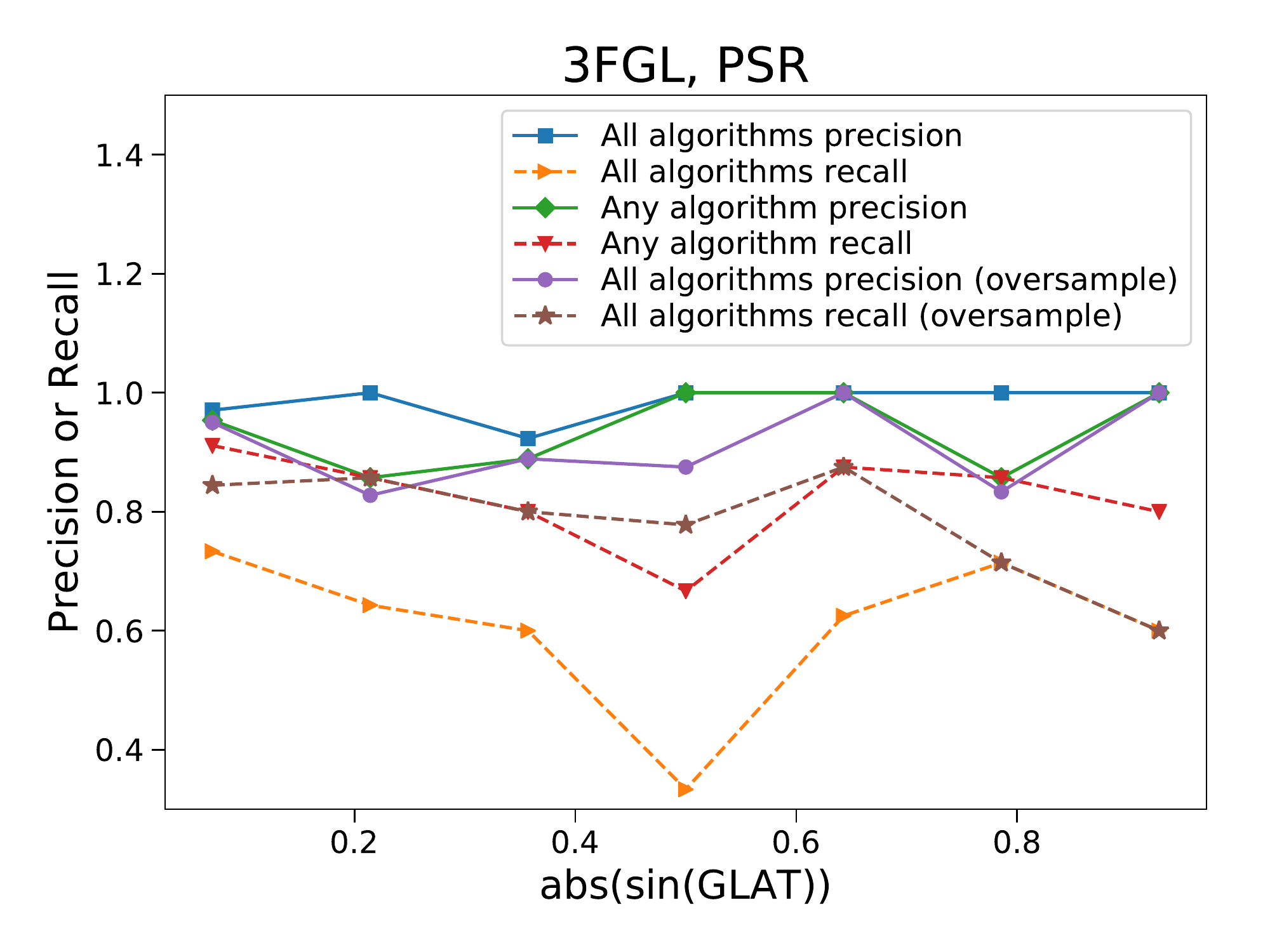}
\caption{Precision and recall for pulsars using all-algorithm and any-algorithm classification for unweighted training data and
all-algorithm classification for oversampling of pulsars in training data (for details, see Sect. \ref{sec:oversampling}).}
\label{fig:prec_recall}
\end{figure}

In Fig. \ref{fig:prec_recall} we show precision and recall for classification of pulsars.
In particular, in the first two lines (solid blue with squares and dashed orange with right triangles) a source is categorized as a pulsar if all four algorithms classify it as a pulsar,
while in lines 3 and 4 (solid green with diamonds and dashed red with down triangles) a source is attributed to the pulsar class, if any of the algorithms classifies it as a pulsar.
It is clear that for lines 1 and 2 the pulsar domain is smaller than for lines 3 and 4, since in the former case, the domain is the intersection of domains for individual algorithms, while in the latter it is the union.
For all-algorithms classification the precision is 100\% for most of latitudes, while the recall is between 40\% and 80\% (i.e., the list of pulsars is generally clean but incomplete).
In case of any-algorithm classification, the recall is increased by about 20\% for most latitudes compared to the all-algorithms classification, but the precision drops by up to 20\% at some latitudes (i.e., the completeness improves at the expense of cleanliness of the sample).
As an alternative to using an any-algorithm classification, one can give larger weights to pulsars or oversample pulsars in the training process (i.e., use the same source several times) so that the numbers of pulsars and AGNs in training are the same.
Provided that in some applications it is beneficial to have as complete a list as possible of pulsar candidates among unassociated sources, we retrained the algorithms using oversampling with the same meta-parameters as in the previous sections.

\begin{figure}[h]
\centering
\includegraphics[width=0.5\textwidth]{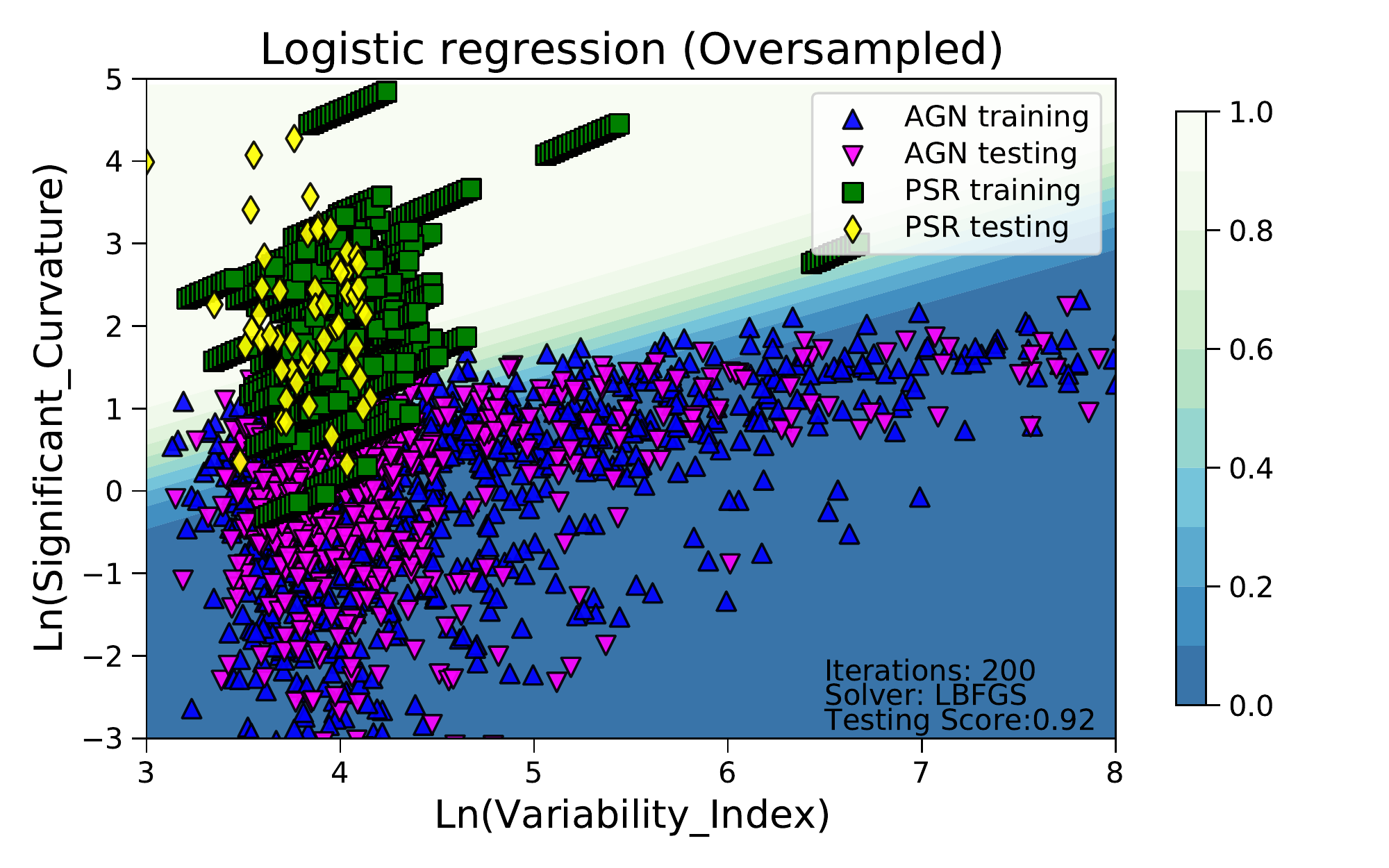} \\
\includegraphics[width=0.5\textwidth]{plots/classification_domains/lr_200_lbfgs.pdf}
\caption{ 
Comparison of classification domains for the LR method with and without oversampling.
Top panel: LR classification domains showing class probabilities for training with oversampling.
The oversampling is illustrated by repeating the pulsar markers with a shift: the number of markers is equal to the number of times the pulsar appears in training.
Bottom panel: Repeat of Fig. \ref{fig:LR_domains} for the convenience of comparison with the oversampled training in the top panel.
In both panels the domains are obtained by averaging over 100 random splits into training and testing samples.
}  
\label{fig:LR_domains_O}
\end{figure}

In general one can either under- or oversample a data set. Undersampling would reduce the number of AGNs to match the number of pulsars. However, since the total number of sources is not very high, we chose to oversample the data. 
For training with oversampling, we copied randomly existing pulsars and added them to the data set until the number of pulsars and AGNs were the same.
Although pulsars in the training data set were redundant, they helped to increase the weight of pulsars in the classification model.
We illustrate the oversampling procedure in Fig. \ref{fig:LR_domains_O} top panel:
the number of times a source appears in training is shown by adding markers with shifts to the right and above the original position of the source (we note that the shift is introduced for presentation only, the parameters of the sources are exactly the same as in the original source).
In the bottom panel of Fig. \ref{fig:LR_domains_O} we repeat Fig.  \ref{fig:LR_domains} in order to compare the classification domains with and without oversampling.
One can see that pulsar domain in the top panel is larger than the pulsar domain in the bottom panel.
As a result, in the top panel more pulsars are classified as pulsars but also more AGNs are falsely classified as pulsars in the intersection region. 
Since the overall number of AGNs is larger than the number of pulsars, the testing accuracy with oversampling is smaller than without oversampling.

The results of training with oversampling are presented  in Fig. \ref{fig:prec_recall},
lines 5 and 6 (solid purple with circles and dashed brown with stars). 
These lines show precision and recall when a source is categorized as a pulsar, if all four algorithms trained with oversampling classify it as a pulsar. The precision and recall in this case are similar to the any-algorithm classification for the training without oversampling.

In order to test the oversampling method, we compare in Appendix \ref{sec:app_O_vs_S} the oversampling-by-repetition with SMOTE \citep{Chawla_2002}. The result of the comparison is that for class probabilities of individual sources, the difference in oversampling is generally smaller than the uncertainty due to the random choice of the training sample, while the differences in population studies are comparable to the differences among the different algorithms.

\section{Probabilistic catalogs based on the 3FGL and 4FGL-DR2 catalogs}
\lb{sec:prob_cats}

In this section we use the ML algorithms optimized in the previous section to construct a probabilistic
classification of sources in the 3FGL and 4FGL-DR2 catalogs.

\subsection{Probabilistic classification of sources in 3FGL and comparison with 4FGL-DR2}
\lb{sec:3FGLprediction1}

We used the following four algorithms for the classification of sources: RF with 50 trees and maximal depth of 6, BDT with 100 trees and maximal depth of 2, NN with 11 neurons, LBFGS solver, and 300 epochs, and LR with LBFGS solver and 200 iterations. 
For training we used the pulsars and AGNs from the 3FGL catalog without missing or unphysical values. 
In addition to original data sets, we performed oversampling of pulsars in order to balance the numbers of pulsars and AGNs.
As a result, we have eight classification methods: four algorithms trained with and without oversampling.

\begin{table}[!h]
    \caption{Testing accuracy of the selected algorithms.}
    \label{tab:selected_algs}
    
\centering
\hspace{-0.2cm}
\resizebox{0.47\textwidth}{!}{
    \tiny
  \centering
    \renewcommand{\tabcolsep}{0.4mm}
\renewcommand{\arraystretch}{1.6}

    \begin{tabular}{c c c c c c c}
    \hline\hline
    Algorithm&Parameters &  Testing&Std. Dev.& Comparison with \\
    & & Accuracy & & 4FGL-DR2 Accuracy \\
    \hline
    RF & 50 trees, max depth 6  &97.37&0.60& 91.09 \\
    RF\_O &   &97.90&0.50& 89.44 \\
    \hline 
    BDT & 100 trees, max depth 2    &   97.65&0.54& 90.43 \\ 
    BDT\_O &     &   97.79&0.51& 91.75 \\
    \hline
    NN & 300 epochs, 11 neurons, LBFGS & 97.29&0.97& 90.10 \\
    NN\_O &  & 94.31&5.13& 87.13 \\
    \hline
    LR & 200 iterations, LBFGS solver & 97.63&0.54& 90.43 \\
    LR\_O &  &93.68&0.99& 85.15 \\
    \hline
    \end{tabular}}
    \tablefoot{Testing accuracy is computed for the classification of 3FGL sources. Comparison with associations in the 4FGL-DR2 catalog
    is presented in the last column. 
    ``\_O'' denotes training with oversampling.}
\end{table}

\begin{figure}[h]
\centering
\includegraphics[width=0.48\textwidth]{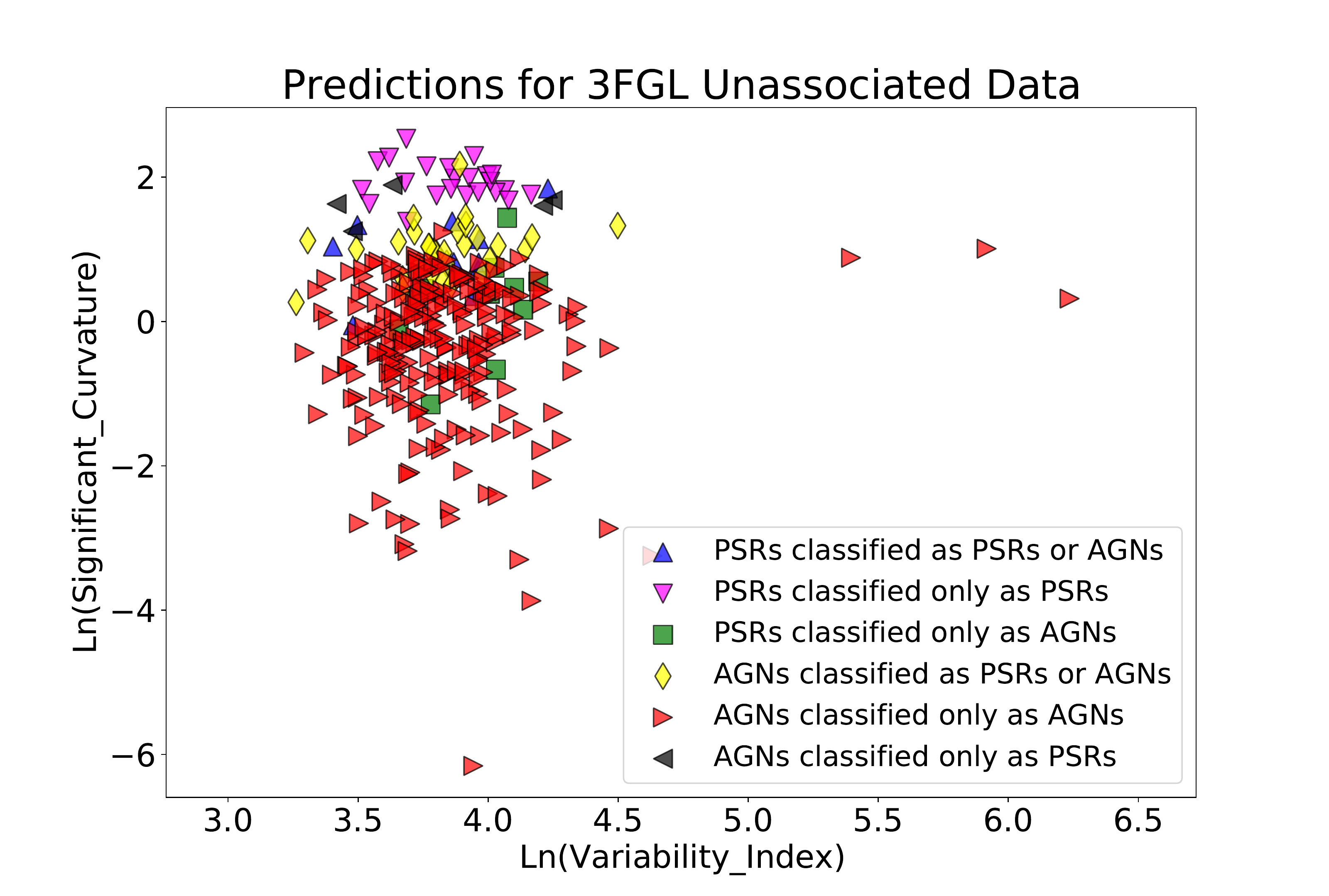}
\caption{Comparison of class prediction for unassociated 3FGL sources with classes in 4FGL-DR2 (for more details, see Sect. \ref{sec:3FGLprediction1}).}
\label{fig:3FGL_vs_4FGL_classes}
\end{figure}

The selected algorithms are summarized in Table \ref{tab:selected_algs}, where oversampling is shown by ``\_O.''
``Average testing accuracy'' is computed by taking 1000 times the 70\% - 30\% split into training and testing samples and averaging over the 
accuracies computed for the testing samples.
In addition, we looked at sources, which are unassociated in 3FGL but have either pulsar or AGN association in 4FGL-DR2: there are 303 such sources.
The accuracy of our prediction for the four selected algorithms with and without oversampling, taking the 4FGL-DR2 classes as the true values, is reported in the column ``Comparison with 4FGL Accuracy.''

As a result of the classification with the eight ML methods,
we created a probabilistic catalog based on the 3FGL sources.
We trained on 70\% of the sources associated with pulsars or AGNs without missing or unphysical values 
(there are thirteen sources with missing or unphysical values in the 3FGL catalog: two unassociated, five AGNs, one pulsar, and five OTHER sources).
We replaced the missing and unphysical values according to the procedure described at the beginning of Sect. \ref{sec:training}.
We calculated the probabilities of classes for testing sources, for sources that are not classified as pulsars or AGNs or have missing or unphysical values, and for unassociated sources.
We repeated the splitting and training 1000 times and report the sample average and standard deviation of the classification probabilities; in other words, we averaged over 1000 values for unassociated sources, sources not classified as AGNs or pulsars, and sources with missing or unphysical values. The average for AGNs and pulsars without missing or unphysical values is over the number of times the sources appeared in the testing sample, which is 300 on average.

In the probabilistic catalogs we added columns with corresponding probabilities for each algorithm and each class; provided that there are eight methods (including oversampling) and two classes, we add 16 columns: eight for unweighted and eight for oversampled training data. The columns with ``\_O'' represent the oversampled probabilities. We also added 16 columns for the standard deviations of probabilities. Although class probabilities and standard deviations for each algorithm are not independent (probabilities add up to 1 and standard deviations are equal for AGN and pulsar classes), we kept the corresponding columns in view of the generalizations to multi-class classification (e.g., the three-class classification in Sect. \ref{sec:3class}).

Table \ref{tab:prob_cat} shows an example of the probabilistic catalog for a few unassociated 3FGL sources.
We notice 
that the last source is classified as a pulsar by BDT and RF algorithms and as an AGN by LR and NN algorithms.
It is therefore an example of a source with mixed classification.

\pgfplotstableread[col sep=comma]{tables/3FGL_unassoc_vs_4FGL_assoc.csv}\loadedtable
\begin{table}
\caption{
Examples of the AGN classification probabilities.}
\label{tab:prob_cat}
\centering
\pgfplotstabletypeset[columns={Source_Name_3FGL,AGN_BDT,AGN_RF,AGN_LR,AGN_NN},
column type=l,
string type,
every head row/.style={before row={\hline\hline & \multicolumn{4}{c}{AGN Probability} \\},after row=\hline,},
every last row/.style={after row=\hline}, 
columns/Source_Name_3FGL/.style={column name=Source\_Name\_3FGL},
columns/AGN_BDT/.style={column name=BDT,numeric type,fixed,precision=3},
columns/AGN_NN/.style={column name=NN,numeric type,fixed,precision=3},
columns/AGN_RF/.style={column name=RF,numeric type,fixed,precision=3},
columns/AGN_LR/.style={column name=LR,numeric type,fixed,precision=3},
skip rows between index={4}{302}
]\loadedtable
\tablefoot{
The probabilities are presented for a few unassociated sources in the 3FGL catalog \citep{2015ApJS..218...23A}. 
We have omitted the oversampled probability columns for brevity.}
\end{table}

For the determination of candidate classes based on the probabilistic classification, we considered the following two conditions:
1) that all algorithms agree (i.e., each algorithm predicts the same class for a source with a more than 50\% probability%
\footnote{We use the 50\% threshold in the two-class case for illustration. Even if the probability of a source being, e.g., a pulsar is larger than 50\% for all eight methods, there is still a large chance of the source being an AGN if the probabilities are around 50\%. 
For this reason, the classes of sources reported in this work derived from the class probabilities should be viewed as candidate classes.
Depending on the application, a higher (lower) threshold can be used for a cleaner (more complete) sample. 
We study the effect of changing the classification threshold on precision and recall for the all-algorithms-agree method in 
Appendix \ref{sec:thres}.
The full catalogs with all probabilities are available online \citep{SOM_material}.
A more confident classification of particular sources can be obtained with multiwavelength studies, 
which are beyond the analysis in this paper.}) and 
2) that the sum of probabilities of a source belonging to a certain class is larger than 7 
(this means that on average the probability is larger than 7/8 = 0.875).
Both of these conditions are stricter than the classification using probabilities for any of the eight algorithms.
For convenience, we added a column in the probabilistic catalogs with the most likely probabilistic classes of sources based on the condition that all algorithms agree on the classification (sources with mixed classification are labeled as ``MIXED'' in this column).
In order to test the performance of the classification conditions we compared the precision and recall for unassociated source in the 3FGL catalog, 
which are associated in the 4FGL-DR2, and also calculated the expected precision and recall based on test samples used in training.
In total there are 340 sources, which are unassociated in 3FGL but have associations in 4FGL-DR2.
The result of classification of the unassociated sources in the 3FGL catalog, which have associations in the 4FGL-DR2 catalog, 
using the condition that all methods agree on the classification are presented in Table~\ref{tab:3FGL_vs_4FGL_2class}.
The MIXED column shows the numbers of sources, for which different algorithms predict different classifications.
Columns show the predictions for the 3FGL unassociated sources, while the rows show the associations in the 4FGL-DR2 catalog.
For the later comparison with the three-class classification, we also added unassociated source in the 3FGL catalog, 
which have associations with sources other than pulsars and AGNs.

We also present in Table~\ref{tab:3FGL_vs_4FGL_2class} the precision and recall estimates from the comparison of the 3FGL and 4FGL-DR2 catalogs (``4FGL assoc'' columns and rows).
For example, the precision for AGNs is the number of true positive predictions (223) over the number of positive predictions (the sum of numbers in the AGN column: (223 + 10 + 8), which gives $223 / 241 \approx 0.93$.
The estimated precision for pulsars is $ 23 / (23 + 5 + 6) \approx 0.68$.
In addition we show the expected precision and recall using the condition that all
methods agree (``all agree'' columns and rows) and the condition that the sum of probabilities is larger than 7 (``$\sum_a p_a > 7$'' columns and rows) calculated using the class probabilities for associated sources reported in the probabilistic catalogs (these probabilities are derived as an average over the test samples).
The precision (recall) estimates using, for example, the condition that all algorithms agree were computed by taking the ratio of associated pulsars in 3FGL, which are also predicted to be pulsars by all eight algorithms, to the number of sources associated with pulsars (to the total number of sources predicted to be pulsars among associated sources) in the 3FGL catalog.
The precision and recall estimates for the $\sum_a p_a > 7$ condition are computed analogously.
This condition is, on average, stricter than the all-agree condition, which resulted in larger precision and smaller recall than in the all-agree case.
In particular, all unassociated 3FGL pulsar candidates, which satisfy the $\sum_a p_a > 7$ condition, are associated with pulsars in the 4FGL-DR2.

We note that the precision and recall calculated with the all-agree condition from the comparison of 3FGL and 4FGL-DR2 catalogs are smaller than the precision and recall determined using the test samples and the all-agree and $\sum_a p_a > 7$ conditions (Table~\ref{tab:3FGL_vs_4FGL_2class}).
In order to understand the reason for the worse performance of classification in comparison of 3FGL predictions with the 4FGL-DR2 associations relative to the expectations, we plot in Fig.~\ref{fig:3FGL_vs_4FGL_classes} the 303 sources unassociated in 3FGL but with pulsar or AGN associations in 4FGL-DR2.
The class at the beginning of the label name in Fig.~\ref{fig:3FGL_vs_4FGL_classes} corresponds to the association in the 4FGL-DR2, while the second half of the labels corresponds to classification of unassociated sources in 3FGL. For example, ``PSRs classified only as PSRs'' shows sources that have a pulsar association in 4FGL-DR2 and all eight methods classified the corresponding unassociated sources in 3FGL as a pulsar. ``PSRs classified as either PSRs or AGNs'' labels sources with pulsar associations in 4FGL-DR2 but the corresponding unassociated sources in 3FGL have both pulsar and AGN classifications by different ML methods.
We notice that misclassified or partially misclassified sources in Fig. \ref{fig:3FGL_vs_4FGL_classes} typically happen on the boundary between the two classes or even inside the opposite class.
Many of these sources also have flags in the 3FGL catalog, such as a potential problem with the background diffuse emission model in the location of the source, which can lead to a poor reconstruction of the source spectrum and, consequently, misclassification of the source.
Thus, we find that the precision and recall calculated from the comparison of the 3FGL and 4FGL-DR2 catalogs give more realistic estimates
of the true precision and recall compared to the estimates determined from the test samples, since the former
take into account possible errors in the reconstruction of source parameters (such as the spectral parameters) as well as the differences in distributions of associated and unassociated sources.

%

\begin{table}[!h]

    \caption{
    Comparison of classes predicted for unassociated sources in the 3FGL catalog
    with associations in the 4FGL-DR2 catalog.}
    \label{tab:3FGL_vs_4FGL_2class}

\centering
\resizebox{0.47\textwidth}{!}{
    \tiny
  
 \renewcommand{\tabcolsep}{0.3mm}
\renewcommand{\arraystretch}{1.5}

    \begin{tabular}{l c c c c c c}
    \hline
    \hline
    4FGL-DR2 class & \multicolumn{3}{c}{3FGL prediction} \ &  Recall\ \ & Recall & Recall\\
      &\ AGN &\ PSR &\ MIXED & (4FGL assoc) & (all agree) & ($\sum_a p_a > 7$) \\
    \hline
    AGN & 223 & 5 &  30 & 0.86 & 0.94 & 0.92 \\ 
    PSR & 10 & 23 &  12  & 0.51 & 0.67 & 0.61 \\ 
    OTHER & 8 & 6 & 23  & -- & -- & --\\ 
    \hline
    Precision (4FGL assoc) & 0.93 & 0.68 & --  \\
    Precision (all agree) & 0.97 & 0.78 & --  \\
    Precision ($\sum_a p_a > 7$) & 0.98 & 0.82 & --  \\
    \hline
    \end{tabular}}
    \tablefoot{
    The classes are predicted using the two-class classification.
    The precision and recall are calculated using the all-algorithms-agree condition and the 4FGL-DR2 associations (4FGL assoc), or the class probabilities of associated 3FGL sources derived from test samples using either the all-algorithms-agree condition (all agree) or the $\sum_a p_a > 7$ condition ($\sum_a p_a > 7$).
    The AGN and pulsar sources are also represented in Fig. \ref{fig:3FGL_vs_4FGL_classes}.}
\end{table}

\begin{table}[!h]

    \caption{Expected numbers of sources among the unassociated 3FGL and 4FGL-DR2 sources.}
    \label{tab:prediction_2and3class}

\centering
\resizebox{0.4\textwidth}{!}{
    \tiny
    \renewcommand{\tabcolsep}{0.3mm}
\renewcommand{\arraystretch}{1.5}

    \begin{tabular}{l  l c c c c}
    \hline
    \hline
    Catalog & Classification &\ AGN &\ PSR &\ OTHER &\ MIXED \\
    \hline
    3FGL & two-class & 599 & 111 & --  &  300 \\
     & {two-class corr} & { 580.0} & { 97.0} & { 56.4}  &  { 276.5} \\
     & three-class & 587 & 53 & 69  &  301 \\
    \hline
    4FGL-DR2 \ & two-class & 878 & 162 & --  &  627 \\
     & {two-class corr} & {826.2} & { 134.5} & { 140.4}  &  {565.9} \\
     & three-class & 739 & 64 & 274  &  590 \\
    \hline
     
    \end{tabular}}
    \tablefoot{Expected numbers of AGNs, pulsars, and OTHER sources as well as sources with mixed classifications
    are derived with the two-class (Sect. \ref{sec:prob_cats}) 
    and three-class (Sect. \ref{sec:3class}) classification.
    The ``two-class corr'' row shows the correction of the two-class classification prediction due to the presence of OTHER sources among 
    the unassociated ones (see Sect. \ref{sec:3FGLprediction1} for details).}
\end{table}

We summarize the results of classification of unassociated 3FGL sources with the two-class classification 
in Table \ref{tab:prediction_2and3class} in the 3FGL two-class row.
The AGN column shows the number of unassociated sources predicted to be AGNs using the all-algorithms-agree condition.
Similarly the PSR column shows the number of unassociated sources where all the algorithms predict the source to more likely be a pulsar.
The MIXED column shows the number of sources with mixed classification (i.e., some algorithms predict that the source is more likely an AGN while the other algorithms predict that it is more likely a pulsar).
We also add the OTHER column in order to compare the results with the three-class classification in Sect. \ref{sec:3class}.
Since there is no OTHER class in the two-class classification, the corresponding entry is empty.
Out of 1010 unassociated sources in 3FGL, 111 are classified as pulsars by all eight methods, 599 are classified as AGNs, and 300 have mixed classifications.

In the two-class corr row of Table \ref{tab:prediction_2and3class}
we show a possible correction of the number of pulsars and AGNs due to the presence of OTHER sources.
Here we assumed that the fraction of AGN-like and pulsar-like sources among the OTHER sources is the same for associated and for unassociated sources.
In particular, if we denote by $N_{\rm AGN}$ the number of unassociated sources with AGN-like probabilistic classification,
by $N_{\rm AGN}^{\rm ass\,OTHER}$ the number of sources with AGN-like classification among associated OTHER sources,
by $N_{\rm ass}$ ($N_{\rm unass}$) the total number of associated (unassociated) sources, then
the number of AGN-like sources among the unassociated ones corrected for the presence of OTHER sources can be estimated as
\bea
\lb{eq:other_correction}
N_{\rm AGN}^{\rm corr} = N_{\rm AGN} - N_{\rm AGN}^{\rm ass\,OTHER} \,\frac{N_{\rm unass}}{N_{\rm ass}}.
\eea
Analogous corrections are applied for the number of unassociated sources with pulsar and with mixed classifications.
If we denote by $N^{\rm ass\,OTHER}$ the total number of associated OTHER sources, then the estimated number of 
OTHER sources among unassociated ones is
\bea
\lb{eq:2class_other}
N^{\rm unass}_{\rm OTHER} = N^{\rm ass\,OTHER} \,\frac{N_{\rm unass}}{N_{\rm ass}}.
\eea
We show this estimate in the OTHER column in the two-class corr row.
We note that since 
$N_{\rm AGN}^{\rm ass\,OTHER} + N_{\rm PSR}^{\rm ass\,OTHER} + N_{\rm MIXED}^{\rm ass\,OTHER} = N^{\rm ass\,OTHER}$,
this estimate is consistent with corrections in Eq. (\ref{eq:other_correction}) for sources classified as AGNs, pulsars, or with mixed classification.
The expected numbers of sources for the 4FGL-DR2 catalog and in the three-class case are calculated in Sects. \ref{sec:4FGLprediction}
and \ref{sec:3class}, respectively.

\subsection{Probabilistic classification of sources in the 4FGL-DR2 catalog}
\lb{sec:4FGLprediction}

In this section we construct a probabilistic classification of sources in the 4FGL-DR2 catalog. The 4FGL-DR2 catalog \citep{2020arXiv200511208B} 
is based on 10 years of \Fermi-LAT data \citep[compared to 8 years of data in the 4FGL catalog,][]{2020ApJS..247...33A}.
It contains 5788 sources, which is 723 sources more than in the 4FGL catalog (all sources in 4FGL are kept in 4FGL-DR2 even if they fall
below the detection threshold with 10 years of data). 
In the 4FGL-DR2 catalog,
3503 sources are associated with AGNs,
271 sources are associated with pulsars,
1658 sources are unassociated (we only look at CLASS1 column in the catalog), 
and the rest 346 sources are OTHER sources, such as pulsar wind nebulae (PWNe), 
supernova remnants (SNRs), and so on.
There are 14 sources in 4FGL-DR2 with missing or unphysical values: four AGNs, one PWN (Crab), and nine unassociated sources.
As in the previous section, we used sources associated with either AGNs or pulsars for training,
which have no missing or unphysical values.
The unphysical and missing values were replaced according to the procedure described at the beginning of Sect. \ref{sec:training}.
We calculated the classification probabilities of AGN and pulsar classes for both the associated and the unassociated sources.

The 4FGL-DR2 catalog has a higher number of features, especially due to the difference in modeling of the spectra compared with the 3FGL catalog. 
We selected 28 of these features plus six hardness ratios HR12, ..., HR67 (the 4FGL-DR2 catalog has seven energy bins)
and looked for correlations among them. 

If any feature is correlated or anticorrelated with a Pearson index of $\pm$0.75 or higher with another feature, then only one of these features was kept. 
The resulting 16 features are:
GLON, GLAT, ln(Pivot\_Energy), ln(Energy\_Flux100), ln(Unc\_Energy\_Flux100), LP\_Index, Unc\_LP\_Index, LP\_beta, LP\_SigCurv, ln(Variability\_Index), and the six hardness ratios.

For the classification of 4FGL-DR2 sources, we confirmed that the parameters used in the 3FGL classification provided an optimal performance also for the 4FGL-DR2 catalog, except for NN, which required more neurons in the hidden layer in the 4FGL-DR2 case.
Therefore, we used the same meta-parameters for the four algorithms as in the construction of the probabilistic catalog based on 3FGL, except for NN where we increased the number of neurons in the hidden layer to 16. Similar to the construction of the 3FGL probabilistic catalog, we used both unweighted training samples and oversampling (i.e., we have eight classification methods).
We retrained the algorithms using the 16 features for the 4FGL-DR2 sources.
The corresponding accuracies are reported in Table \ref{tab:selected_algs2}.

\begin{table}[!h]
    \caption{Testing accuracy of the algorithms for the associated 4FGL-DR2 sources. }
    \label{tab:selected_algs2}

\centering
    \tiny
    \renewcommand{\tabcolsep}{0.4mm}
\renewcommand{\arraystretch}{1.6}

    \begin{tabular}{ c c c c }
    \hline
    \hline
    Algorithm&Parameters &  Testing&Std. Dev.\\
    & & Accuracy\ &  \\
    \hline
    RF& 50 trees, max depth 6  &97.87 & 0.36\\
    RF\_O   &&97.56&0.39 \\
    \hline
    BDT & 100 trees, max depth 2    &   97.63 &0.39\\
    BDT\_O&&97.72&0.38\\
    \hline
    NN & 300 epochs, 16 neurons, LBFGS  & 97.41 & 0.47\\
    NN\_O&&95.48&0.66\\
    \hline
    LR & 200 iterations, LBFGS solver & 97.80&0.38\\
    LR\_O&&96.03&0.53\\
    \hline
     
    \end{tabular}
    \tablefoot{``\_O'' denotes training with oversampling.}
\end{table}

\begin{table}[!h]
    \caption{Expected precision and recall for the classification of 4FGL-DR2 sources.}
    \label{tab:prec_recall_4FGL}

\centering
    \tiny
  
 \renewcommand{\tabcolsep}{0.3mm}
\renewcommand{\arraystretch}{1.5}

    \begin{tabular}{l c c c c}
    \hline
    \hline
    & classification &\ \  AGN &\ \   PSR &\ \   OTHER \\
    \hline
    two-class precision & all     agree            & 0.96 &  0.71 & --  \\ 
                                & $\sum_a p_a > 7$ & 0.97 &  0.78  & -- \\
    \hline
    two-class recall       & all     agree            & 0.95 &  0.70 & --  \\ 
                                & $\sum_a p_a > 7$ & 0.94 &  0.64  & -- \\
    \hline
    three-class precision & all     agree            & 0.98 & 0.89  & 0.80  \\ 
                                & $\sum_a p_a > 7$ & 0.99 &  0.94  & 0.85 \\
    \hline
    three-class recall       & all     agree            & 0.94 &  0.62 & 0.38  \\ 
                                & $\sum_a p_a > 7$ & 0.87 &  0.30  & 0.06 \\
    \hline
    \end{tabular}
    \tablefoot{The expected precision and recall are calculated
    for two classification conditions: all algorithms agree (all agree) and the sum of probabilities for the eight algorithms 
    is larger than 7 ($\sum_a p_a > 7$) (for details on the two- and three-class classification, see Sects. \ref{sec:4FGLprediction} and 
    \ref{sec:3class}, respectively). }
\end{table}

Analogously to the 3FGL catalog, we used the agreement among the algorithms for the probabilistic classification of sources.
We report the corresponding precision and recall in Table \ref{tab:prec_recall_4FGL}, where we also show precision and recall
for the $\sum_a p_a > 7$ condition and for the three-class classification described in Sect. \ref{sec:3class}.
Using the agreement among the algorithms condition,
we calculated the expected numbers of pulsars and AGNs among the 1658 unassociated sources in 4FGL-DR2 without missing values 
(see 4FGL-DR2 part of Table \ref{tab:prediction_2and3class}).
The definition of rows is the same as in the 3FGL catalog two-class classification in Sect. \ref{sec:3FGLprediction1}.

Finally, we looked at sources that are unassociated in both 3FGL and 4FGL-DR2 (using ``ASSOC\_FGL'' as an identifier for 3FGL sources). Out of 303 such sources%
\footnote{The 303 unassociated 4FGL-DR2 sources correspond to 302 unassociated 3FGL sources because there are two 4FGL-DR2 sources that are associated with one 3FGL source.},
40 sources are predicted to be pulsars using 3FGL features and 75 sources are predicted to be pulsars using 4FGL-DR2 features. This leads to 29 sources which are predicted by all eight methods to be pulsars for features taken from both the 3FGL and 4FGL-DR2 catalogs. 
For convenience, we save these 29 pulsar candidates as a separate file (``3FGL\_4FGL-DR2\_Candidates\_PSR'' in the supplementary online materials; \citealt{SOM_material}). Out of these 29 sources classified as pulsars, 4 sources have counterparts in Parkes survey \citep{Camilo2015} within 2 arc minutes (see Table \ref{tab:parkes}). The data for the Parkes association candidates is also added to the ``3FGL\_4FGL-DR2\_Candidates\_PSR'' file.

\pgfplotstableread[col sep=comma]{tables/3fgl_unassoc_predictions_matches_with_Parkes_2015_1.csv}\loadedtable
\begin{table}[h]
\caption{
Connection of unassociated 3FGL and 4FGL-DR2 sources classified as pulsars with the Parkes pulsar survey.}
\label{tab:parkes}
\centering
\pgfplotstabletypeset[columns={Source_Name_4FGL,GLON_4FGL,GLAT_4FGL,Separation_Parkes},
column type=l,
string type,
every head row/.style={before row={\hline \hline},after row=\hline,},
every last row/.style={after row=\hline},
columns/Source_Name_4FGL/.style={column name=Source\_Name\_4FGL},
columns/GLON_4FGL/.style={column name=GLON,numeric type,fixed,precision=1},
columns/GLAT_4FGL/.style={column name=GLAT,numeric type,fixed,precision=1},
columns/Separation_Parkes/.style={column name=Sep (arcsec),numeric type,fixed,precision=1}
]\loadedtable
\vspace{2mm}
\tablefoot{
Galactic longitude (GLON) and latitude (GLAT) coordinates are taken from the 4FGL-DR2 catalog. 
The corresponding separation in arcseconds from the Parkes pulsars \citep{Camilo2015} 
is shown in the ``Sep (arcsec)'' column.}
\end{table}

\section{Three-class classification}
\lb{sec:3class}

One of the caveats of the analysis with two classes is that there are associated sources which do not belong to the AGN or pulsar classes. 
These sources have the following labels: unk, spp, glc, snr, gal, sbg, GAL, sfr, bin, SNR, HMB, LMB, css, PWN, pwn, hmb, SFR, BIN, lmb, and NOV.
We collect all associated sources, which do not belong to AGN and pulsar classes, into a new class, which we label as OTHER.
Since in two-class classification we train algorithm to classify sources only into AGN and pulsar classes, OTHER sources are also classified as either AGN or pulsar.
This introduces a bias in the estimates of the number of AGNs and pulsars among unassociated sources.
One possibility to correct this bias is to assume that the fraction of OTHER sources among associated and unassociated sources are the same (Eq. (\ref{eq:other_correction})).
This correction can be applied for the total number of sources or for the number of sources in some window of parameters,
for example, in a flux bin or in a range of latitudes and longitudes.
This is a straightforward calculation but it has some limitations. 
In particular, it implicitly assigns one probability for each AGN and one probability for each pulsar of belonging to the OTHER class inside a selected range of parameters.
For a small range of parameters the variance of this estimate can be very large due to a small number of associated OTHER sources in this parameter range.
As we show in Sect. \ref{sec:pop_studies}, this correction depends on the choice of the variable used for binning; for example,
the overall correction with latitude bins is not equal to the correction with longitude bins.

In this section we discuss the construction of probabilistic catalogs with multi-class classification (three-class classification in our case).
We start with the construction of the probabilistic catalog based on 3FGL by adding the class OTHER, which includes all associated sources without AGN or pulsar associations: there are 113 such sources in 3FGL (108 OTHER sources are without missing or unphysical values).
We used the same 11 features as in the two-class classification: the only difference is that we used cos(GLON) instead of GLON.
The reason is that the LR and NN methods have a significantly worse performance than RF and BDT methods when GLON is used,
but, as we show below, all four methods have comparable accuracy if cos(GLON) is used.
This may be due to a discontinuity in the GLON variable. 
We performed optimization of the meta-parameters for the four ML algorithms with the three classes.
In the calculation of accuracy we determined the probabilistic class as the class with the maximum probability among the three classes,
in some cases the maximal probability can be less than 0.5, but it is always above one-third.

\begin{figure}[h]
\centering
\includegraphics[width=0.5\textwidth]{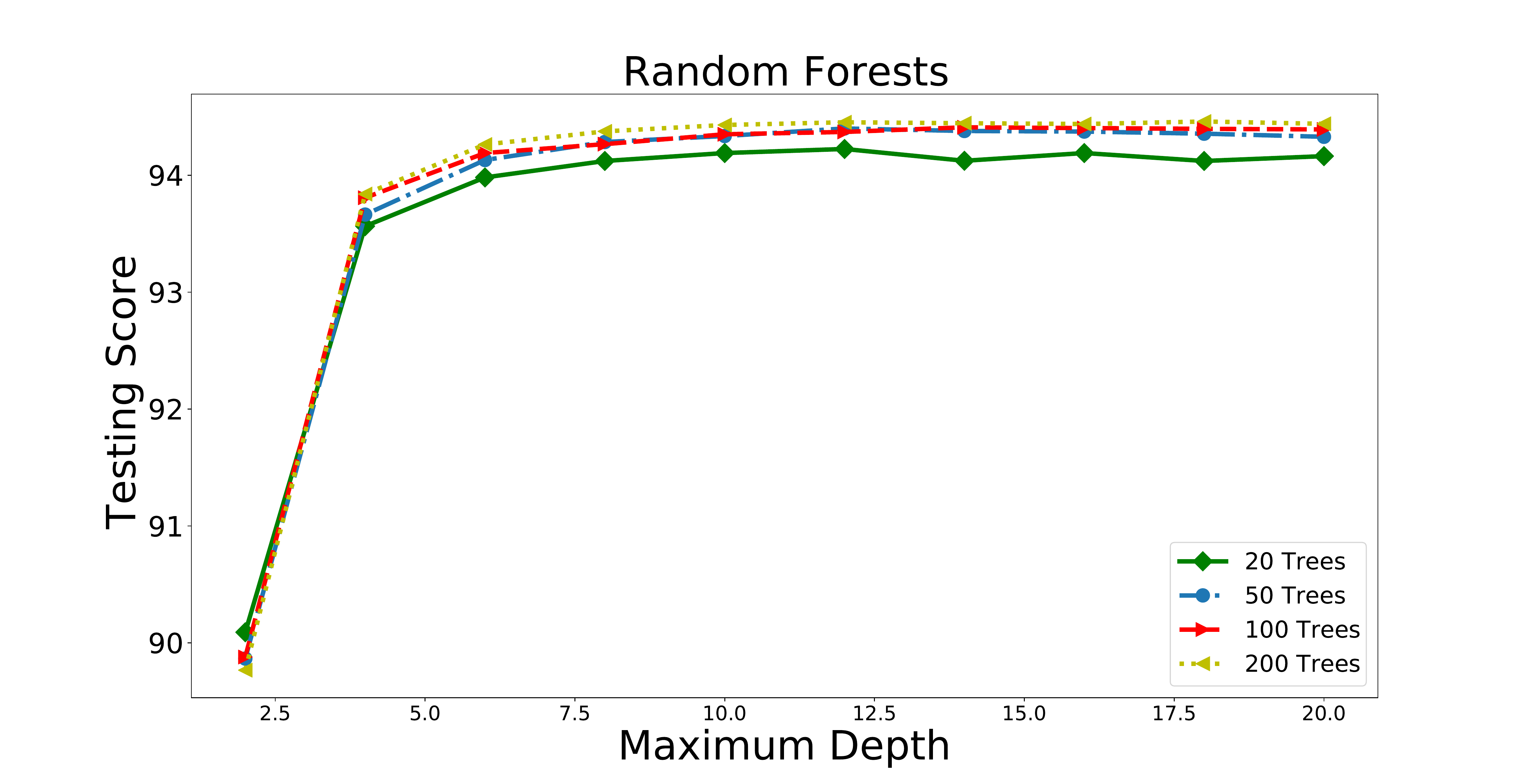}\\
\includegraphics[width=0.5\textwidth]{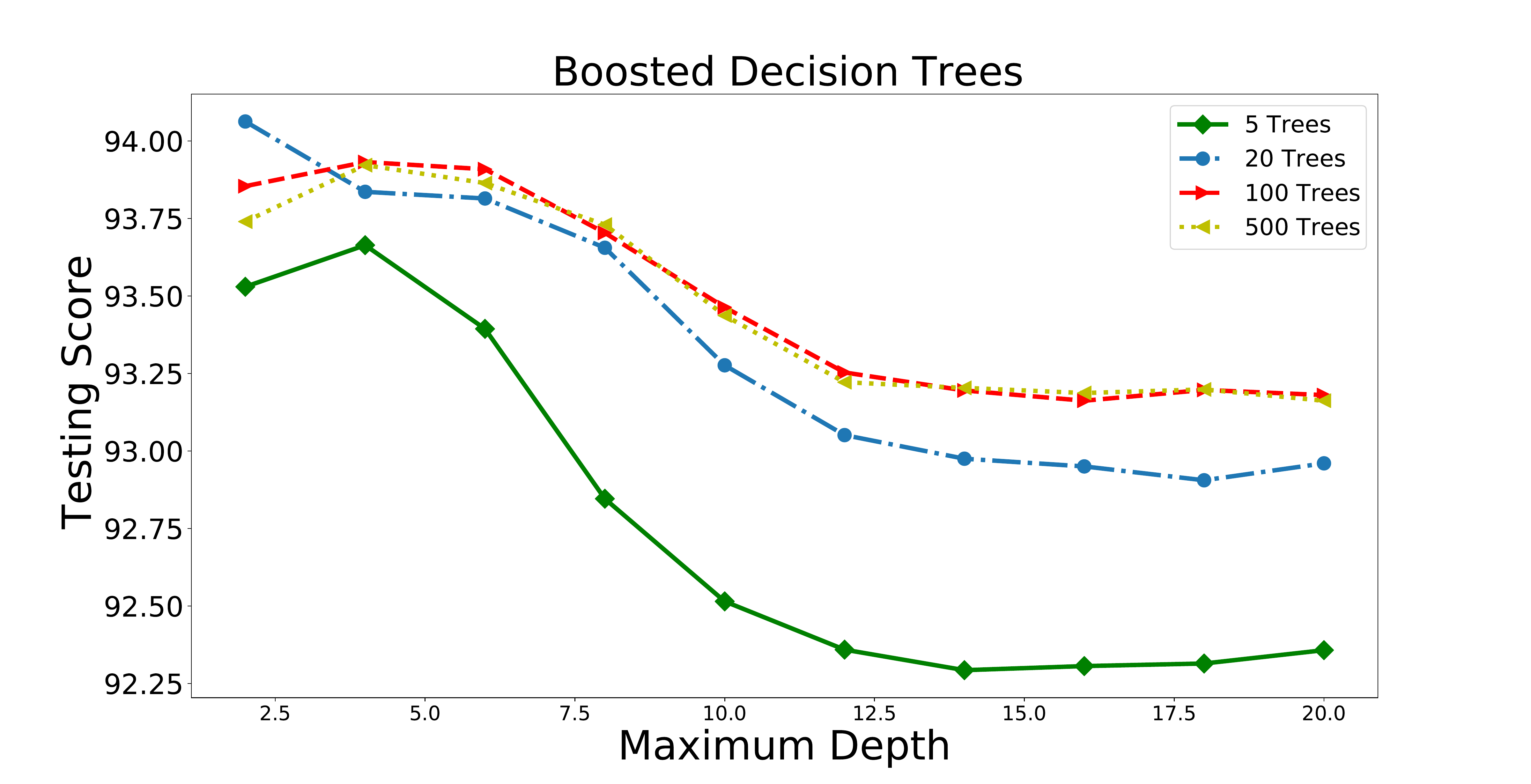}
\includegraphics[width=0.5\textwidth]{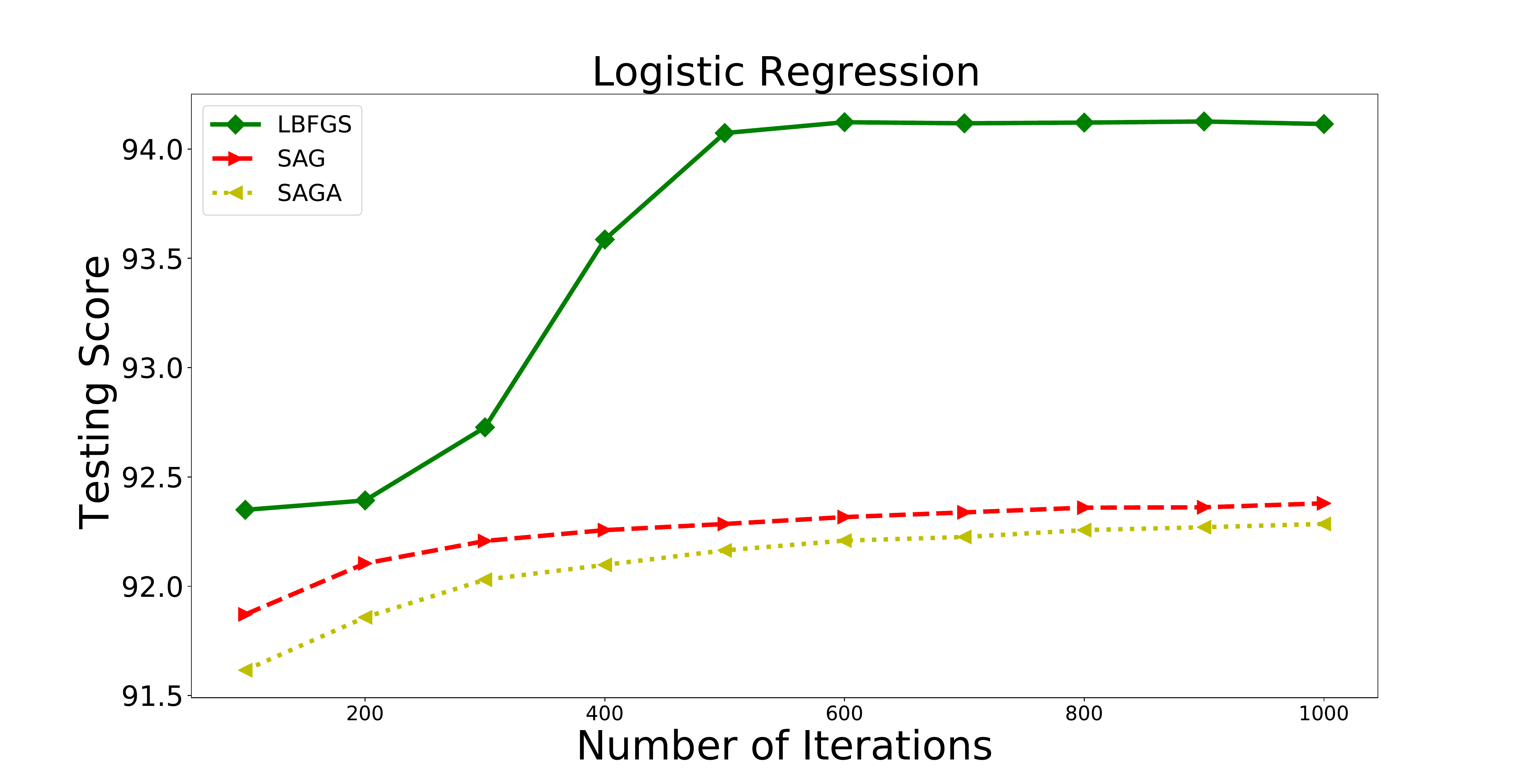}
\caption{Accuracy of the three-class classification with the RF, BDT, and LR  methods. LR does not have a LIBLINEAR solver here, since LIBLINEAR 
does not support multi-class loss.
}
\label{fig:tree_multi}
\end{figure}

\begin{figure}[h]
\centering
\includegraphics[width=0.5\textwidth]{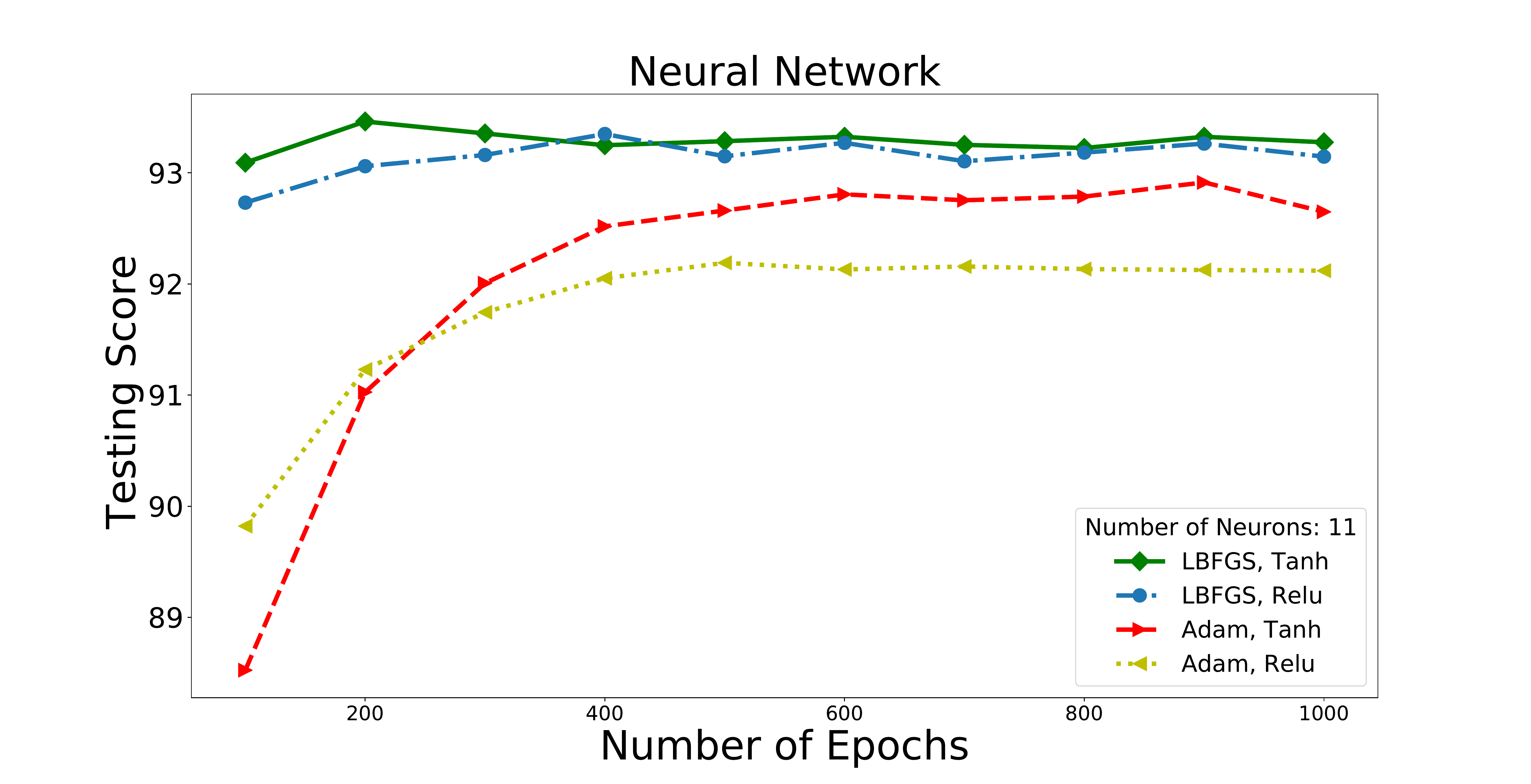}\\
\includegraphics[width=0.5\textwidth]{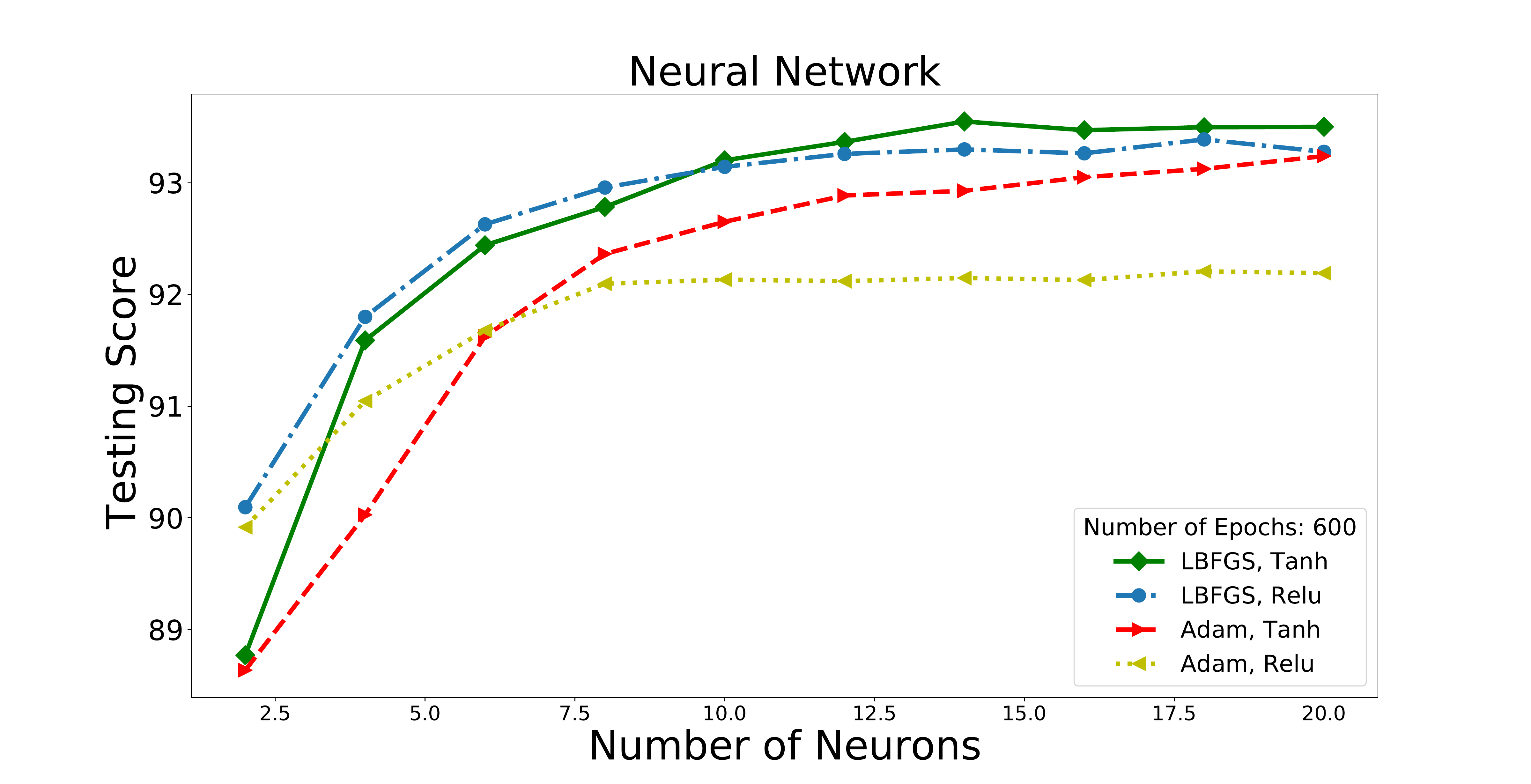}
\caption{Accuracy of the NN classification as a function of the number of epochs and of the number of neurons
 for the three-class classification. 
 }
\label{fig:nets_multi}
\end{figure}

The dependence of accuracy on meta-parameters of the algorithms is shown in Figs. \ref{fig:tree_multi} and \ref{fig:nets_multi}.
We see that for the tree-based algorithms, the optimal parameters are similar to the two-class classification: 50 trees with depth 6 for RF and 100 trees with depth 2 for BDT 
provide close to optimal performance at a minimal cost in complexity (depth of the trees).
The main difference for NN and LR algorithms is that more steps are needed for convergence, especially in the case of oversampling. 
In the following we used 600 epochs for NN and 500 iterations for LR instead of 300 epochs and 200 iterations, respectively, in the two-class case.
For NN, the accuracy stops increasing above about ten neurons in the hidden layer (in the following we used 11 neurons for classification: the same as in the two-class case).
For oversampling, we used the oversampling factors $\sqrt{\frac{\text{\# AGN}}{\text{\# PSR}}}$ and $\sqrt{\frac{\text{\# AGN}}{\text{\# OTHER}}}$ for the pulsar and OTHER classes, respectively (compared to the $\frac{\text{\# AGN}}{\text{\# PSR}}$ oversampling factor in the two-class case).
The reason for the smaller oversampling factors is to avoid overweighting the two relatively small pulsar and OTHER classes.

\begin{figure}[h]
\centering
\includegraphics[width=0.46\textwidth]{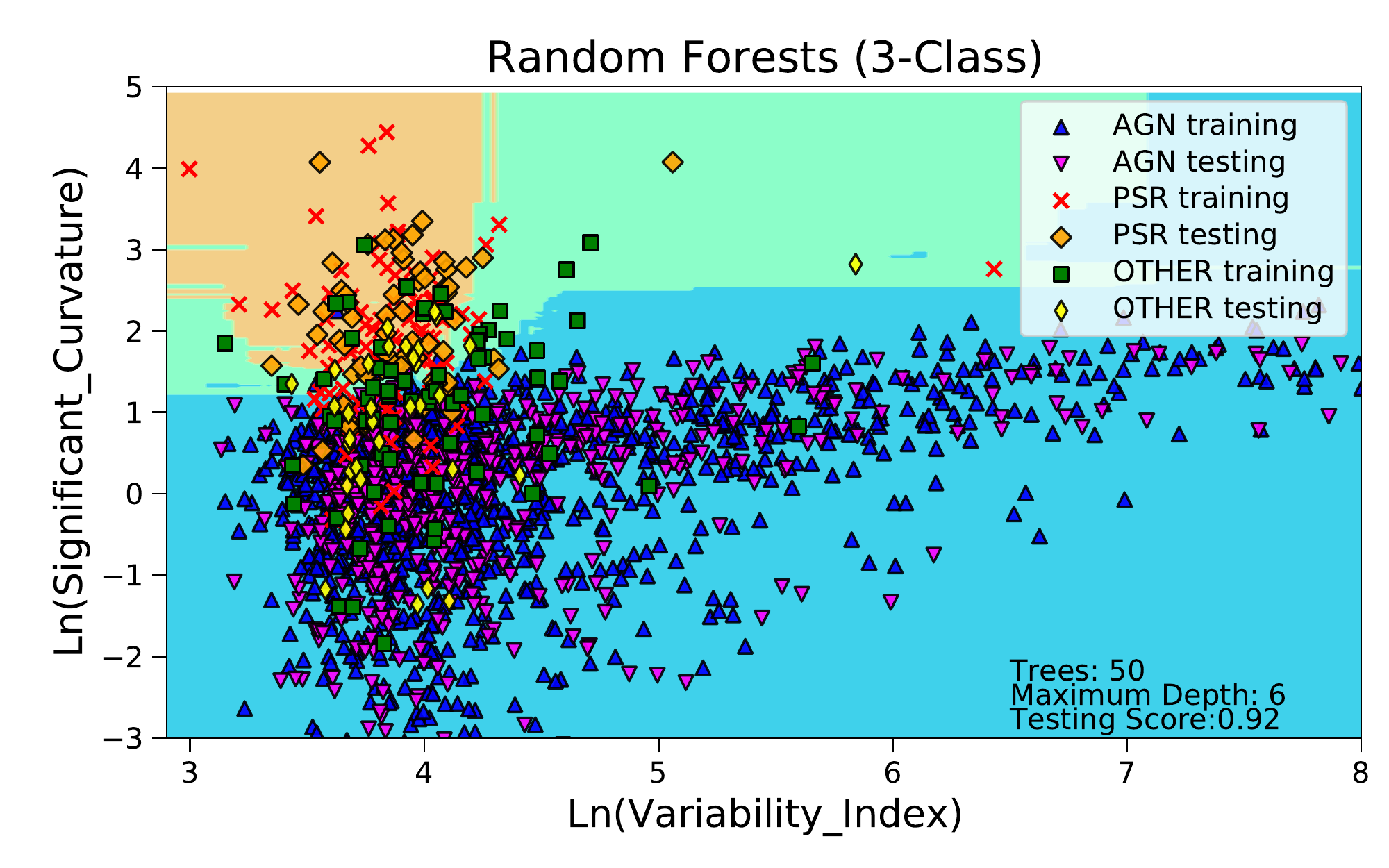}
\caption{Classification domains for RF in the three-class classification.
The sources in the blue, green, and brown areas are attributed to AGN, OTHER, and pulsar classes, respectively.
}
\label{fig:RF_domains_3class}
\end{figure}

We show an example of domains in the three-class case in Fig. \ref{fig:RF_domains_3class}.
A class domain is determined by the class with the largest probability.
Since in the three-class case there are two independent probabilities, which are difficult to show with a single color bar,
we present only the domains represented by three different colors: brown for pulsar, green for OTHER, and blue for AGN classes.
The corresponding training and testing data are shown by red crosses and brown rotated squares for pulsar, by green squares and yellow diamonds for OTHER,
and by blue and purple triangles for AGN.
The classification domains were averaged over 100 realizations of splitting the data into training and testing samples.
One of these splittings is shown in the figure for illustration.

The accuracies of our chosen models for classification of the 3FGL sources are presented in Table \ref{tab:selected_algs_multi}.
As in the two-class case, the accuracies were averaged over 1000 realizations of splitting the data into training and testing samples.
We notice that accuracies presented in Table \ref{tab:selected_algs} are calculated relative to AGN and pulsar classes only. If we take into account that all OTHER sources
are misclassified in this case, then the testing accuracy is reduced by about 5\% (the fraction of OTHER sources among associated sources in 3FGL),
while the accuracy of comparison with 4FGL-DR2 is reduced by about 10\% (there are 37 unassociated sources in 3FGL with OTHER class associations in 4FGL-DR2,
while there are in total 340 unassociated sources in 3FGL with associations in 4FGL-DR2).
Thus the testing accuracy of 93-94\% in Table \ref{tab:selected_algs_multi} provides at least a 1-2\% improvement over the accuracy in Table \ref{tab:selected_algs},
after taking into account the misclassification of OTHER sources in the two-class case.
A similar improvement is seen for the accuracy of classification of unassociated sources in 3FGL with corresponding 4FGL-DR2 associations.

Similarly to the two-class classification, we used the condition that all algorithms agree to determine the candidate probabilistic classes of sources.
In the end of this section we also use a stricter condition that the sum of probabilities is larger than 7 in order to determine lists of most likely pulsar and OTHER source candidates.
A comparison of predicted classes (using the all-algorithms-agree condition) for the 3FGL unassociated sources in the three-class classification case with the classes of the corresponding associated sources in the 4FGL-DR2 catalog is presented in Table~\ref{tab:3FGL_vs_4FGL_3class}.
It is interesting to note that there are fewer sources with mixed classification in this case relative to the two-class classification in 
Table~\ref{tab:3FGL_vs_4FGL_2class}.
Also, the number of correct predictions in the three-class case is 263 (out of 340 sources), while in the two-class case there are 246 correct predictions.
We also show in this table the expected precision and recall for the all-algorithms-agree and $\sum_a p_a > 7$
conditions.
We note that, despite adding an additional class, the precision of the AGN and pulsar classifications in 3FGL
is higher in the three-class case (Table~\ref{tab:3FGL_vs_4FGL_3class}) compared to the two-class case
(Table~\ref{tab:3FGL_vs_4FGL_2class}), but the recall is smaller for pulsars and similar or larger for AGNs in the three-class case.

\begin{table}[!h]
    \caption{Testing accuracy of the selected algorithms for the three-class classification of 3FGL sources.}
    \label{tab:selected_algs_multi}

\centering
\hspace{-0.2cm}
\resizebox{0.47\textwidth}{!}{
    \tiny
  \centering
    \renewcommand{\tabcolsep}{0.4mm}
\renewcommand{\arraystretch}{1.6}

    \begin{tabular}{c c c c c c c}
    \hline
    \hline
    Algorithm&Parameters &  Testing\ &Std. Dev.& Comparison with \\
    & & Accuracy\ & & 4FGL-DR2 Accuracy \\
    \hline
    RF & 50 trees, max depth 6  & 93.96 & 0.85 & 85.00 \\
    RF\_O &   & 94.38 & 0.76 & 85.00 \\ 
    \hline
    BDT & 100 trees, max depth 2    &   93.72 & 0.83 & 83.24 \\
    BDT\_O &     &   93.83 & 0.80 & 85.29 \\
    \hline
    NN & 600 epochs, 11 neurons, LBFGS & 93.17 & 1.05 & 83.53 \\
    NN\_O &&  92.51 & 1.34 & 81.76 \\
    \hline
    LR & 500 iterations, LBFGS solver & 93.93 & 0.88 & 83.24 \\
    LR\_O &   & 93.01 & 0.96 & 83.24 \\
    \hline
    \end{tabular}}
    \tablefoot{
    Comparison with associations in the 4FGL-DR2 catalog
    is presented in the last column.
    ``\_O'' denotes training with oversampling.}
\end{table}

\begin{table}[!h]
    \caption{
    Comparison of classes predicted for unassociated sources in the 3FGL catalog using three-class classification
    with associations in the 4FGL-DR2 catalog.}
    \label{tab:3FGL_vs_4FGL_3class}

\centering
\resizebox{0.45\textwidth}{!}{
    \tiny
 \renewcommand{\tabcolsep}{0.3mm}
\renewcommand{\arraystretch}{1.5}

    \begin{tabular}{l c c c c c c c}
    \hline
    \hline
    4FGL-DR2 class & \multicolumn{4}{c}{3FGL prediction} & Recall\ \ & Recall & Recall \\
      &\ AGN &\ PSR &\ OTHER &\ MIXED & (4FGL assoc) & (all agree) & ($\sum_a p_a > 7$)\\
    \hline
    AGN & 238 & 2 &  1 & 17 & 0.92 & 0.97 & 0.91\\ 
    PSR & 12 & 17 &  0 & 16 & 0.38 & 0.58 & 0.28 \\ 
    OTHER & 6 & 5 & 8 & 18 & 0.22 & 0.29 & --\\ 
    \hline
    Precision (4FGL assoc) & 0.93 & 0.71 & 0.89  \\
    Precision (all agree) & 0.99 & 0.89 & 0.75  \\
    Precision ($\sum_a p_a > 7$) & 0.99 & 0.98 & --  \\
    \hline
    \end{tabular}}
    \tablefoot{For the definition of precision and recall labels, see Table \ref{tab:3FGL_vs_4FGL_2class}.}
\end{table}

The three-class classification of 4FGL-DR2 sources was performed analogously to the three-class classification of the 3FGL sources.
The differences between the 4FGL-DR2 and 3FGL three-class classifications are similar to the differences in the two-class classification of 3FGL and 4FGL-DR2 sources:
we used 16 features instead of 11 features in the 3FGL case (the features are the same as in the two-class classification of 4FGL-DR2 sources in Sect. \ref{sec:4FGLprediction} but with GLON replaced by cos(GLON))
and we have 16 neurons in the hidden layer of the NN method. Furthermore, for LR we used 1000 iterations as it gives better performance for oversampled cases.
The corresponding accuracies are reported in Table \ref{tab:selected_algs_4fgl_multi}.
In comparing the accuracies with the two-class classification in Table \ref{tab:selected_algs2}, 
one has to take into account that there are 346 OTHER sources among 4116 associated sources in 4FGL-DR2, which is about 8.4\%.
Since all OTHER sources are ``misclassified'' by the two-class classification, the three-class classification provides an improvement of about 2-4\% compared to the two-class classification.

The expected precision and recall for the classification of sources using agreement among all algorithms and the
$\sum_a p_a > 7$ conditions are presented in Table \ref{tab:prec_recall_4FGL}.
We note that for the 4FGL-DR2 sources the
precision in the three-class case is better than the precision in the two-class case, while the recall is better in the two-class case.
Another comparison of the performance of the two- and three-class classifications is provided in Appendix \ref{sec:reliability} where we present the reliability diagrams.
The conclusion is that the reliability diagrams for the three-class are similar to the reliability diagrams for the two-class case, which includes only AGN and pulsar sources, with an additional advantage of the three-class case that in the two-class classification one needs to account for the contribution of the OTHER sources among the unassociated ones, while in the three-class case the contribution of the OTHER sources is included in the model.

\begin{table}[!h]
    \caption{Testing accuracy of the four selected algorithms for the three-class classification of 4FGL-DR2 sources.}
    \label{tab:selected_algs_4fgl_multi}

\centering
\hspace{-0.2cm}
\resizebox{0.47\textwidth}{!}{
    \tiny
    \renewcommand{\tabcolsep}{0.4mm}
\renewcommand{\arraystretch}{1.6}
    \begin{tabular}{c c c c c c}
    \hline
    \hline
    Algorithm&Parameters &  Testing&Std. Dev.\\
    & & Accuracy\ &  \\
    \hline
    RF & 50 trees, max depth 6  &92.91&0.66\\
    RF\_O &   &92.83&0.63 \\
    \hline
    BDT & 100 trees, max depth 2    &   92.51&0.67 \\
    BDT\_O &     &   92.27&0.67 \\
    \hline
    NN & 600 epochs, 16 neurons, LBFGS & 91.86&0.72\\
    NN\_O &  & 90.26&0.83\\
    \hline
    LR & 1000 iterations, LBFGS solver & 92.63&0.67 \\
    LR\_O &  &92.22&0.69\\
    \hline
     
    \end{tabular}}
    \tablefoot{``\_O'' denotes training with oversampling.}
\end{table}

The numbers of unassociated sources classified by all 8 methods as AGNs, pulsars, and OTHER sources for the 3FGL and 4FGL-DR2 catalogs are presented in Table \ref{tab:prediction_2and3class} in the three-class rows.
For each algorithm the most probable class of the source is determined by the class with the largest probability.
The MIXED column shows the number of sources with different classification results for different algorithms.

Classification of \Fermi-LAT 4FGL sources into three classes has been considered earlier by, for example, \cite{2021RAA....21...15Z}, 
who primarily use a two-step classification procedure, where in the first step AGNs are separated from the rest of sources and in the second step the remaining sources are split into pulsars and OTHER sources.
\cite{2021RAA....21...15Z} have also tested a simultaneous classification of sources into three classes (AGN, pulsars, and OTHER),
but the results are inconsistent for the two ML algorithms used by \cite{2021RAA....21...15Z} (RF and NN).
In particular, the number of OTHER sources predicted by NN is zero.
In our case, the predictions of various algorithms are relatively consistent with each other.
For example, in the 3FGL (4FGL-DR2) catalog all 8 methods classify 69 (271) unassociated sources as OTHER.
Also, 8 out of 37 unassociated 3FGL sources, which are associated with OTHER sources in 4FGL-DR2, are classified by all eight algorithms as
OTHER (see Table \ref{tab:3FGL_vs_4FGL_3class}).
We also checked that the reliability diagrams for the OTHER class in the three-class classification look reasonable
(see Fig. \ref{fig:rel_OTHER}).

We also used the three-class classification to create lists of most likely pulsar and OTHER sources among the unassociated
sources in the 4FGL-DR2 catalog.
In Sect. \ref{sec:4FGLprediction} we determined a list of 29 pulsar candidates by requiring that 
the unassociated sources are predicted to be pulsars by all 8 methods both in the 3FGL and in the 4FGL-DR2 catalogs.
As one can see from Table \ref{tab:selected_algs_multi} and also from the comparison of Tables \ref{tab:3FGL_vs_4FGL_3class}
and \ref{tab:3FGL_vs_4FGL_2class}, the $\sum_a p_a > 7$ condition provides a better precision than the 
agreement among the algorithms condition.
For this reason, in this section we used the $\sum_a p_a > 7$ condition to create lists of pulsar and OTHER candidates
among the unassociated 4FGL-DR2 sources.
In the 3FGL case, all pulsar candidates satisfying the $\sum_a p_a > 7$ condition are already associated with pulsars in 4FGL-DR2,
while there are no OTHER candidates satisfying the $\sum_a p_a > 7$ condition among the unassociated sources. 

There are 6 unassociated 4FGL-DR2 sources
with the sum of pulsar class probabilities in the three-class classification larger than 7.
The pulsar candidates are shown in Table \ref{tab:psr_candidates_3class}. 
All of these sources are also among the list of 29 pulsar candidates determined in the two-class classification using both 3FGL and 4FGL-DR2 
features in Sect. \ref{sec:4FGLprediction}. 
For convenience, we also add the sums of pulsar-like probabilities for the three-class classification of the 4FGL-DR2 sources
in the ``3FGL\_4FGL-DR2\_Candidates\_PSR'' file.
Three sources (4FGL J1539.4-3323, 4FGL J0933.8-6232,  and 4FGL J2112.5-3043) 
have also possible associations in the Parkes survey (Table \ref{tab:parkes}).

\pgfplotstableread[col sep=comma]{tables/4FGLDR2_PSR.csv}\loadedtable
\begin{table}
\caption{\label{tab:psr_candidates_3class}
Unassociated 4FGL-DR2 sources with the sum of pulsar class probabilities for all eight ML methods in the three-class classification larger than 7. }
\centering
\hspace{-0.1cm}
\tiny
\resizebox{0.47\textwidth}{!}{
\pgfplotstabletypeset[columns={Source_Name_4FGL,GLON,GLAT,PSR_TOTAL_4FGL,PSR_TOTAL_3FGL,Category_Prob},
column type=c,
string type,
every head row/.style={before row={\hline \hline},after row=\hline,},
every last row/.style={after row=\hline},
columns/Source_Name_4FGL/.style={column name=Source\_Name\_4FGL, column type=l},
columns/GLON/.style={column name=GLON,numeric type,fixed,precision=2},
columns/GLAT/.style={column name=GLAT,numeric type,fixed,precision=2},
columns/PSR_TOTAL_4FGL/.style={column name=$P_{\rm tot\ PSR}^{\rm 4FGL-DR2}$, numeric type, fixed, precision=2},
columns/PSR_TOTAL_3FGL/.style={column name=$P_{\rm tot\ PSR}^{\rm 3FGL}$, numeric type, fixed, precision=2},
columns/Category_Prob/.style={column name=3FGL cat.},
]\loadedtable
}
\normalsize
\tablefoot{All of these sources are also unassociated sources in the 3FGL catalog. 
$P_{\rm tot\ PSR}^{\rm 4FGL-DR2}$ ($P_{\rm tot\ PSR}^{\rm 3FGL}$) represents the sum of pulsar class probabilities 
in the three-class classification for the 4FGL-DR2 (3FGL) catalog. The 
``3FGL cat.'' column is the probabilistic category based on the three-class classification for the corresponding 3FGL source.}
\end{table}

\pgfplotstableread[col sep=comma]{tables/4FGLDR2_Candidates_OTHER.csv}\loadedtable
\begin{table}
\caption{
OTHER-class candidates among the unassociated 4FGL-DR2 sources.
}
\label{tab:other_candidates_3class}

\centering
\hspace{-0.5cm}
\tiny
\resizebox{0.47\textwidth}{!}{
\pgfplotstabletypeset[columns={Source_Name_4FGL,GLON,GLAT,OTHER_TOTAL,Category_Prob_3FGL},
column type=c,
string type,
every head row/.style={before row={\hline \hline},after row=\hline,},
every last row/.style={after row=\hline},
columns/Source_Name_4FGL/.style={column name=Source\_Name\_4FGL, column type=l},
columns/GLON/.style={column name=GLON,numeric type,fixed,precision=2},
columns/GLAT/.style={column name=GLAT,numeric type,fixed,precision=2},
columns/OTHER_TOTAL/.style={column name=$P_{\rm tot\ OTHER}^{\rm 4FGL-DR2}$,numeric type,fixed,precision=2},
columns/Category_Prob_3FGL/.style={column name=3FGL cat.},
]\loadedtable
}
\normalsize
\vspace{2mm}
\tablefoot{
For the selected sources the sum of OTHER-class probabilities of all eight ML methods in the three-class classification is larger than 7. 
See Table \ref{tab:psr_candidates_3class} for the description of columns.
Sources with missing values in the 3FGL columns are not detected in the 3FGL.
We also save these OTHER-class 
candidates in the supplementary online materials as ``4FGL-DR2\_Candidates\_OTHER\_3classes'' \citep{SOM_material}.
}
\end{table}

There are 30 unassociated 4FGL-DR2 sources
with the sum of OTHER-class probabilities in the three-class classification larger than 7.
The OTHER-class candidates are shown in Table \ref{tab:other_candidates_3class}.
Out of the 30 sources only two sources have no flags in 4FGL, and only eight sources have an association with the previous FGL catalogs (column name ASSOC\_FGL). 
The following sources had additional associations in the SIMBAD database
ordered by decreasing sum of OTHER-class probabilities:

\textbf{4FGL J1800.2-2403c}: This source has the largest sum of OTHER-class probabilities but no entry in Simbad. 
However, the source 1FGL J1800.5-2359c \citep{2010ApJS..188..405A} is associated with two sources in 4FGL-DR2:
4FGL J1800.7-2355, which has OTHER association in the region of the SNR W28 \citep{2020MNRAS.495.2909R}, and 4FGL J1800.2-2403c.

\textbf{4FGL J1847.7-0125}: This source is within 1 arcminute of the candidate young stellar object (YSO) SSTOERC G031.2256+00.1711 \citep{2017ApJ...839..108S}.

\textbf{4FGL J1843.7-0326}: This source is associated with 3FGL J1843.7-0322 and is near the HESS source HESS J1843-033, next to the SNR G28.6-0.1 \citep{2018A&A...612A...1H}. This source is
among the 120 unassociated sources according to significance (>10) in the list of \citet{2016ApJ...820....8S} where the RF and LR methods predicted it to be a young pulsar based on the two-class classification. 
In our 3FGL three-class catalog, this source has a MIXED classification.

\textbf{4FGL J1556.8-5242c}: This source is within 1 arcminute of the candidate YSO 2MASS J15564953-5241450 \citep{2008AJ....136.2413R}.

\textbf{4FGL J1631.7-4826c} (3FGL J1632.4-4820; \citealt{2015ApJS..218...23A}): This source is within 30 arcseconds of the dark cloud (nebula) SDC G335.894-0.184 \citep{2016A&A...590A..72P}.

\textbf{4FGL J1626.0-4917c} (3FGL J1626.2-4911 -- \citealt{2015ApJS..218...23A}; 3FHL J1626.3-4915 -- \citealt{2017ApJS..232...18A}): 
This source is associated with the HESS source HESS J1626-490. It is also one of the 27 sources shortlisted by \citet{2020MNRAS.495.1093H}, 
who used ML methods to select pulsar candidates from the 3FHL catalog. It is also classified as OTHER based on the 3FGL values.

\textbf{4FGL J1849.4-0117} (3FGL J1849.5-0124c; \citealt{2015ApJS..218...23A}): This source is in the region of Galactic mini starburst W43 studied by \citet{2020A&A...640A..60Y} and is also within 1 arcminute of the candidate YSO SSTOERC G031.5367-00.1555. It has a MIXED classification based on the 3FGL values.

\textbf{4FGL J1109.4-6115e} (3FGL J1111.9-6038; \citealt{2015ApJS..218...23A}): This source is associated with the extended galactic source FGES J1109.4-6115 \citep{2017ApJ...843..139A}, near the speculated star forming region (SFR) 4FGL J1115.1-6118, in the region of the young massive stellar cluster NGC 3603 \citep{2020ApJ...897..131S}. It has a MIXED classification with the 3FGL values.

\textbf{4FGL J1850.2-0201}: This source is also in the region of the starburst W43 \citep{2020A&A...640A..60Y}.

\textbf{4FGL J1801.8-2358}: This source is associated with HESS J1800-240A and 2FHL J1801.7-2358. It is located south of the SNR W28 \citep{2020MNRAS.495.2909R}.

\textbf{4}\textbf{FGL J1855.8+0150}: This source is in the region of SNR W44 \citep{2020ApJ...896L..23P}

\textbf{4FGL J1742.0-3020}: This source is within 1 arcminute of the molecular cloud [MML2017] 777 \citep{2017ApJ...834...57M}.

\textbf{4FGL J1904.7+0615}: This source is within 1 arcminute of the bubble [SPK2012] MWP1G040012-001102 \citep{2012MNRAS.424.2442S}.

\section{Application of probabilistic catalogs for population studies}
\lb{sec:pop_studies}

\subsection{Number of sources as a function of flux}
\lb{sec:dNdS}

In this section we show how probabilistic catalogs can be used for population studies.
One of the most important questions in gamma-ray astronomy is the contribution of point sources, 
for example, AGNs, to the extragalactic gamma-ray flux 
\citep[e.g.,][]{2010ApJ...720..435A, 2011ApJ...738..181M, 2016PhRvL.116o1105A, 2016ApJS..225...18Z, 2016ApJ...826L..31Z, 2016ApJ...832..117L, 2018ApJ...856..106D}:
if most of the extragalactic emission is explained by point sources, then one can put stringent constraints, 
for example, on  dark matter annihilation or decay into gamma rays 
\citep{2015ApJ...800L..27A, 2015PhRvD..91l3001D, 2015JCAP...09..008F, 2015PhR...598....1F, 2017ChPhC..41d5104L} or 
on the evaporation of primordial black holes \citep{2010PhRvD..81j4019C}.
In particular, it is important to understand the contribution to the population of AGNs from the unassociated sources.
A probabilistic catalog provides an answer to the question: how many sources among the unassociated ones are expected to belong to different classes, such as pulsars or AGNs. 
One can calculate the total expected number of AGNs or pulsars among the unassociated sources, or calculate the contribution as a function of one or more parameters.
In this subsection we present the estimated numbers of AGNs, pulsars, and, in case of three-class classification, OTHER sources as a function of their flux.
In the following subsection we also discuss the distributions of sources as functions
of Galactic latitude and longitude.

\begin{figure*}[h]
\centering
\includegraphics[width=0.45\textwidth]{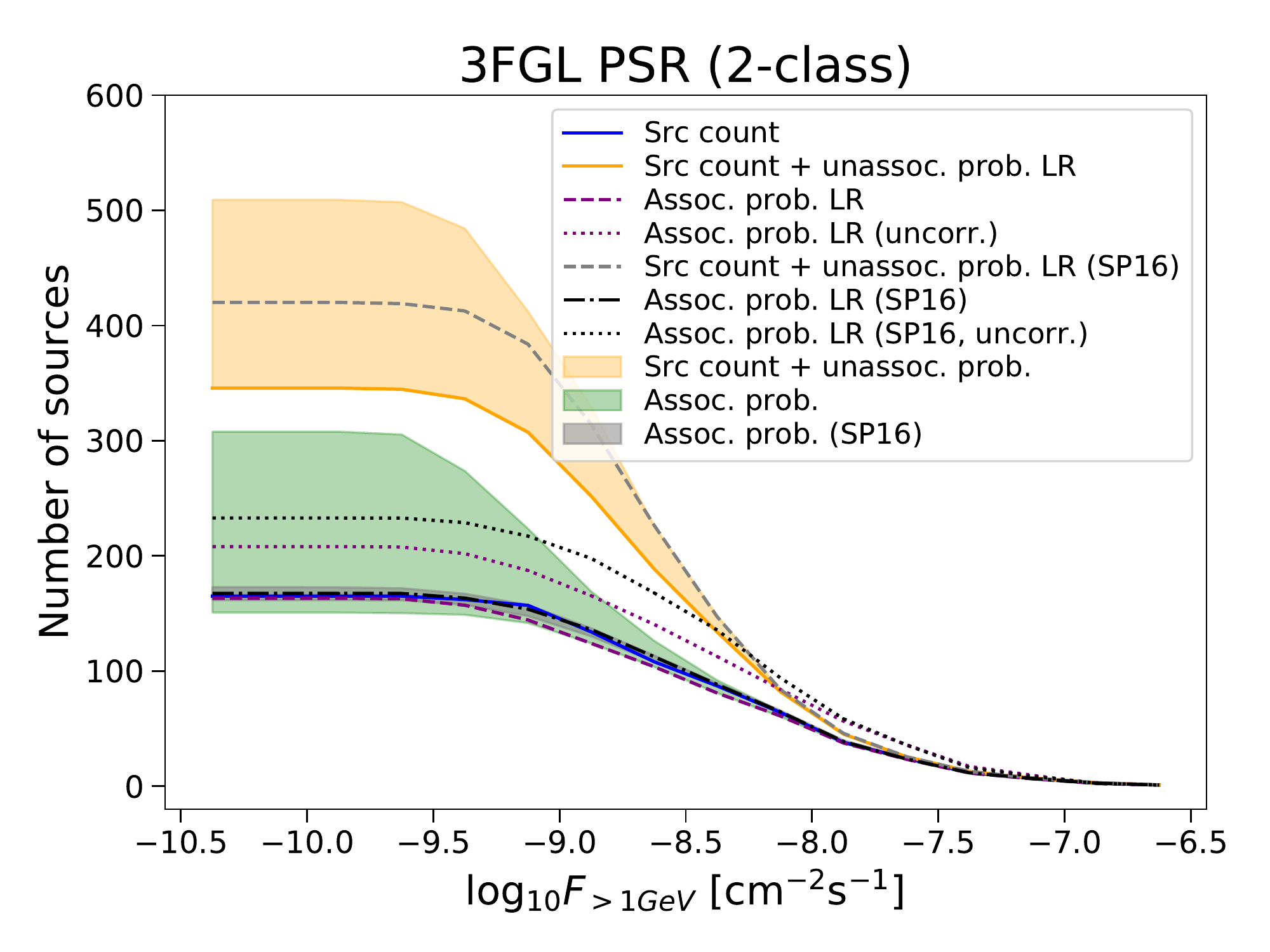}
\includegraphics[width=0.45\textwidth]{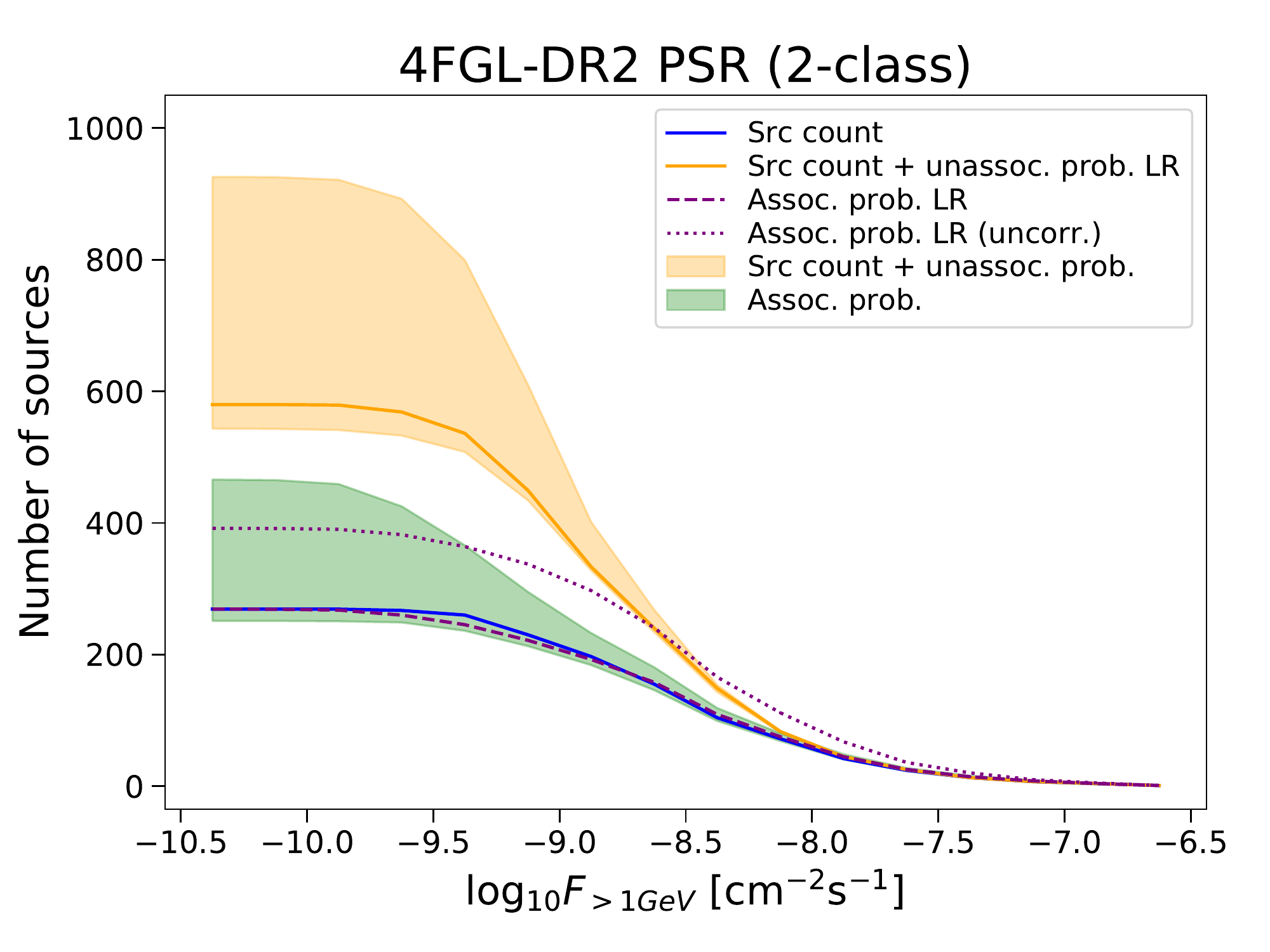} \\
\includegraphics[width=0.45\textwidth]{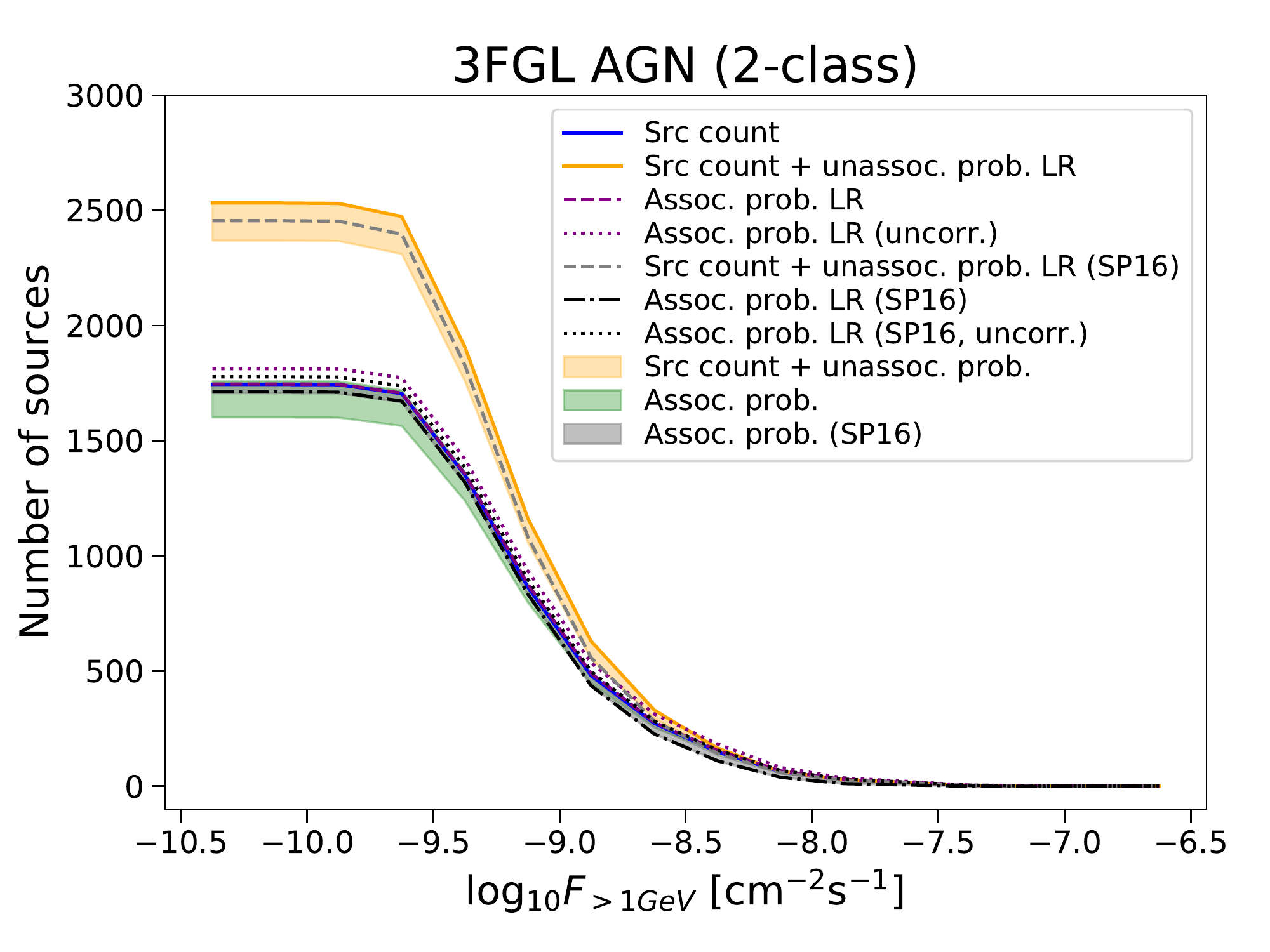}
\includegraphics[width=0.45\textwidth]{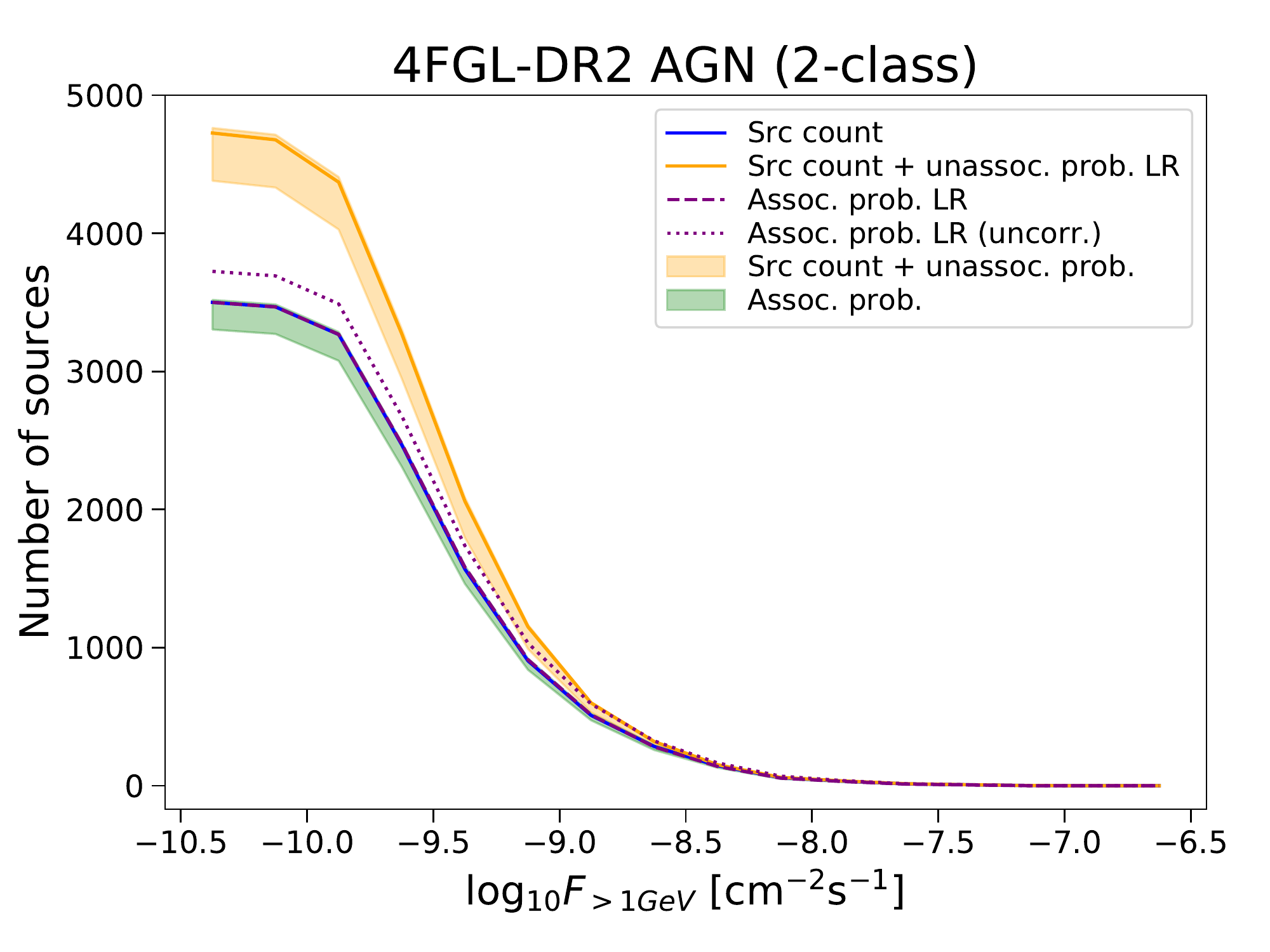}
\caption{Cumulative number of sources as a function of their flux. 
Solid blue lines represent associated 3FGL and 4FGL-DR2  sources; green bands show the envelope of sums of class probabilities for associated sources for the eight ML methods corrected for the presence of OTHER sources; solid orange lines (bands) show the sum of class probabilities for the LR model (the envelope of the eight ML methods), corrected for the presence of OTHER sources and added to the source count of associated sources; purple dashed (dotted) lines show the sum of class probabilities for associated sources for the LR method without oversampling and corrected (uncorrected) for the presence of OTHER sources; gray dash-dotted (dotted) lines show the sums of class probabilities from \cite{2016ApJ...820....8S} using LR corrected (uncorrected) for the presence of OTHER sources; and gray bands show the envelope of sums of class probabilities for associated sources for LR and RF methods from \cite{2016ApJ...820....8S} corrected for the presence of OTHER sources (for details, see Sect. \ref{sec:dNdS}).}  
\label{fig:logN_logS}
\end{figure*}

In Fig. \ref{fig:logN_logS} we show the cumulative number of AGNs and pulsars with a flux above 1 GeV larger than the
value on the x-axis.
Solid blue lines show the actual counts of sources (AGNs or pulsars) in the 3FGL and 4FGL-DR2 catalogs.
As a consistency check of the method, we calculated the AGN- and pulsar-like probabilities for associated sources.
The sum of probabilities (uncorrected for sources other than AGNs and pulsars) for LR algorithm are shown by dotted purple lines.
In order to correct the expected number of AGNs among associated sources for AGN-like probabilities in OTHER sources, 
we subtracted the corresponding AGN-like probabilities in each flux band:
\bea
\lb{eq:assoc_ev}
N_{\rm AGN}^{\rm ass}  = \sum_{i \in \rm ass} p^i_{\rm AGN}\,\, - \sum_{i \in \rm ass\,other} p^i_{\rm AGN}.
\eea
The corrected sums of probabilities for LR method are shown by dashed purple lines.
The green bands show the envelope of the sums of corrected probabilities for the eight methods used in this paper.
We see that the counts of associated sources, AGNs and pulsars, are consistent with the expected number of associated sources
calculated from the class probabilities of associated sources.
This conclusion is not very surprising since we used associated sources for training the ML algorithms.
We note that the correction for OTHER sources is important for consistency of the sum of probabilities and the number of associated sources.
We also compared the sums of probabilities for the 3FGL associated sources in \cite{2016ApJ...820....8S}.
The sum of probabilities for associated sources in the LR case uncorrected for OTHER sources are shown by dotted black line,
while the sums corrected for OTHER sources are shown by black dash-dotted lines.
The gray band is the envelope of the two methods (LR and RF) used by \cite{2016ApJ...820....8S}.
We see that the sum of probabilities for AGNs and pulsars overpredicts the source counts in 3FGL, 
while correction for OTHER sources makes the prediction for associated sources consistent with the source counts.

The predictions for the number of AGNs and pulsars among the unassociated sources corrected for OTHER sources 
added to the 3FGL and 4FGL-DR2  source counts are shown by solid orange lines for the LR model.
The orange bands show the corresponding envelopes for the eight ML methods.
We assume that the fractional contribution of OTHER sources is the same for associated and unassociated sources in the different flux bands.
Thus, the correction for the presence of OTHER sources is calculated similarly to the associated sources in Eq. (\ref{eq:assoc_ev}),
but we adjust for the fact that there are fewer unassociated than associated sources (i.e., 
the correction is assumed to be proportionally smaller).
In particular, the number of AGNs among unassociated sources in a flux band $\Delta F$ is estimated as
\bea
\lb{eq:unassoc_ev}
N_{\rm AGN}^{\rm unass} = \sum_{i \in \rm unass} p^i_{\rm AGN}\,\, - \sum_{i \in \rm ass\,other} p^i_{\rm AGN} \cdot 
\frac{N_{\rm unass}}{N_{\rm ass}}
,\eea
where all probabilities and the number of sources are computed for sources with flux inside $\Delta F$.
The first term is the sum of AGN-like probabilities among the unassociated sources,
while the second term is the sum of AGN-like probabilities among associated OTHER sources rescaled by the total number
of unassociated and associated sources in this flux band.
The expected number of pulsars among the unassociated sources is calculated analogously.
The corresponding sums of associated source counts plus the expected number of sources calculated with the LR method of \cite{2016ApJ...820....8S} 
and corrected for OTHER sources are shown by dashed gray lines.

The expected numbers of pulsars and AGNs among the unassociated sources
are summarized in Table \ref{tab:expected_counts_prob_3FGL} for the 3FGL catalog and 
in Table \ref{tab:expected_counts_prob_4FGL-DR2} for the 4FGL-DR2 catalog.
The row two-class shows the expectations for the two-class classification uncorrected for the presence of OTHER sources.
The row two-class corr total shows the expectations corrected for OTHER sources using Eq. (\ref{eq:unassoc_ev}) without any binning,
while two-class corr F-bins shows the correction with the flux bins as in Fig. \ref{fig:logN_logS}.
We see that corrections using total source counts and source counts binned in flux are similar and relatively small.
In the following section we also calculate the correction using latitude and longitude bins.

\begin{figure*}[h]
\centering
\includegraphics[width=0.45\textwidth]{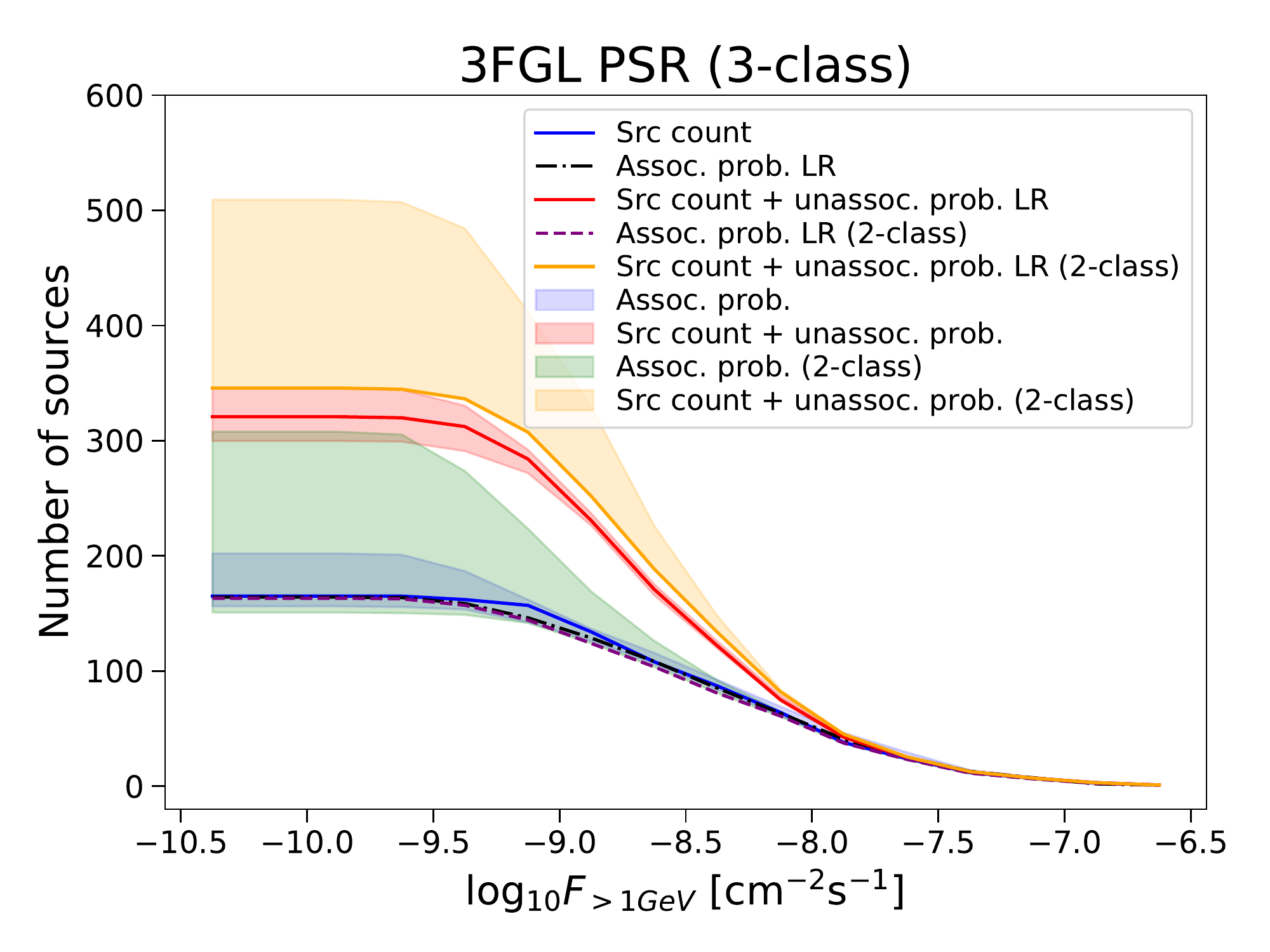}
\includegraphics[width=0.45\textwidth]{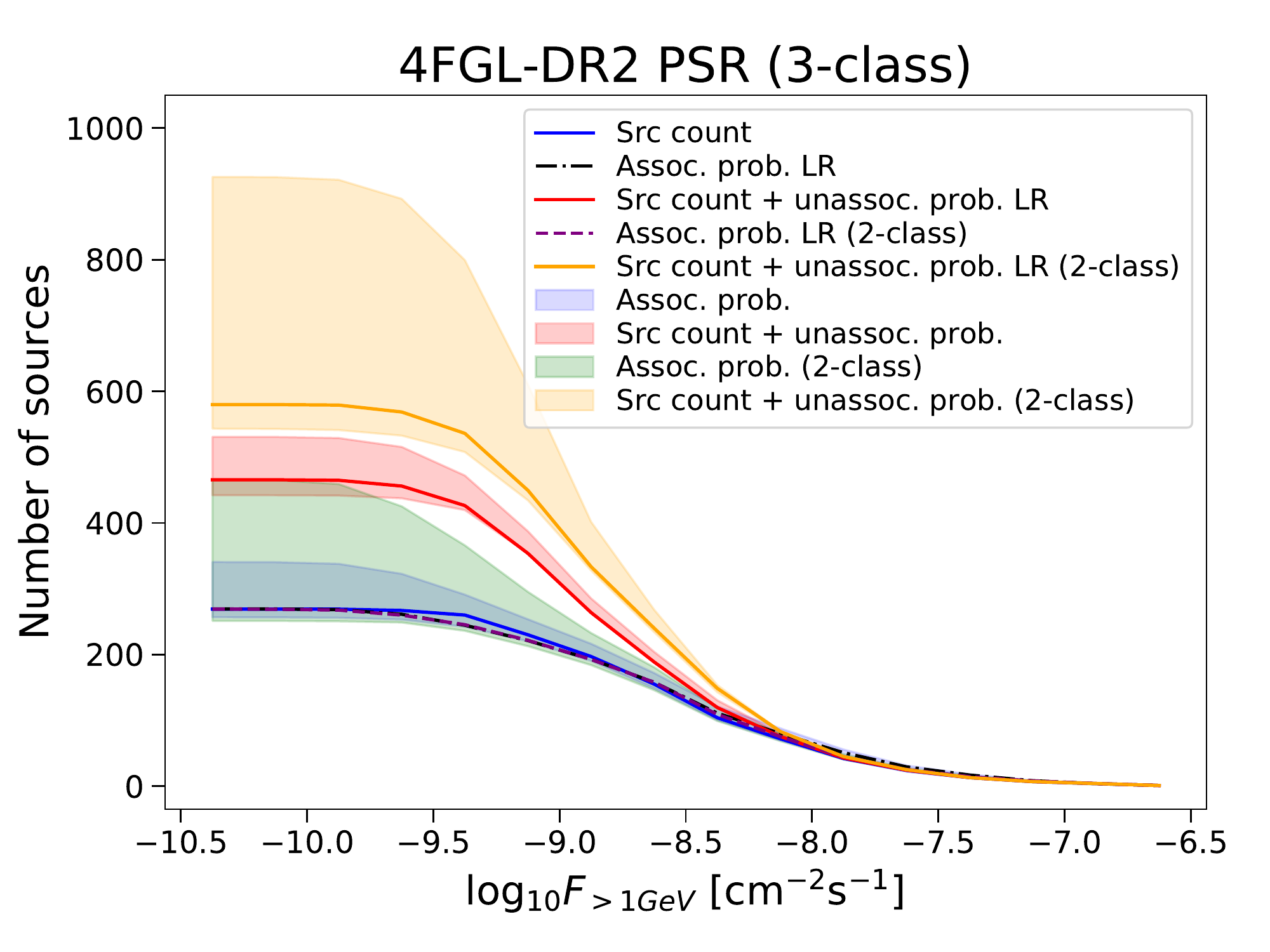} \\
\includegraphics[width=0.45\textwidth]{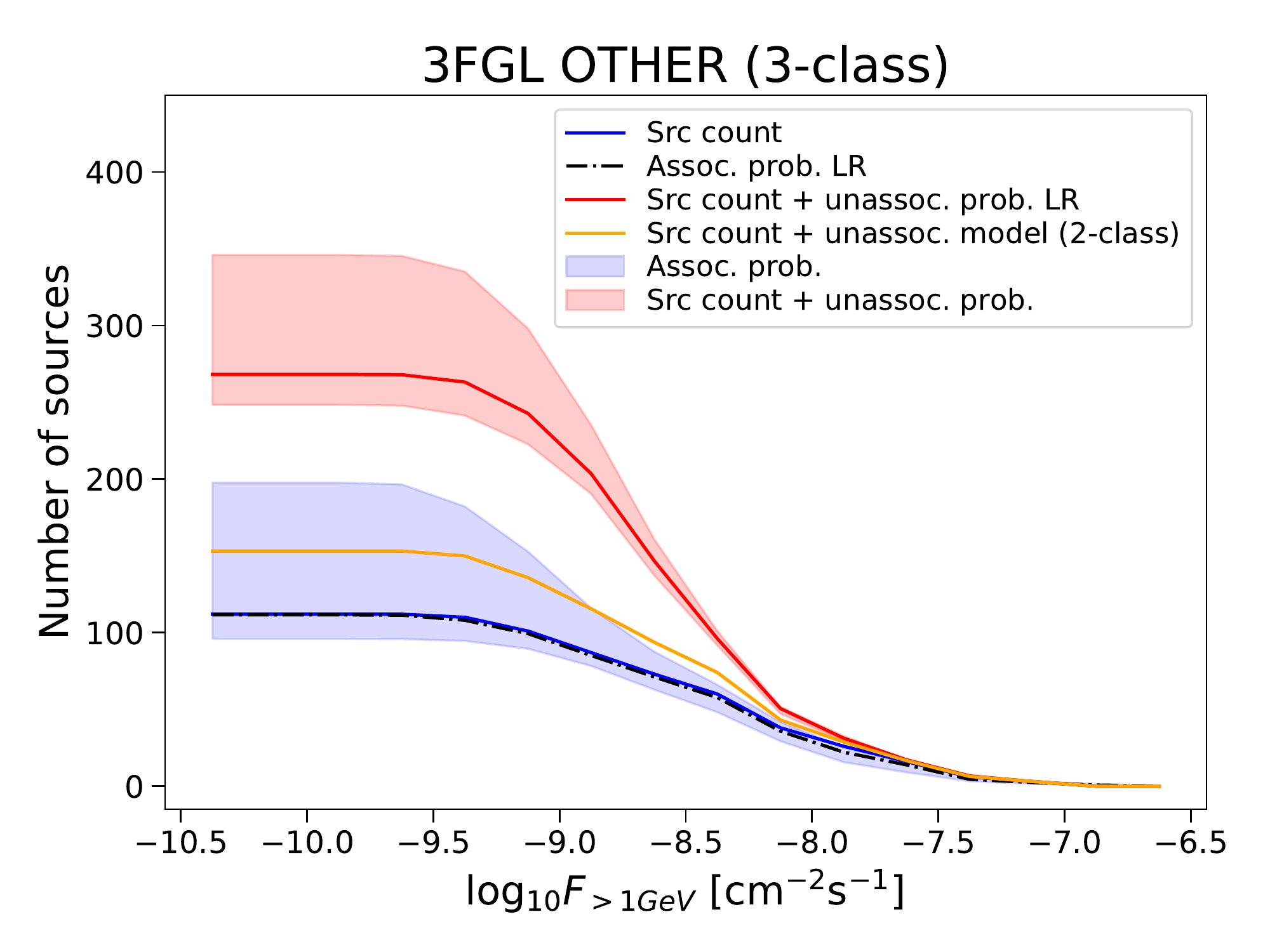}
\includegraphics[width=0.45\textwidth]{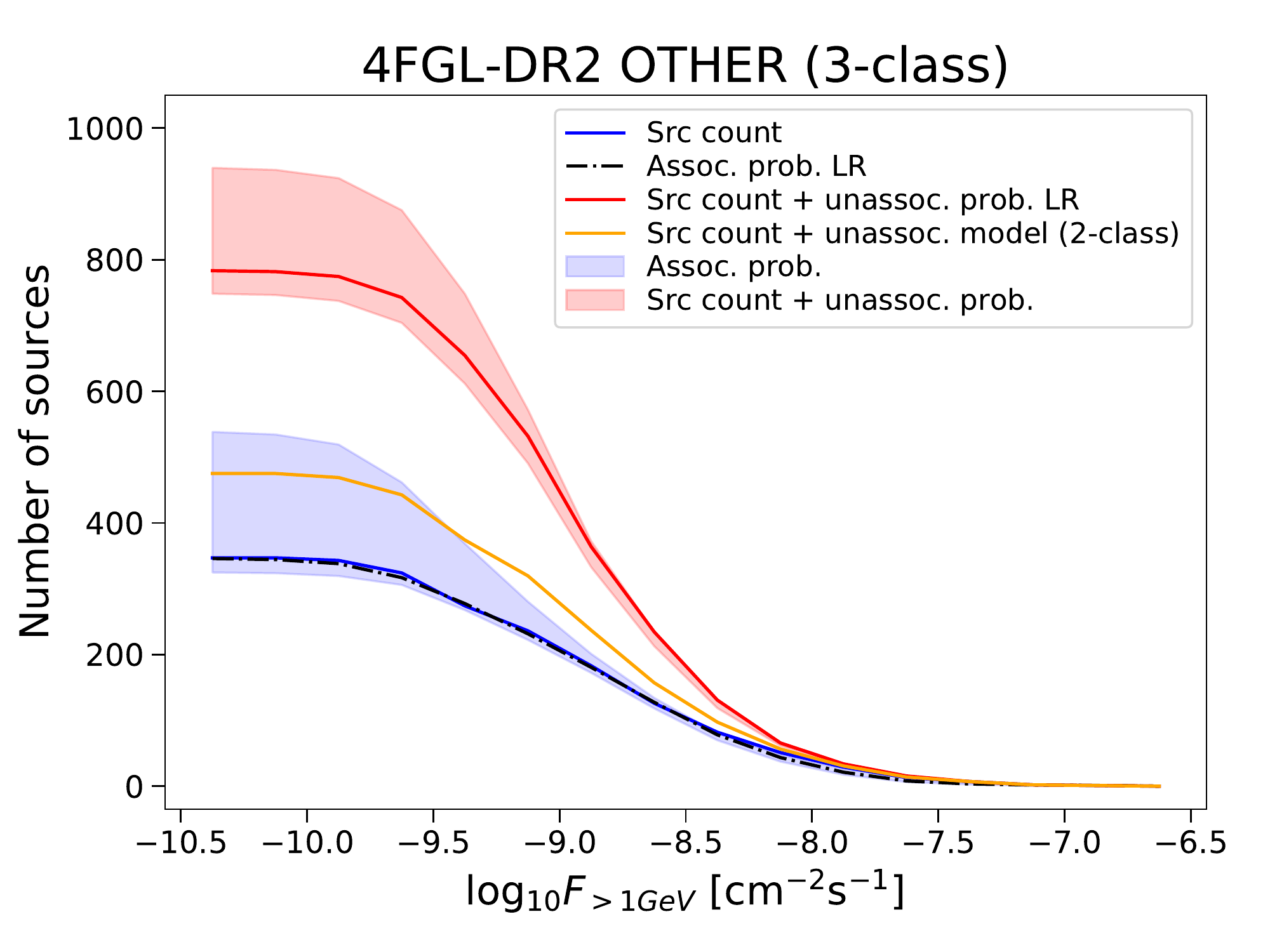}
\caption{Cumulative number of sources as a function of their flux. 
The solid blue lines represent the associated 3FGL and 4FGL-DR2  sources; the blue bands show the envelope of sums of class probabilities for associated sources for the eight ML methods;  
the purple dash-dotted lines show the sum of class probabilities for associated sources for the LR method without oversampling; the solid red lines (bands) show the sum of class probabilities for the LR model (the envelope of the eight ML methods) added to the source count of associated sources; the purple dashed lines show the sum of class probabilities for associated sources for the LR method without oversampling in the two-class classification corrected for the presence of OTHER sources; the solid orange lines (bands) show the sum of class probabilities for the LR model (the envelope of the eight ML methods), corrected for the presence of OTHER sources and added to the source count of associated sources (the band and the line are the same as in Fig. \ref{fig:logN_logS}); green bands in the pulsar plots show the envelope of sums of class probabilities for associated sources for the eight ML methods in the two-class classification corrected for the presence of OTHER sources (the same as in Fig. \ref{fig:logN_logS}); and solid orange lines in the OTHER plots show OTHER source counts plus an estimate of the number of OTHER sources among unassociated ones using Eq. (\ref{eq:unas_other}) (for details, see Sect. \ref{sec:dNdS}).
}  
\label{fig:logN_logS_3classes}
\end{figure*}

The numbers of pulsars and OTHER sources in the three-class classification as a function of flux are shown in Fig. \ref{fig:logN_logS_3classes}.
As in the two-class case, solid blue line shows the counts of associated sources in the corresponding classes.
Red and blue bands show the envelops of the expected number of associated sources and the number of associated sources plus the expected numbers of sources among unassociated ones, respectively.
In the pulsar plots, we also replot the envelops of expected numbers of associated pulsars in the two-class classification
(shown by the green band) and the associated pulsar counts plus predictions for unassociated sources by orange bands corrected for the presence of OTHER sources among unassociated ones.
We notice that the bands for the three-class case are narrower in the pulsar plots than the bands for the two-class case.
In part this is due to less oversampling in the three-class case. 
We also note that the red band lies almost entirely below the orange one.
This is due to a possible underestimation of the contribution of OTHER sources to the pulsar class among unassociated sources
in the two-class case.
In particular, in the bottom panels of Fig. \ref{fig:logN_logS_3classes} we show the estimated total number of OTHER sources in the two-class case by the orange line.
Since we do not have probabilities for the OTHER sources in this case, we estimate the number of OTHER sources among unassociated ones 
simply by rescaling the number of associated OTHER sources in each energy band as
\bea
\lb{eq:unas_other}
N_{\rm OTHER}^{\rm unass} = N_{\rm OTHER}^{\rm ass} \frac{N^{\rm unass}}{N^{\rm ass}}
.\eea
We notice that this estimate agrees with the correction of the estimated numbers of AGNs and pulsars in the two-class case due to presence of OTHER sources since
\bea
\nonumber
N_{\rm OTHER}^{\rm ass} &=& \sum_{i \in \rm ass\,other} p^i_{\rm AGN} + \sum_{i \in \rm ass\,other} p^i_{\rm PSR} \\
&=& \sum_{i \in \rm ass\,other} (p^i_{\rm AGN} + p^i_{\rm PSR})
\eea
provided that $p^i_{\rm AGN} + p^i_{\rm PSR} = 1$ for each source in the two-class case.
The estimated total number of OTHER sources in the two-class case (orange line) is significantly below the red band derived from the sum of probabilities of the OTHER class in the three-class case.
The model in Eq. (\ref{eq:unas_other}) depends on binning: as we show in the following section, for latitude binning this two-class
estimate of the number of OTHER sources among unassociated ones agrees with the three-class estimate.
The reason is that for flux bins, the ratio $\frac{N^{\rm unass}}{N^{\rm ass}}$ is dominated by AGNs at high latitudes and it is always smaller than 1, while at low latitudes, where most of OTHER sources are, the ratio $\frac{N^{\rm unass}}{N^{\rm ass}}$ is about 1.

\begin{figure}[h]
\centering
\includegraphics[width=0.45\textwidth]{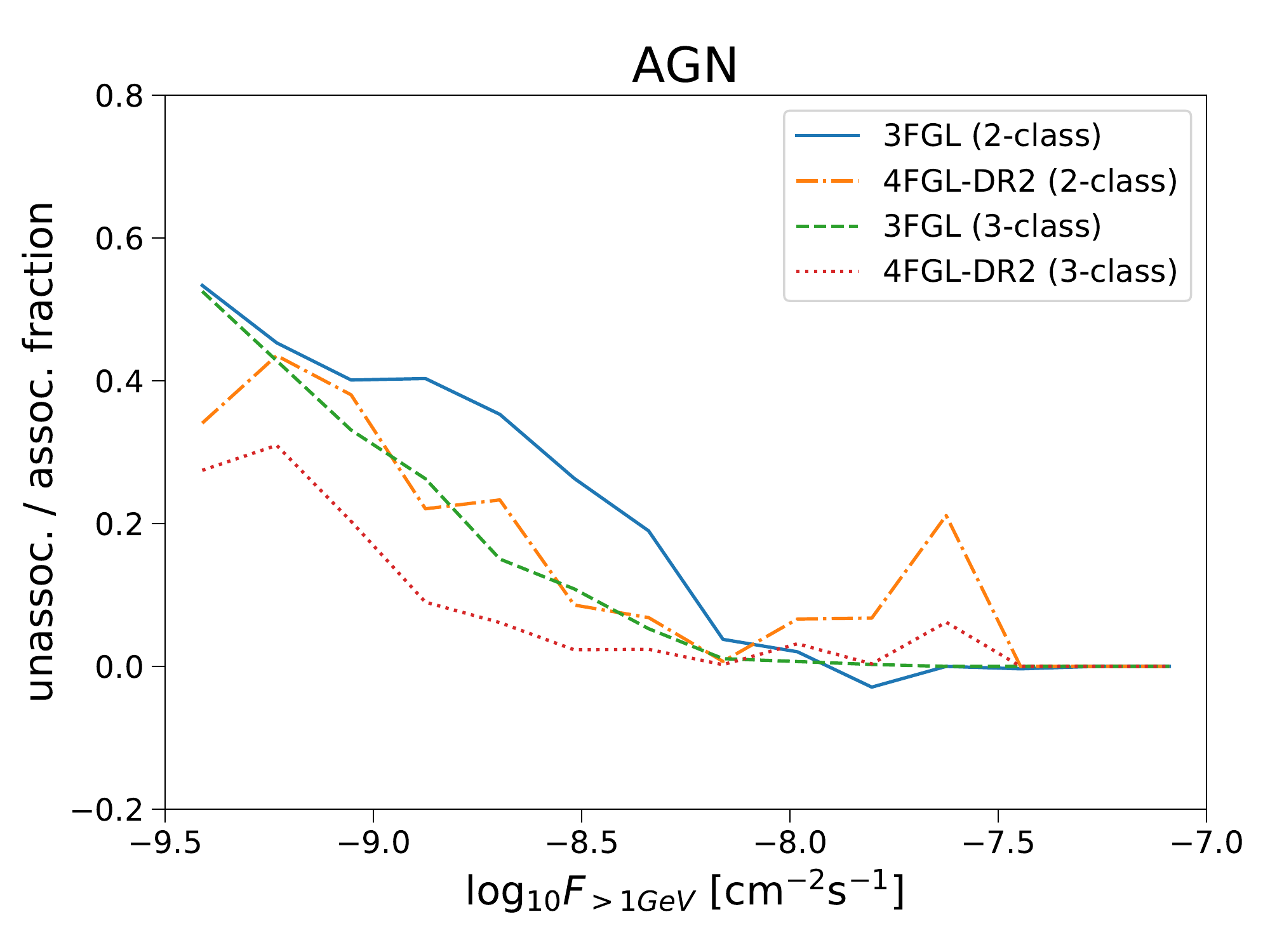} \\
\includegraphics[width=0.45\textwidth]{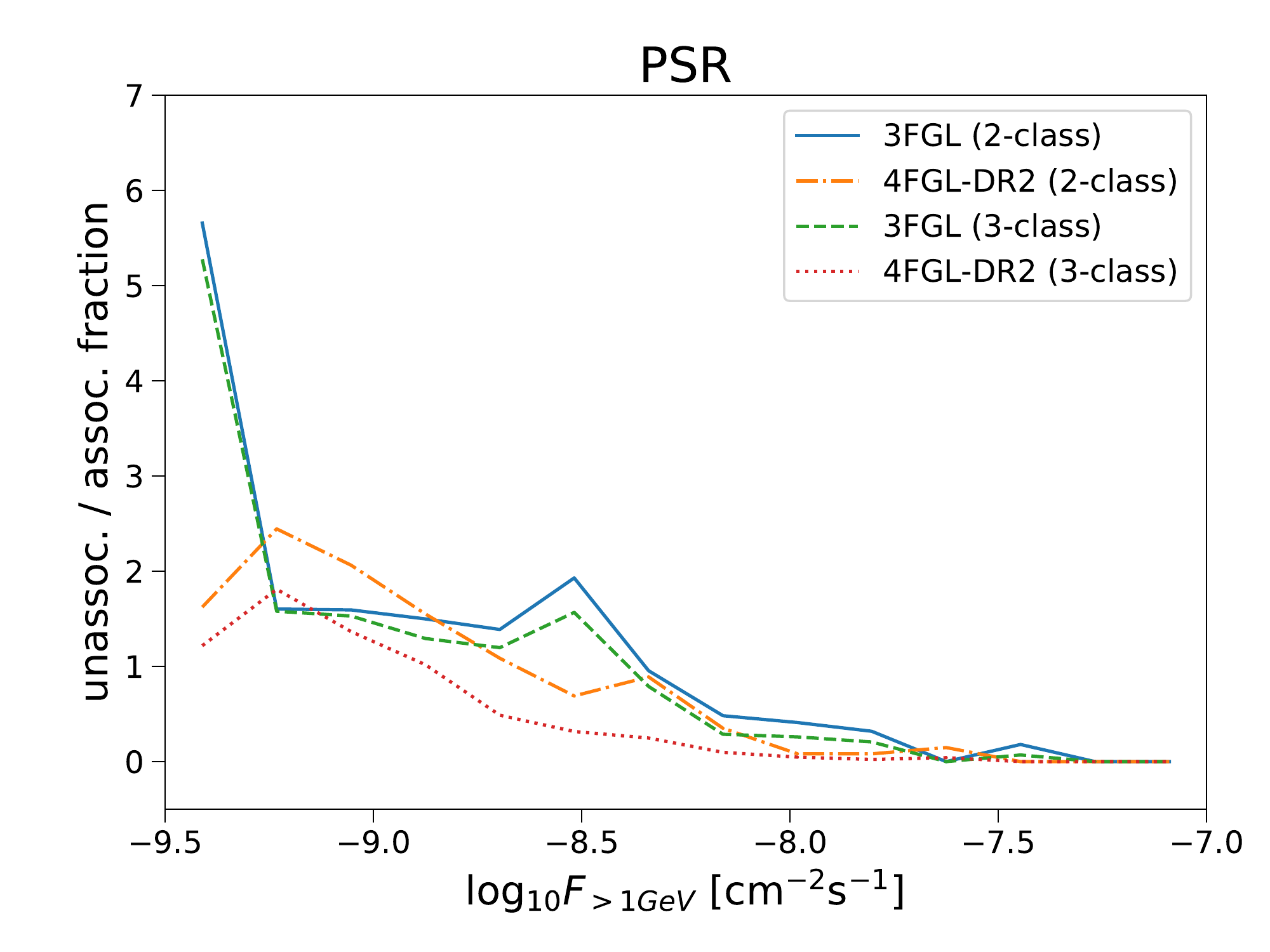} \\
\includegraphics[width=0.45\textwidth]{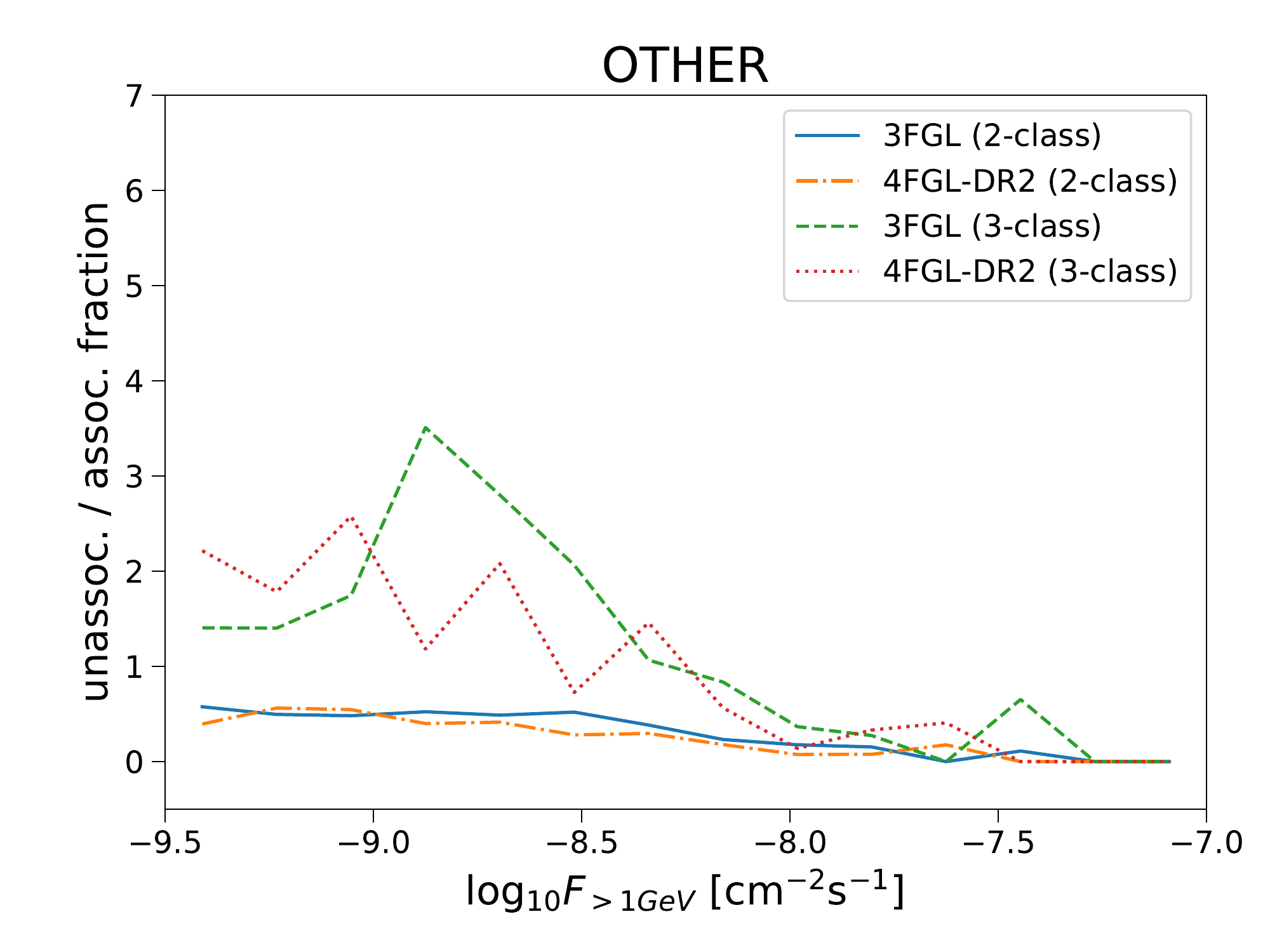}
\caption{Ratio of the number of AGNs, pulsars, and OTHER sources among unassociated sources estimated with LR without oversampling
to the counts of associated sources, respectively.
AGN and pulsar estimates are corrected for the presence of OTHER sources using Eq. (\ref{eq:unassoc_ev}),
while the number of OTHER sources among unassociated ones in the two-class case is estimated using Eq. (\ref{eq:unas_other}).
} 
\label{fig:unass_vs_ass_frac}
\end{figure}

We note that the probabilistic classification mostly affects sources with small fluxes.
In Fig. \ref{fig:unass_vs_ass_frac} we plot the ratio of the expected number of sources of a certain class among unassociated sources
computed according to Eq.~(\ref{eq:unassoc_ev}) with the LR algorithm (without oversampling) to the number of associated sources in this class.
The ratio generally increases as the flux decreases.
Negative values (e.g., at high fluxes for AGNs) are due to subtraction of probabilities for the OTHER associated sources.
For OTHER sources in the two-class case we use estimates in Eq. (\ref{eq:unas_other}).
As discussed above, we see that these estimates are significantly below the estimated numbers of OTHER sources among unassociated ones in the three-class classification (for the LR method without oversampling in this case).

\subsection{Latitude and longitude profiles}
\lb{sec:lat-lon-profiles}

\begin{figure*}[h]
\centering
\includegraphics[width=0.45\textwidth]{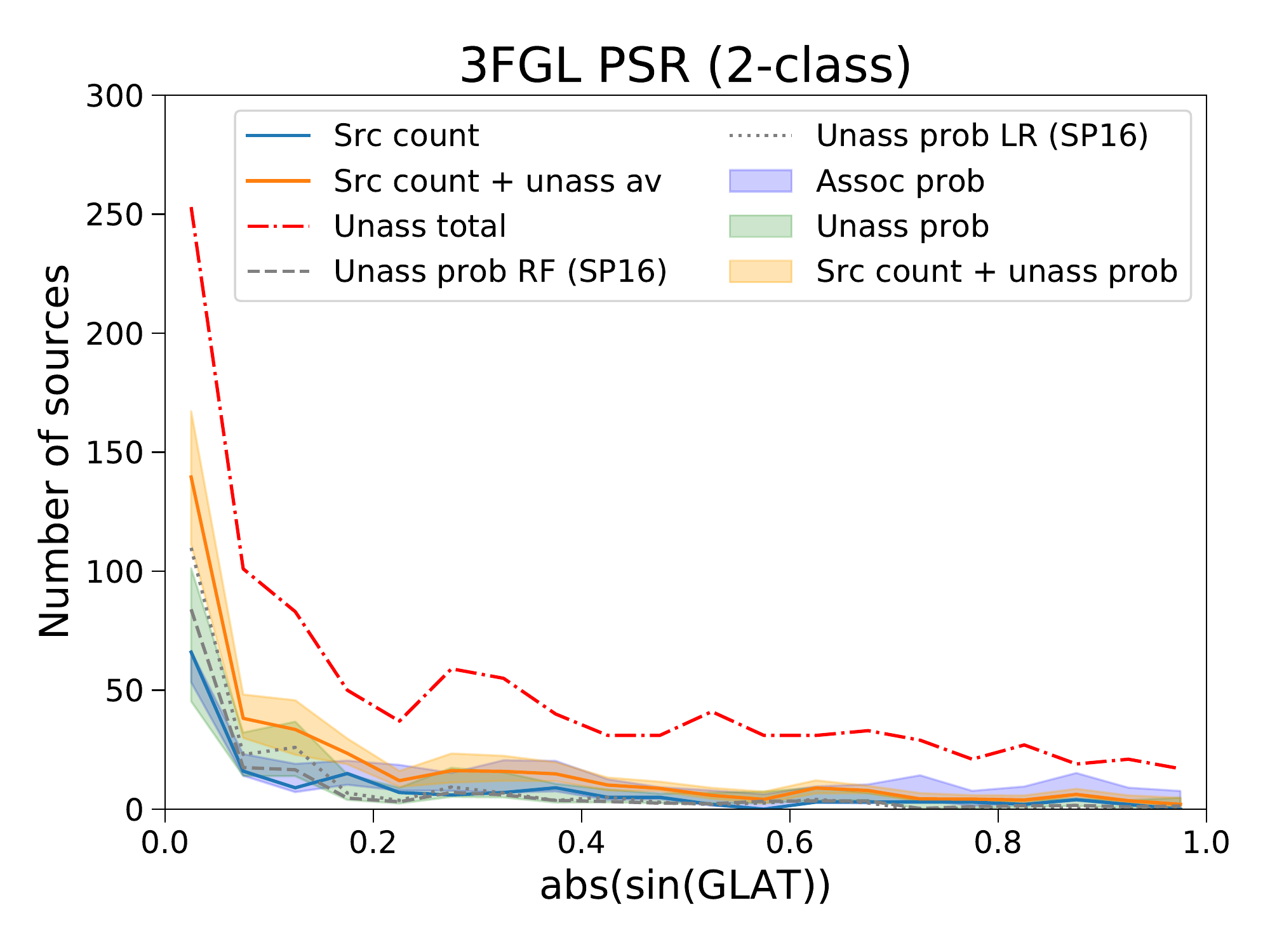}
\includegraphics[width=0.45\textwidth]{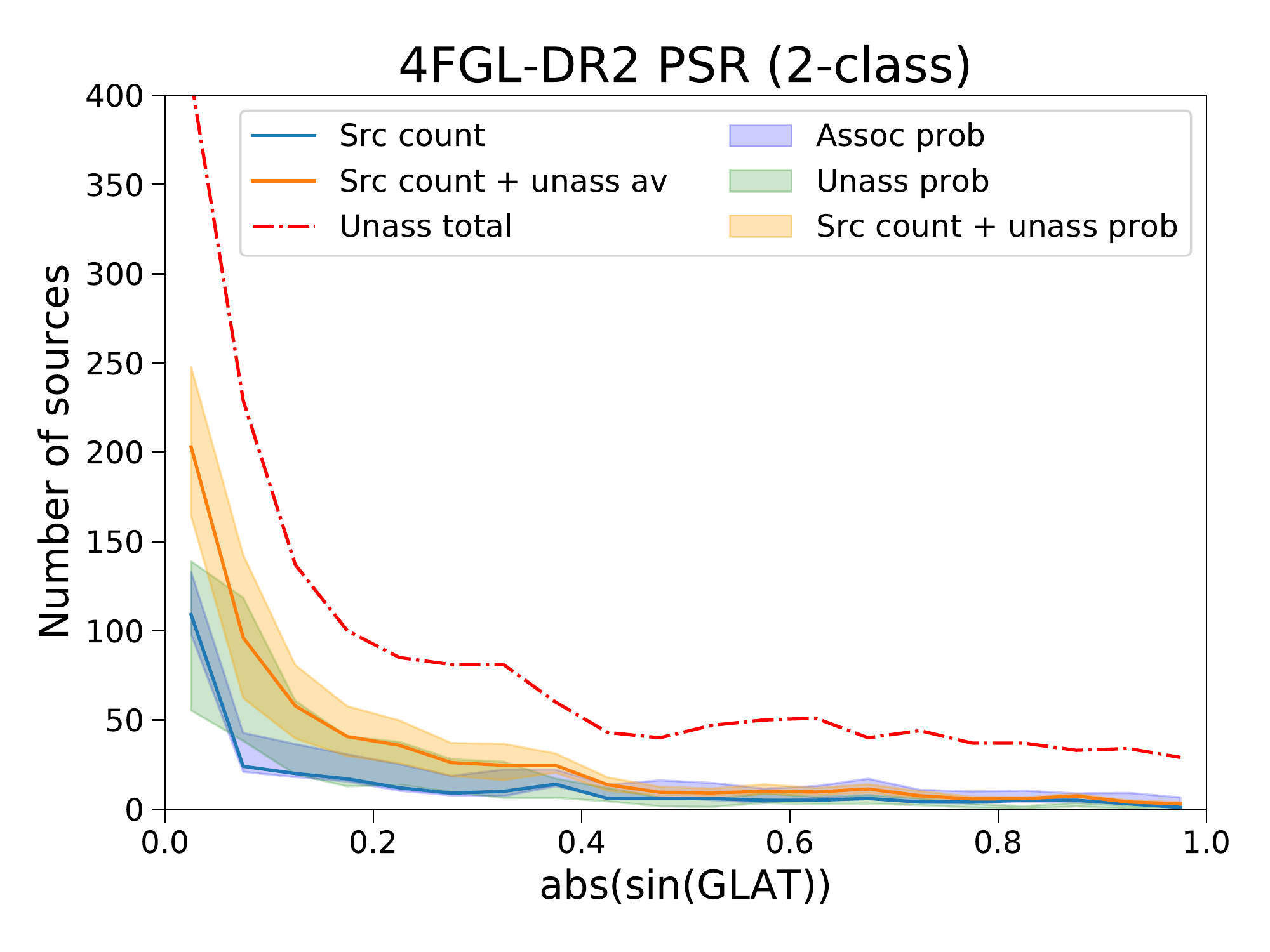} \\
\includegraphics[width=0.45\textwidth]{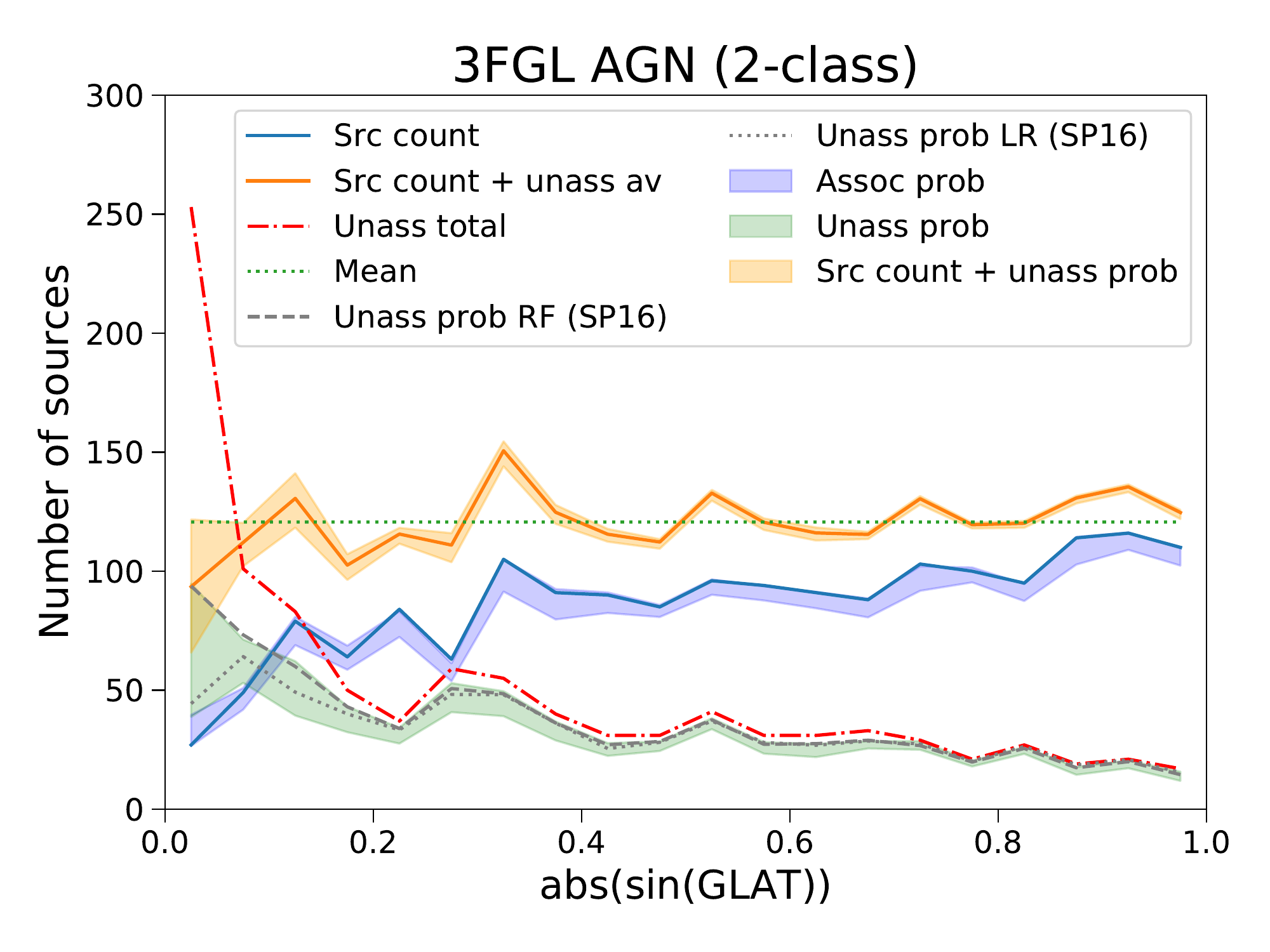}
\includegraphics[width=0.45\textwidth]{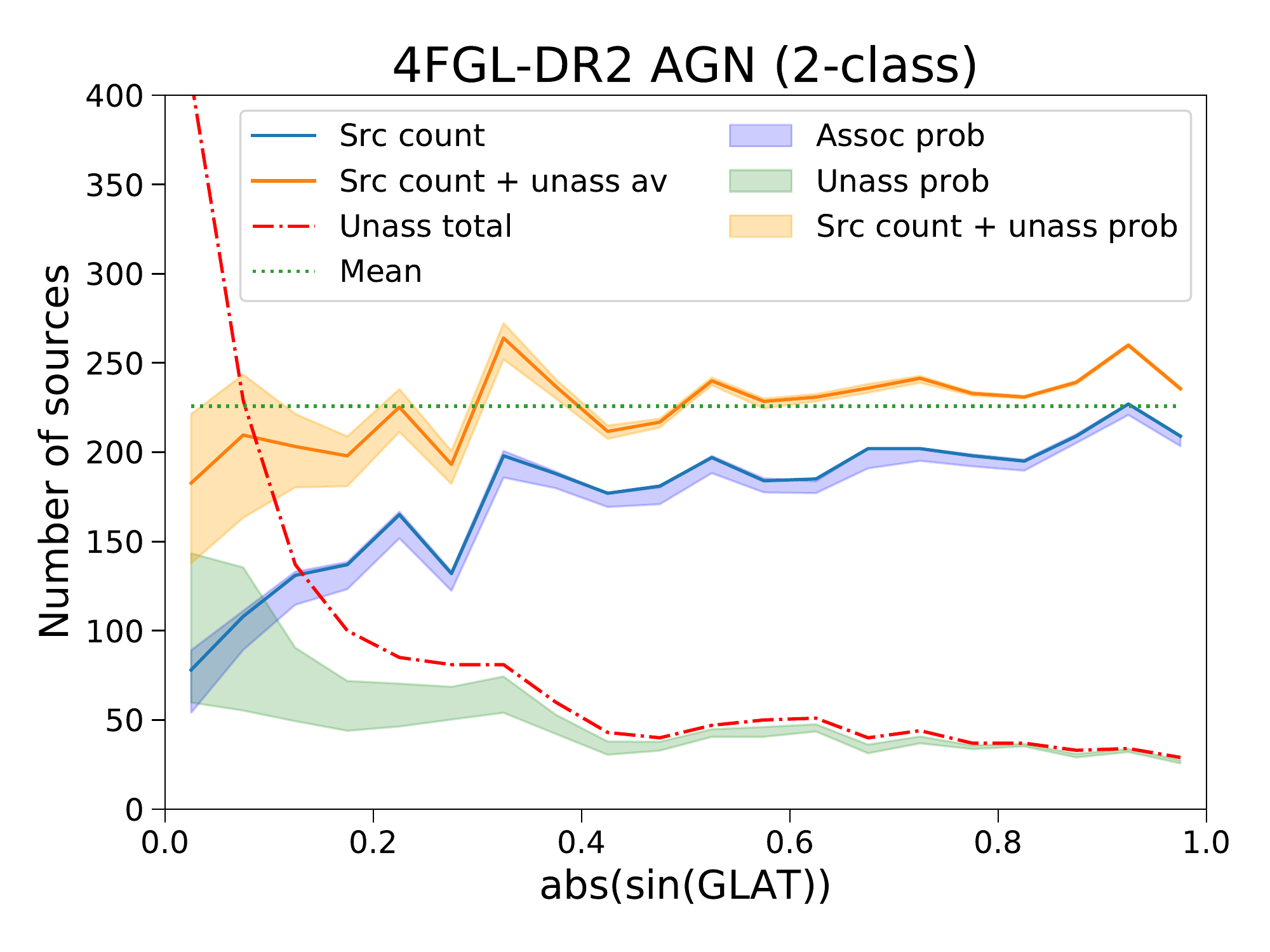}
\caption{Latitude profiles of source counts in the case of two-class classification. 
Solid blue lines represent associated 3FGL and 4FGL-DR2  sources; red dash-dotted lines show counts of all unassociated sources; 
blue bands show the envelope of sums of class probabilities for associated sources for the eight ML methods with and without oversampling
corrected for the presence of OTHER sources; 
green bands show the envelope of sums of class probabilities for unassociated sources for the eight ML methods corrected for the presence of OTHER sources; 
solid orange lines (bands) show the average (envelope) of sums of class probabilities for the eight ML methods added to the source count of associated sources; 
green dotted lines on the AGN plots show the mean of the orange solid lines; and gray dashed (dotted) lines show the RF (LR) sums of class probabilities from \cite{2016ApJ...820....8S} (for details, see Sect. \ref{sec:lat-lon-profiles}). }  
\label{fig:lat_profile}
\end{figure*}

\begin{figure*}[h]
\centering
\includegraphics[width=0.45\textwidth]{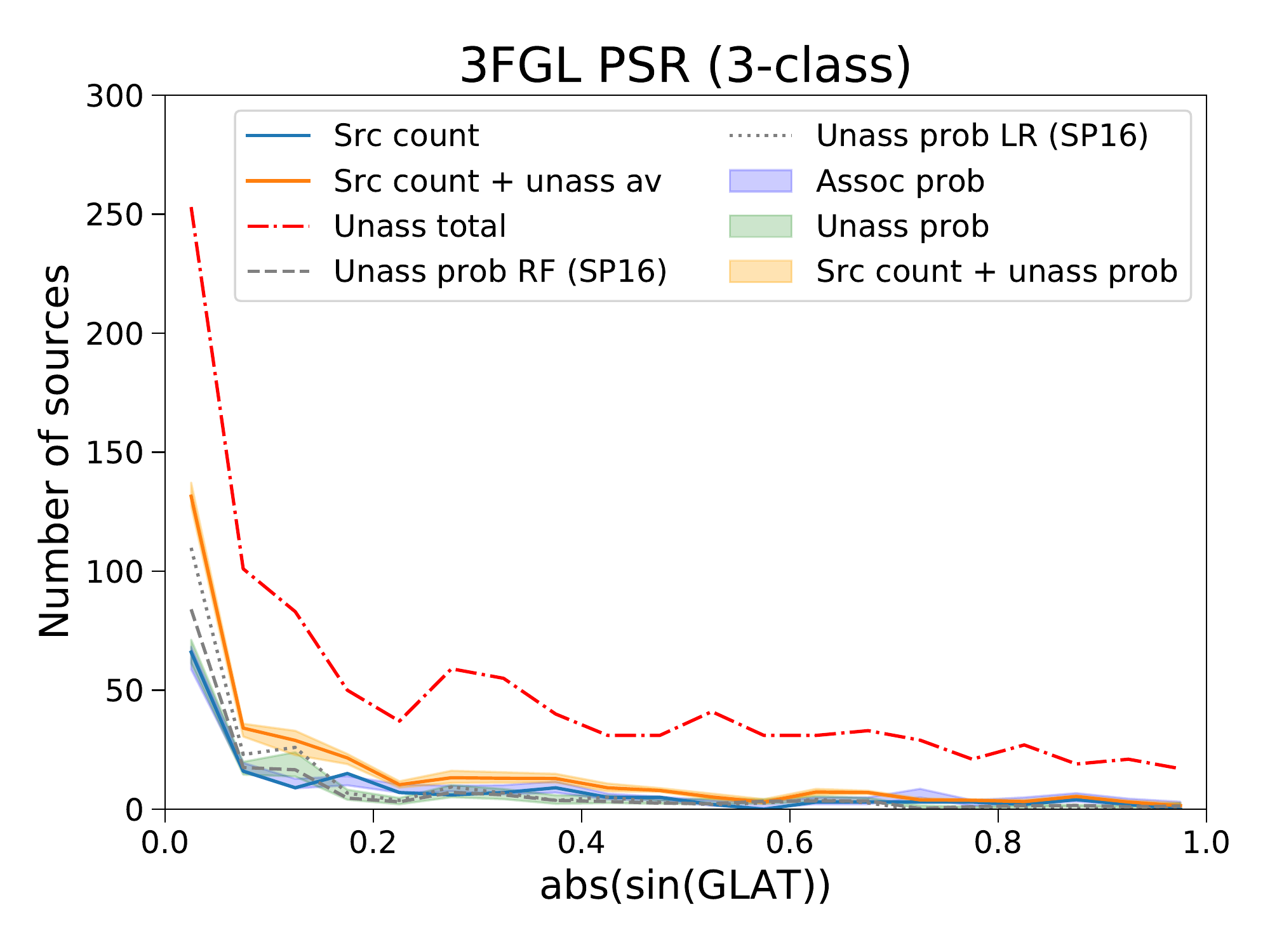}
\includegraphics[width=0.45\textwidth]{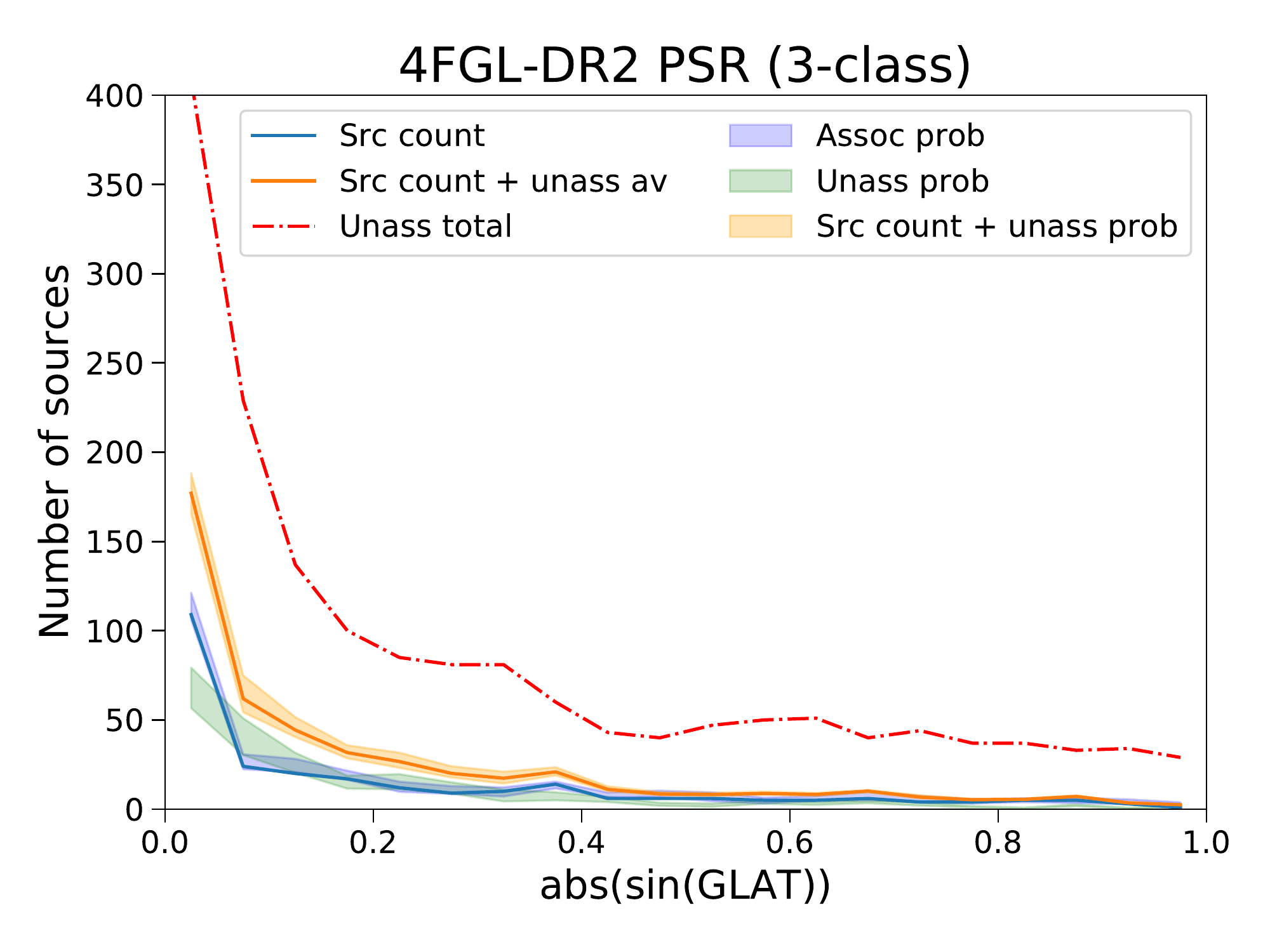} \\
\includegraphics[width=0.45\textwidth]{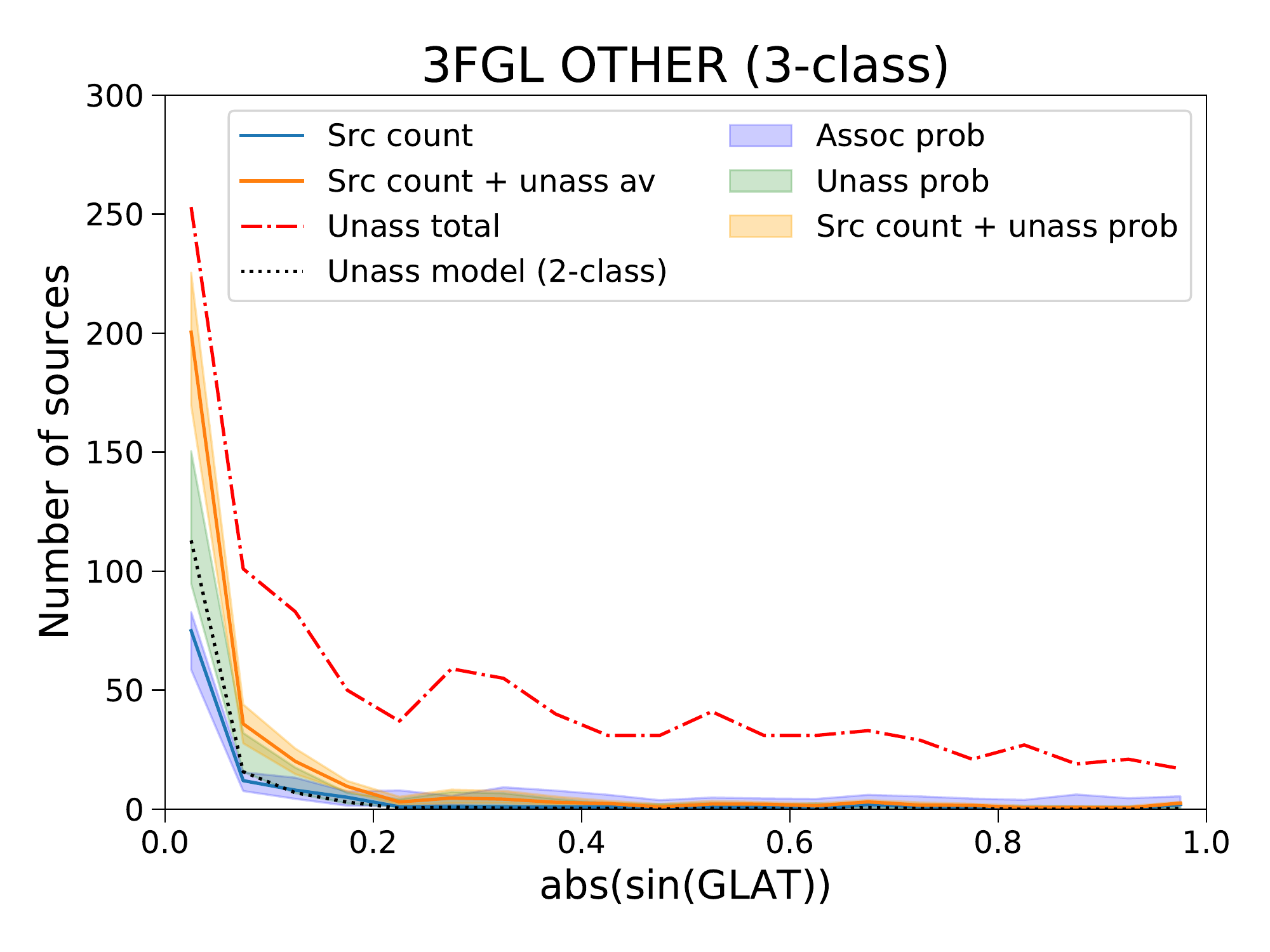}
\includegraphics[width=0.45\textwidth]{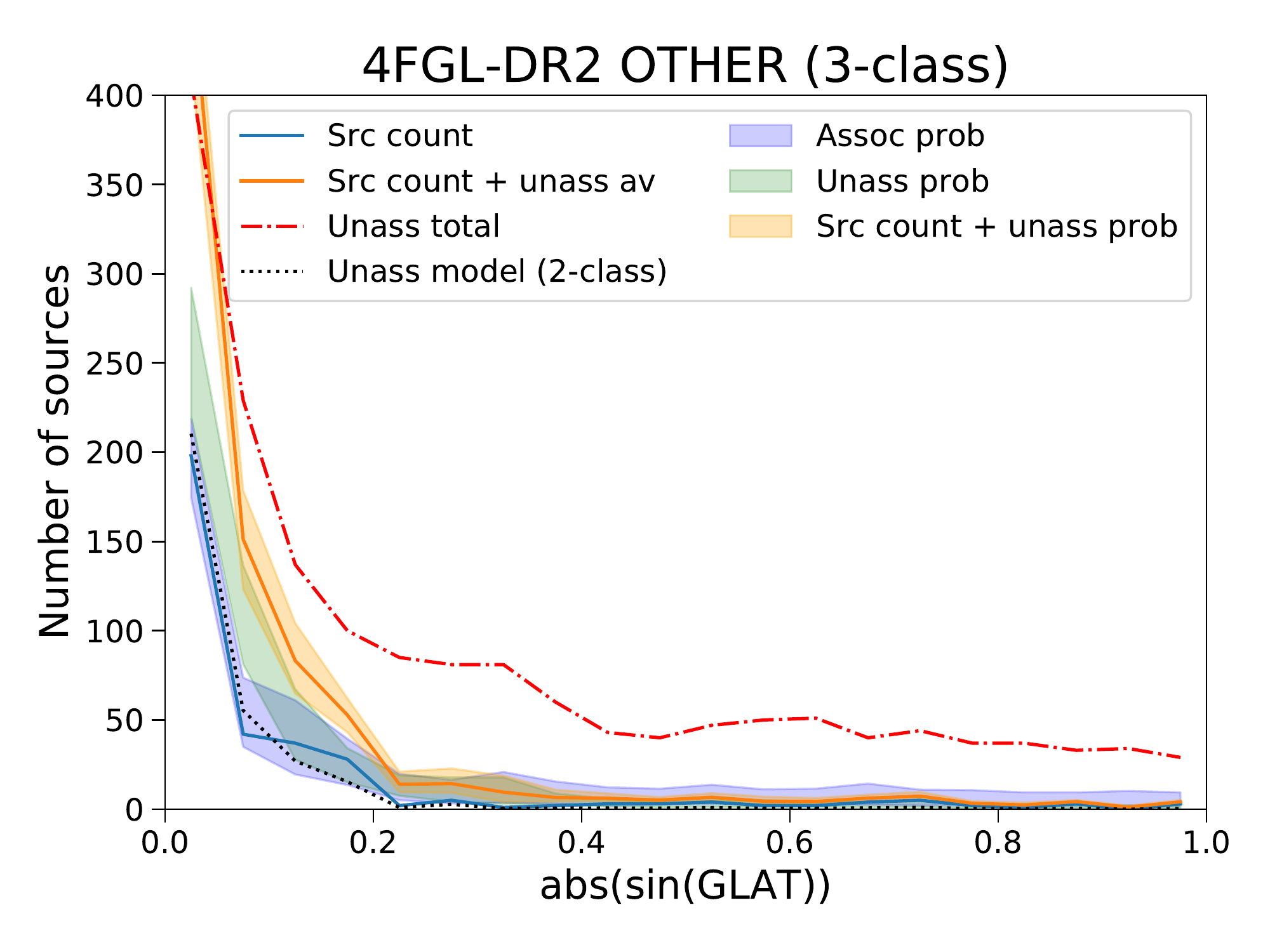} \\ 
\includegraphics[width=0.45\textwidth]{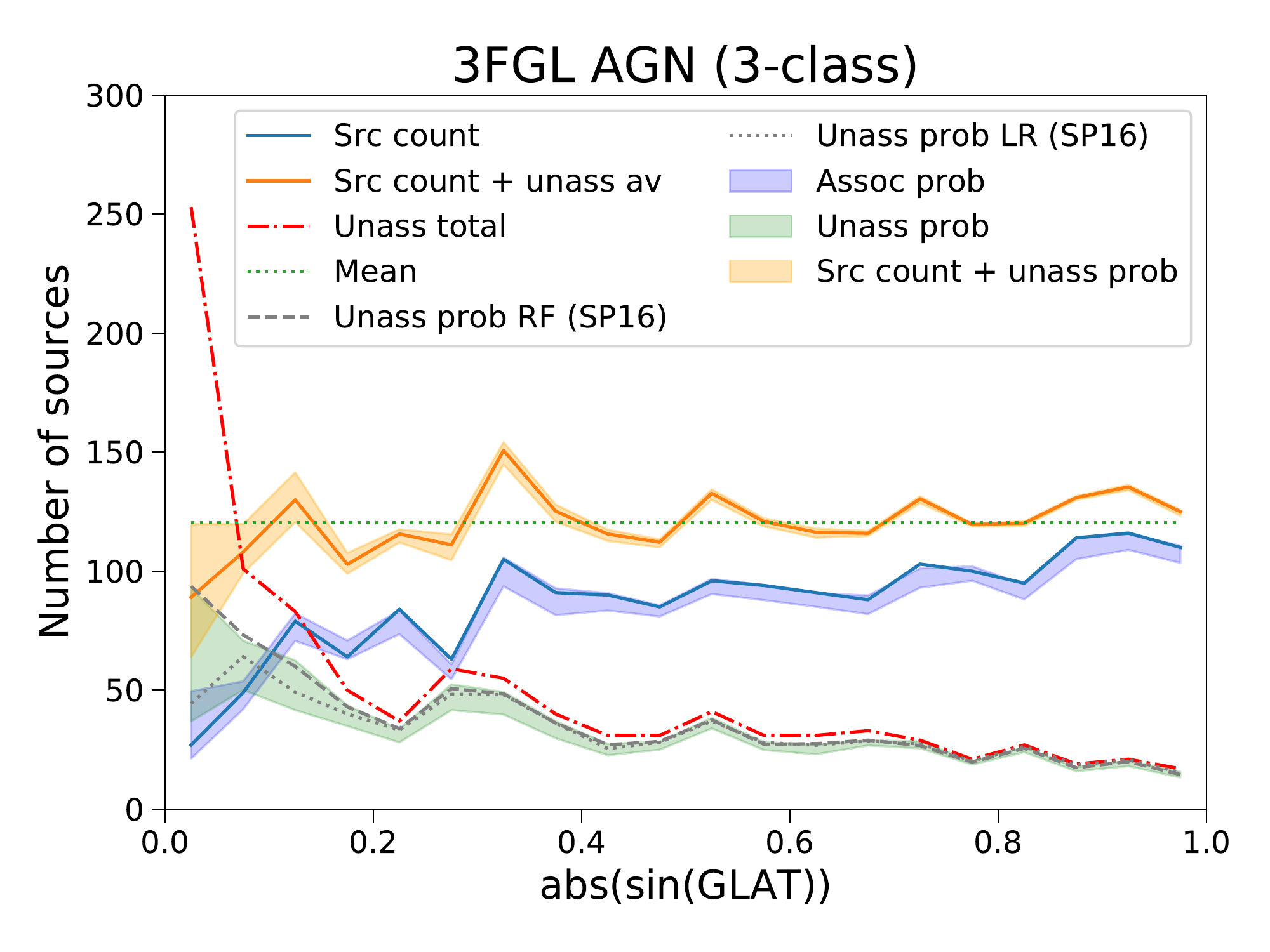}
\includegraphics[width=0.45\textwidth]{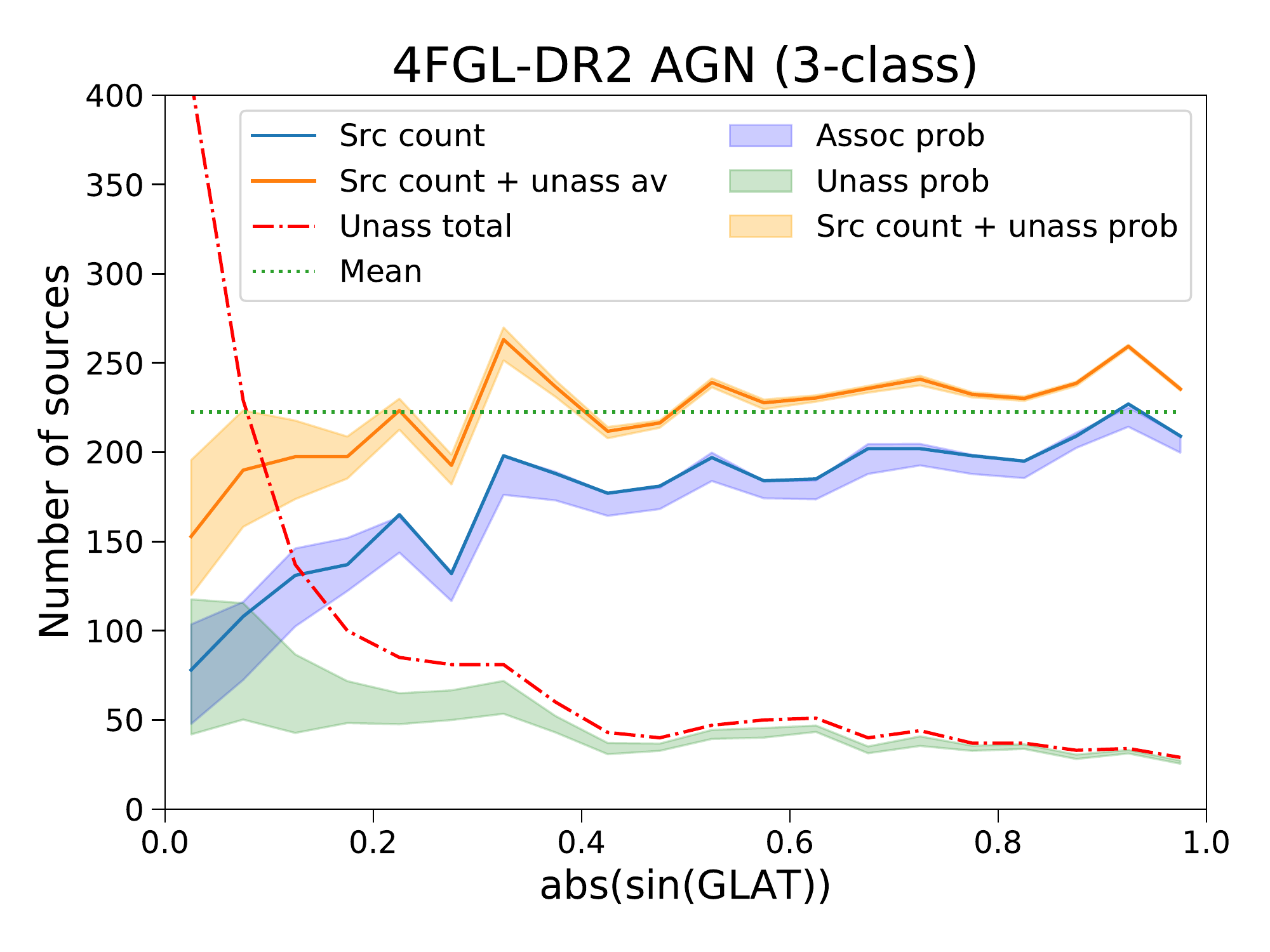}
\caption{Latitude profiles of source counts in the case of three-class classification (for the definition of the labels, see Fig. \ref{fig:lat_profile}).
The black dashed lines in the plots of the OTHER class show the model for the number of OTHER sources among the unassociated ones
in Eq. (\ref{eq:unas_other}) for latitude bins.}
\label{fig:lat_profile_3class}
\end{figure*}

\begin{figure*}[h]
\centering
\includegraphics[width=0.45\textwidth]{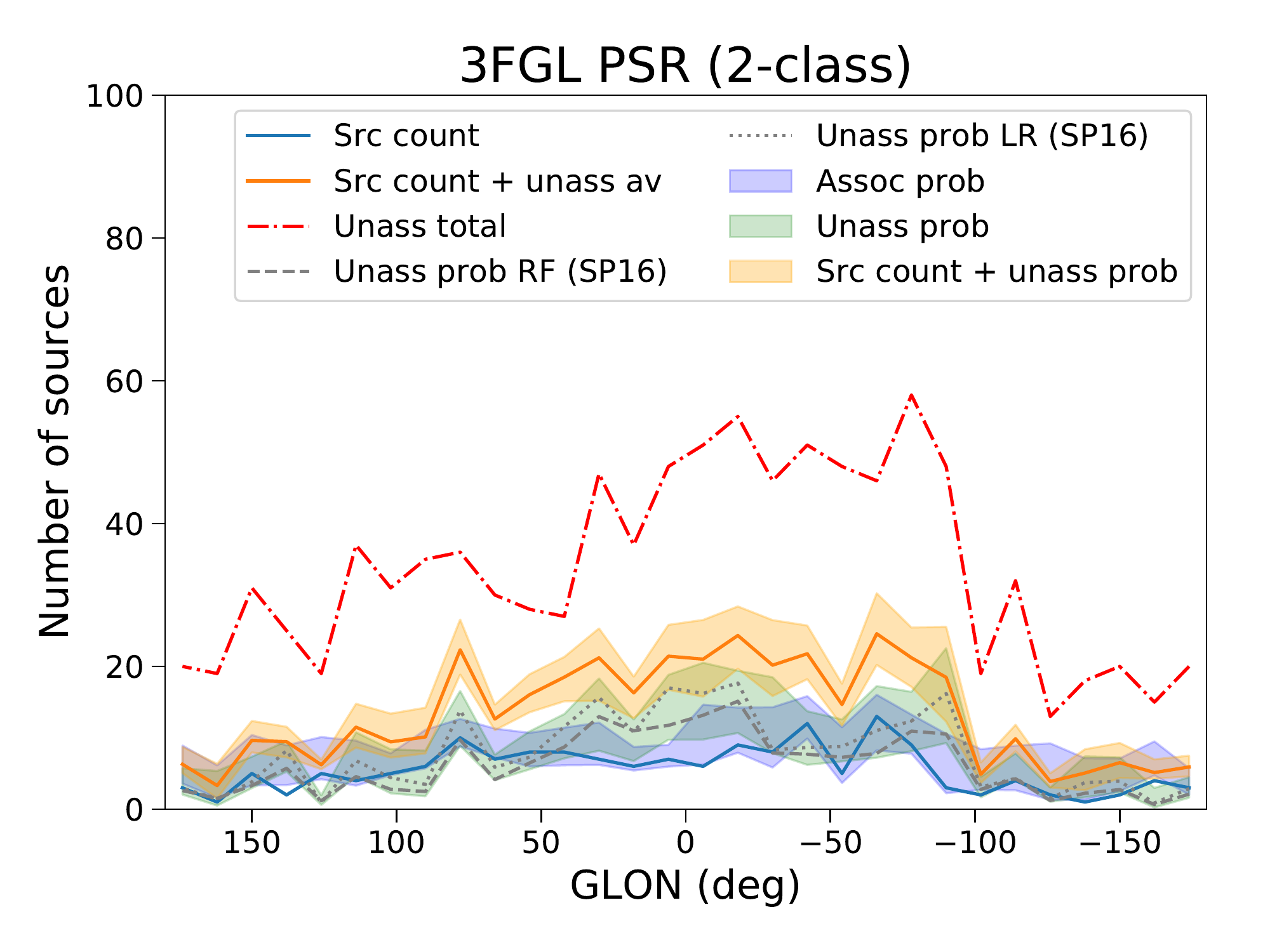}
\includegraphics[width=0.45\textwidth]{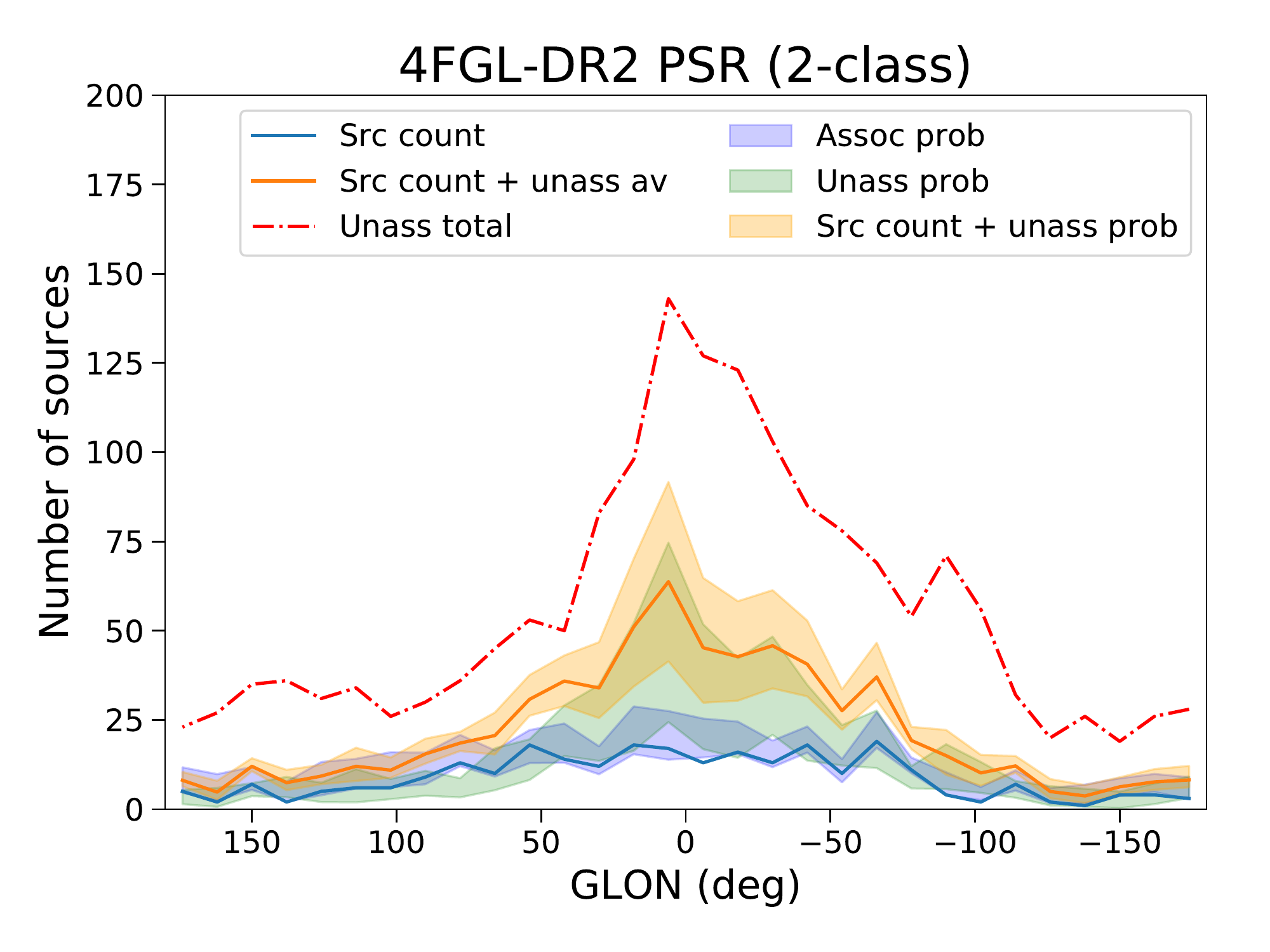} \\
\includegraphics[width=0.45\textwidth]{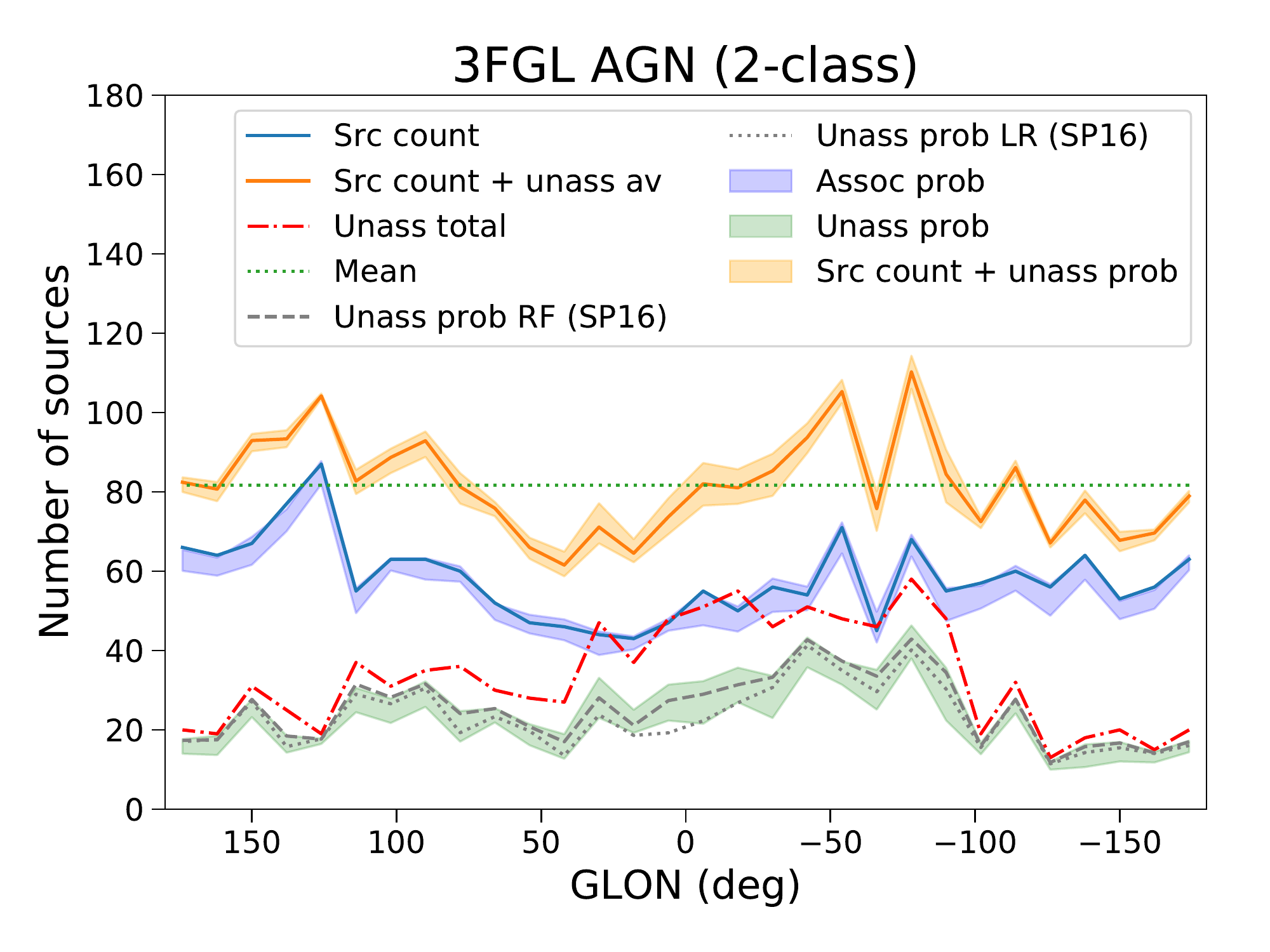}
\includegraphics[width=0.45\textwidth]{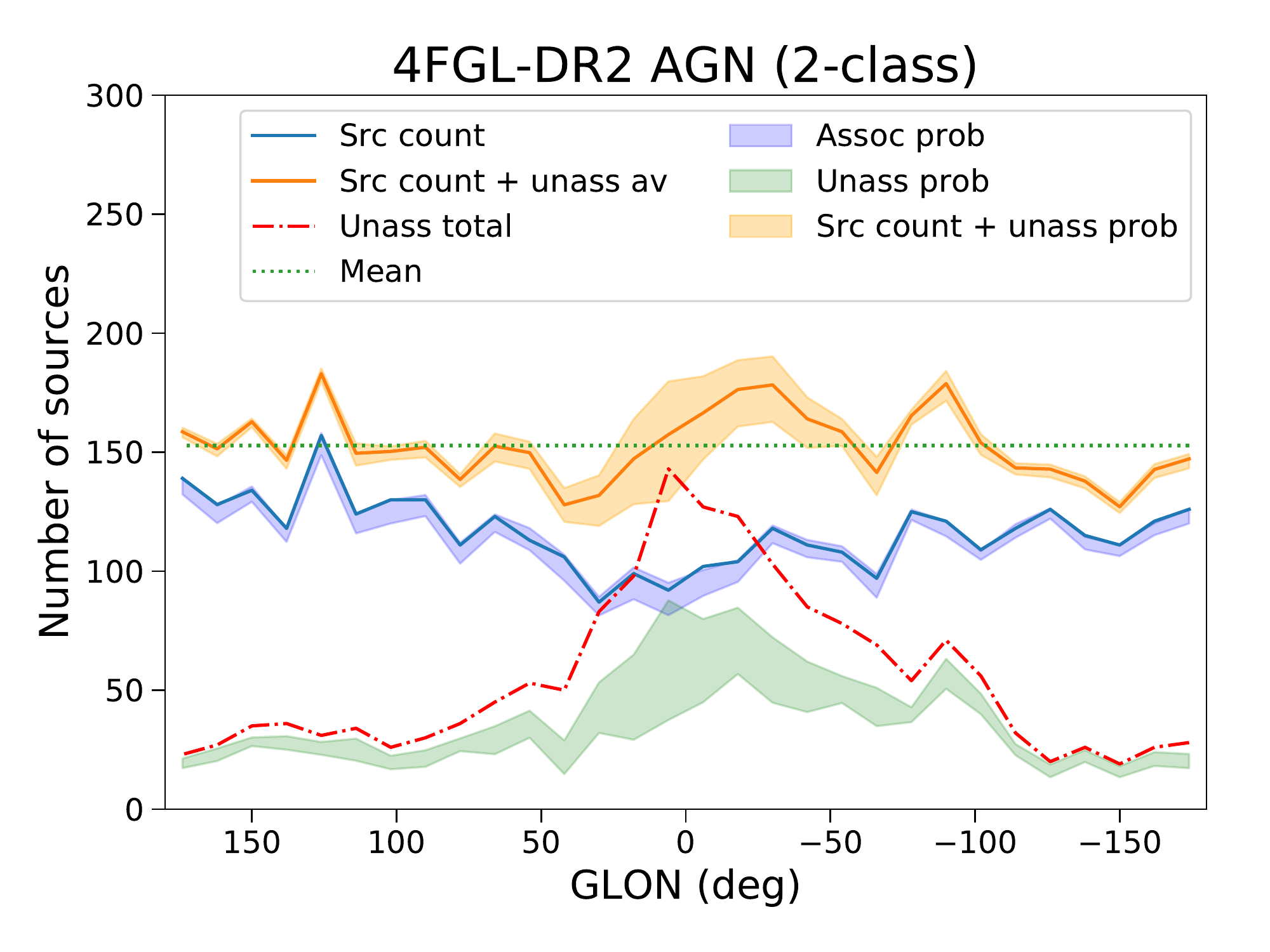}
\caption{Longitude profiles of source counts in the case of two-class classification (for the definition of the labels, see Fig. \ref{fig:lat_profile}).}  
\label{fig:lon_profile}
\end{figure*}

\begin{figure*}[h]
\centering
\includegraphics[width=0.45\textwidth]{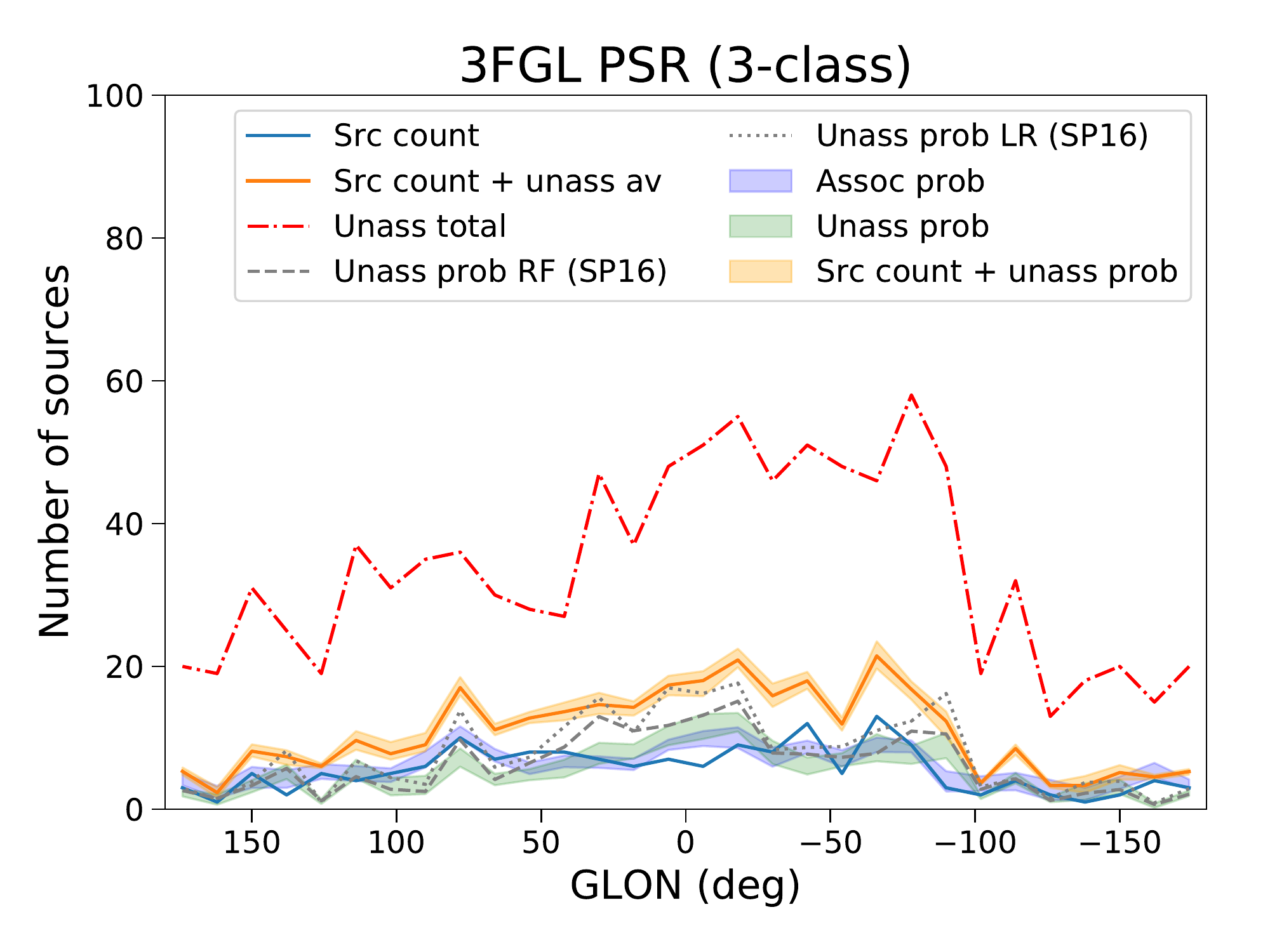}
\includegraphics[width=0.45\textwidth]{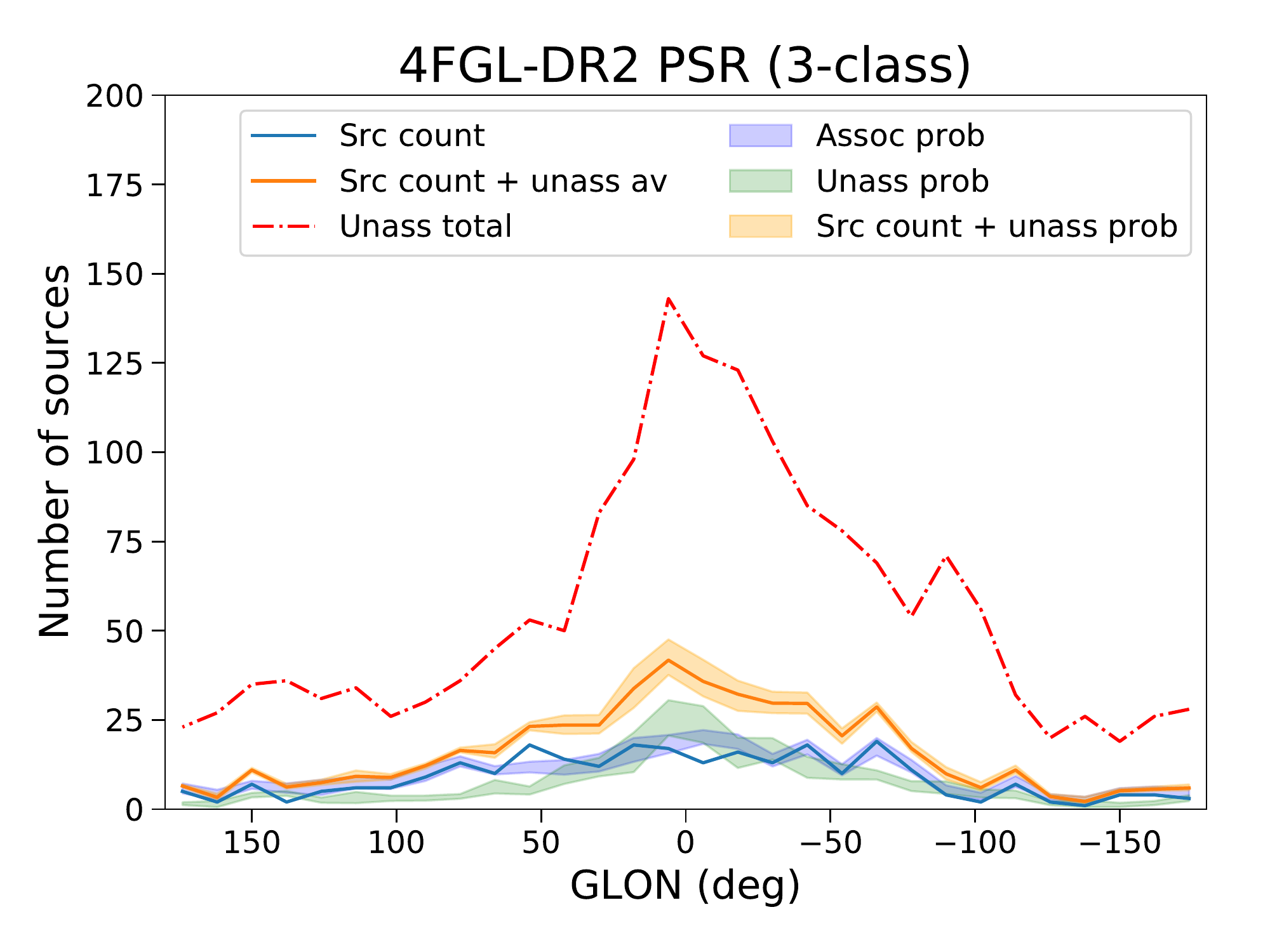} \\
\includegraphics[width=0.45\textwidth]{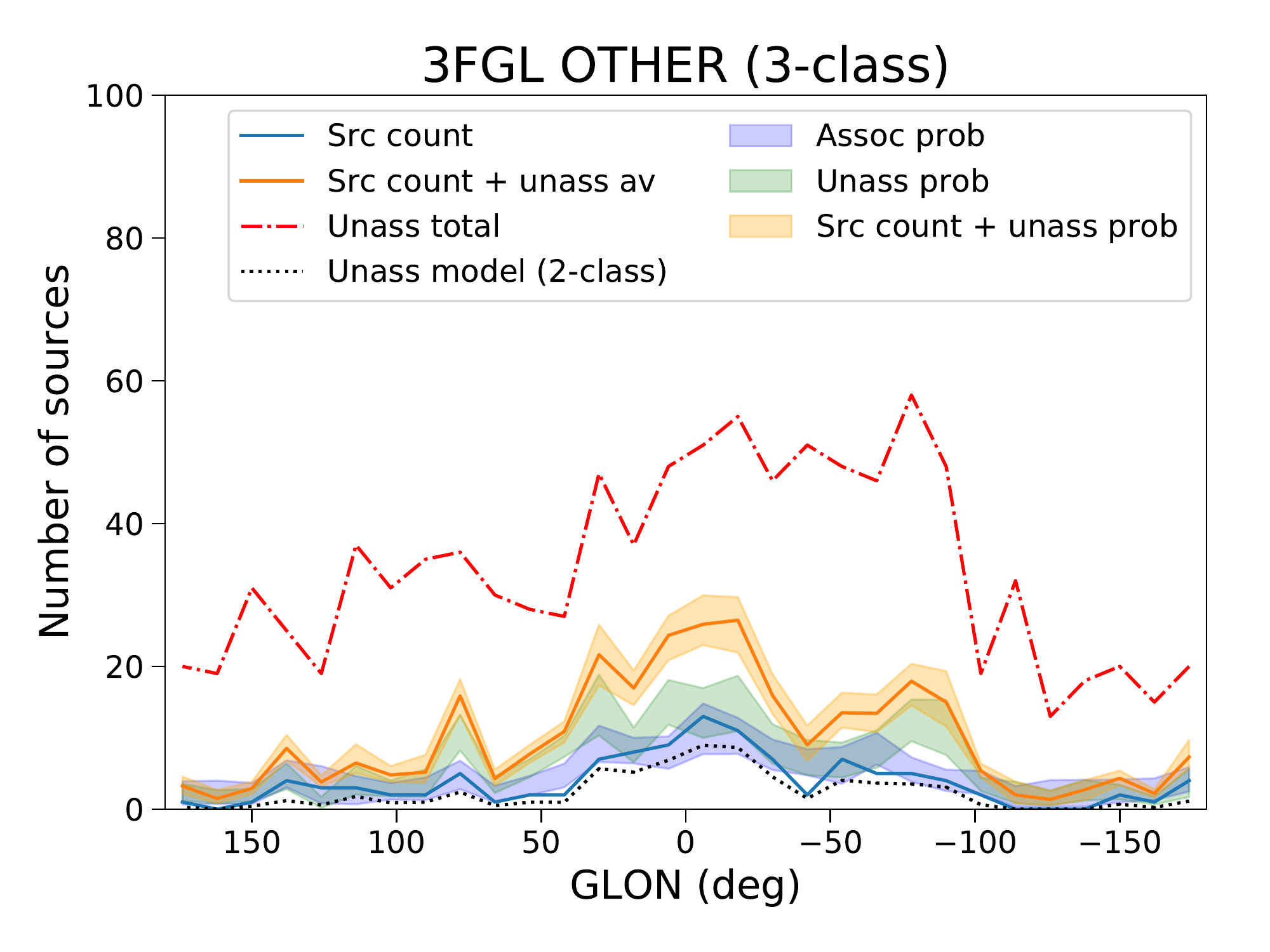}
\includegraphics[width=0.45\textwidth]{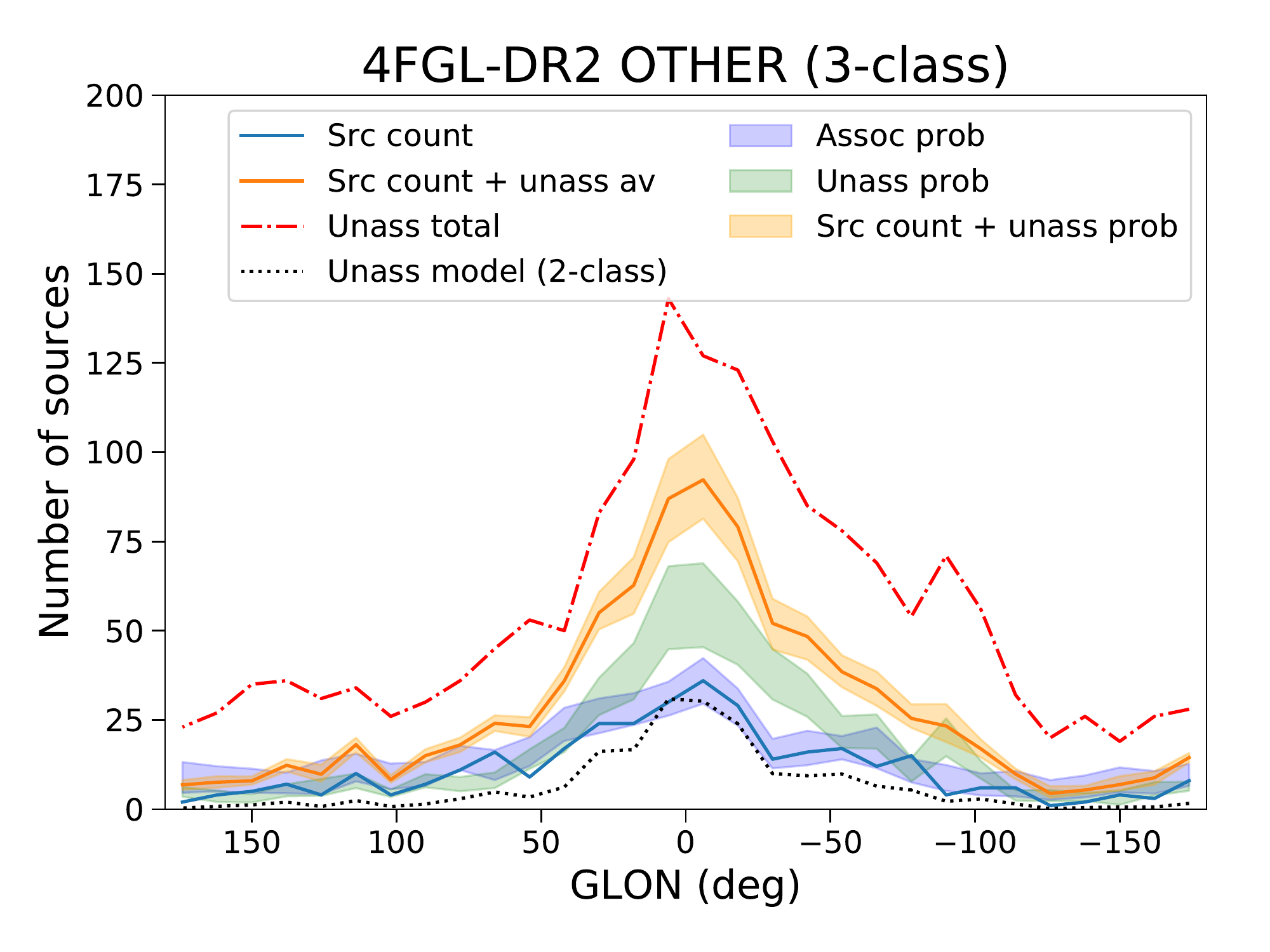} \\
\includegraphics[width=0.45\textwidth]{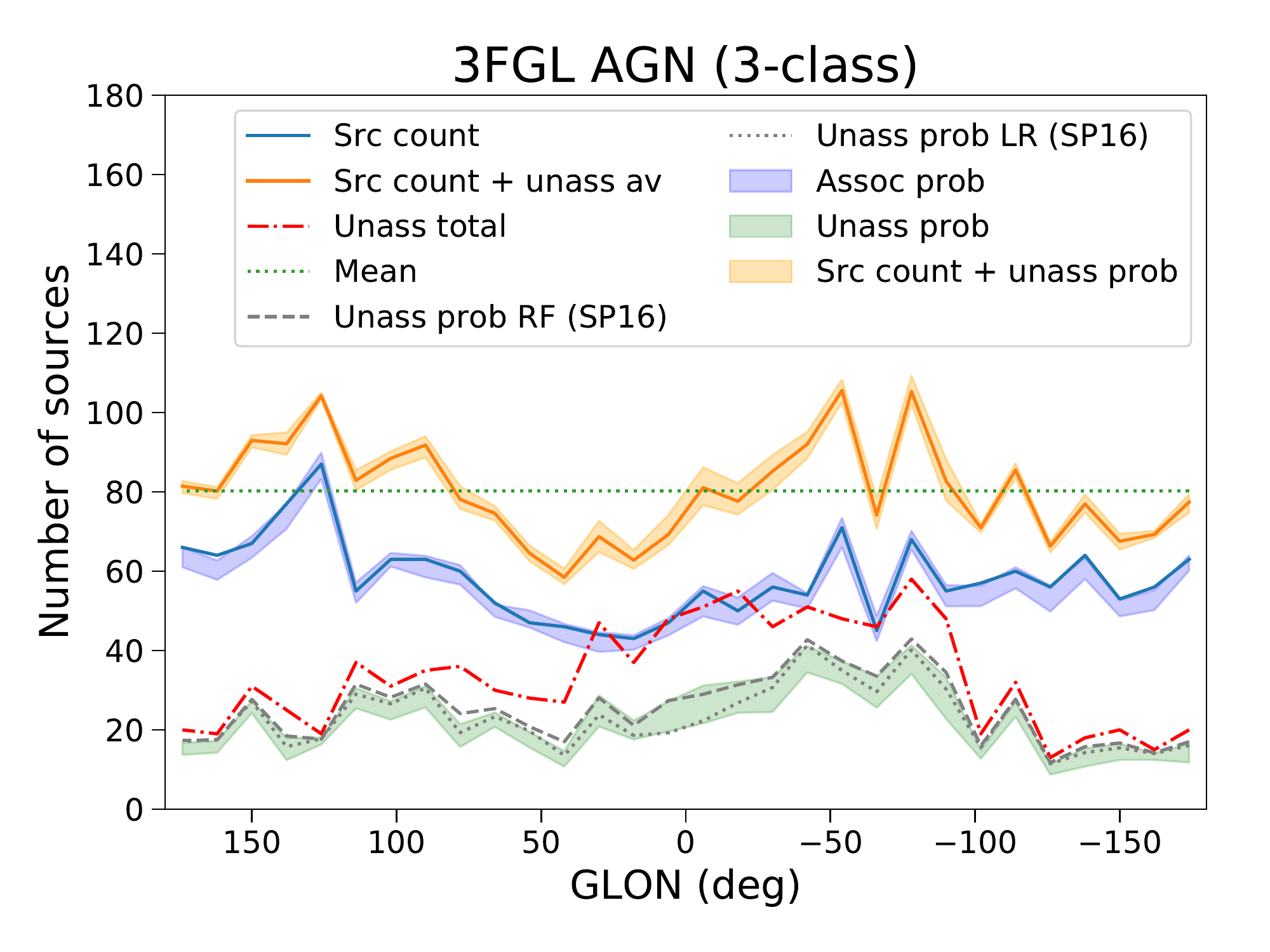}
\includegraphics[width=0.45\textwidth]{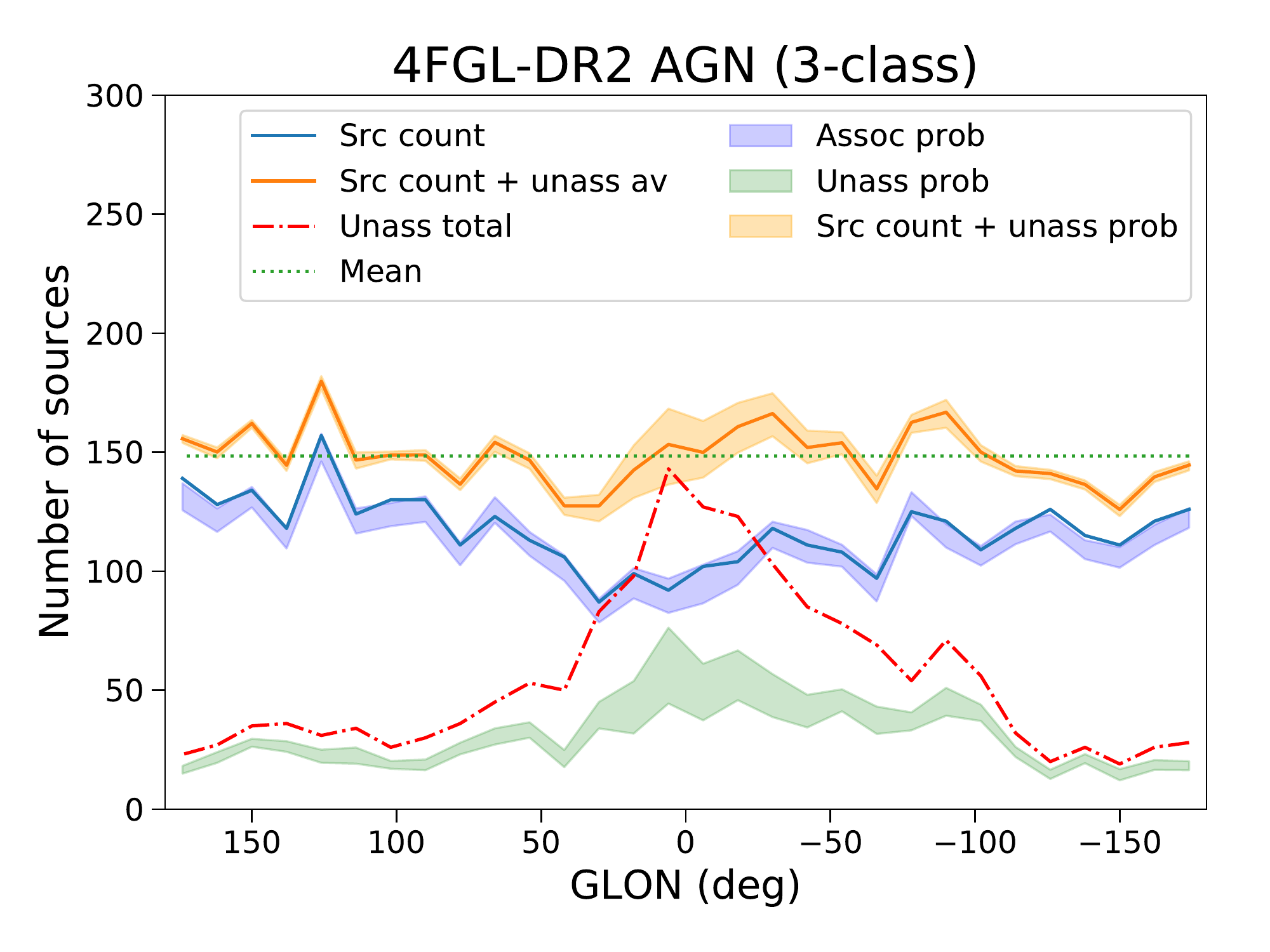}
\caption{Longitude profiles of source counts in the case of three-class classification (for the definition of the labels, see Figs. \ref{fig:lat_profile} and \ref{fig:lat_profile_3class}).}  
\label{fig:lon_profile_3class}
\end{figure*}

In this section we show the Galactic latitude and longitude profiles of the distributions of associated and unassociated sources.
In Figs. \ref{fig:lat_profile} and \ref{fig:lat_profile_3class} we present the source counts as a function of ${\rm abs(sin(GLAT))}$ 
for two-class and three-class classifications, respectively.
We used 20 bins (i.e., each bin corresponds to a solid angle of $4 \pi / 20$). 
Solid blue lines show counts of associated sources in 3FGL and 4FGL-DR2  catalogs.
The total counts of unassociated sources are shown by red dash-dotted lines.
Green bands show the envelopes of sums of probabilities for classification of unassociated sources into AGN, pulsar, and, in case of three classes, OTHER classes for the eight ML methods corrected in the case of two classes for the presence of OTHER sources.
Black dotted line in plots of the OTHER class show the model for the number of OTHER sources among the unassociated ones
in Eq. (\ref{eq:unas_other}) for latitude bins.
We see that in the latitude profile, the two-class model for the contribution of OTHER sources to the unassociated ones is generally consistent with the estimate of the number of OTHER sources in the three-class model.

The classifications of unassociated 3FGL sources by \cite{2016ApJ...820....8S} are shown by gray dashed (RF) and dotted (LR) lines.
The numbers of unassociated sources classified as AGN, pulsar, or OTHER grow toward the Galactic plane (GP).
Within $\approx 3^\circ$ from the GP the expected number of pulsars is about the same as the number of AGNs among unassociated sources, while at high latitudes, most of the unassociated sources are classified as AGNs.
It is interesting to note that, according to Table \ref{tab:feat_imp}, GLAT is one of the least important features for the RF and BDT algorithms.
It can be a posteriori explained by the fact that the density of AGNs is so high that even in the GP the expected number of AGNs is comparable to the expected number of pulsars.

Orange shaded areas show the sum of the source counts and the expected number of sources for the eight methods (both with and without oversampling).
The average among the eight methods added to the counts of associated sources is shown by solid orange line 
(for AGNs we also show the mean of these points by dotted green line).
We find that the number of associated AGNs is decreasing toward the GP, the expected number of AGNs among unassociated sources is increasing toward the GP, so that the sum of the two is relatively uniform as a function of Galactic latitude.

In Figs. \ref{fig:lon_profile} and \ref{fig:lon_profile_3class} we show plots analogous to Figs. \ref{fig:lat_profile} and \ref{fig:lat_profile_3class} for Galactic longitudes (we use 30 bins).
We note that there is a significant increase in the number of unassociated sources in the 4FGL-DR2  catalog for $|\ell | \lesssim 50^\circ$.
In the two-class classification about half of the unassociated sources at these longitudes are attributed to pulsars and half to AGNs.
In the three-class classification at least one-third of the unassociated sources in this range of longitudes is attributed to the OTHER class.
This comes mostly at the expense of reducing the predicted number of pulsars.
It is interesting to note that the two-class model for the OTHER sources calculated using Eq. (\ref{eq:unas_other}) for longitude bins
(black dotted line in the OTHER plots) is significantly below the bands of three-class expectations, whereas for the latitude bins the two-class model
is consistent with the bands.
The reason is that the ratio of unassociated with associated sources, which we use to estimate the number of OTHER sources among the unassociated ones, depends on binning.
In particular, this ratio is about 1 for low latitudes, which leads to an estimate that at low latitudes the number of OTHER sources among unassociated ones is similar to the number of associated OTHER sources.
But for longitude and flux binning the ratio of unassociated with associated sources is smaller than 1, which leads to a prediction (in the two-class case) that the number of OTHER sources among unassociated ones is smaller than the number of associated OTHER sources.

We summarize the expected numbers of sources in the two-class and three-class models in Table \ref{tab:expected_counts_prob_3FGL} for the 3FGL and in Table \ref{tab:expected_counts_prob_4FGL-DR2} for the 4FGL-DR2 catalogs.
In the two-class case we also show the correction for the presence of OTHER sources among unassociated ones.
We make the correction using the total numbers of unassociated and associated sources, as well as binning in flux from Fig. \ref{fig:logN_logS} (F-bins), in latitudes from Fig. \ref{fig:lat_profile} (Lat-bins), and in longitudes from Fig. \ref{fig:lon_profile} (Lon-bins).
We see that predictions for the number of sources using latitude binning in the two-class case is closest to the three-class case.
Even better agreement would likely be achieved if one uses a simultaneous binning in flux, latitudes, longitudes and, possibly, other variables.
If, in addition, one allows bins of variable size and non-rectangular boundaries in multidimensional space, then one will likely end up with one of the ML algorithms for classification.
Indeed, ML algorithms are designed to provide an optimal binning, which maximizes the separation of classes and makes predictions for data with unknown labels (e.g., unassociated sources) based on counts of samples in bins with known labels (e.g., associated sources).
Thus the three-class prediction for classification of unassociated sources is likely more accurate than the two-class classification with correction for OTHER sources based on ad hoc binning of one of the variables.
The two-class classification may still be useful in situations when high recall is necessary for either pulsars or AGNs:
since OTHER sources are mixed with pulsars and AGNs (cf. Fig. \ref{fig:RF_domains_3class}), 
some of pulsars and AGNs can be mistakenly classified as OTHER in the three-class case.

\begin{table}[!h]
\caption{Expected counts of sources among unassociated 3FGL sources.}
\label{tab:expected_counts_prob_3FGL}

\centering
\resizebox{0.47\textwidth}{!}{
\begin{tabular}{l c c c }
\hline
\hline
Classification &  AGN & PSR & OTHER \\
\hline
two-class  & $740.3^{+75.8}_{-97.6}$ & $269.7^{+97.6}_{-75.8}$ & --  \\
two-class corr total  & $711.6^{+70.6}_{-90.9}$ & $242.0^{+90.9}_{-70.6}$ & 56.4 \\
two-class corr F-bins  & $717.3^{+70.8}_{-92.7}$ & $251.5^{+92.7}_{-70.8}$ & 41.1 \\
two-class corr Lat-bins  & $670.2^{+64.9}_{-80.9}$ & $196.5^{+80.9}_{-64.9}$ & 143.3 \\
two-class corr Lon-bins   & $705.7^{+68.3}_{-89.8}$ & $234.8^{+89.8}_{-68.3}$ & 69.4 \\
three-class   &$664.0^{+74.9}_{-65.2}$ & $158.3^{+22.3}_{-23.6}$ & $187.6^{+46.4}_{-51.3}$ \\
\hline
\end{tabular}}
\tablefoot{
The ``two-class corr total (F-bins, Lat-bins, Lon-bins)'' row shows the two-class expectations corrected for the presence of OTHER sources
using the ratio of unassociated and associated OTHER sources for the total counts (using flux, latitude, or longitude bins)
in Eq. (\ref{eq:unassoc_ev}).
The number of OTHER sources in the two-class case is estimated in Eq. (\ref{eq:unas_other}).}
\end{table}

\begin{table}[!h]
\caption{Expected counts of sources among unassociated 4FGL-DR2 sources.}
\label{tab:expected_counts_prob_4FGL-DR2}

\centering
\resizebox{0.47\textwidth}{!}{
\begin{tabular}{l c c c }
\hline
\hline
Classification &  AGN & PSR & OTHER \\
\hline
two-class  & $1184.4^{+174.3}_{-244.9}$ & $482.6^{+244.9}_{-174.3}$ & --  \\
two-class corr total  & $1104.2^{+160.0}_{-221.5}$ &$422.4^{+221.5}_{-160.0}$ & 140.4 \\
two-class corr F-bins  & $1103.4^{+159.3}_{-223.3}$ & $433.5^{+223.3}_{-159.3}$ & 128.1  \\
two-class corr Lat-bins  & $1013.2^{+142.6}_{-191.0}$ &$334.6^{+191.0}_{-142.6}$ & 319.2 \\
two-class corr Lon-bins   & $1081.0^{+155.3}_{-211.6}$ & $390.0^{+211.6}_{-155.3}$ & 196.0 \\
three-class   & $948.0^{+122.7}_{-135.8}$ &$214.9^{+47.0}_{-41.8}$ & $504.1^{+88.9}_{-102.6}$ \\
\hline
\end{tabular}}
\tablefoot{
For the description of rows, see Table \ref{tab:expected_counts_prob_3FGL}.}
\end{table}

\section{Conclusions}
\lb{sec:conclusions}

In this paper we determined the probabilities of classification of sources in the 3FGL and 4FGL-DR2 \Fermi-LAT catalogs
into two (AGNs and pulsars) and three (AGNs, pulsars, and OTHER) classes.
The probabilities were calculated with eight different ML methods: RF, BDT, LR, and NN -- each algorithm with and without oversampling during training.
The algorithms were trained and tested with associated sources.
We optimized the meta-parameters of the algorithms, such as the depth of the trees, the number of trees, the number of neurons, and so on, to avoid any overfitting of data while providing a good accuracy of classification.

The testing accuracies of the classification of associated sources in the 3FGL catalog for the four algorithms in the two-class case without (with) oversampling are about  97\% (between 93\% and 97\%).
We also checked the accuracy of the classification by selecting the unassociated sources in 3FGL that have associations in 4FGL-DR2.
If we take the 4FGL-DR2 associations as the true classes, then the accuracies of the two-cass classification in this subset of sources 
without (with) oversampling are between 90\% and 91\% (85\% and 92\%).
We find that most of the misclassified sources in this comparison have spectral parameters in 3FGL that are typical of the opposite class 
(Fig. \ref{fig:3FGL_vs_4FGL_classes}); in other words, the misclassification may be due to problems with reconstructing the spectrum of the sources.

In the three-class classification, the testing accuracies for the 3FGL catalog are between 92\% and 94\% (both with and without oversampling), while comparison with 4FGL-DR2 gives accuracies between 82\% and 85\%. For the 4FGL-DR2 classifications, the testing accuracy is between 90\% and 93\% (both with and without oversampling). If one takes into account that all OTHER sources are misclassified in the two-class case, then the three-class case provides an improvement in accuracy of 1\% to 5\%.

We created four catalogs with probabilistic classifications of sources: based on 3FGL and on 4FGL-DR2 with two- and three-class classifications.
For each source and for each class we report class probabilities for each of the eight ML methods (with and without oversampling). 
We also provide individual standard deviations for all classification probabilities by using sample average over selection of training and testing data sets.
We report the classification probabilities not only for the unassociated sources, but also for the associated ones, which can be used to find outliers.
The full probabilistic catalogs for two- and three-class classifications of the 3FGL and 4FGL-DR2 catalogs are available online \citep{SOM_material}.
An advantage of such a probabilistic classification is that a threshold on probability for selecting, for example, pulsar candidates can be chosen by the user based on their needs.
For example, in a search for new pulsars, one can select a low threshold in order to avoid missing possible pulsars.
In a derivation of an average property of the class, such as spectral index or cutoff energy, one can select a high threshold in order to avoid contamination from the other classes.

We discuss two examples of applications of probabilistic catalogs: determination of the most likely classes of unassociated sources (which can be used for searches of new class members, such as AGNs or pulsars) and population studies using class probabilities that include both associated and unassociated sources.
For the determination of the candidate classes of unassociated sources, we used two conditions: agreement among algorithms (i.e., each algorithm predicts the same class as the most likely for a source) and that the sum of probabilities for the eight classification methods is above 7.
In order to evaluate the performance of these classification conditions, we estimated the precision and recall using associated sources and tested the estimations using unassociated 3FGL sources that have associations in the 4FGL-DR2 catalog.
Both precision and recall for the classification of unassociated 3FGL sources with 4FGL-DR2 associations 
are smaller than the estimates based on testing samples.
For instance, the precision for AGNs (pulsars) estimated from the comparison of the 3FGL and 4FGL-DR2 catalogs is about 93\% (68\%),
compared to the precision of 97\% (78\%) determined from the testing samples for the two-class classification of 3FGL sources (Table~\ref{tab:3FGL_vs_4FGL_2class}).
A similar reduction in precision is observed for the three-class classification of the 3FGL sources.
We notice that many misclassified sources in the 3FGL versus 4FGL-DR2 comparison (Fig. \ref{fig:3FGL_vs_4FGL_classes}) were outside of the expected class domains when 4FGL-DR2 features were used instead of the 3FGL features. This shows that in many cases such a misclassification is due to errors or uncertainties in the input data rather than issues in the classification methods. Such errors in the input data provide an irreducible uncertainty in the analysis. As a result, we find that precision and recall determined from the comparison of the 3FGL and 4FGL-DR2 catalogs provide a more realistic estimate of the true precision and recall than the values based on testing samples.
Although a comparison of the 4FGL-DR2 catalog with a newer \Fermi-LAT catalog is not possible at the moment, 
a similar reduction in precision and recall compared to the values determined from testing samples in the 4FGL-DR2 case can be expected.

We used the all-algorithms-agree method for the general assignment of the candidate classes to the sources (for convenience, we include these classes in the catalogs) and also to create a list of 29 pulsar candidates based on unassociated sources classified as pulsars by all algorithms in both the 3FGL and 4FGL-DR2 catalogs using two-class classification.
We find that the expected precision for the ``sum of probabilities above 7'' condition is larger than the ``all-methods-agreement'' condition (albeit at the expense of generally smaller recall), and thus the sum of probabilities above 7 condition is stricter.
In particular, all pulsar candidate sources in 3FGL satisfying this condition are associated with pulsars in 4FGL-DR2, while there are no OTHER sources in 3FGL satisfying this condition. In the three-class classification based on the 4FGL-DR2 catalog, there are six pulsar candidates and 30 OTHER source candidates among unassociated sources satisfying  the sum of probabilities above 7 condition. 
We report these sources in Tables \ref{tab:psr_candidates_3class} and \ref{tab:other_candidates_3class}, respectively (digital versions of the tables are available in the supplementary online materials; \citealt{SOM_material}). We discuss possible associations of the pulsar candidates with pulsars from the Parkes survey \citep{Camilo2015} and OTHER source candidates with sources in the SIMBAD database.

As a second example of the application of the probabilistic catalogs, we performed population studies using class probabilities for both associated and unassociated sources.
In particular, we derived the expected number of sources in the catalog as a function of their flux.
As a consistency check, we compared the counts of associated sources to the sums of probabilities for the associated sources.
We find that correcting for the contribution of OTHER sources in the two-class case plays an important role in the estimation of the expected number of sources in a particular class.
We found the total expected number of AGNs and pulsars in the 3FGL and 4FGL catalogs by adding the class probabilities for the unassociated sources in the two- and three-class cases to the source counts of associated sources and correcting in the two-class case for the contribution of other classes in the unassociated sources.
In particular, we find that the total expected number of pulsars is about two times larger than the number of associated pulsars.

We also plotted the counts of associated sources and the expected numbers of AGNs, pulsars, and OTHER sources among unassociated sources
as a function of Galactic latitude and longitude.
We find that the number of associated AGNs decreases toward low latitudes, while the expected number of AGNs among unassociated sources increases; therefore, the sum of the two is relatively uniform, as expected for extragalactic sources.

We perform the checks of the classification probabilities using reliability diagrams in Appendix \ref{sec:reliability}. 
We find that the performance of the three-class classification is similar to the two-class classification if we only take AGNs and pulsars in the two-class case into account. For all associated sources, the two-class classification overestimates the number of pulsars due to the presence of the
OTHER sources, while in the three-class classification case the OTHER sources are included in the model and the reliability diagrams are reasonably close to the perfect calibration line.

\subsection*{Acknowledgements}

The authors would like to thank Jean Ballet, Isabelle Grenier, Pablo Saz Parkinson, and the anonymous referee for valuable comments and suggestions.
The work of DM was in part supported by BMBF under the ErUM-Data project ``Innovative Digital Technologies for Research on Universe and Matter'' (grant number 05H18WERC1) and by DFG grant MA 8279/2-1.
We would like to acknowledge the use of the following software:
Astropy \citep[\url{http://www.astropy.org},][]{2013A&A...558A..33A}, 
matplotlib \citep{Hunter:2007}, 
scikit-learn [\url{https://scikit-learn.org/stable/about.html}], 
TOPCAT \citep{2005ASPC..347...29T}, and Imbalanced-learn \cite{JMLR:v18:16-365}.
This research has also made use of the SIMBAD database,
operated at CDS, Strasbourg, France \citep{2000A&AS..143....9W}.

\bibliography{ML_3FGL_papers}  

\begin{appendix}
\section{Tests of additional meta-parameters}
\lb{sec:app}

In this appendix we discuss tests of some meta-parameters, which had a relatively little effect on the 
accuracy of the algorithms. For these tests we used the two-class classification in the 3FGL catalog.

\begin{figure}[h]
\centering
\includegraphics[width=\twopicsp\textwidth]{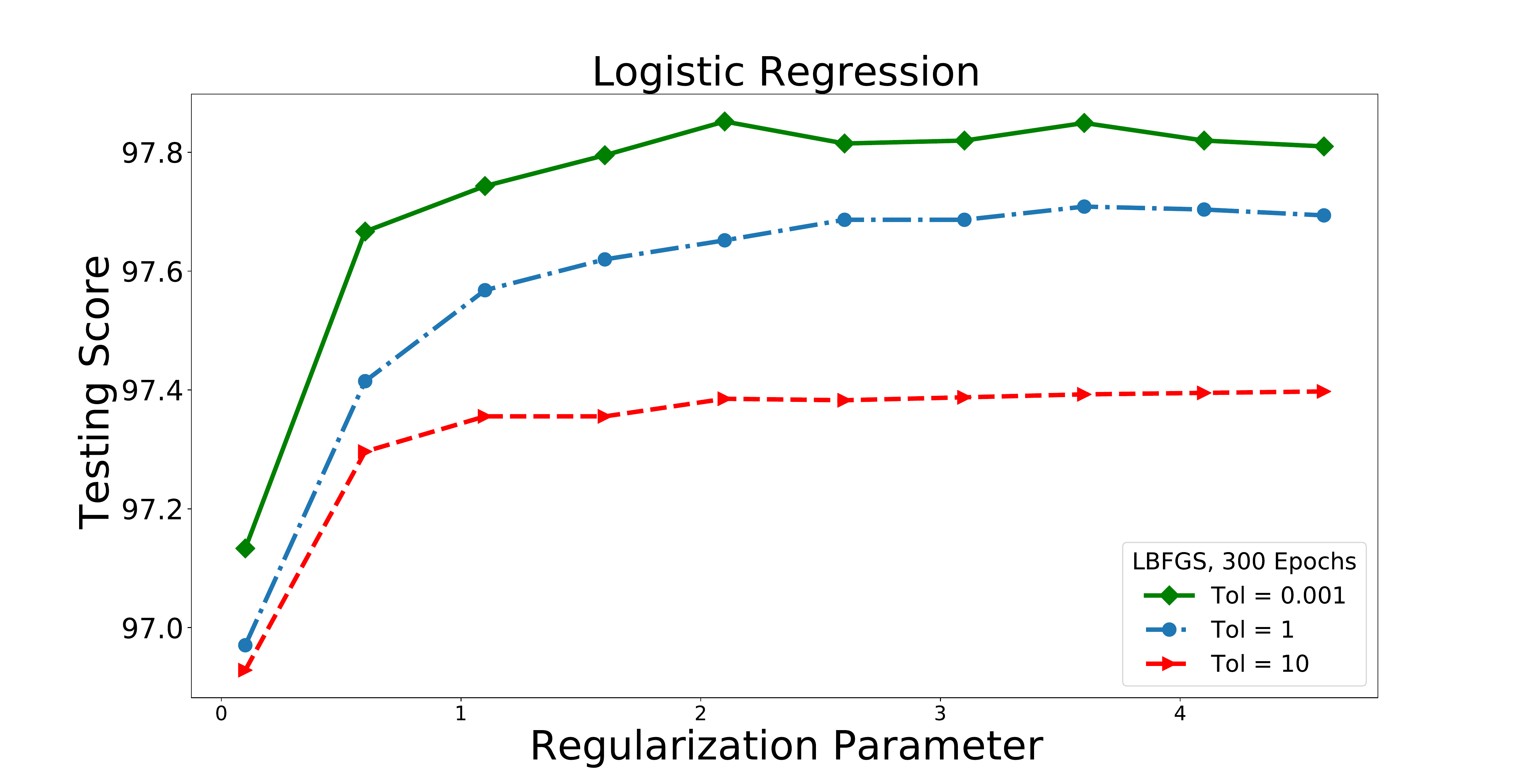}
\caption{Dependence of LR on tolerance and regularization. 
}
\label{fig:LR_tol_reg}
\end{figure}

For the LR algorithm, we tested two additional meta-parameters: regularization and tolerance. 
The effect of the choice of these parameters on accuracy is less than 1\% (Fig. \ref{fig:LR_tol_reg}). 
Therefore, we used the default values for these parameters (tolerance is $10^{-4}$ and regularization parameter is 1)
both in the two-class and in the three-class classifications.

\begin{figure}[h]
\centering
\includegraphics[width=\twopicsp\textwidth]{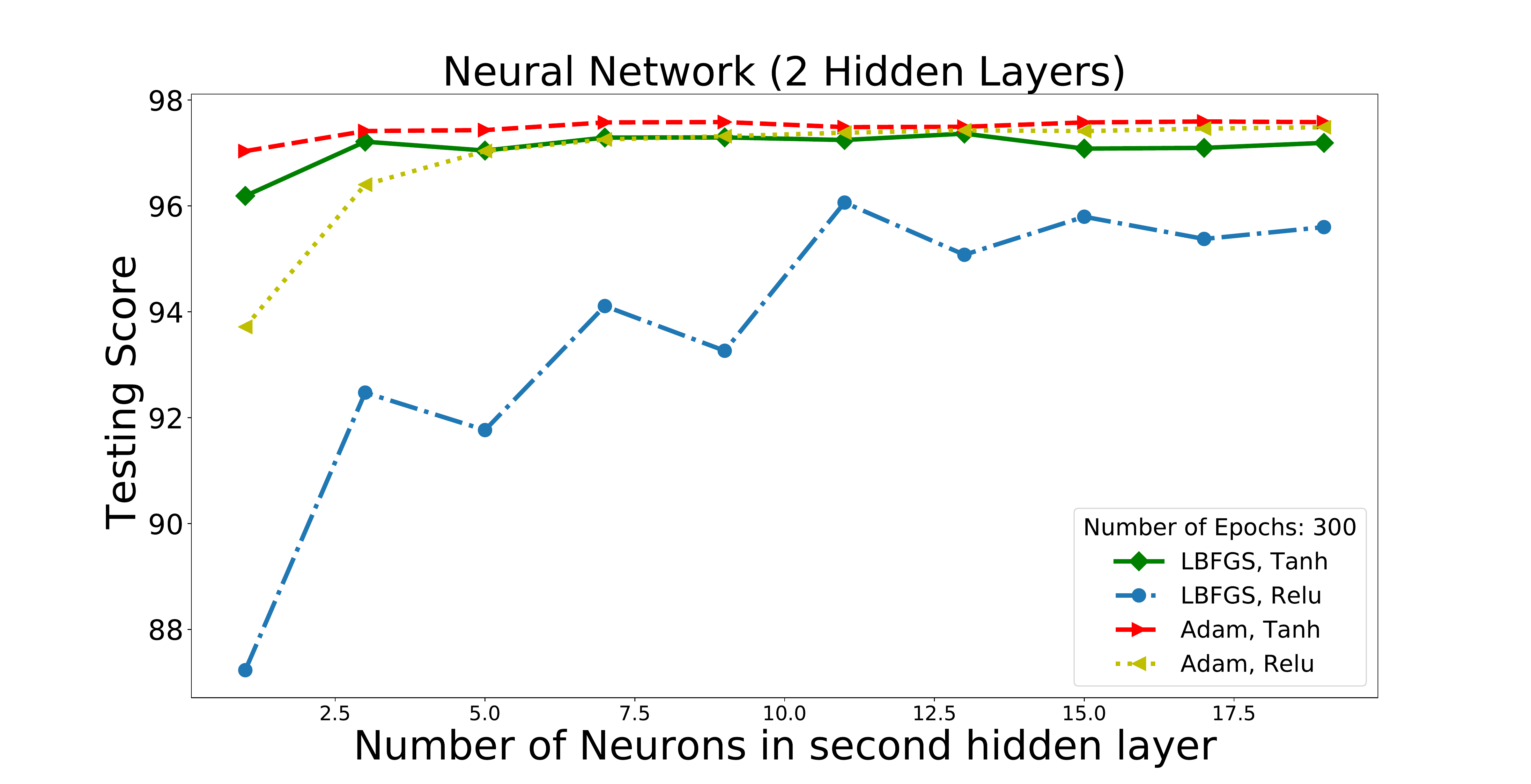}
\caption{Dependence of NN on the number of neurons in the second hidden layer, for 11 neurons in the first hidden layer.
}
\label{fig:nn_nn}
\end{figure}

In Fig. \ref{fig:nn_nn} we show the effect of adding the second hidden layer in the NN algorithm.
The difference between the best accuracies of the NN with one hidden layer (cf. Table \ref{tab:selected_algs})
and  the NN with the additional hidden layer is less than 1\%.

\pgfplotstableread[col sep=comma]{tables/features/3fglassocfeaturesAGNPSRnewfeats.csv}\tablea
\begin{table}[h]
\caption{Statistics of features used for the two-class probabilistic classification of the 3FGL sources.
\label{tab:3FGL_features}}
\centering
\resizebox{0.45\textwidth}{!}{
\pgfplotstabletypeset[
columns={Name,Mean,SD,Minimum,Maximum},
column type=c,
string type,
every head row/.style={before row=\hline\hline,after row=\hline,},
every last row/.append style={after row={\hline} },
every first column/.style={column type/.add={}{}},
every last column/.style={column type/.add={}{}},
columns/Name/.style={column name=Feature Name,string replace*={_}{\textunderscore}},
columns/Mean/.style={column name=Mean,column type=c,numeric type,fixed,precision=2},
columns/SD/.style={column name=Standard Deviation,numeric type,fixed,precision=2},
columns/Minimum/.style={column name=Minimum,numeric type,fixed,precision=2},
columns/Maximum/.style={column name=Maximum,numeric type,fixed,precision=2},
skip rows between index={11}{25}
]{\tablea}
}
\vspace{2mm}
\end{table}

\pgfplotstableread[col sep=comma]{tables/features/4fgldr2agnpsrfeatures.csv}\tableaf
\begin{table}
\caption{Statistics of features used for the two-class probabilistic classification of the 4FGL-DR2 sources.
\lb{tab:4FGL_features}}
\centering
\resizebox{0.45\textwidth}{!}{
\pgfplotstabletypeset[
columns={Name,Mean,SD,Minimum,Maximum},
column type=c,
string type,
every head row/.style={before row=\hline\hline,after row=\hline,},
every last row/.append style={after row={\hline} },
every first column/.style={column type/.add={}{}},
every last column/.style={column type/.add={}{}},
columns/Name/.style={column name=Feature Name,string replace*={_}{\textunderscore}},
columns/Mean/.style={column name=Mean,column type=c,numeric type,fixed,precision=2},
columns/SD/.style={column name=Standard Deviation,numeric type,fixed,precision=2},
columns/Minimum/.style={column name=Minimum,numeric type,fixed,precision=2},
columns/Maximum/.style={column name=Maximum,numeric type,fixed,precision=2},
]{\tableaf}
}
\end{table}

\begin{table}[!h]
\caption{Feature importances for the classification of 4FGL-DR2 sources for
the RF and BDT algorithms.
}
\label{tab:feat_imp2}

\tiny
\centering
\renewcommand{\tabcolsep}{1mm}
\renewcommand{\arraystretch}{1}

\begin{tabular}{c c c}
\hline
\hline
Feature & RF: 50, 6& BDT: 100, 2\\
\hline
{ $\ln$(LP\_SigCurv)}&  0.297  & 0.465   \\
{LP\_beta}&0.151&0.109\\
{ $\ln$(Variability\_Index)} &0.085& 0.253   \\
$\ln$(Unc\_Energy\_Flux100)& 0.081&0.059  \\
$\ln$(Energy\_Flux100) & 0.076&0.008   \\
HR56&0.071& 0.015  \\
Unc\_LP\_Index & 0.067&0.009  \\
HR34& 0.035&0.005  \\
$\ln$(Pivot\_Energy)&0.031&0.006\\
HR23 &0.025& 0.005     \\
 LP\_Index& 0.015&0.016  \\
HR67&0.015&0.010\\
HR45&0.015&0.006\\
GLON&0.013&0.017\\
HR12&0.009&0.004\\
GLAT&0.007&0.003\\
\hline
\end{tabular}
\tablefoot{RF algorithm: 50 trees with maximal depth 6; BDT algorithm: 100 trees with maximal depth 2.
The features are ordered by decreasing importance in the case of the RF algorithm.
}
\end{table}

We summarize features and their statistics,
which we used for probabilistic classification of sources in the 3FGL and 4FGL-DR2 catalogs
in Tables \ref{tab:3FGL_features} and \ref{tab:4FGL_features}, respectively. 
We show the feature importances for the two-class classification of 4FGL-DR2 sources in Table \ref{tab:feat_imp2}.
Similar to feature importances for the 3FGL two-class classification reported in Table \ref{tab:feat_imp}, 
some of the most important features are significance of curvature in the spectrum, variability index, energy flux above 100 MeV and its uncertainty.

\section{Comparison of oversampling methods}
\lb{sec:app_O_vs_S}

\begin{table}[!h]
\caption{
Mean and standard deviation of the significance of differences between 
oversampling-by-repeating and SMOTE class probabilities for the 3FGL catalog.
}
\label{tab:OvsS_3FGL}

\tiny
\centering
\renewcommand{\tabcolsep}{1mm}
\renewcommand{\arraystretch}{1.3}

\begin{tabular}{c c c c c }
\hline
\hline
Classes&\multicolumn{2}{c}{two-class}&\multicolumn{2}{c}{three-class}\\
Method & Mean&Std.&Mean&Std.\\
\hline
RF& -0.431 & 0.461&0.114&0.343\\
\hline
BDT&-0.161&0.426 &-0.102&0.337\\
\hline
NN&0.285&0.419&0.224&0.254\\
\hline
LR&0.437&0.615&0.357&0.325\\
\end{tabular}
\vspace{2mm}
\tablefoot{The significance of the probability difference, $\Delta$, is defined in Eq. (\ref{eq:OS_diff}).
The mean and the standard deviation of $\Delta$
are computed with the pulsar-like probabilities for all sources in the 3FGL catalog.
}
\end{table}

\begin{table}[!h]
\caption{
Mean and standard deviation of the significance of differences between 
oversampling-by-repeating and SMOTE class probabilities for the 4FGL-DR2 catalog.
}
\label{tab:OvsS_4FGL}

\tiny
\centering
\renewcommand{\tabcolsep}{1mm}
\renewcommand{\arraystretch}{1.3}

\begin{tabular}{c c c c c}
\hline
\hline
Classes&\multicolumn{2}{c}{two-class}&\multicolumn{2}{c}{three-class}\\
Method & Mean&Std.&Mean&Std.\\
\hline
RF& -0.335 & 0.489&-0.491&0.464\\
\hline
BDT&-0.120&0.581 &0.007&0.423\\
\hline
NN&0.472&0.430&0.171&0.145\\
\hline
LR&0.531&0.732&0.347&370\\
\end{tabular}
\vspace{2mm}
\tablefoot{The significance of the probability difference, $\Delta$, is defined in Eq. (\ref{eq:OS_diff}).
The mean and the standard deviation of $\Delta$
are computed with the pulsar-like probabilities for all sources in the 4FGL-DR2 catalog.
}
\end{table}
\begin{figure}[h!]
\centering
\includegraphics[width=0.5\textwidth]{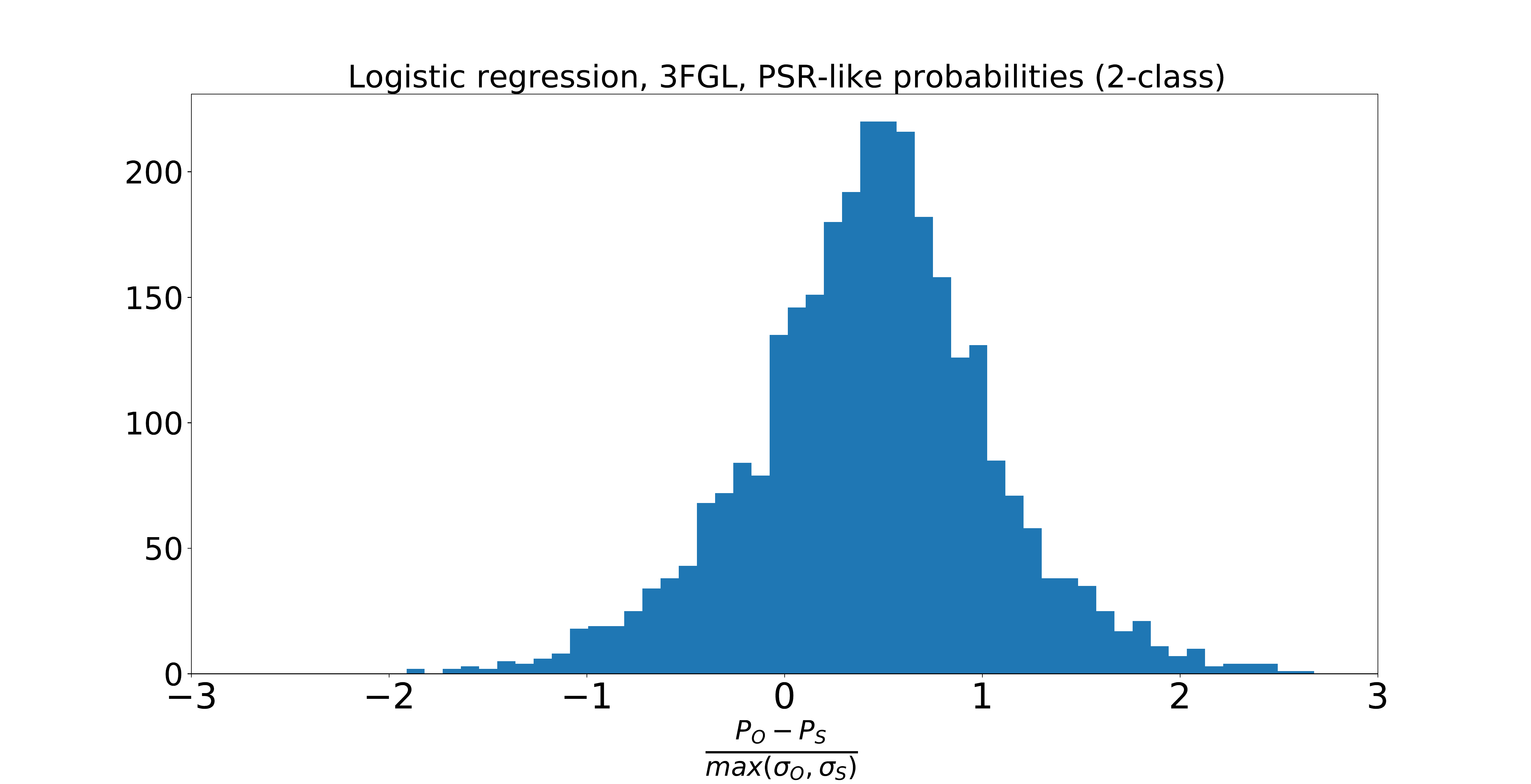}
\includegraphics[width=0.5\textwidth]{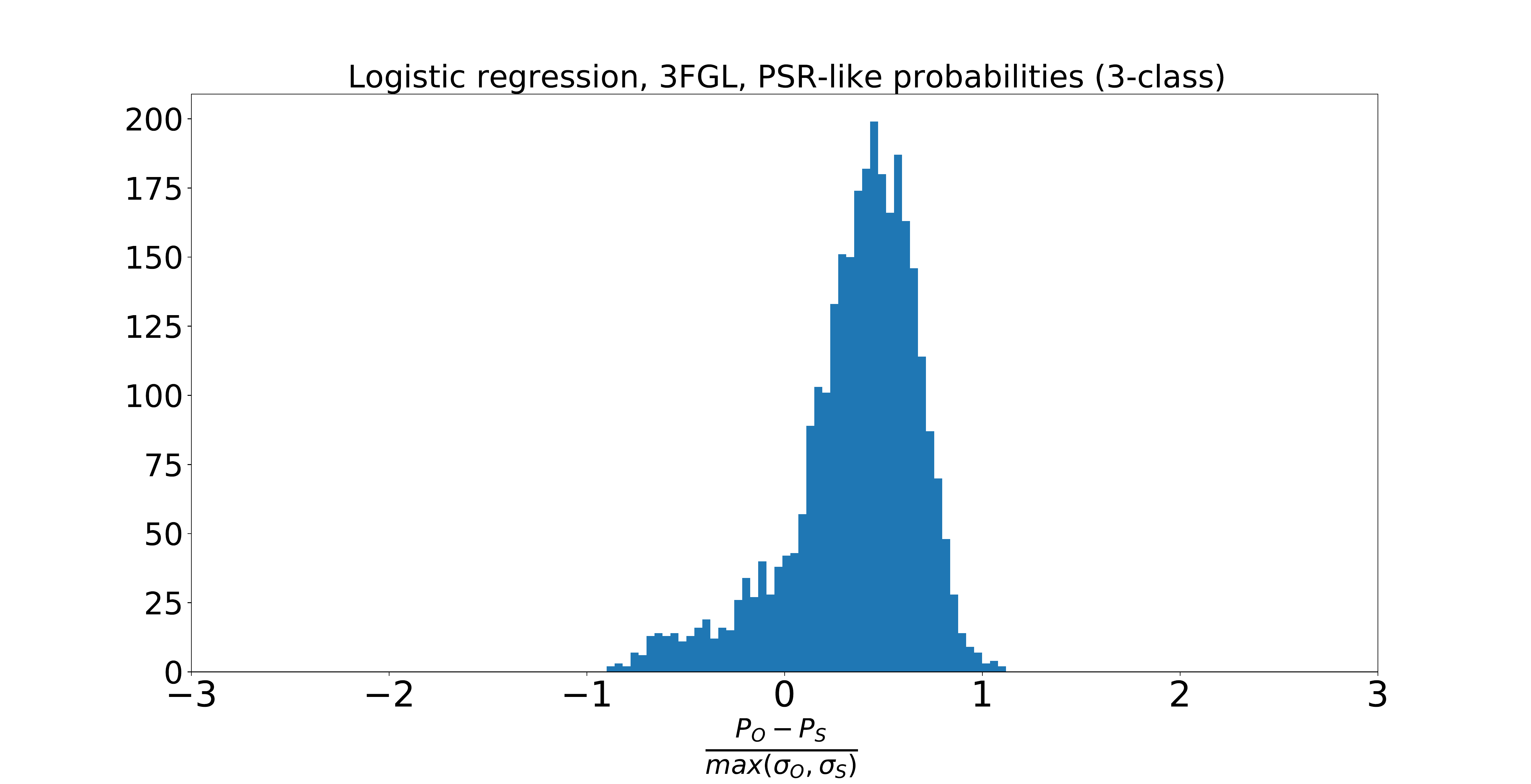}
\caption{Distribution of the difference  of probabilities derived with oversampling-by-repeating ($P_O$) and SMOTE ($P_S$) 
relative to the standard deviations due to the random choice of training samples
for LR in the two-class (top) and three-class (bottom) classifications of the 3FGL sources.
}
\label{fig:OvsS_3FGL_PSR}
\end{figure}

\begin{figure}[h!]
\centering
\includegraphics[width=0.5\textwidth]{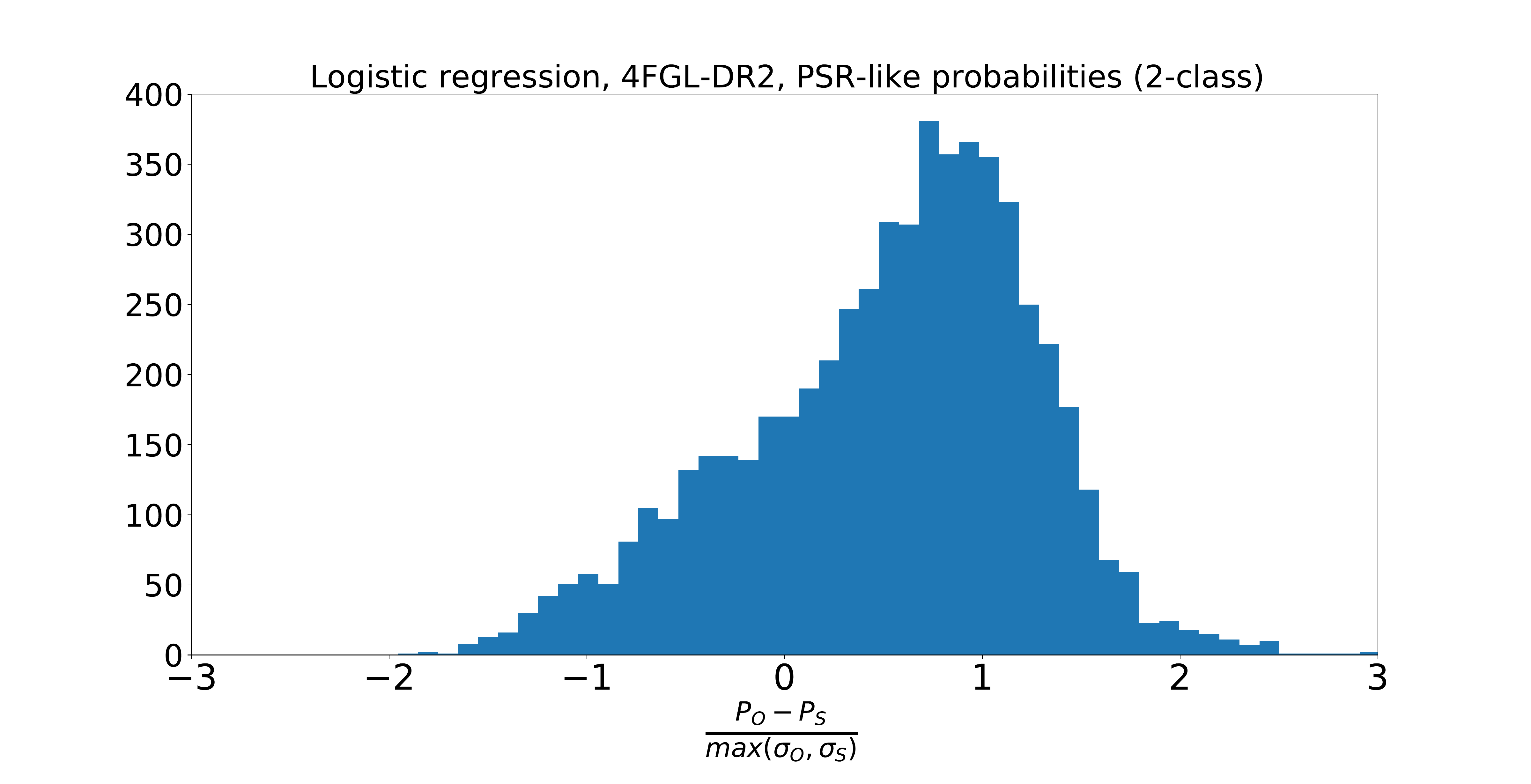}
\includegraphics[width=0.5\textwidth]{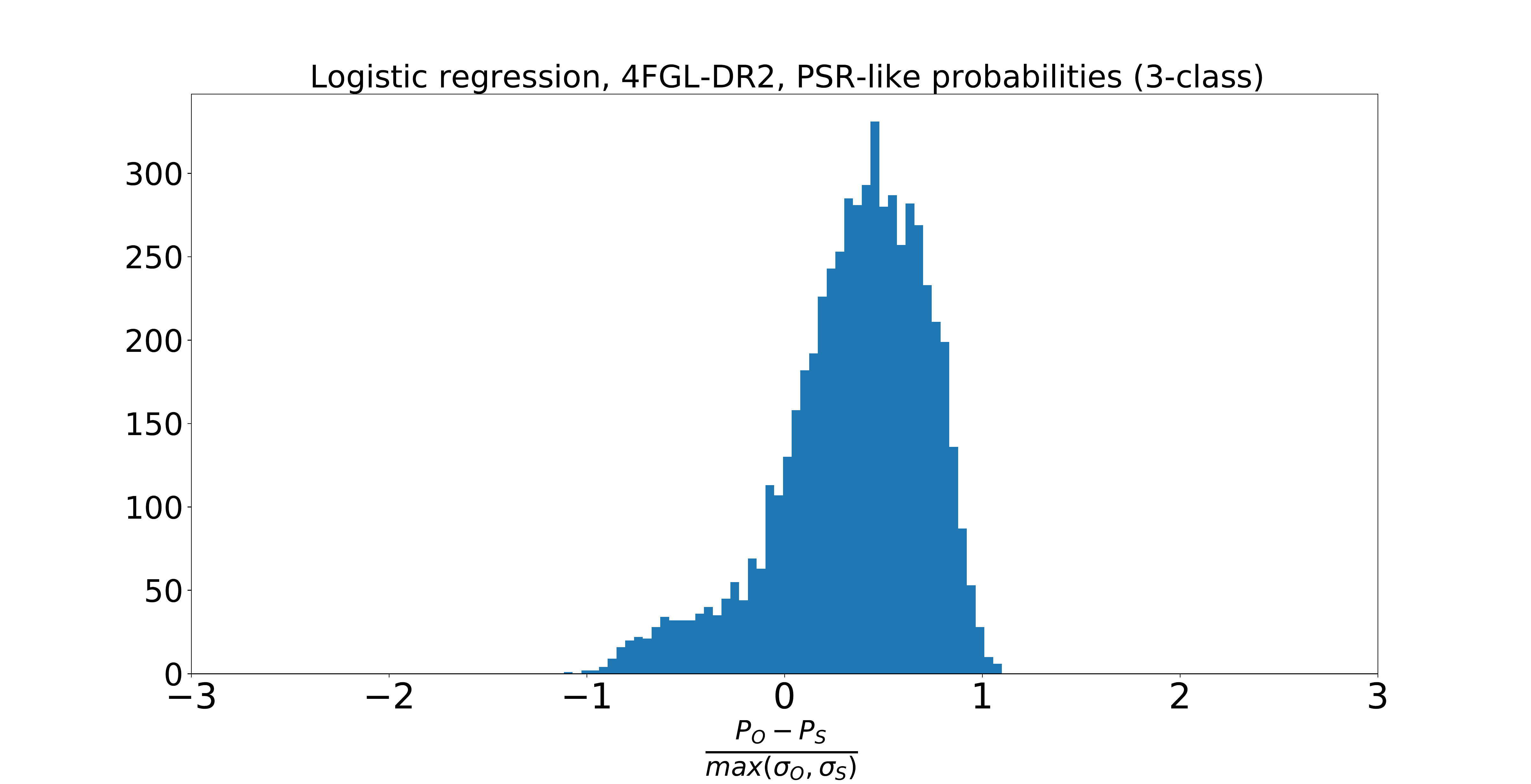}
\caption{Same as Fig. \ref{fig:OvsS_3FGL_PSR} but for the 4FGL-DR2 catalog.
}
\label{fig:OvsS_4FGL_PSR}
\end{figure}

\begin{figure}[h!]
\centering
\includegraphics[width=0.5\textwidth]{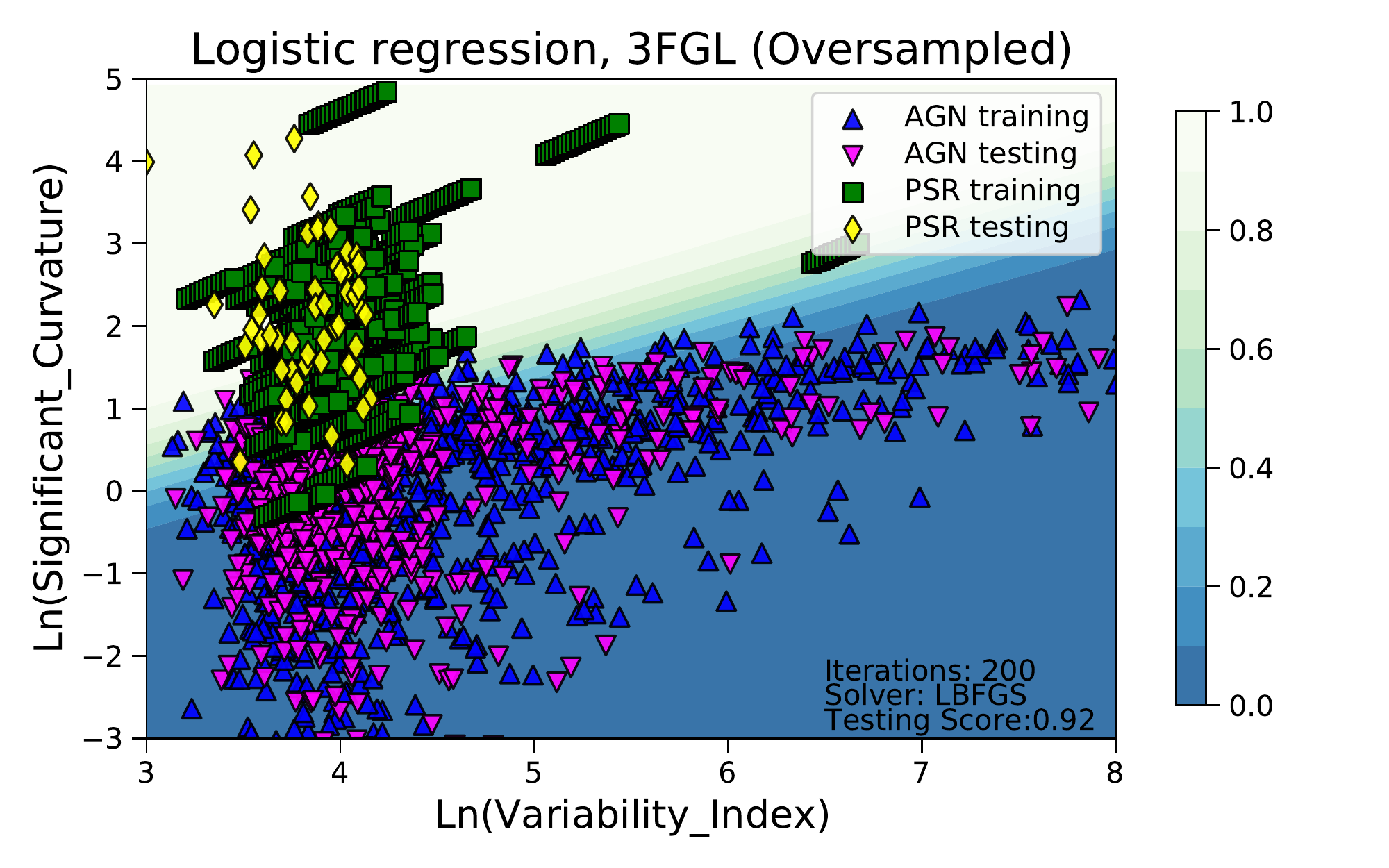}
\includegraphics[width=0.5\textwidth]{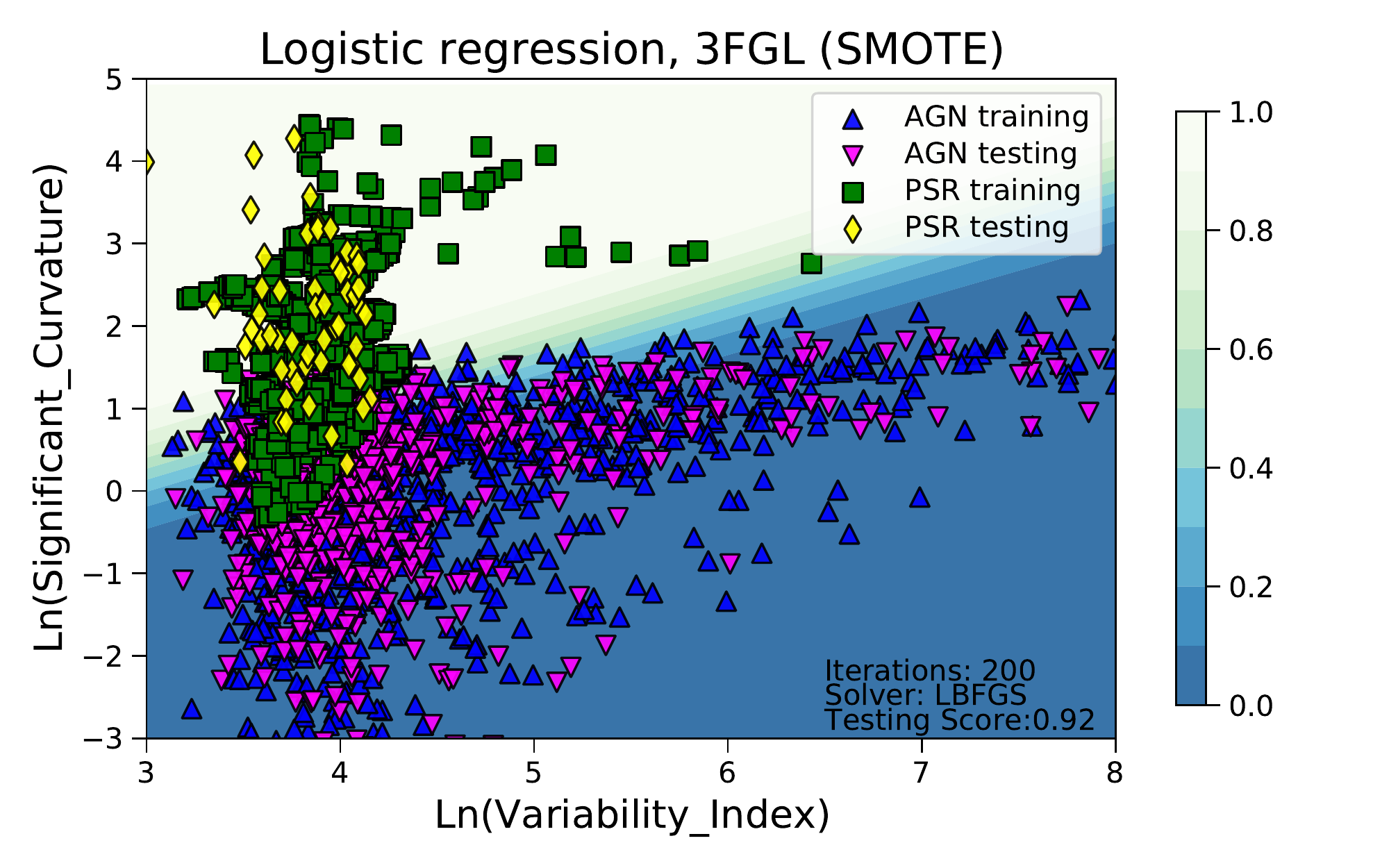}
\caption{Comparison of classification domains for the two-class classification of the 3FGL sources
using LR with oversampling-by-repeating sources (top) and SMOTE (bottom).
}
\label{fig:domains_smote_over}
\end{figure}

\begin{figure*}[h]
\centering
\includegraphics[width=0.45\textwidth]{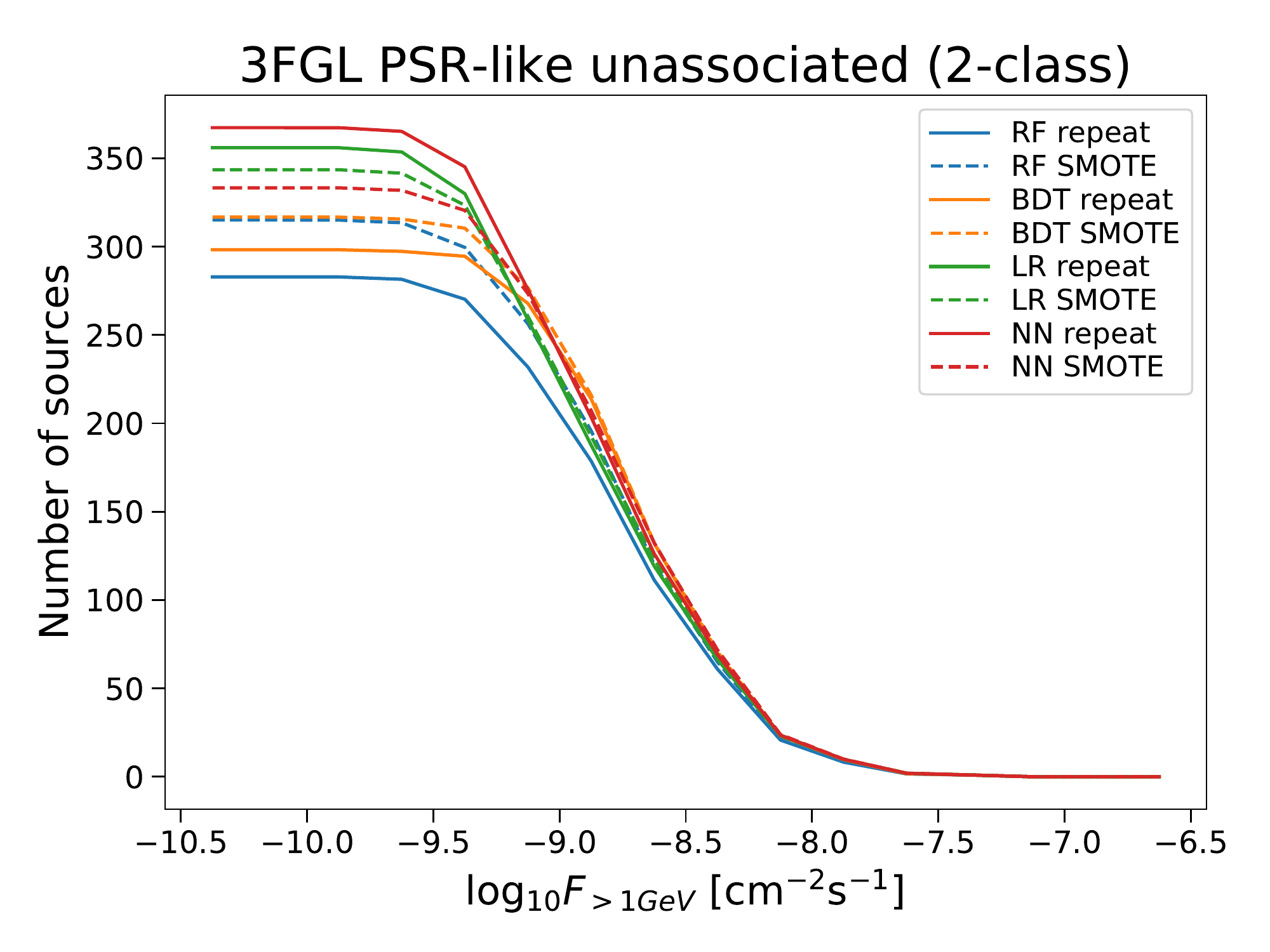}
\includegraphics[width=0.45\textwidth]{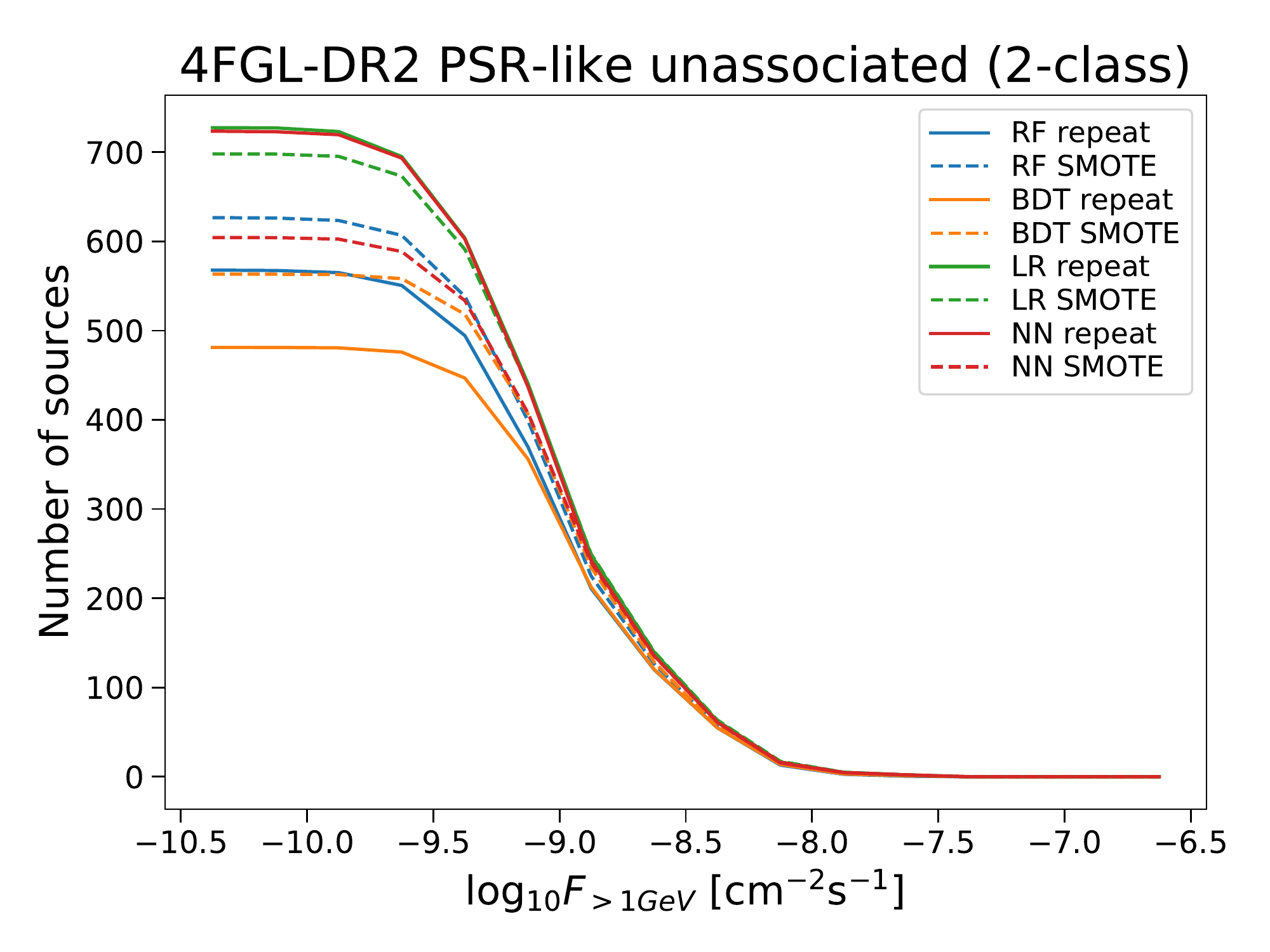}\\
\includegraphics[width=0.45\textwidth]{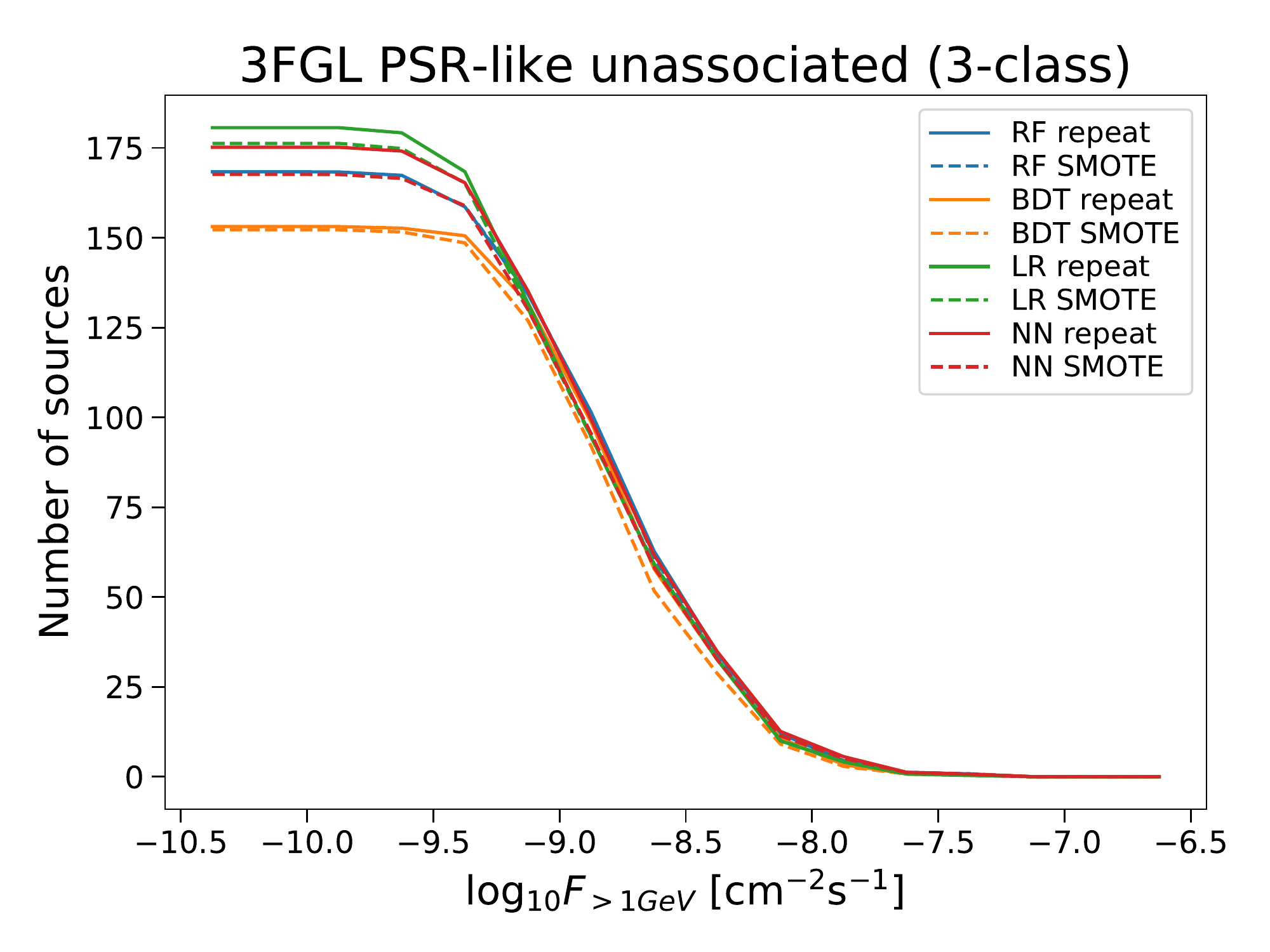}
\includegraphics[width=0.45\textwidth]{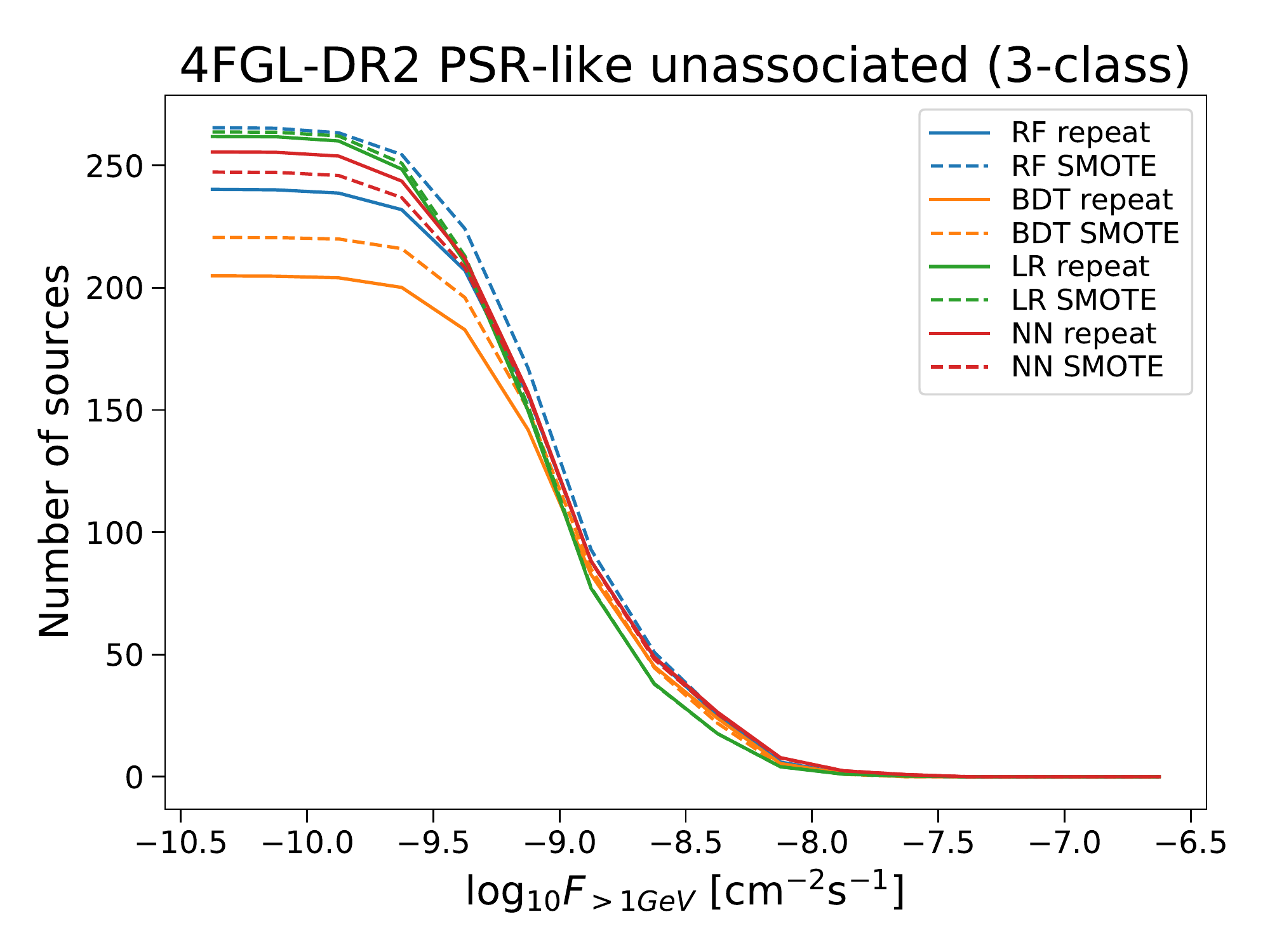}\\
\includegraphics[width=0.45\textwidth]{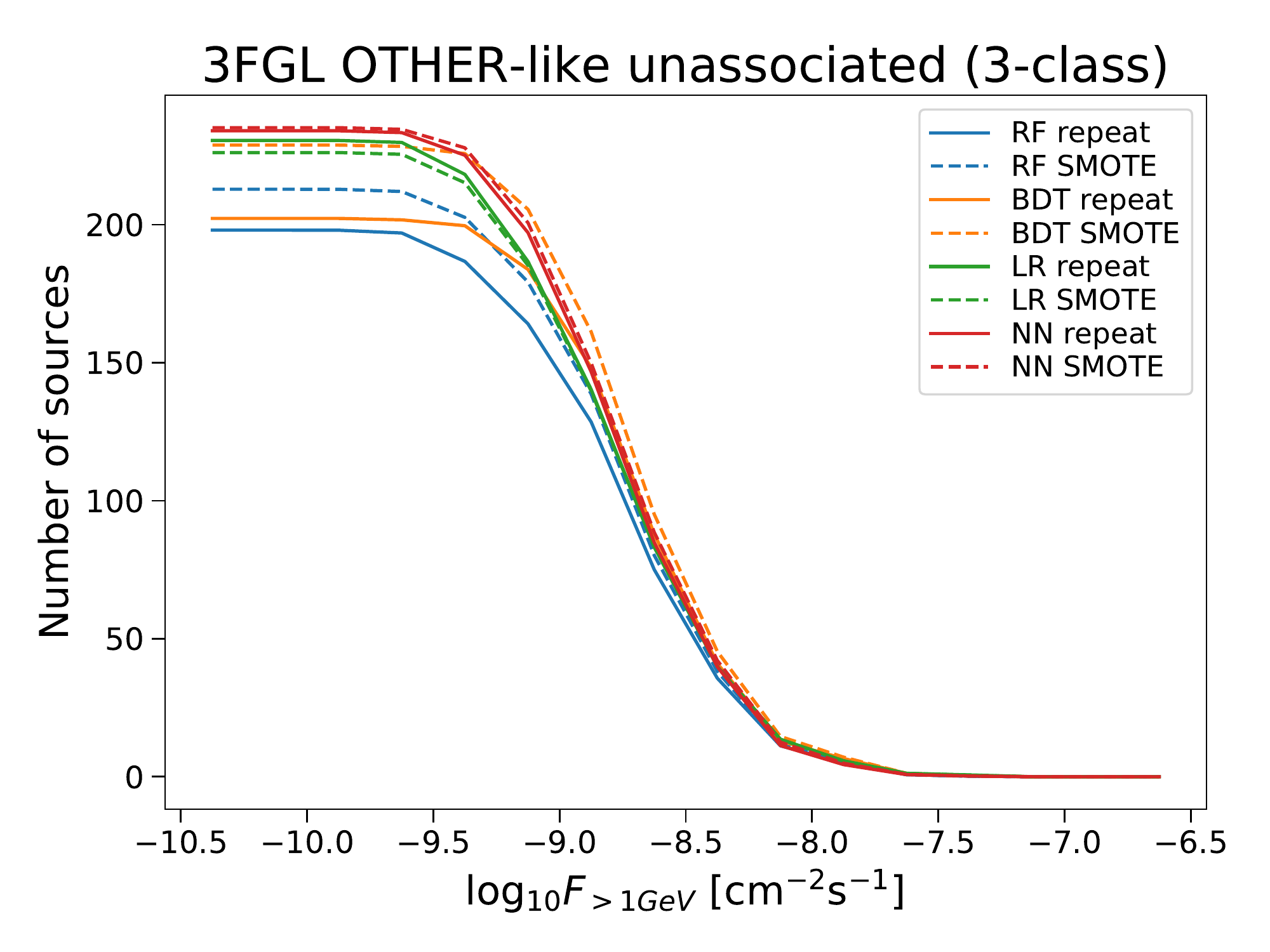}
\includegraphics[width=0.45\textwidth]{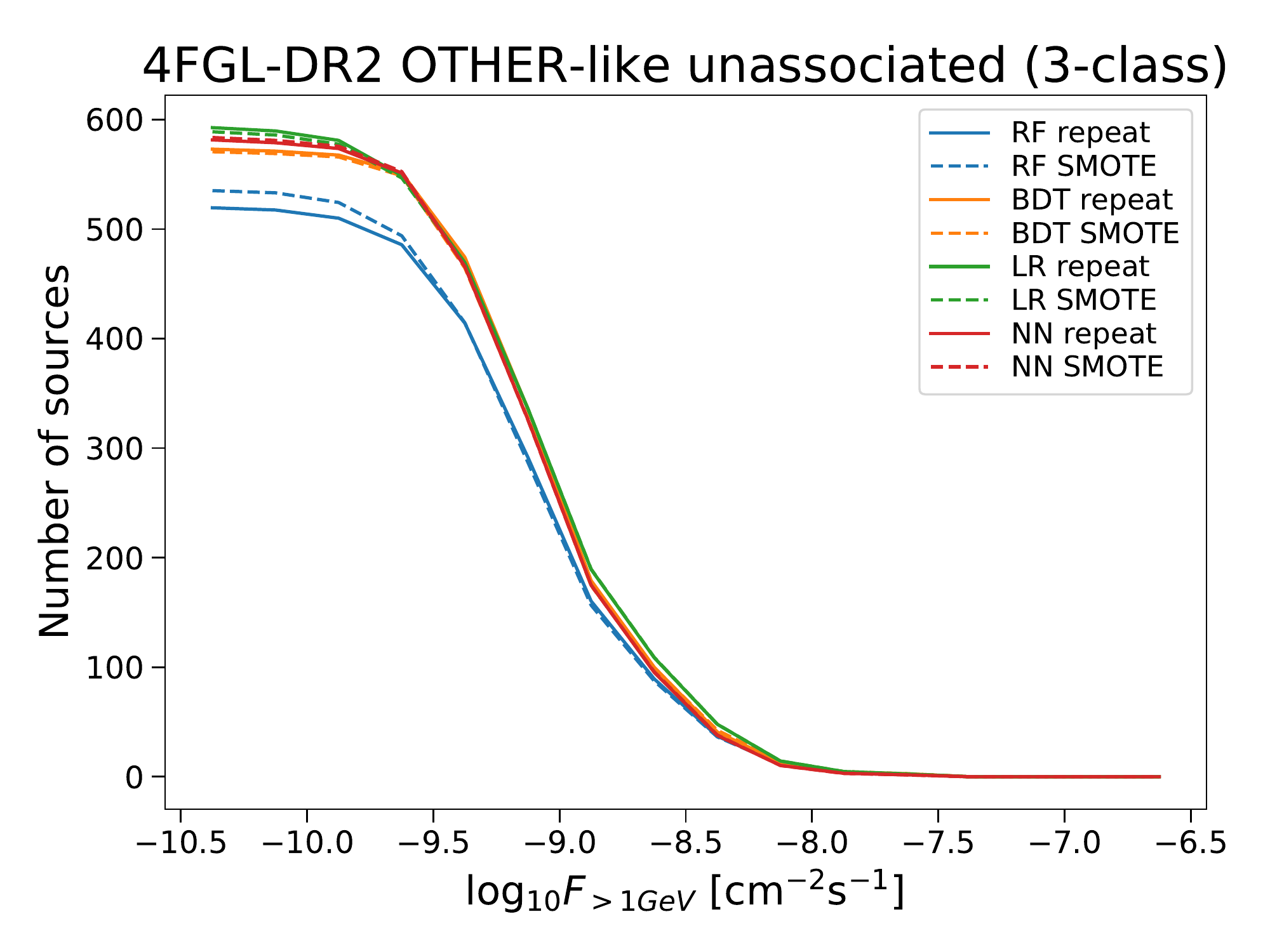}
\caption{Comparison of source count distributions as a function of flux for the estimated number of pulsar and OTHER sources
among the unassociated sources in the two- and three-class cases for oversampling-by-repeating and SMOTE.
The number of pulsar-like sources in the two-class case is not corrected for the presence of OTHER sources.
The changes due to using the SMOTE oversampling technique are comparable to the difference among the algorithms
for oversampling-by-repeating.
}
\label{fig:S_vs_O_NlogS}
\end{figure*}

In this appendix we compare the method of oversampling by repeating sources described in Sect. \ref{sec:oversampling} 
with SMOTE \citep{Chawla_2002}. We used SMOTE from the Imbalanced-learn library, which is based on an implementation of \citet{Chawla_2002}. 
We estimated the probabilities of classification of all sources in the 3FGL and 4FGL-DR2 catalogs using the same algorithms and meta-parameters as in Sect. \ref{sec:oversampling}.
The only difference in the SMOTE case was the oversampling technique.
First, we compared the difference in class probabilities for individual sources relative to the statistical uncertainty due to the random choice of the training samples.
In particular, we calculated the difference in probabilities of belonging to, for example, the pulsar class relative to the maximal standard deviation
between oversampling-by-repeating and SMOTE

\be
\lb{eq:OS_diff}
\Delta = \frac{P_O - P_S}{max(\sigma_{O},\sigma_{S})},
\ee
where $P_O$ ($P_S$) is the probability for the oversampling-by-repeating (SMOTE) and $\sigma_O$ ($\sigma_S$) is the corresponding standard deviation. 
The mean and the standard deviation of $\Delta$ for two- and three-class cases in the 3FGL and 4FGL-DR2 catalogs are presented in 
Tables \ref{tab:OvsS_3FGL} and \ref{tab:OvsS_4FGL}, respectively.
We note that in the three-class case we use less oversampling than in the two-class case, namely, the oversampling factor in the three-class case is equal to the square root of the ratio of the number of sources, while in the two-class case it is equal to the ratio of the number of associated AGNs to the number of associated pulsars.
This is the reason for the smaller bias and standard deviation of the difference in the three-class case relative to the two-class case.
Overall, the differences for individual probabilities are smaller than the uncertainties due to randomness of training.
The LR algorithm has some of the largest differences.
We plot the histogram of the $\Delta$ for the pulsar-like probabilities in the 3FGL (4FGL-DR2) catalogs in Fig. \ref{fig:OvsS_3FGL_PSR}
(\ref{fig:OvsS_4FGL_PSR}).

The difference between oversampling-by-repeating and SMOTE is illustrated in Fig. \ref{fig:domains_smote_over}.
The domains were determined by averaging over 100 random choices of the training data.
One of these choices is shown on the plots: in this case, both oversampling-by-repeating and SMOTE have the same 
training and testing samples.
In the oversampling-by-repeating the training sources were oversampled by simply repeating the sources,
while in SMOTE new sources were created by randomly placing sources in the parameter space along the lines connecting a source
and one of its nearest neighbors. In our implementation we chose one out of the five nearest neighbors.

Although the mean and the standard deviations for the probabilities of the individual sources are smaller than the 
statistical uncertainties of the probabilities, the presence of the bias can have a significant effect when we sum the probabilities, for example, in population studies.
In order to check this effect we compare the source count distributions as a function of flux for oversampling-by-repeating and SMOTE 
in Fig. \ref{fig:S_vs_O_NlogS}.
We show the expected number of pulsars among unassociated sources for the two- and three-class cases using the 3FGL and 4FGL-DR2 catalogs.
We also show the expected number of OTHER sources in the three-class case.
The difference can indeed be significant for some of the algorithms, but the change is comparable or smaller than the difference among the algorithms.
The classification probabilities in the two- and three-class cases with SMOTE for the 3FGL and 4FGL-DR2 catalogs 
are available online in the supplementary online materials \citep{SOM_material}.

\section{Choice of the classification threshold}
\lb{sec:thres}

\begin{figure*}[h!]
\centering
\includegraphics[width=0.45\textwidth]{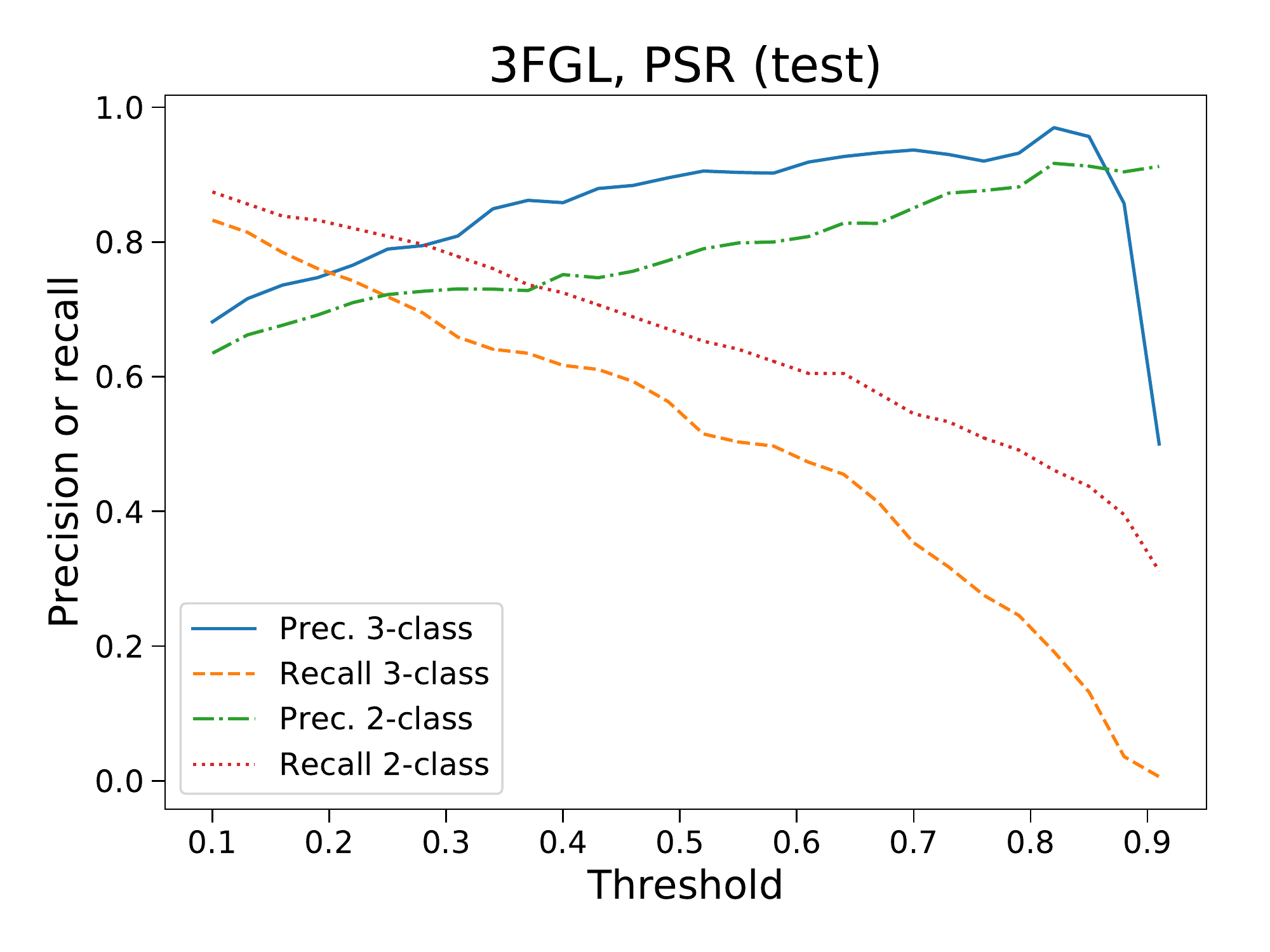}
\includegraphics[width=0.45\textwidth]{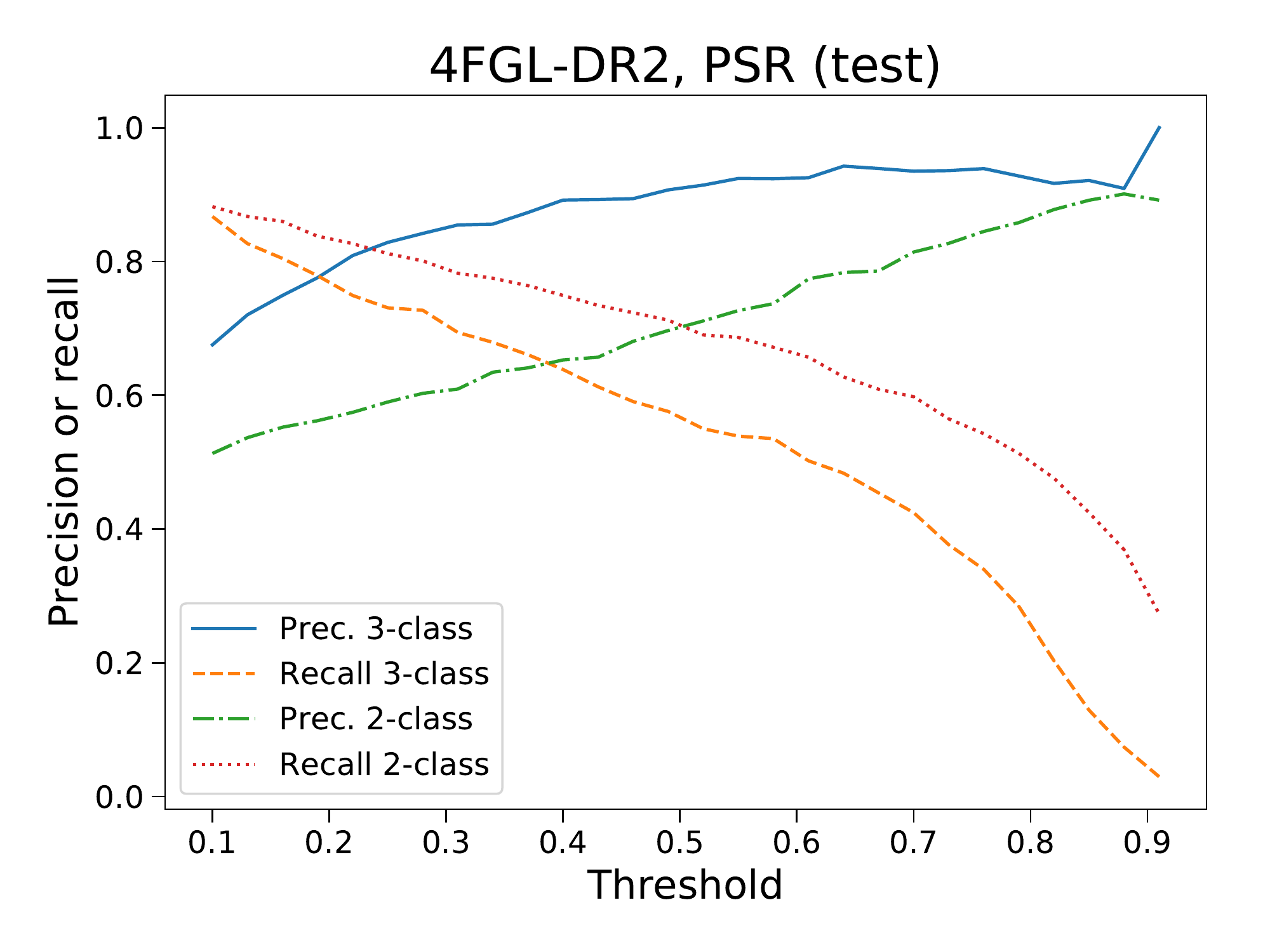} \\ 
\includegraphics[width=0.45\textwidth]{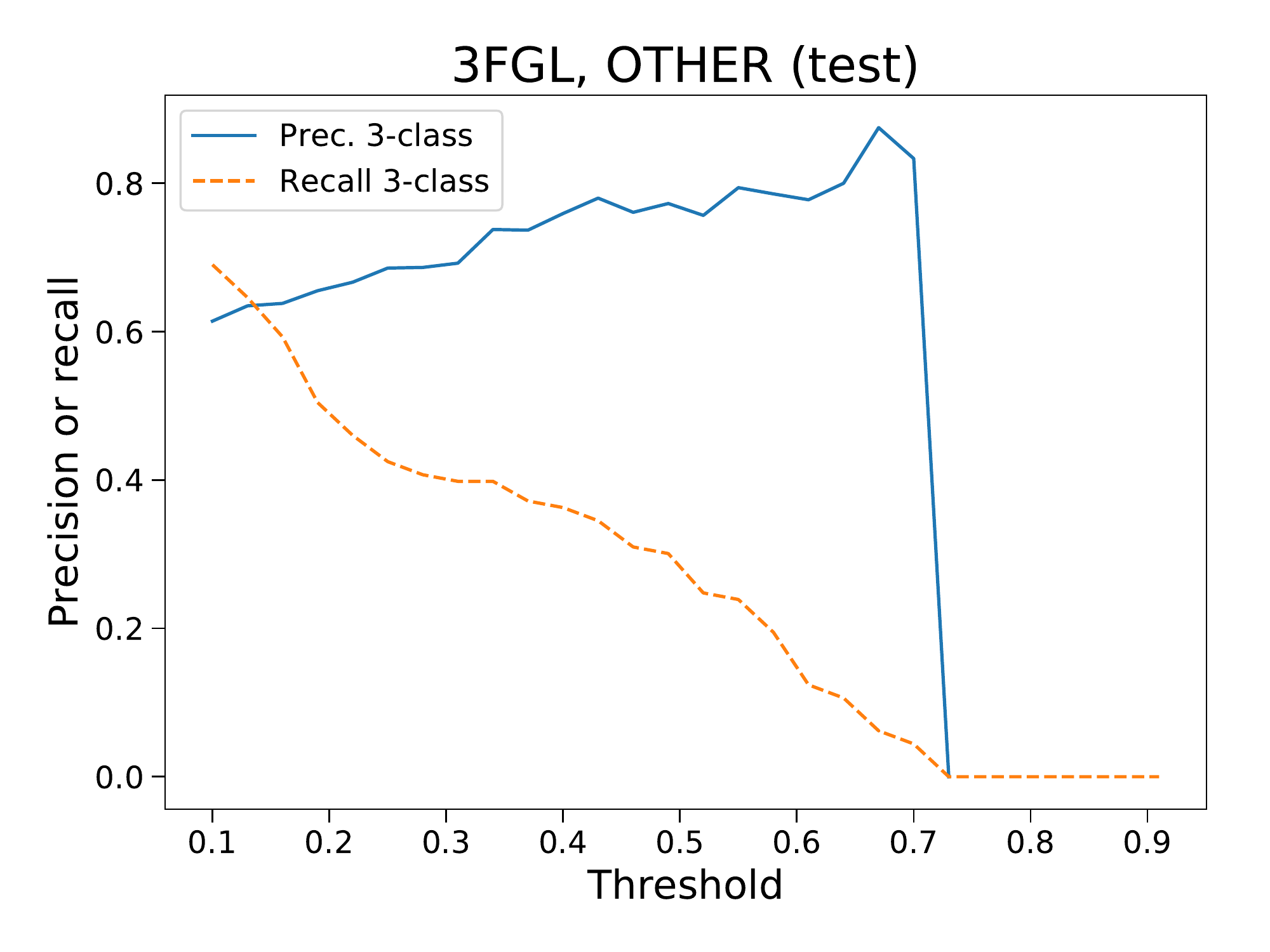}
\includegraphics[width=0.45\textwidth]{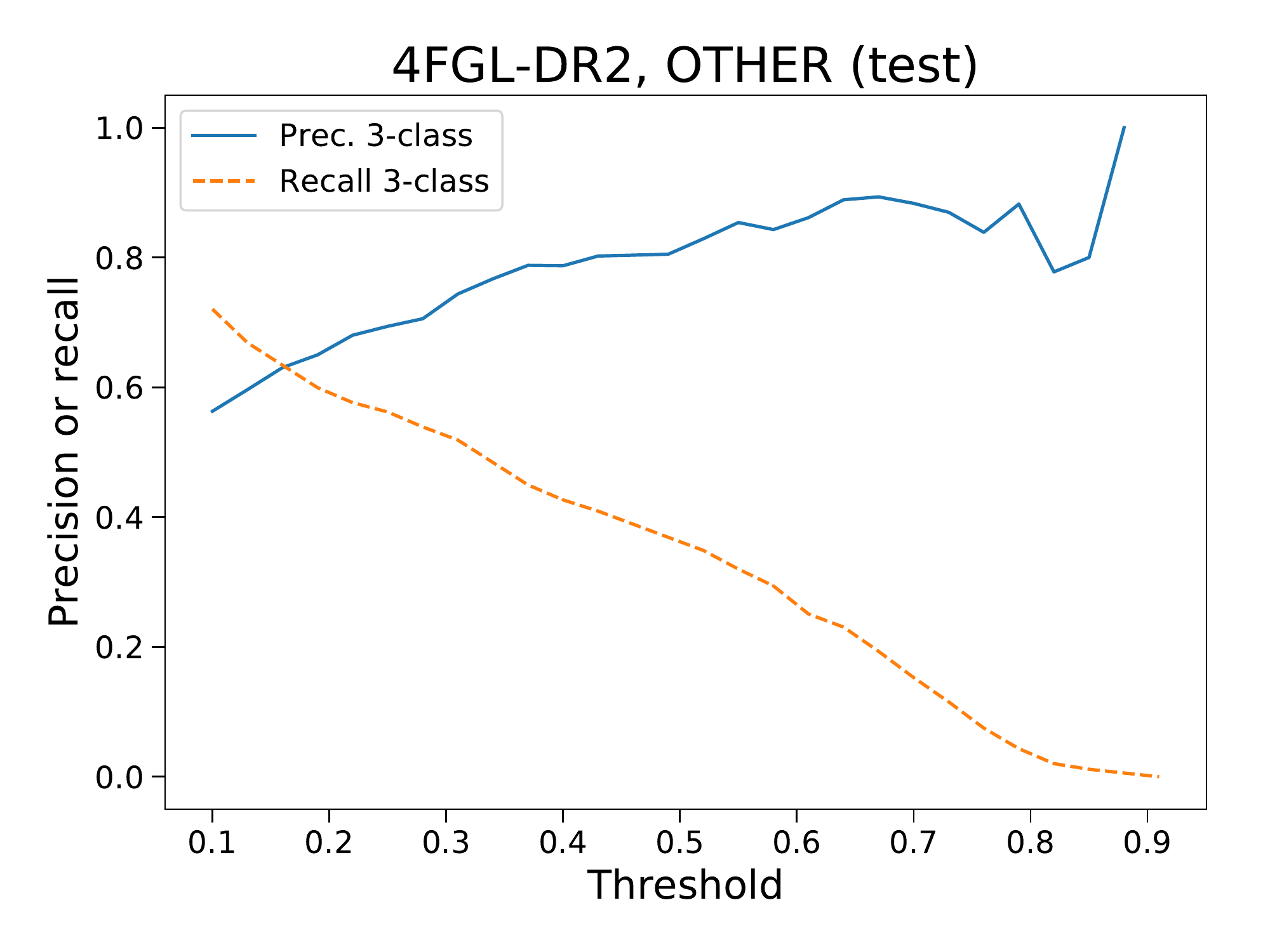}
\caption{Precision and recall for different choices of the threshold for the classification of sources by individual algorithms.
The all-algorithms-agree method is used for the final classification. 
In these estimates, we use associated sources; the corresponding class probabilities are calculated by averaging over the probabilities when the sources are included in testing samples. Above the threshold of 0.73, no sources were classified as OTHER in the three-class classification of the 3FGL sources. The corresponding points in the precision curve in the bottom-left panel are absent.
}
\label{fig:thres}
\end{figure*}

\begin{figure*}[h!]
\centering
\includegraphics[width=0.45\textwidth]{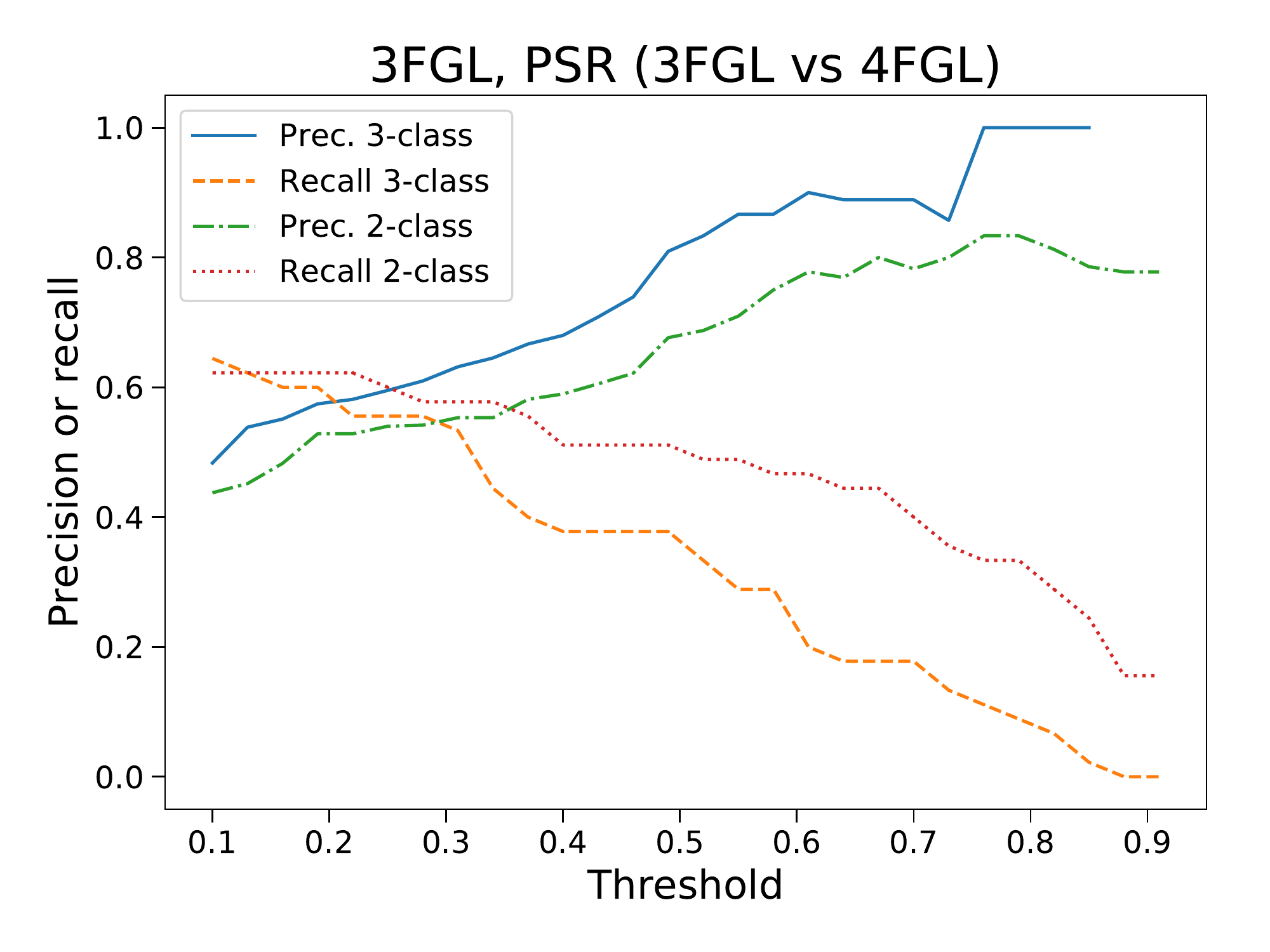}
\includegraphics[width=0.45\textwidth]{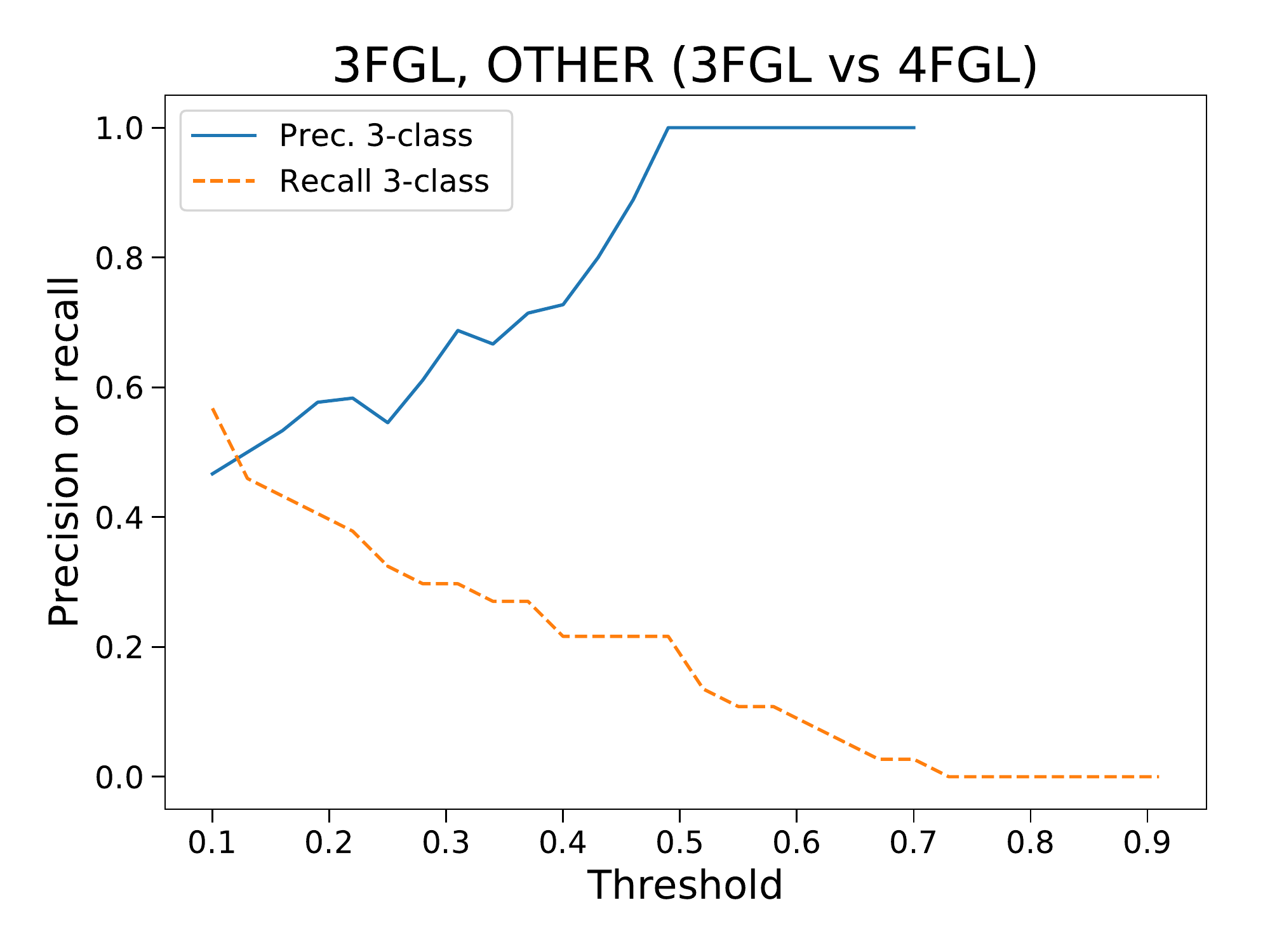}
\caption{Precision and recall for different choices of the threshold for the classification of sources by individual algorithms.
The all-algorithms-agree method is used for the final classification. 
The precision and recall were calculated for unassociated 3FGL sources that have associations in the 4FGL-DR2 catalog.
The classes in the 4FGL-DR2 catalog were considered as the true classes.
The points on some precision curves are absent for high thresholds when there are no sources classified as pulsars (OTHER) in the left (right) panel in the three-class case.
}
\label{fig:thres_3_vs_4}
\end{figure*}

In the paper we used the largest probability rule to classify sources by individual algorithms.
In particular, this means that in the two-class case a source is classified as an AGN or a pulsar if the corresponding probability
is larger than 50\%. Similarly, in the three-class case a source may be classified based on a minimum required 33.3\% probability.
Classification by the largest probability is a common approach in ML and we used it
for the calculation of accuracy of individual algorithms and for the optimization of the meta-parameters. 
However, one can choose other thresholds for the classification depending on the goals of the analysis.
For instance, if a list of high-probability pulsars is required one could use a higher threshold, 
which is expected to reject most of the false candidates but real pulsars would be rejected as well. 
In general, one expects a higher precision for higher threshold
at the expense of lower recall.
Lower threshold would increase the recall (one would miss fewer true candidates), 
but decrease the precision (there will be more false positives).

In this appendix we study the effect of changing the threshold for the probabilistic classification of sources by the individual
algorithms on the overall precision and recall of classification using the agreement of all eight algorithms.
In Fig. \ref{fig:thres} we show precision and recall for the two- and three-class classification of pulsar and OTHER sources in the 3FGL and 4FGL-DR2 catalogs.
In the calculation we used the probabilistic classification of associated sources described in Sect. \ref{sec:3FGLprediction1}: we performed 1000 random splits into training and testing samples and determined the class probabilities for a source by averaging the probabilities when the source is included in the test samples.
We note that with increasing threshold the recall is decreasing while the precision is generally increasing  (except for a few high threshold values in the 3FGL pulsar and OTHER three-class classification, where the number of candidates is very small).
The precision in the three-class case is generally better than in the two-class case (see the top panels of Fig. \ref{fig:thres}),
while the recall is better in the two-class case.
There are a few points for high threshold values where no associated 3FGL sources were classified as OTHER.
In this case the precision is undetermined due to division by zero and the corresponding points are absent in the precision curve
on the lower left panel of Fig. \ref{fig:thres}.

In Fig. \ref{fig:thres_3_vs_4} we show the precision and recall for different thresholds for classification of unassociated sources in the 3FGL catalog that have associations in the 4FGL-DR2 catalog. In general the precision and recall in this case are smaller than the estimates in Fig. \ref{fig:thres}.  The estimates in Fig. \ref{fig:thres_3_vs_4} are likely more realistic than in the test samples case, since they also take into account possible difficulties in reconstructing the properties of the sources, such as the spectrum, which can affect the probabilistic classification of the sources.
\newpage
\section{Reliability diagrams}
\lb{sec:reliability}

\begin{figure*}[h!t]
\centering
\includegraphics[width=0.45\textwidth]{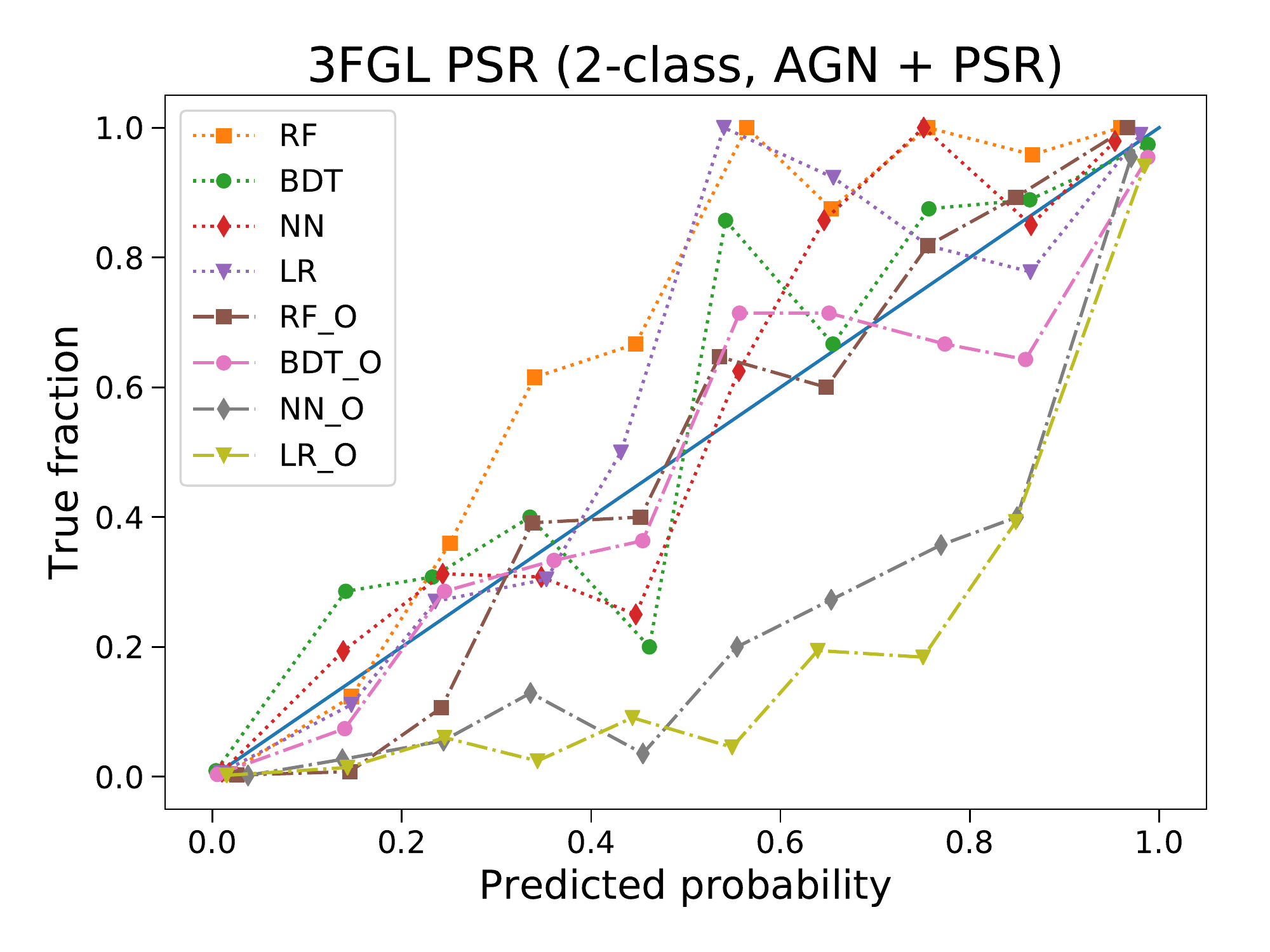}
\includegraphics[width=0.45\textwidth]{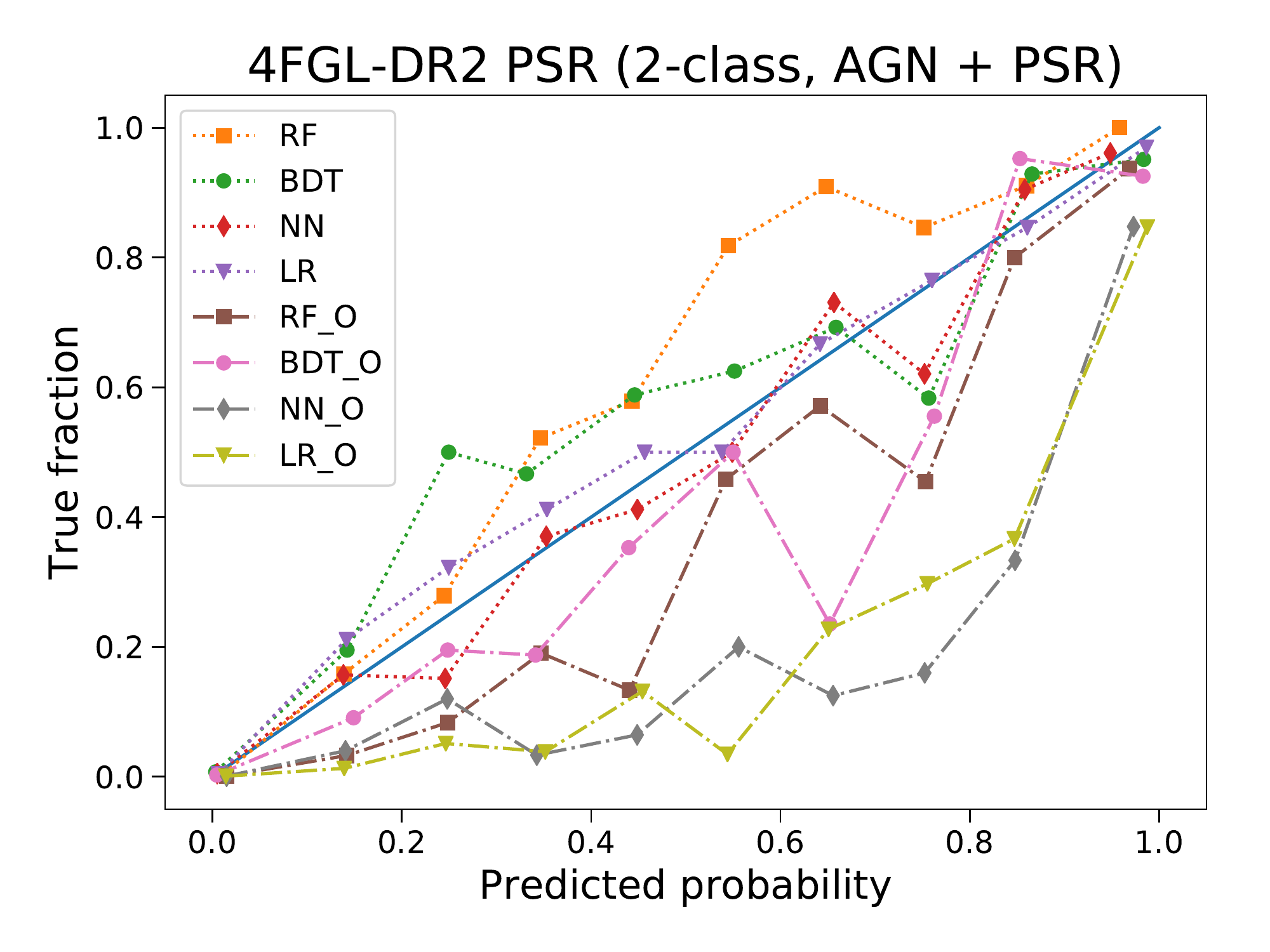} \\ 
\includegraphics[width=0.45\textwidth]{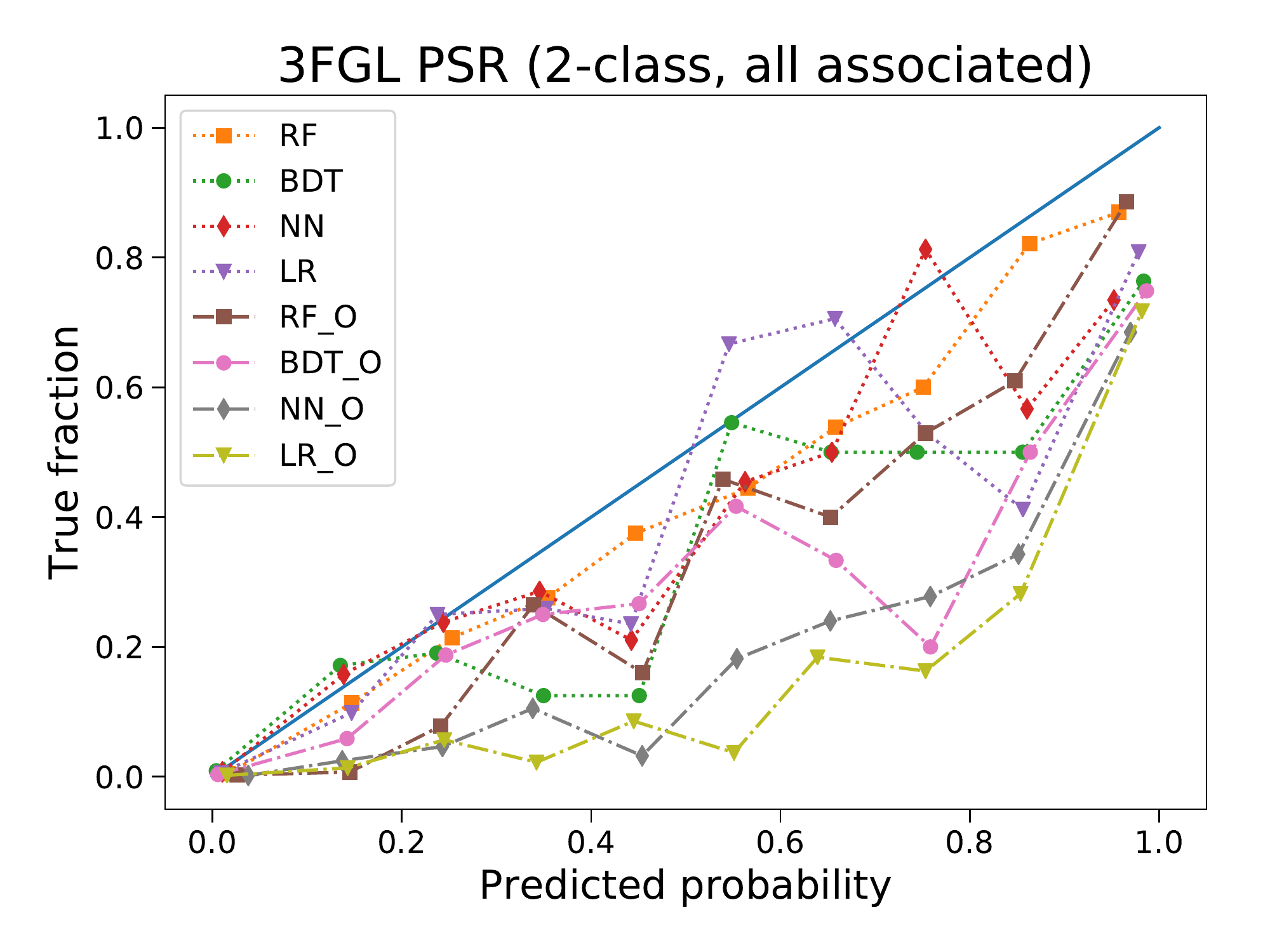}
\includegraphics[width=0.45\textwidth]{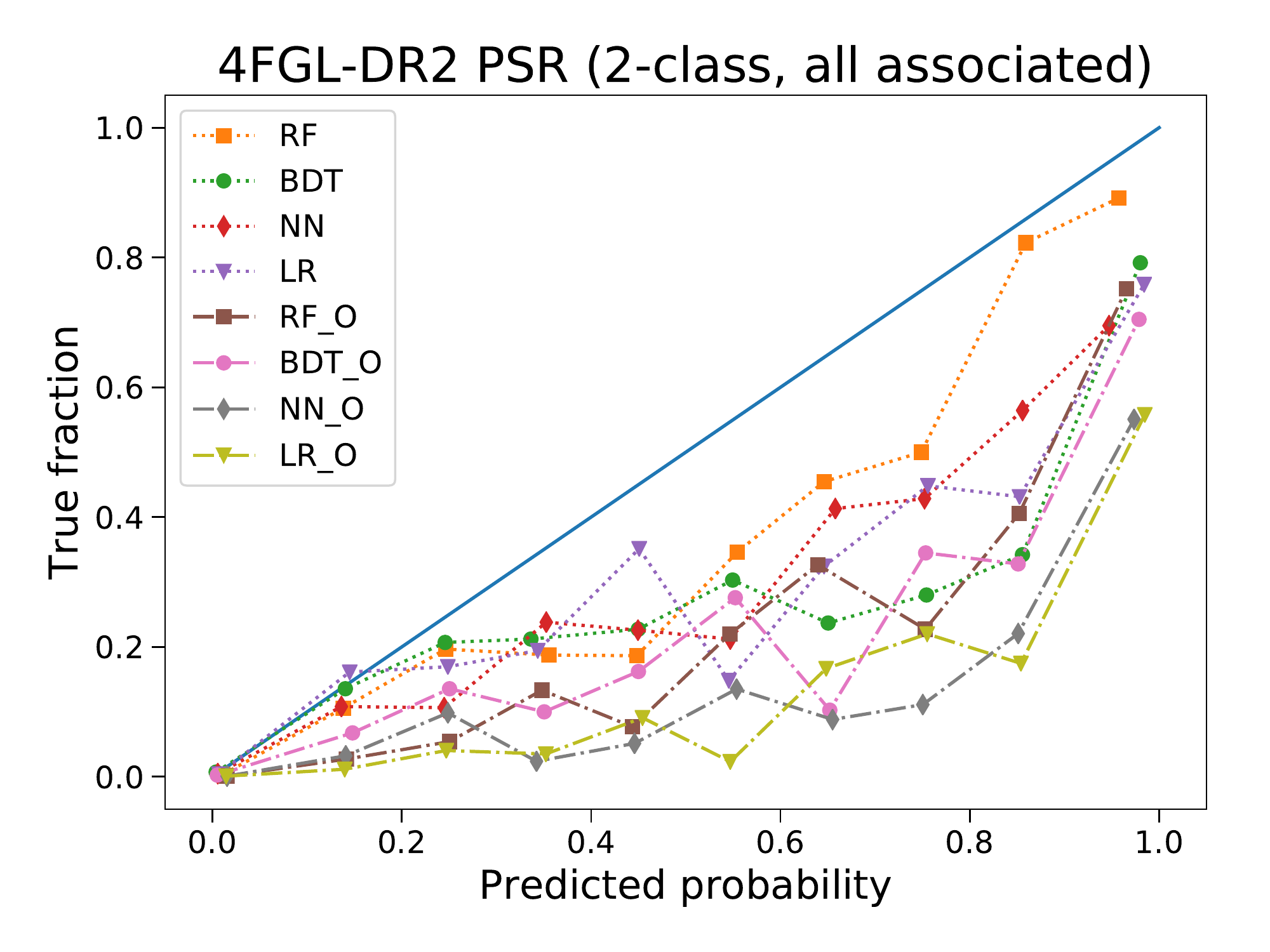} \\ 
\includegraphics[width=0.45\textwidth]{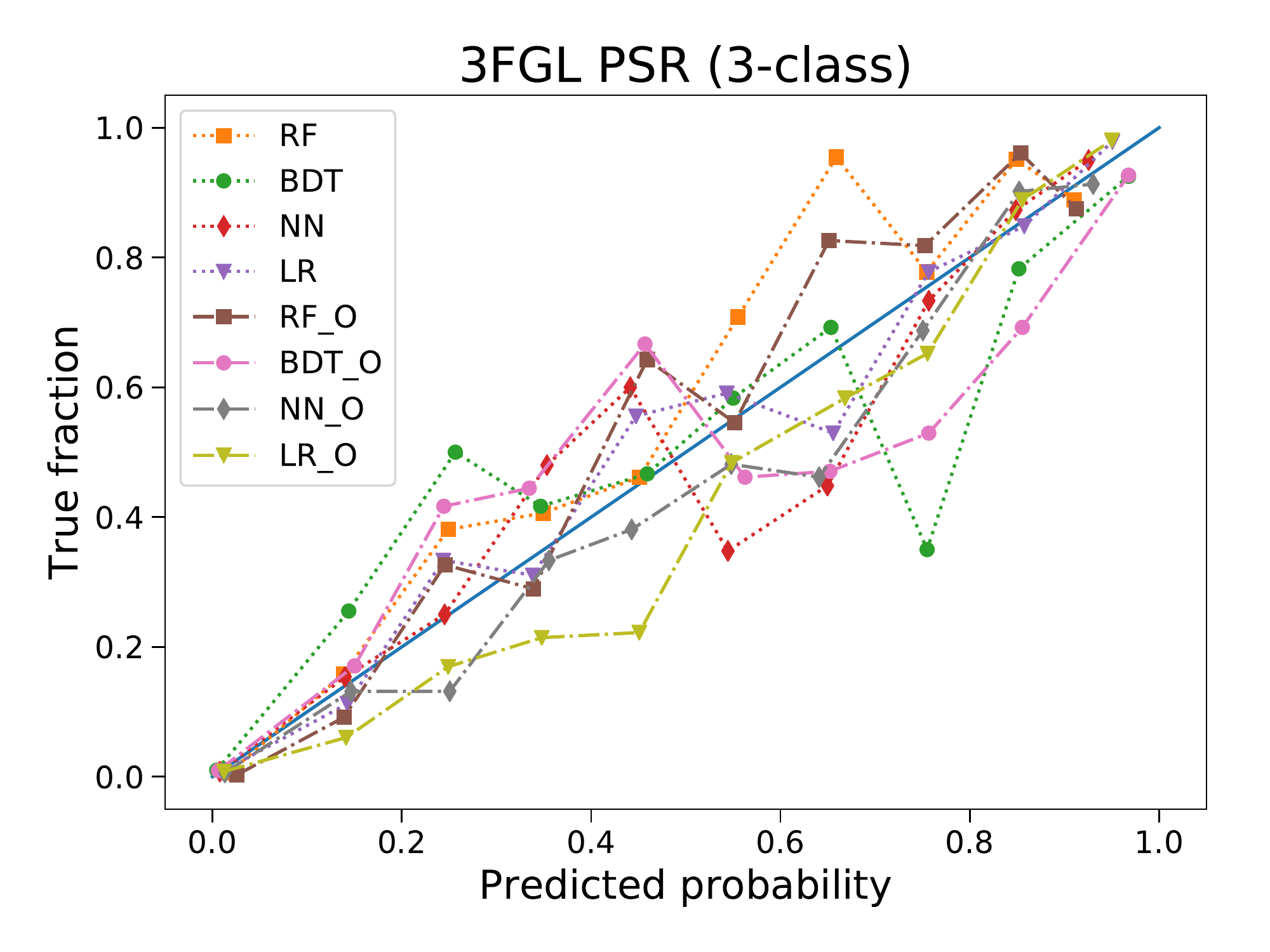}
\includegraphics[width=0.45\textwidth]{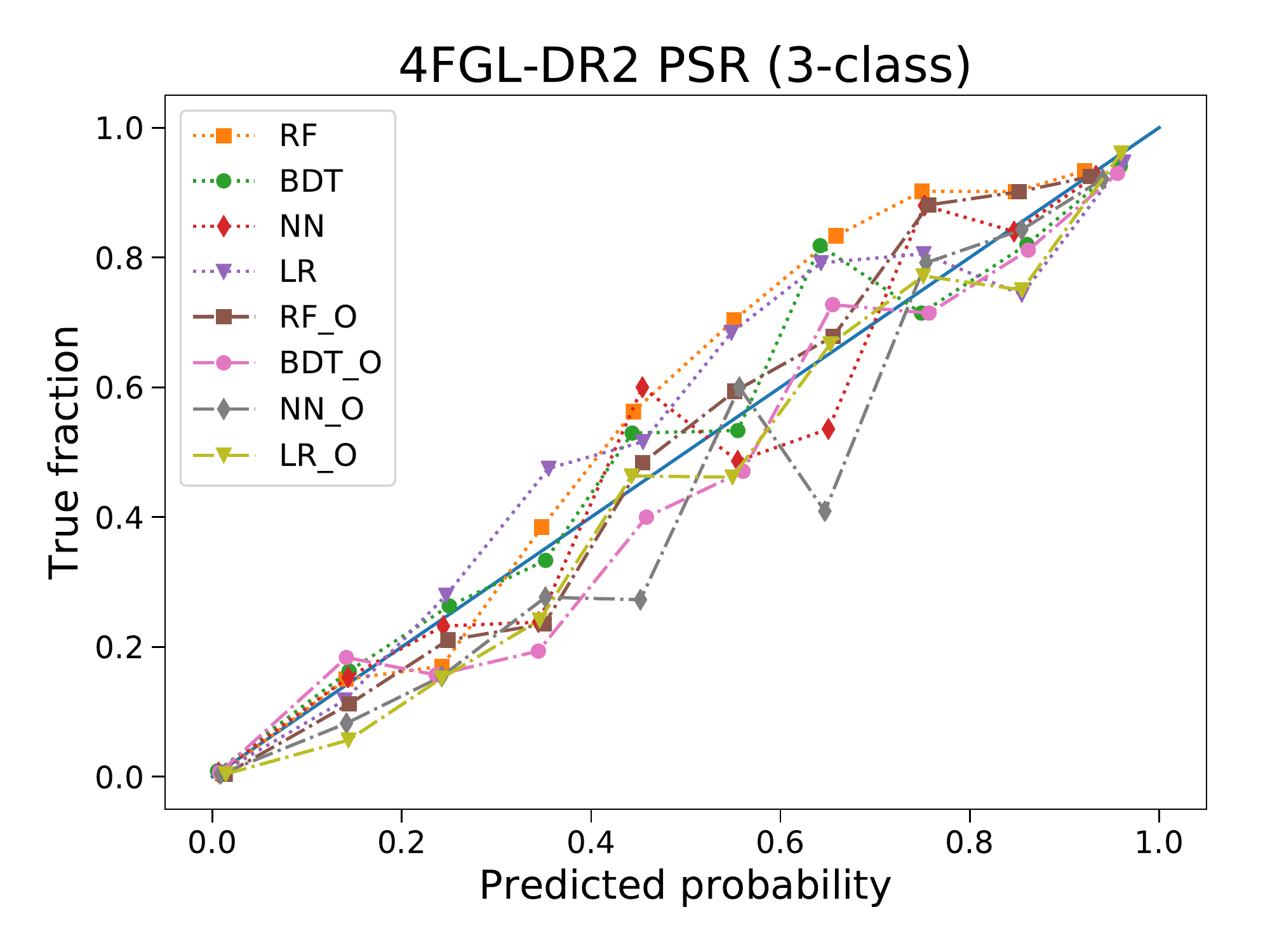}
\caption{Reliability diagrams for the pulsar class. Top panels: Two-class classification 
taking only AGN and pulsar associated 3FGL and 4FGL-DR2 sources into account.
Middle panels: Same as the top panels but taking all associated 3FGL and 4FGL-DR2 sources  into account.
Bottom panels: Three-class classification of 3FGL and 4FGL-DR2 sources.
}
\label{fig:rel_PSR}
\end{figure*}

\begin{figure*}[ht]
\centering
\includegraphics[width=0.45\textwidth]{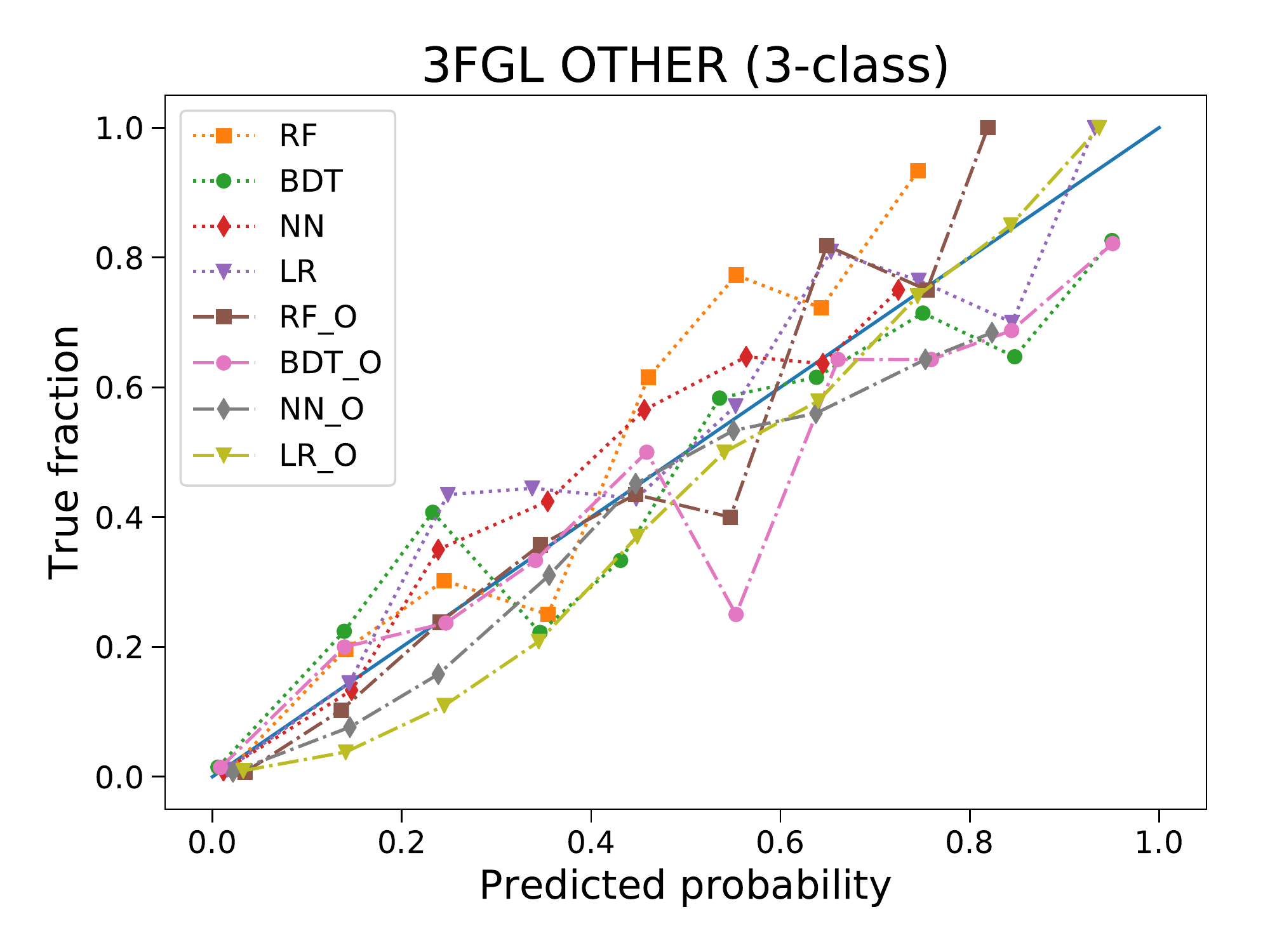}
\includegraphics[width=0.45\textwidth]{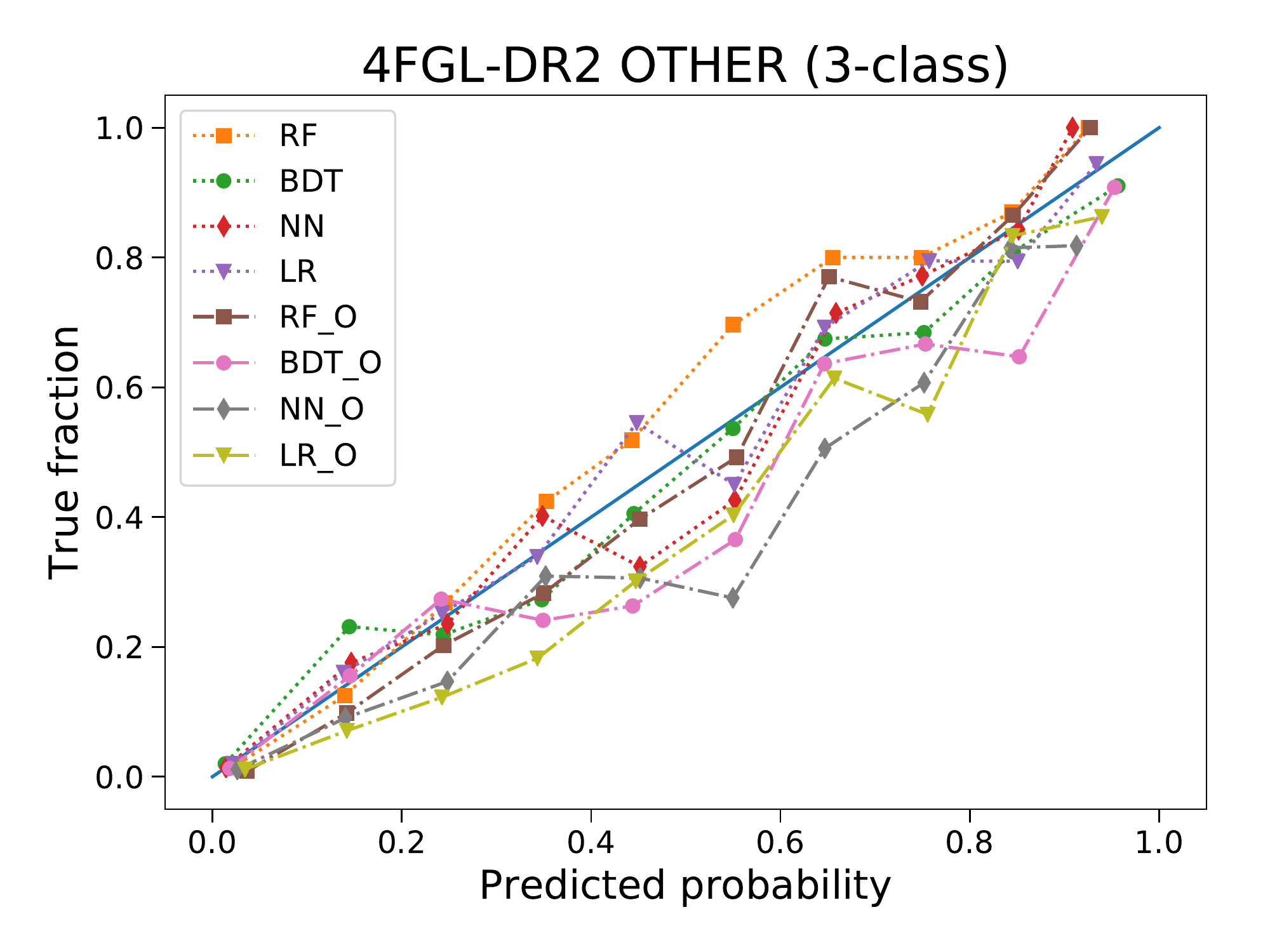}
\caption{Reliability diagrams for the OTHER class in the case of the three-class classification of the 3FGL and 4FGL-DR2 sources.
}
\label{fig:rel_OTHER}
\end{figure*}

One of the characteristics of a classification algorithm is the reliability diagram (also known as the calibration curve),
where one compares predicted probabilities with the true fractions of correct classifications.
For the calculations we used the ``calibration\_curve'' function implemented in scikit-learn.

In Fig. \ref{fig:rel_PSR} we show the reliability diagram for pulsar classification in the two- and three-class cases
for the 3FGL and 4FGL-DR2 catalogs.
The predicted pulsar-like probabilities for the associated sources are separated into ten equally spaced bins between 0 and 1, 
the x-values are the average predicted probabilities in each bin,
the y-values are the fractions of associated pulsars among the sources with predicted probabilities 
in the corresponding bins.
The solid lines at $45^\circ$ show the perfect calibration, when the expected probabilities are on average equal to the 
fraction of true classifications.
If the curve is above (below) the perfect calibration line, then the expected probabilities are smaller (larger) than the true fraction; in other words, the algorithm underestimates (overestimates) the true fraction.

In the top panels of Fig. \ref{fig:rel_PSR} we show the reliability diagrams for the two-class classification
where we take into account only sources in pulsar and AGN classes.
One can see that without oversampling, some algorithms tend to underestimate the true fraction (e.g., RF and LR in the 3FGL two-class case),
while with oversampling, the algorithms generally overestimate the true fraction of pulsars (e.g., NN\_O and LR\_O for 3FGL and all oversampling algorithms for 4FGL-DR2) -- this behavior is not unexpected, since in the oversampling case we artificially increase the number of sources in the smaller class.

In the middle panels of Fig. \ref{fig:rel_PSR} we show the reliability diagrams  for the two-class classification
when we add the OTHER sources.
In this case all algorithms underestimate the true fraction of pulsars, this is due to the presence of additional sources,
none of which are pulsars.
It shows that the two-class classification is likely overestimating the number of pulsars among the unassociated sources
due to the presence of the OTHER sources.
Thus some correction or calibration is needed.

The bottom panels of Fig. \ref{fig:rel_PSR} show the reliability diagrams in the three-class case.
One can see that the performance of the algorithms is not worse than in the two-class case on the top panels
(the better performance of the oversampling cases can be in part attributed to fewer sources added in the oversampling
for the three-class case).
In Fig. \ref{fig:rel_OTHER} we show the reliability diagrams for the OTHER class in the three-class classification.
One can see that the performance of the algorithms is also good in this case, although the OTHER class is the smallest class
and has different types of sources, which can in principle lead to confusion with other classes
and poor performance of classification.

Overall, we find that the reliability of predictions in the three-class case is similar to the performance
in the two-class case when only pulsar and AGN classes are taken into account.
In addition, we expect a similar performance for the unassociated sources since the OTHER class is included in the 
calculation of the reliability diagrams (i.e., no correction is needed, contrary to the two-class case). 
We also note that, although a particular algorithm can be above or below the perfect calibration curve, 
the envelope of the predictions contains the perfect calibration curve (i.e., the envelope of the predictions gives a reasonable
estimate of the modeling uncertainty).

\end{appendix}

\end{document}